\documentclass[useAMS,usenatbib]{mn2e}
\usepackage{graphicx}
\usepackage[fleqn]{amsmath}
\usepackage{threeparttable}
\usepackage{ulem}
\usepackage[colorlinks=true,linkcolor=red,citecolor=blue,urlcolor=magenta]{hyperref}
\usepackage[usenames,dvipsnames]{color} 

\pdfoutput=1

\newcommand{\aap}{A\&A}

\newcommand{\apjl}{{ApJL}}
\newcommand{\apjs}{ApJS}

\newcommand{\ao}{{Appl. Opt.}}

\title[SPIRE FTS calibrators]{Systematic characterisation of the \textit{Herschel} SPIRE Fourier Transform Spectrometer \thanks{{\it Herschel} is an ESA space observatory with science instruments provided by European-led Principal Investigator consortia and with important participation from NASA.}}
\author[R. Hopwood et al.]{R. Hopwood$^{1}$\thanks{E-mail: r.hopwood@imperial.ac.uk}, 
E. T. Polehampton$^{2, 3}$, 
I. Valtchanov$^{4}$, 
B. M. Swinyard$^{2, 5}$,
\newauthor
T. Fulton$^{3}$,
N. Lu$^{7}$,
N. Marchili$^{11}$,
M. H. D. van der Wiel$^{3, 14}$,
D. Benielli$^{9}$,
\newauthor
P. Imhof$^{8, 3}$,
J.-P. Baluteau$^{9}$,
C. Pearson$^{2, 6, 13}$,
D. L. Clements$^{1}$,
M. J. Griffin$^{10}$,
\newauthor
T. L. Lim$^{2}$,
G. Makiwa$^{3}$, 
D. A. Naylor$^{3}$,
G. Noble$^{12, 1}$,
E. Puga$^{4}$,
L. D. Spencer$^{3}$\\
$^{1}$Department of Physics, Imperial College London, Prince Consort Road, London SW7 2AZ, UK \\
$^{2}$RAL Space, Rutherford Appleton Laboratory, Chilton, Didcot, Oxfordshire, OX11 0QX, UK \\
$^{3}$Institute for Space Imaging Science, Department~of Physics \& Astronomy, University of Lethbridge, 4401 University Drive, \\
Lethbridge, Alberta, T1K 3M4, Canada \\
$^{4}$European Space Astronomy Centre, Herschel Science Centre, ESA, 28691 Villanueva de la Ca\~nada, Spain\\
$^{5}$Department of Physics and Astronomy, University College London, Gower St, London WC1E 6BT, UK \\
$^{6}$Department of Physical Sciences, The Open University, Milton Keynes, MK7 6AA, UK\\
$^{7}$NASA Herschel Science Center, MS 100-22, California Institute of Technology, Pasadena, CA 91125, USA\\
$^{8}$Blue Sky Spectroscopy, Lethbridge, AB, T1J 0N9, Canada\\
$^{9}$Laboratoire d'Astrophysique de Marseille (LAM), Universit\'e d'Aix-Marseille \& CNRS, UMR7326, 13388 Marseille Cedex 13, France \\
$^{10}$School of Physics and Astronomy, Cardiff University, The Parade, Cardiff, CF24  3AA, UK\\
$^{11}$IAPS-INAF, Via Fosso del Cavaliere 100, I-00133 Roma, Italy\\
$^{12}$Department of Physics \& Astronomy, University of British Columbia, 6224 Agricultural Road, Vancouver, BC V6T 1Z1, Canada\\
$^{13}$Oxford Astrophysics, Denys Wilkinson Building, University of Oxford, Keble
Rd, Oxford OX1 3RH, UK\\
$^{14}$Centre for Star and Planet Formation, Niels Bohr Institute and Natural History Museum of Denmark, University of Copenhagen,\\
\O ster Voldgade 5--7, DK-1350 \mbox{Copenhagen~K}, Denmark}

\begin{document}

\date{Accepted.. Received..; in original form ..}

\pagerange{\pageref{firstpage}--\pageref{lastpage}} \pubyear{2013}

\maketitle

\label{firstpage}

\begin{abstract}
A systematic programme of calibration observations was carried out to monitor the performance of the SPIRE FTS instrument on board the {\it Herschel} Space Observatory. Observations of planets (including the prime point-source calibrator, Uranus), asteroids, line sources, dark sky, and cross-calibration sources were made in order to monitor repeatability and sensitivity, and to improve FTS calibration. We present a complete analysis of the full set of calibration observations and use them to assess the performance of the FTS. Particular care is taken to understand and separate out the effect of pointing uncertainties, including the position of the internal beam steering mirror for sparse observations in the early part of the mission. The repeatability of spectral line centre positions is $<$5\,km\,s$^{-1},$ for lines with signal-to-noise ratios $>$40, corresponding to $<$0.5--2.0\% of a resolution element. For spectral line flux, the repeatability is better than 6\%, which improves to 1--2\% for spectra corrected for pointing offsets. The continuum repeatability is 4.4\% for the SLW band and 13.6\% for the SSW band, which reduces to $\sim$1\% once the data have been corrected for pointing offsets. Observations of dark sky were used to assess the sensitivity and the systematic offset in the continuum, both of which were found to be consistent across the FTS detector arrays. The average point-source calibrated sensitivity for the centre detectors is 0.20 and 0.21\,Jy [1\,$\sigma$; 1 hour], for SLW and SSW. The average continuum offset is 0.40\,Jy for the SLW band and 0.28\,Jy for the SSW band.
\end{abstract}



\section{Introduction}\label{sec:intro}

The ESA {\it Herschel} Space Observatory \citep[{\it Herschel};][]{pilbratt10} conducted observations of the far infrared and sub-millimetre (submm) sky from an orbit around the Sun-Earth L2 Lagrangian point, over a period of four years (May 2009--April 2013). After a six-month period of commissioning and performance demonstration, {\it Herschel} entered its routine operation phase, during which science observations were made, with a small proportion of the observing time devoted to a systematic programme of calibration observations. The Spectral and Photometric Imaging REceiver \citep[SPIRE;][]{Griffin10} was one of three focal plane instruments on board \textit{Herschel}, and consisted of an imaging photometric camera and an imaging Fourier Transform Spectrometer (FTS). The FTS provided simultaneous frequency coverage across a wide band in the submm with two bolometer arrays: SLW (447--990\,GHz) and SSW (958--1546\,GHz). The bolometric detectors \citep{Turner2001} operated at $\sim$300\,mK with feedhorn focal plane optics, giving sparse spatial sampling over an extended field of view \citep{Dohlen2000}.
The basic calibration scheme and the calibration accuracy for the FTS is described by \citet[][]{Swinyard2014}. 

This paper details the programme of systematic calibration observations designed to monitor the SPIRE/FTS performance and consistency during the entire {\it Herschel} mission, and to allow end-to-end performance and calibration to be improved. 
Care is taken to understand and separate out the effect from pointing uncertainties, which includes the position of the internal beam steering mirror for sparse observations in the early part of the mission.

All FTS spectra are presented as a function of frequency, in GHz, in the local standard of rest (LSR) reference frame. The designations of the two central detectors for the SSW and SLW arrays are \texttt{SSWD4} and \texttt{SLWC3} respectively.

In Section~\ref{sec:calProg} the FTS calibration sources are introduced. Section~\ref{sec:dataProcessing} details the data reduction procedures that were applied beyond the standard pipeline. Section~\ref{sec:repeatabilityLineSources} investigates spectral line fitting of FTS data and the repeatability of observations of line sources. The continuum spectra of the line sources are considered in Section~\ref{sec:continuum}.  Comparisons to planet and asteroid models are made in Section~\ref{sec:ratios}, and with SPIRE photometer observations in Section~\ref{sec:specPhot}. The importance of an extensive set of FTS observations of the SPIRE dark sky field is discussed in Section~\ref{sec:darkSky}, which are used to assess the FTS sensitivity and uncertainty on the continuum for all detectors. The findings are summarised in Section~\ref{sec:summary}. Tables summarising the FTS observations used are provided in Appendix~\ref{app:tables}.

\section{FTS routine-phase calibrators}\label{sec:calProg}

\subsection{Overview}

The primary calibrators for the SPIRE FTS are the planet Uranus for point sources, and for extended sources the {\it Herschel} telescope itself, observed against a region of dark sky. See \citet{Swinyard2014} for more details on the calibration. In addition, a set of secondary calibrators was selected in order to monitor the spectral line calibration and line shape, the continuum calibration and shape, the frequency calibration, measure the spectral resolution and determine the instrument stability by assessing repeatability and trends with time. These calibration sources include a range of point-like targets, extended sources, planets and asteroids. The secondary calibrators were regularly observed during the {\it Herschel} mission, with targets chosen so that at least one calibrator in each category was visible on each FTS observing day, ensuring coverage of all instrument modes (i.e. sparse and mapping observations, and nominal and bright source modes). In order to create a more complete set of data for cross-calibration with PACS and HIFI, further sources were added during the mission for regular monitoring. Tables detailing the FTS observations for each source can be found in Appendix~\ref{app:tables}. 
A summary of the FTS calibration sources with the number of times that they were observed is provided in Table~\ref{tab:sourceCount}. The mapping mode calibration observations of the Orion Bar are not included, as these have already been discussed by \citet{Benielli2014}.

All of the calibration sources were most commonly observed with 4 repetitions for both high resolution (HR) and low resolution (LR) modes, bar Hebe, Pallas, Juno and Saturn. Each repetition is two scans of the Spectrometer Mechanism, one forward and one reverse. The integration time for each scan is 66.6 seconds for HR and 6.4 seconds for LR. The number of repetitions per observation are provided in the individual tables (B1--B20) listing the repeated observations for each calibration source. For each of the sources included in Table~\ref{tab:sourceCount}, the total integration time for all HR and all LR observations are provided in the final two columns.
11.5\% of the total FTS observing time was dedicated to sparse and mapping calibration observations, which translates to 254.5 hours on sparse and 15 hours on mapping, and includes observations of dark sky.

\subsection{Operating modes}

The spectral resolution of FTS observations was determined by the scan distance of the Spectrometer Mechanism (SMEC) mirror \citep{spireHandbook}. Initial observations of all the calibration targets were made using a dedicated calibration resolution (CR) setting, which involved scanning the SMEC over its entire range. Initially it was expected that both HR and LR spectra could be extracted from CR, by using a subset of the mirror travel distance. However, it was found that observing in HR and LR separately provided more consistent results, and so from {\it Herschel} operational day (OD) 1079 onwards all calibration observations were made in either HR or LR. In this paper all CR observations are processed as HR.

An additional medium resolution (MR) mode was available at the start of the mission, but not extensively used and never fully calibrated. Any MR observation is now processed as LR for the archive. The optical path difference (OPD) range and corresponding spectral resolution are given in table~\ref{tab:opd}.

\begin{table*}
\caption{Summary of the FTS routine-phase calibrators. The numbers of observations for sparse high resolution (HR), sparse low resolution (LR), mapping (Map) and other special calibration observations are given in the final column. RA and Dec are J2000. For a simple brightness comparison the 250$\mu$m SPIRE photometer values are given in the S$_{\rm PSW}$ column. For the line sources and stars, these values are simply the average peak value of the respective point-source-calibrated maps, used in Section~\ref{sec:specPhot} and detailed in Table~\ref{tab:specMatchPhot}. The  range given for each asteroid is the minimum to maximum photometry from Lim et al. (in preparation). For the planets, the minimum and maximum of the peaks are given for all the small and large photometer maps available, which showed no processing issues. 25 maps of Uranus were used and 52 maps of Neptune. The number of Spectrometer observations taken are given in the four ``Number of observations'' columns, where ``Map'' is the total of intermediate and fully sampled, and SC are special calibration observations, i.e. non-standard configurations. The final two columns provide the total  integration time in hours for the HR and LR observations per source.}
\medskip
\begin{center}
\begin{tabular}{lcccccccccc}
\hline\hline
 & & & & & \multicolumn{4}{c}{Number of observations} & \multicolumn{2}{c}{Integration} \\
Source & RA & Dec & Type & S$_{\rm PSW}$[Jy] & HR & LR & Map & SC & Int$_{\rm HR}$[hr] & Int$_{\rm LR}$[hr] \\ \hline
AFGL2688$^{\rm *}$ & 21:02:18.78 & +36:41:41.2 & proto planetary nebular & 120 & 23 & 10 & 0 & 0 & 7.8 & 1.5 \\
AFGL4106$^{\rm *}$ & 10:23:19.47 & --59:32:04.9 & post red supergiant & 13 & 30 & 11 & 0 & 0 & 11.1 & 1.8 \\
CRL618$^{\rm * \dagger}$ & 04:45:53.64 & +36:06:53.4 & proto planetary nebular & 51 & 21 & 5 & 1 & 1 & 6.0 & 0.6 \\
NGC7027$^{\rm *}$ & 21:07:01.59 & +42:14:10.2 & planetary nebular & 24 & 31 & 11 & 3 & 0 & 11.5 & 1.3 \\
NGC6302 & 17:13:44.21 & --37:06:15.9 & planetary nebular & 51 & 8 & 5 & 0 & 0 & 2.8 & 1.1 \\
CW Leo$^{\rm \dagger\dagger}$ & 09:47:57.41 & +13:16:43.6 & variable star & 69 & 9 & 3 & 0 & 0 & 3.3 & 0.3 \\
IK Tau & 03:53:28.84 & +11:24:22.6 & variable star & 5 & 1 & 0 & 0 & 0 & 0.8 & -- \\
Omi Cet & 02:19:20.79 & --02:58:39.5 & variable star & 7 & 2 & 1 & 0 & 0 & 1.3 & 0.4 \\
R Dor  & 04:36:45.59 & --62:04:37.8 & Semi-regular pulsating Star & 11 & 4 & 2 & 0 & 0 & 2.2 & 0.3 \\
VY CMa & 07:22:58.33 & --25:46:03.2 & red supergiant & 37 & 8 & 5 & 0 & 0 & 2.6 & 1.4 \\
W Hya & 13:49:02.00 & --28:22:03.5 & Semi-regular pulsating Star & 8 & 1 & 0 & 0 & 0 & 0.8 & --- \\
Ceres & --- & --- & Asteroid & 14--38 & 13 & 7 & 0 & 3 & 5.0 & 1.0 \\
Cybele & --- & --- & Asteroid & 1--2 & 0 & 1 & 0 & 0 & --- & 0.6 \\
Europa & --- & --- & Asteroid & 2--5 & 1 & 2 & 0 & 0 & 0.6 & 1.2 \\
Hebe & --- & --- & Asteroid & 1--3 & 3 & 2 & 0 & 0 & 1.7 & 0.3 \\
Hygiea & --- & --- & Asteroid & 2--8 & 8 & 6 & 0 & 0 & 3.0 & 1.0 \\
Juno & --- & --- & Asteroid & 1--6 & 3 & 2 & 0 & 0 & 1.8 & 1.4 \\
Pallas & --- & --- & Asteroid & 5--12 & 8 & 5 & 0 & 0 & 3.8 & 0.8 \\
Thisbe & --- & --- & Asteroid & 1--3 & 0 & 1 & 0 & 0 & --- & 0.5 \\
Vesta & --- & --- & Asteroid & 10--17 & 13 & 5 & 0 & 1 & 7.5 & 1.2 \\
Neptune & --- & --- & Planet & 127--152 & 21 & 7 & 2 & 5 & 12.8 & 1.2 \\
Uranus & --- & --- & Planet & 300--353 & 21 & 10 & 4 & 7 & 10.3 & 1.4 \\
Dark field & 17:40:12.00 & +69:00:0.00 & Dark & --- & 125 & 34 & 17 & 5 & --- & --- \\ \hline
\end{tabular}
\end{center}
\begin{tablenotes}[normal,flushleft]
\item $^{\rm *}${One of the four main FTS line sources.}
\item $^{\rm \dagger}${Also known as AFGL618.}
\item $^{\rm \dagger\dagger}${Also known as IRC+10216, and has intrinsic FIR/submm domain variability \citep{Cernicharo2014}.}

\end{tablenotes}
\label{tab:sourceCount}
\end{table*}

\begin{table}
\caption{Optical path difference (OPD) range and spectral resolution (Resolution) for the FTS resolution modes. For an unresolved line, the spectral resolution is the distance from the peak to the first zero crossing. Note that in the current pipeline, CR is processed as HR and MR is processed as LR.}
\medskip
\begin{center}
\begin{tabular}{lcc}
\hline\hline
Mode & OPD range [cm] & Resolution [GHz] \\ \hline
LR & -0.555--0.560 & 24.98 \\
MR & -2.395--2.400 & 7.200 \\ 
HR & -0.555--12.645 & 1.184 \\
CR & -2.395--12.645 & 1.184 \\ \hline 
\end{tabular}
\end{center}
\label{tab:opd}
\end{table}
\subsection{SPIRE dark field}

\begin{figure}
\begin{center}
\includegraphics[width=\hsize]{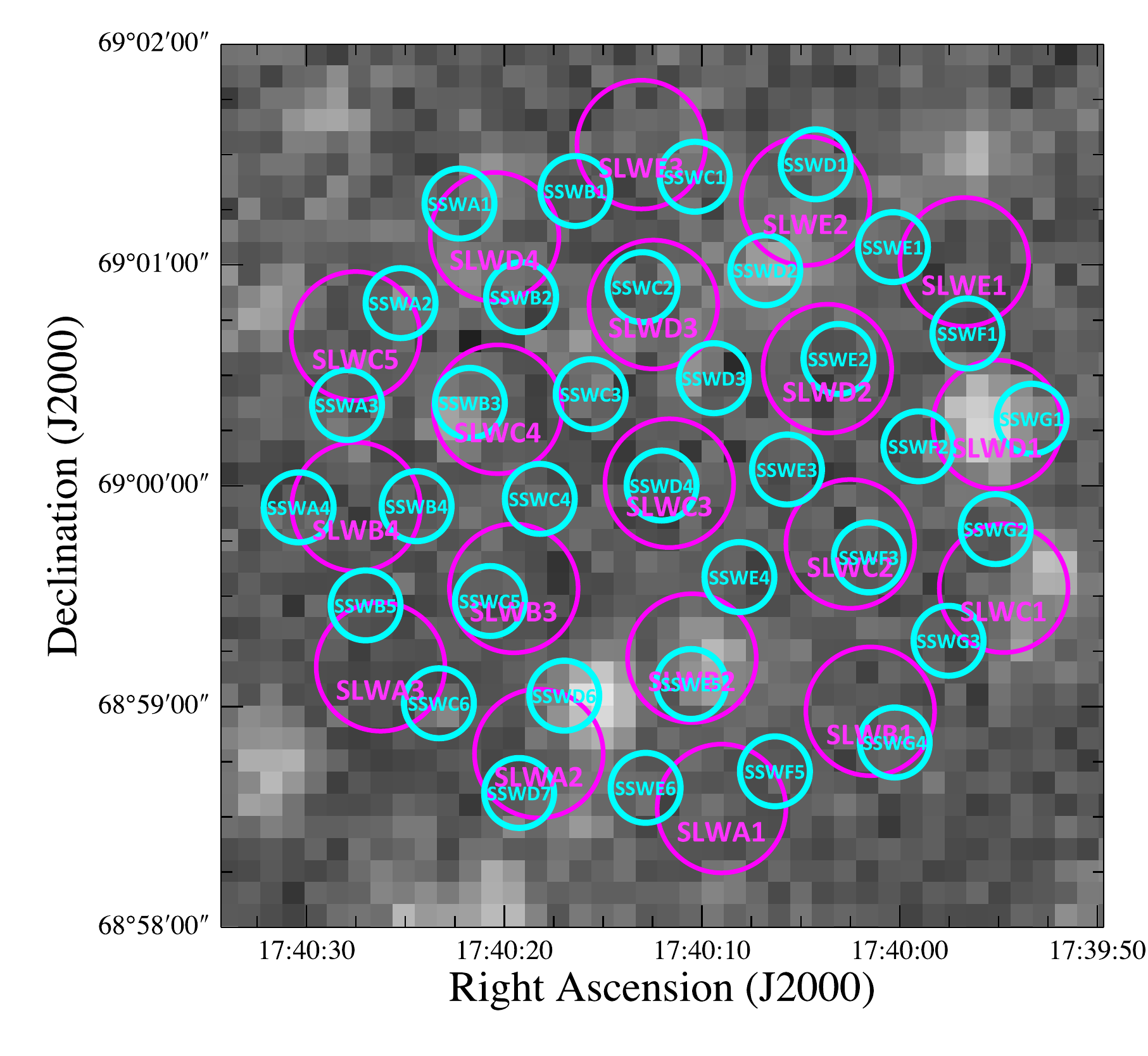}
\end{center}
\caption{Example sparse FTS footprint overlaid on a photometer short wavelength (PSW; 250\,$\mu$m) map of the SPIRE dark sky region. Several faint sources are evident, but all are $<$\,40\,mJy, except for the peak located near to SSWE6. As the FTS footprint rotates depending on the date, the relative position of this 40\,mJy peak changes over the {\it Herschel} mission, but is always below the FTS detection limit, as discussed in section \ref{sec:darkSky}.}
\label{fig:darkSkyField}
\end{figure}

To provide regular measurements of the {\it Herschel} telescope itself, which must be precisely removed from all observations, a dark region of sky centred on RA:17h40m12s and Dec:+69d00m00s (J2000) was selected to be the prime dark field for SPIRE. The selection process required the region to be of low cirrus, with no SPIRE-bright sources and to be visible at all times. The North Ecliptic Pole has low dust emission and satisfied the visibility criteria. During the performance verification period, test photometer mapping observations of the chosen region confirmed its suitability. As shown in Fig.~\ref{fig:darkSkyField}, there is a 40 mJy (at 250\,$\mu$m) source towards the edge of the FTS footprint, but stacking all sparse dark sky observations taken during the mission gives no obvious detection. The stacking results are briefly discussed in Section~\ref{sec:darkSky}, but are beyond the scope of this paper and will be presented in full in a future publication.

\subsection{Line sources}\label{sec:lineSources}
\begin{figure*}
\begin{center}
\makebox[\textwidth][l]{
\includegraphics[trim = 8mm 8mm 8mm 8mm, clip,width=0.33\textwidth]{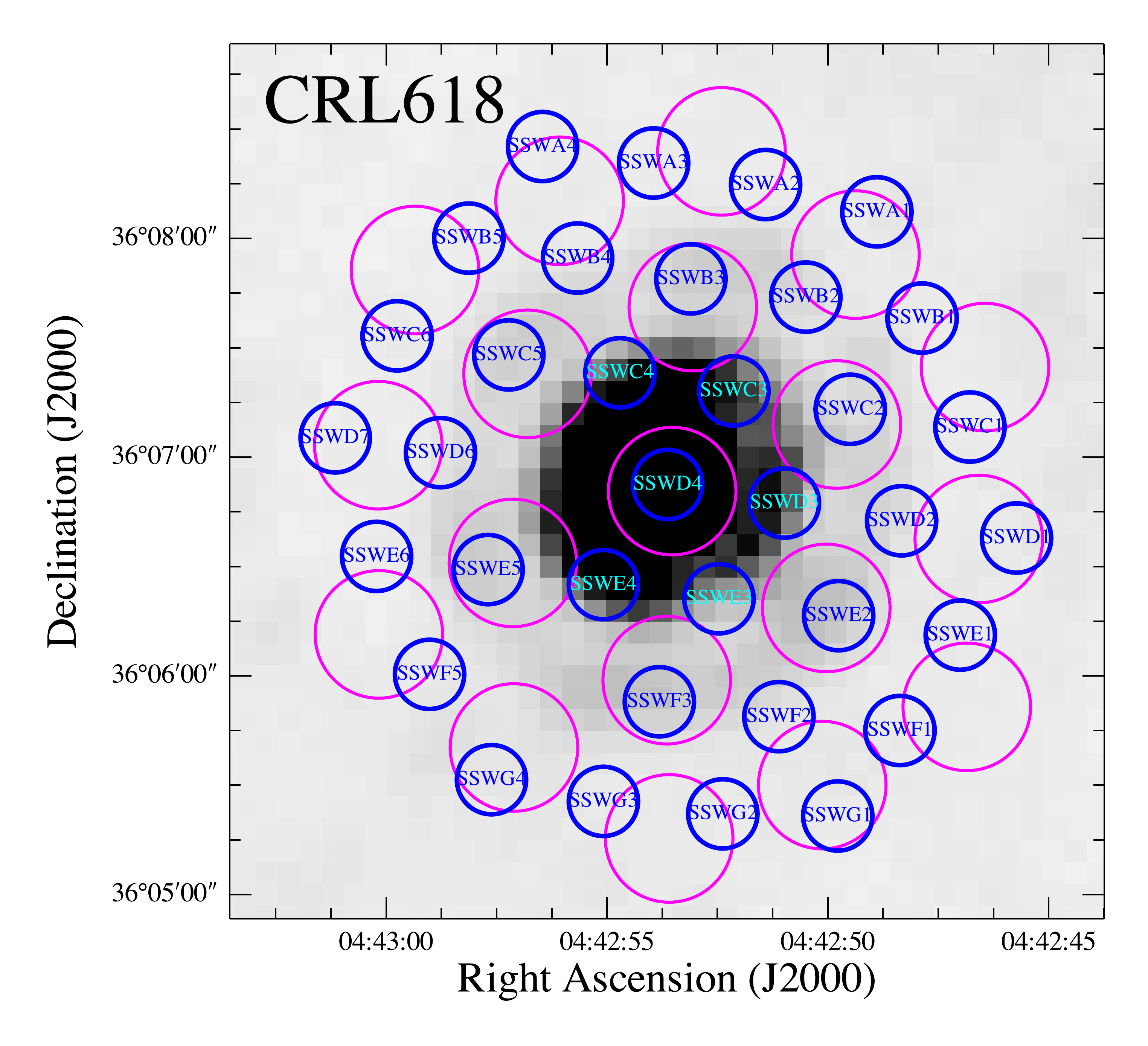}
\includegraphics[trim = 8mm 8mm 8mm 8mm, clip,width=0.33\textwidth]{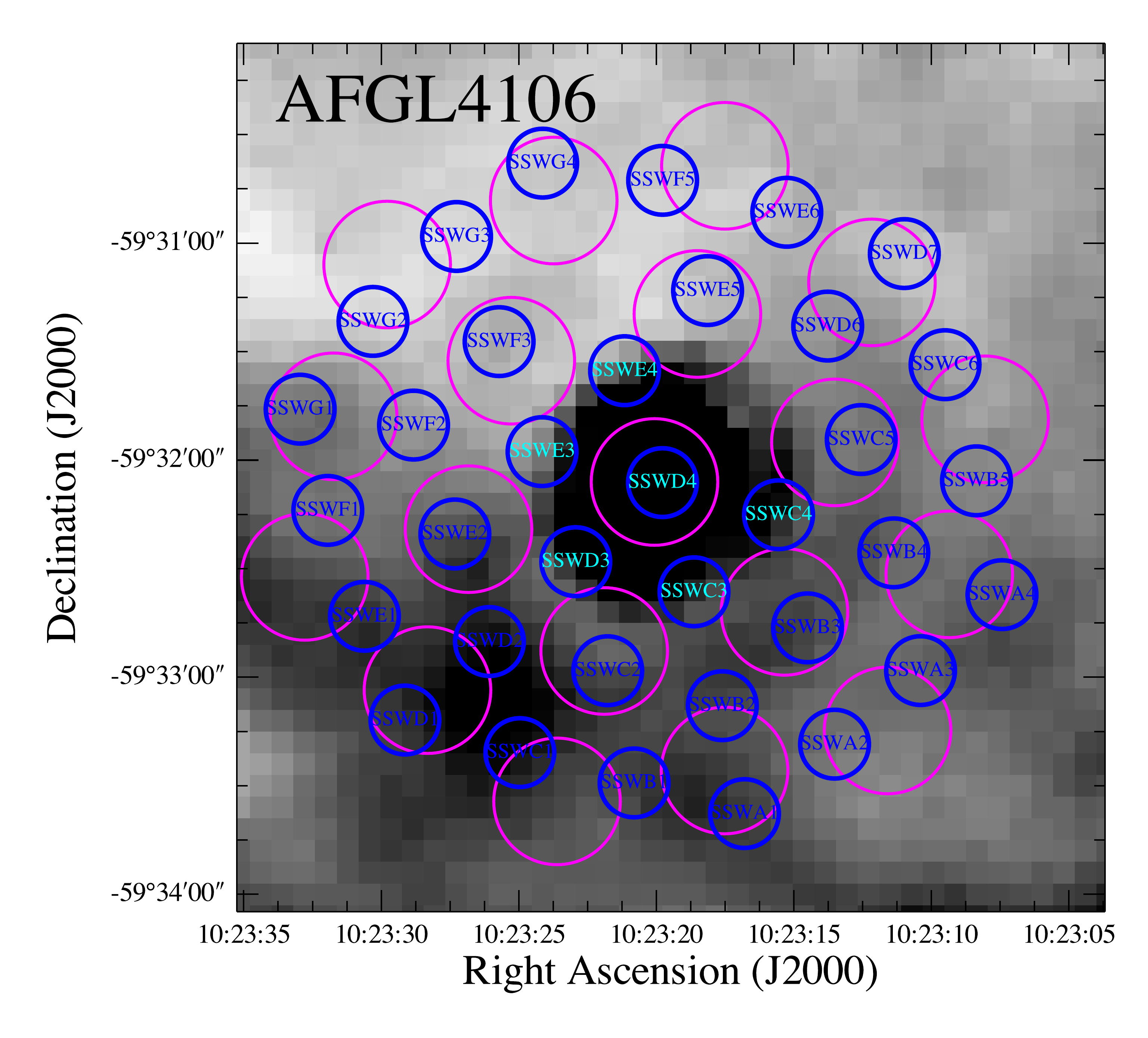}
\includegraphics[trim = 8mm 8mm 8mm 8mm, clip,width=0.33\textwidth]{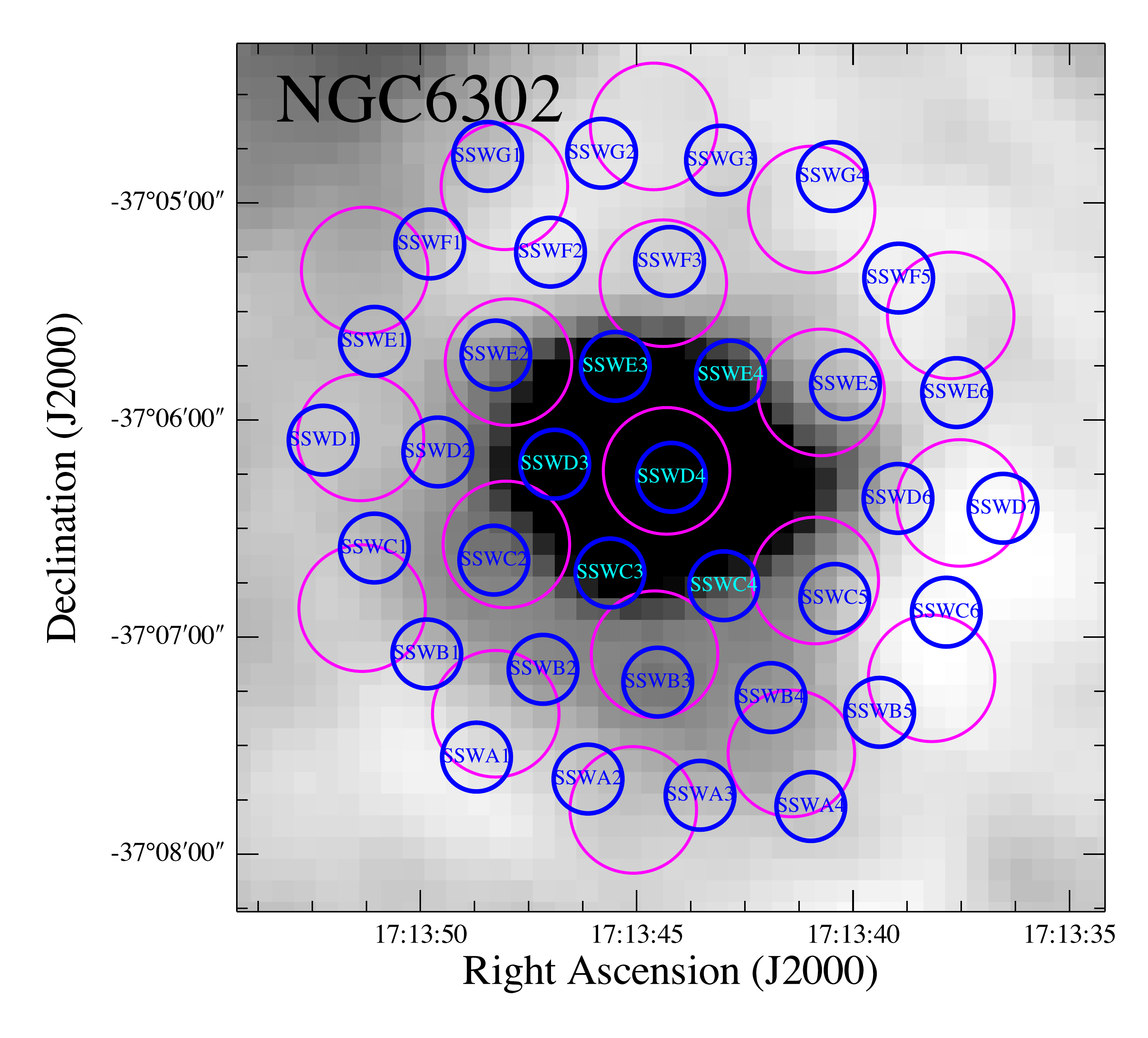}
}
\end{center}
\caption{Example FTS footprints for three of the repeated calibration sources, overlaid on {\it Herschel} SPIRE short wavelength photometer maps. CRL618 is a point-like source. AFGL4106 is a point-like source embedded in an extended background of Galactic cirrus and NGC6302 is a semi-extended source. A single FTS observation is used for each plot, but the footprint will be rotated for observations of the same source taken at different times in the mission.}
\label{fig:footprints}
\end{figure*}

\begin{figure}
\begin{center}
\includegraphics[trim = 5mm 5mm 5mm 0mm, clip=False,width=1.0\hsize]{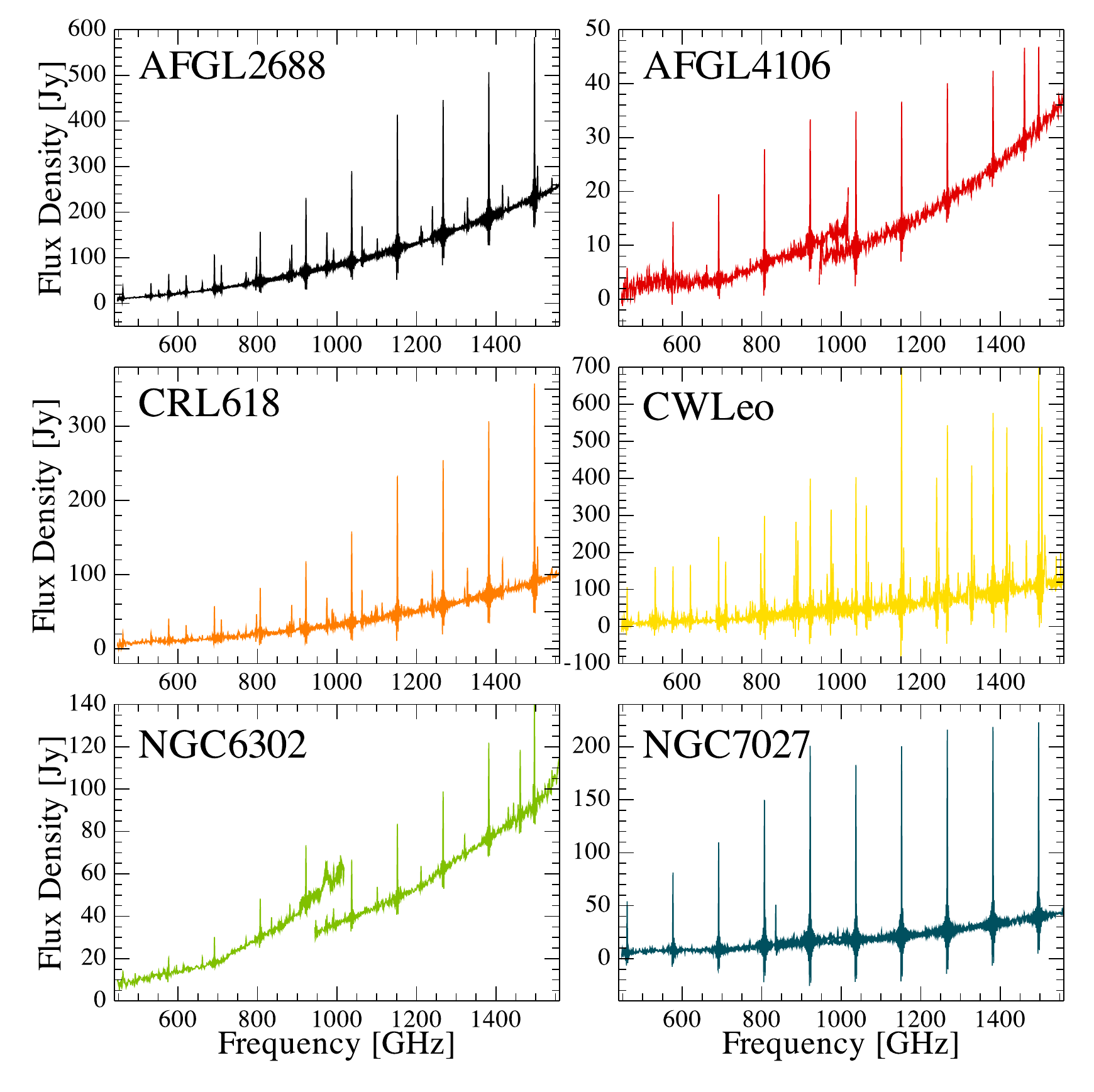}
\end{center}
\caption{Example point-source-calibrated spectra for the FTS line sources. Two sources show a discontinuity between the frequency bands. For AFGL4106 this is due to the point-like source sitting in an extended background, whereas for NGC6302 the difference is due to its partially extended morphology. If uncorrected, the latter issue can significantly affect spectral line flux measurements.}
\label{fig:lineSourcesEg}
\end{figure}

AFGL2688, AFGL4106, CRL618 and NGC7027 are four spectral-line rich sources, chosen to provide continuous visibility coverage throughout the mission. They are all planetary nebulae and post-AGB stars, which matched the necessary criteria well, i.e. they  are relatively bright and point-like within the SPIRE beam, with strong CO lines \citep[e.g.][]{Wesson2010, Wesson2011} and a wealth of ancillary observations. CRL618 and AFGL2688 are two commonly used secondary calibrators at the JCMT \citep{Jenness02} and AFGL2688 was used as a secondary calibrator for BLAST, a balloon experiment with an instrument based on the design of SPIRE \citep{Pascale2008}. An image of the FTS detector array over a corresponding SPIRE photometer short wavelength (PSW; 250\,$\mu$m) map is shown for CRL618 and AFGL4106 in Fig.~\ref{fig:footprints} and example spectra of all four of these sources in Fig.~\ref{fig:lineSourcesEg}. Note that the beam size varies between 31--42\arcsec\,across SLW and 16--20\arcsec\,across SSW \citep[see][]{Makiwa2013}. Many other spectral lines have been identified in FTS observations of AFGL2688, CRL618 and NGC7027. These line sources are summarised in the following sections and Table~\ref{tab:sourceCount}, with all associated observations detailed in the tables in Appendix~\ref{app:tables}.

\subsubsection{AFGL2688}\label{sec:lineSources:AFGL2688}

AFGL2688 is a very young proto-planetary nebula, with several main components: a hot fast wind extending up to 0.3\arcsec\,from the central star; intermediate layers, where the fast wind meets the envelope, out to 3\arcsec; a medium velocity wind up to 15\arcsec; and the AGB circumstellar remnant which extends to 25\arcsec\,\citep{Herpin02}. The higher-J CO lines originate from the hotter layers towards the centre of the nebula. 
The source size, as measured at 450 and 850\,$\mu$m at the JCMT, is $\sim$5\arcsec\,\citep{Jenness02}.

The SPIRE observations appear to be picking up continuum emission from the central region, as the spectrum appears point-like, i.e. no step is observed between the two bands, where the beam size changes, therefore there is no extension, e.g. see \citet[][]{Wu2013}.
There were four observations with significantly lower flux in SSW, which is assigned to pointing offset. Three of these observations were offset because of an error in the commanded coordinates used. 

\subsubsection{CRL618}\label{sec:lineSources:CRL618}

CRL618 (or AFGL618) is a very young proto-planetary nebula that initiated its post-AGB phase about 100 years ago \citep{KwokBignell1984}. It contains a compact HII region around the star of $0.4\arcsec\times0.1\arcsec$ \citep{KwokBignell1984}, a high velocity outflow to $\pm$2.5\arcsec, a slow axial component at $\pm$6\arcsec\,and a roughly circular outer halo \citep{SanchezContreras2004}. The halo component contributes to the core part of the line profile and is more important for low-J CO lines. The high-velocity outflow contributes to a much broader line component, which is more important for higher energy CO transitions \citep{SoriaRuiz2013} and manifests itself in the SPIRE spectrum as a clear line broadening towards higher frequencies in the SSW band, where the CO lines become partially resolved.

The continuum observed in the SPIRE spectrum appears point-like and is therefore likely to originate from the compact core. CRL618 is also known to be unresolved by the JCMT at 450 and 850\,$\mu$m, out to a 20\arcsec\,radius \citep[][]{Jenness02}. The CO lines likely have differing contributions, with more extended emission contributing to the lower-J lines. 

\subsubsection{NGC7027}\label{sec:lineSources:NGC7027}

NGC7027 is a young planetary nebula \citep[e.g.][]{volkKwok97} and consists of a central HII region with a bipolar outflow covering a region $\sim$15\arcsec\,in diameter \citep[e.g.][]{Deguchi1992, Huang2010}, surrounded by a more extended CO envelope that is approximately 35\arcsec\,in diameter \citep[e.g.][]{Phillips1991}, with $J$=1--0 measurements showing this extends out to $\sim$60\arcsec\,diameter \citep[]{Fong2006}. Higher energy levels of CO are populated at smaller radii towards the central hot region \citep{Jaminet1991, Herpin02}. The continuum emission as observed by the SPIRE photometer and spectrometer can be explained by a diameter of $\sim$15\arcsec.
Despite the slightly extended nature of the source, the SPIRE FTS observations show high signal-to-noise ($>$\,200) CO lines, with the least blending of the four main line sources. 

\subsubsection{AFGL4106}\label{sec:lineSources:AFGL4106}

AFGL4106 is fainter and has fewer previous observations than the other three sources, but was added to the programme to fill a gap in coverage, and ensure at least one calibration source was available on every SPIRE FTS observing day. It is an evolved massive star surrounded by a dust shell in the post-red supergiant phase, and known to be part of a binary system \citep{Molster1999}. It has ongoing mass-loss, and is surrounded by a bow shaped emission complex that extends to 5--10\arcsec\,from the star \citep{vanLoon99}. In the mid-infrared, the dust distribution is clumpy with a size of $3.4\times3.3$\arcsec\,\citep{Lagadec2011}.

Infrared Astronomical Satellite \citep[IRAS;][]{Neugebauer1984} 100\,$\mu$m observations \citep{MivilleLagache2005} show this source to be embedded in a region of Galactic cirrus, which is evident in the SPIRE photometer data (see Fig.~\ref{fig:footprints}). Once this extended emission is subtracted from the spectra on the centre detectors, the source appears point-like, i.e. the remaining continuum emission is coming from the central part of the shell, as seen with the aforementioned mid-infrared observations.

\subsection{Planets and asteroids}\label{sec:planetsAsteroidSummary}

\begin{figure}
\begin{center}
\includegraphics[width=\hsize]{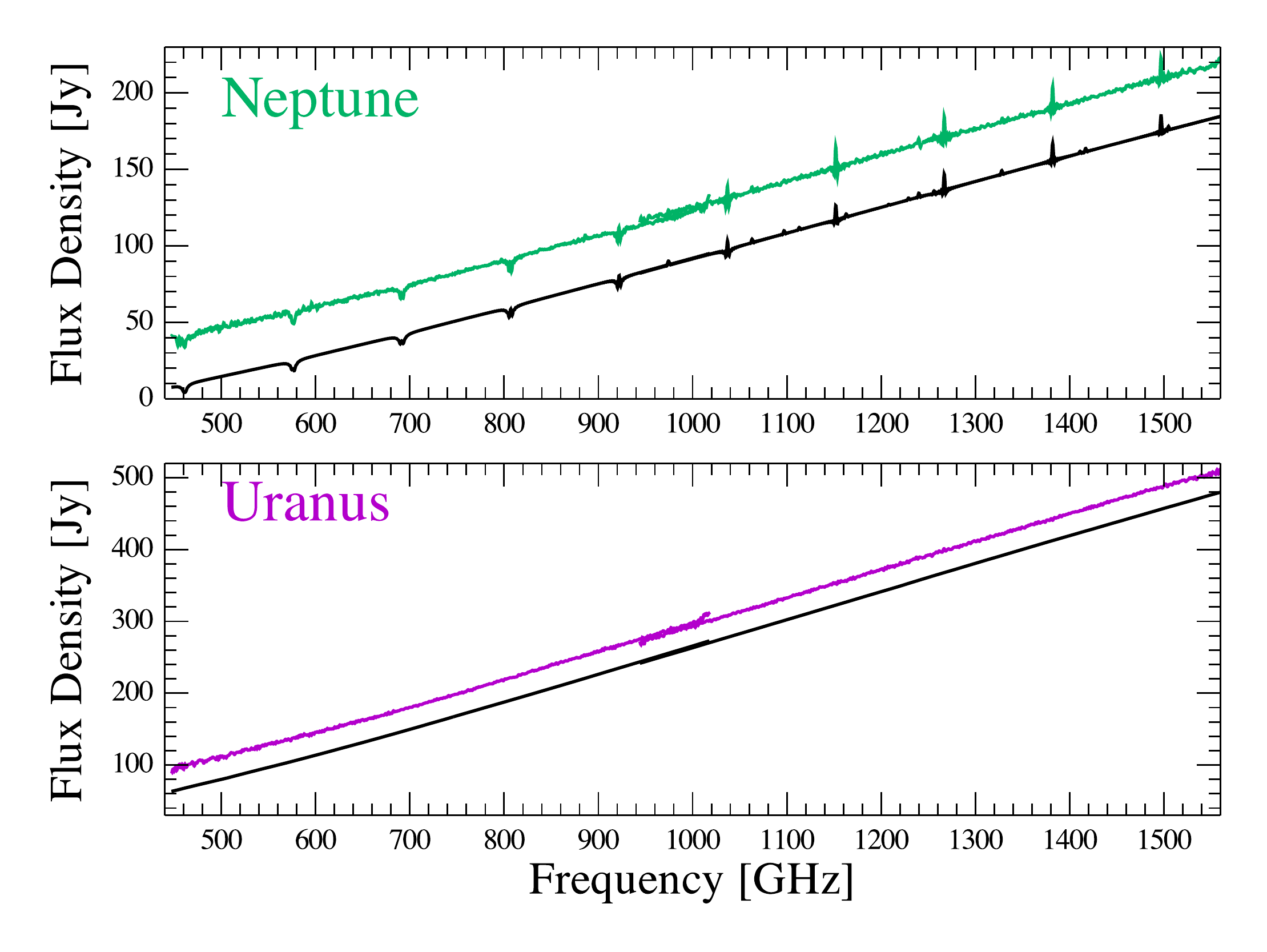}
\end{center}
\caption{Example FTS point-source-calibrated spectra for Neptune and Uranus. The corresponding model is plotted in black and offset for clarity.}
\label{fig:planetEg}
\end{figure}

The planet Uranus is used as the primary point-source calibrator for the FTS. See \citet{Swinyard2014} for details of the observations and model used. It was observed regularly during the mission, including observations centred on different detectors. These observations are summarised in Table~\ref{tab:UranusObs} 
and an example spectrum is shown in Fig.~\ref{fig:planetEg}.

The planet Neptune is used as the primary flux calibrator for the SPIRE photometer \citep{Bendo13} and was also regularly observed with the FTS. The model used for Neptune is described in \citet{Swinyard2014}. FTS observations of Neptune are summarised in Table~\ref{tab:NeptuneObs} 
and an example spectrum is shown in Fig.~\ref{fig:planetEg}.

Two additional planets, Mars and Saturn, were observed in order to monitor the calibration in  bright-source mode \citep[see][]{Lu14}. 

\begin{figure}
\begin{center}
\includegraphics[width=\hsize]{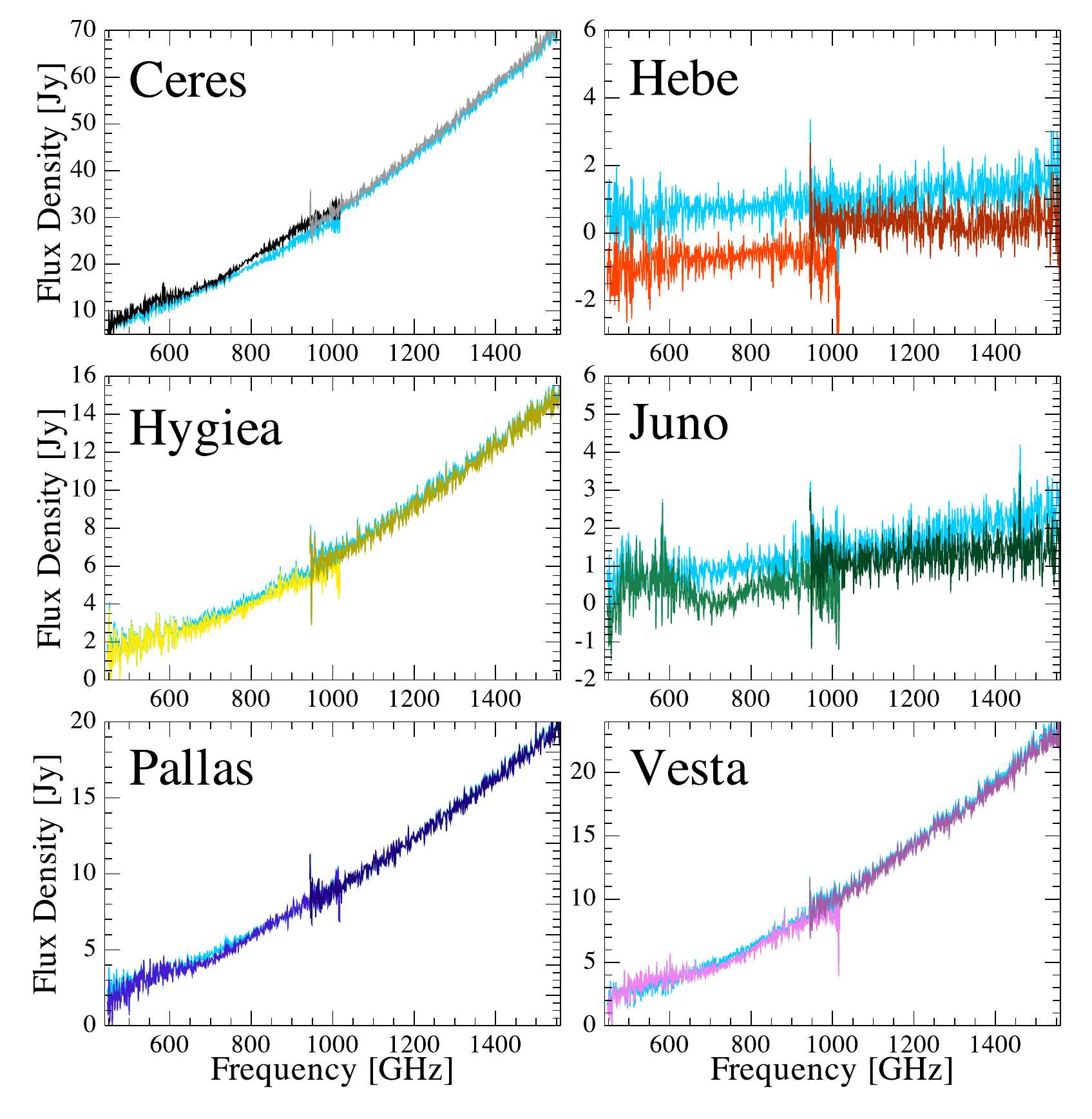}
\end{center}
\caption{Example point-source-calibrated spectra for the FTS asteroids. The examples after background subtraction are plotted in light blue, which show a significant improvement in the continuum shape for the faintest two sources, Hebe and Juno.}
\label{fig:asteroidEg}
\end{figure}

A number of asteroids were also included for regular observation, as these cover a fainter flux range than both the planets and the four main line sources, thus allowing the fainter end of the FTS capabilities to be monitored.
Asteroids have been successfully used as secondary calibrators on previous missions, for example ISOPHOT \citep{Schulz02}, AKARI \citep{Kawada2007} and Spitzer-MIPS \citep{Stansberry2007}. In order to compare observations on different days, the thermophysical models of \citet{Muller13, Muller02} were used. The asteroids observed (with number of HR/LR observations given in brackets) were: Ceres (13/7), Pallas (8/5), Vesta (13/5), Hebe (3/2), Hygiea (8/5), and Juno (3/0). Example HR spectra for each of these asteroids are shown in Fig.~\ref{fig:asteroidEg}. In addition, there were some MR observations made early in the mission for the following asteroids: Cybele (1), Europa (2), Hygiea (1), Juno (2) and Thisbe (1), which are processed as LR and therefore the number of MR observations (given in the brackets) are included in the total LR observations for these sources in Table~\ref{tab:sourceCount}. 

\subsection{Cross calibration targets}\label{sec:crossCalSummary}

A number of additional spectral line rich targets were added to the calibration programme to allow comparison with HIFI and PACS observations. These were: NGC6302 (9HR/5LR), see Fig.~\ref{fig:footprints}, CW Leo (9HR/3LR), R Dor (4HR/2LR), and VY CMa (8HR/5LR). Several other late additions to the programme were only measured once or twice: Omi Cet (2HR), W Hya (1HR), and IK Tau (1HR). These observations are summarised by Table~\ref{tab:othersObs}, and their comparison with HIFI and PACS will be detailed in Puga et al. (in preparation).

\section{Data processing}\label{sec:dataProcessing}

All FTS data presented were reduced with the standard {\it Herschel} Interactive Processing Environment \citep[HIPE;][]{Ott10} pipeline (Fulton et al. in preparation), or standard pipeline tasks where required, with HIPE version 13 and \texttt{spire\_cal\_13\_1} calibration tree used for the main results presented. There can be a number of considerations for FTS data on top of a standard reduction, and some of these are explored in this section.

\subsection{Pointing considerations}\label{sec:pointCorrBsm}

As discussed in \citet{Swinyard2014}, the accuracy of the {\it Herschel} pointing is one of the largest sources of uncertainty on line flux measurements and point-source continuum for the SSW array. The SLW array is less affected by pointing, due to its larger beam size. The \textit{Herschel} absolute pointing error (APE) is the offset of the actual telescope pointing from the commanded position. Due to several improvements in the satellite pointing system (see \citealt{SanchezPortal2014} for more details), the initial APE of 2\arcsec\,(68\% confidence interval) was improved to $\sim$\,1\arcsec\,for observations in the second half of the mission (after OD\,866, 27 September 2011). Once at the commanded position, the telescope stability, i.e. the Relative Pointing Error (RPE), was within 0.3\arcsec, which has a negligible effect compared to the other sources of uncertainty associated with FTS data. 

In addition to the APE, another source of complications for FTS pointing comes from the fact that the Beam Steering Mirror (BSM) rest position for sparse-sampled observations was not at the nominal rest position before OD\,1011 (18 February 2012). In the remainder of this sub-section we discuss this 1.7\arcsec\ shift in the BSM rest position, its impact of on FTS data, and how to account and correct for \textit{systematic} pointing offset.

\subsubsection{BSM position}\label{sec:bsm}

The 1.7\arcsec\ difference in BSM position divides FTS sparse observations into two mission epochs -- observations taken \textit{``Before''} OD\,1011 and those taken on or \textit{``After''} this OD. This division into two epochs is not necessary for intermediate and fully sampled observations, as these were always observed with the BSM at its nominal rest position.

It is important to note that the 1.7\arcsec\,BSM shift is automatically taken into account in the point-source calibration, as the primary calibrator was also measured in the shifted position during the {\it Before} epoch.

\begin{figure*}
\centering
\makebox[\textwidth][l]{
\includegraphics[trim = 8mm 15mm 8mm 5mm, clip,width=0.152\textwidth]{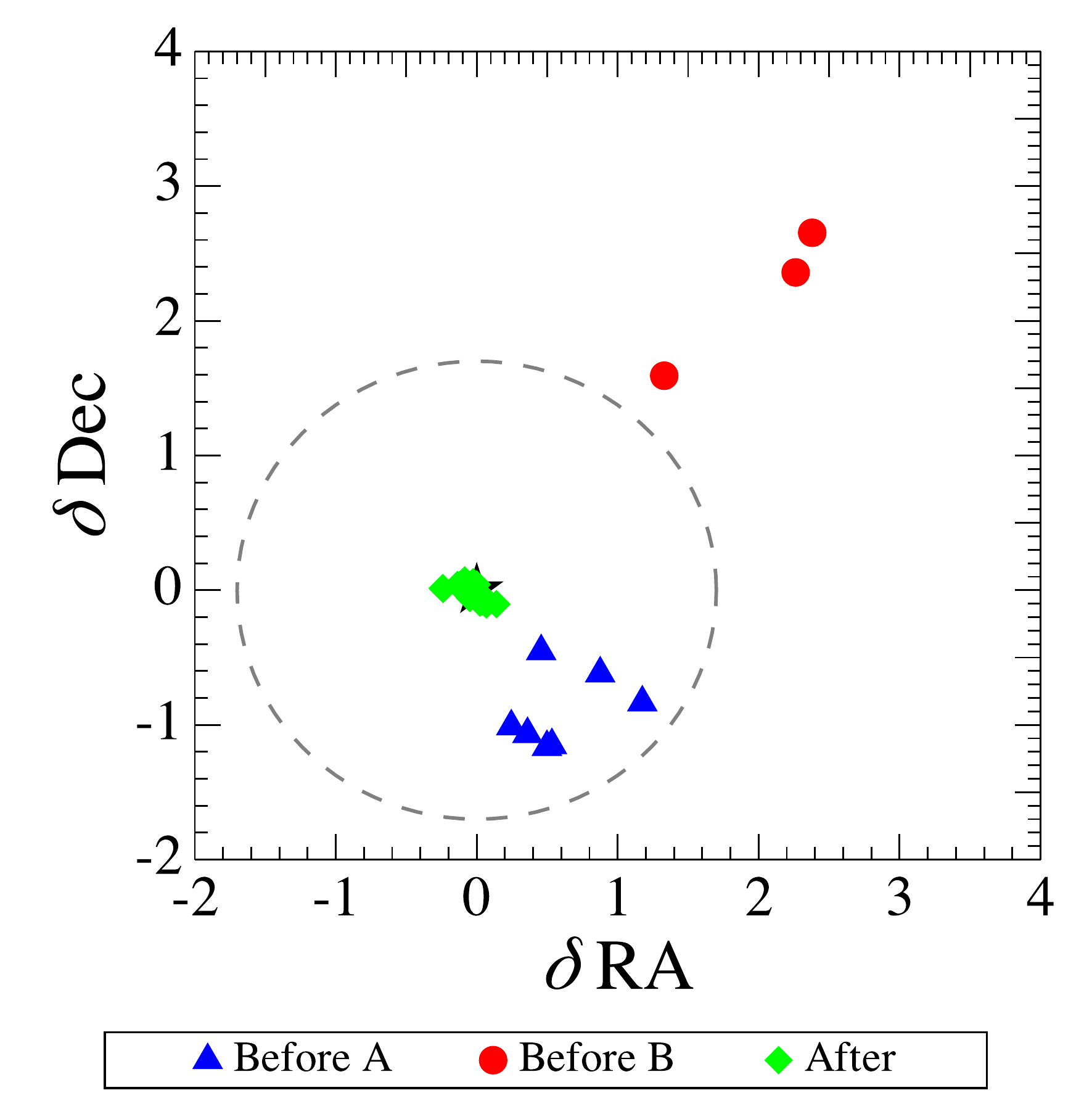}
\includegraphics[trim = 8mm 15mm 8mm 5mm, clip,width=0.333\textwidth]{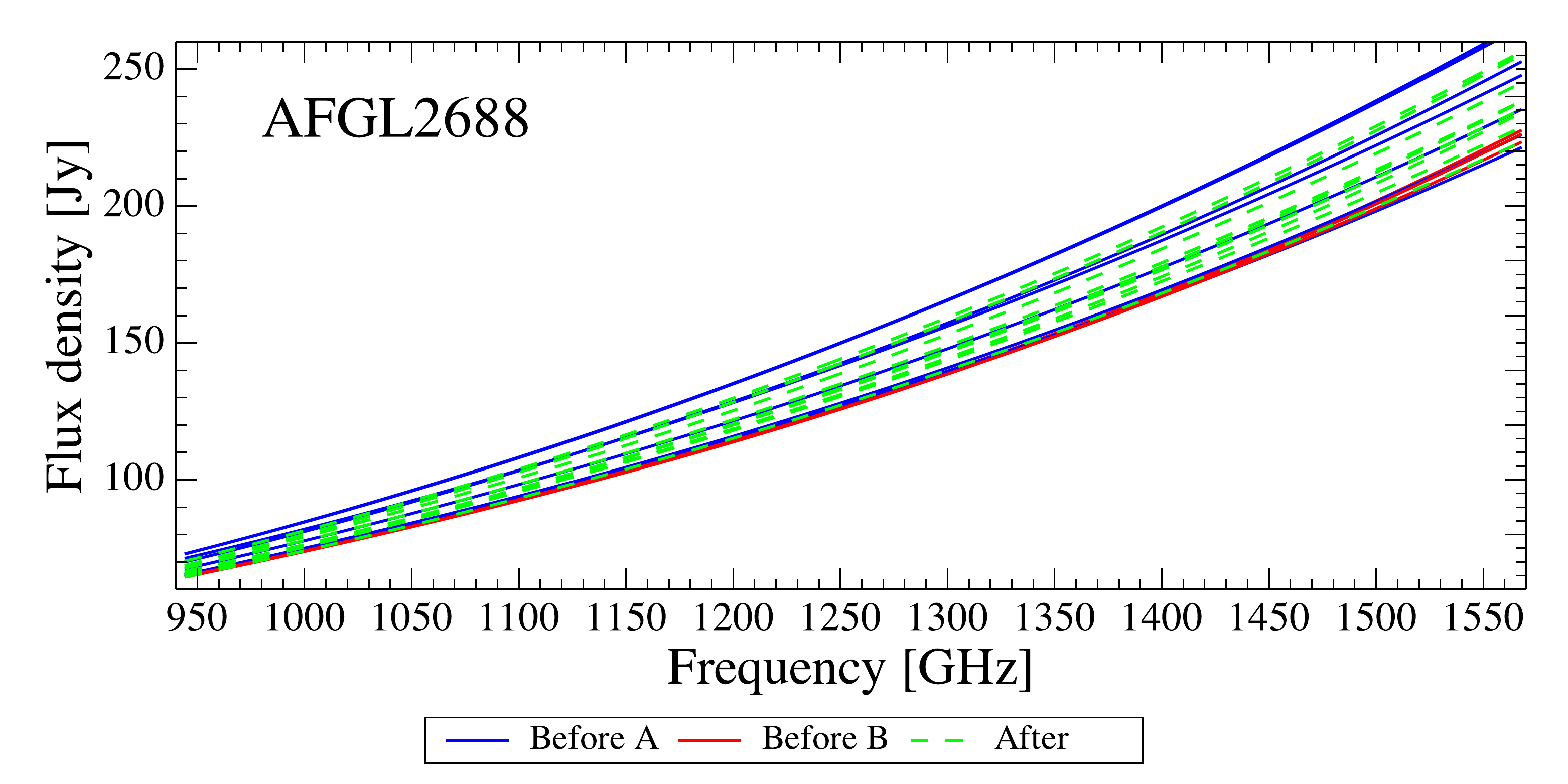}
\includegraphics[trim = 8mm 15mm 8mm 5mm, clip,width=0.152\textwidth]{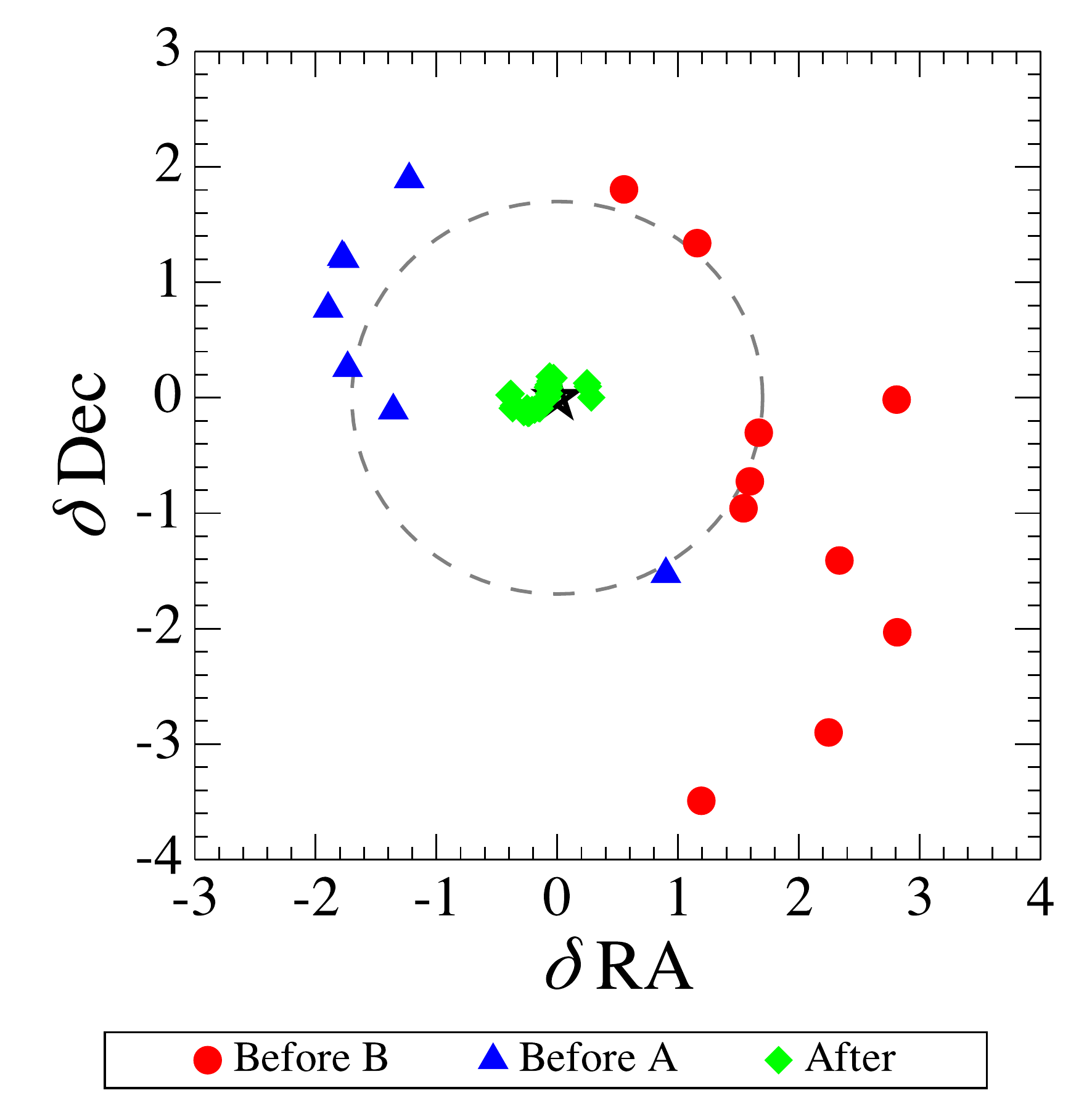}
\includegraphics[trim = 8mm 15mm 8mm 5mm, clip,width=0.333\textwidth]{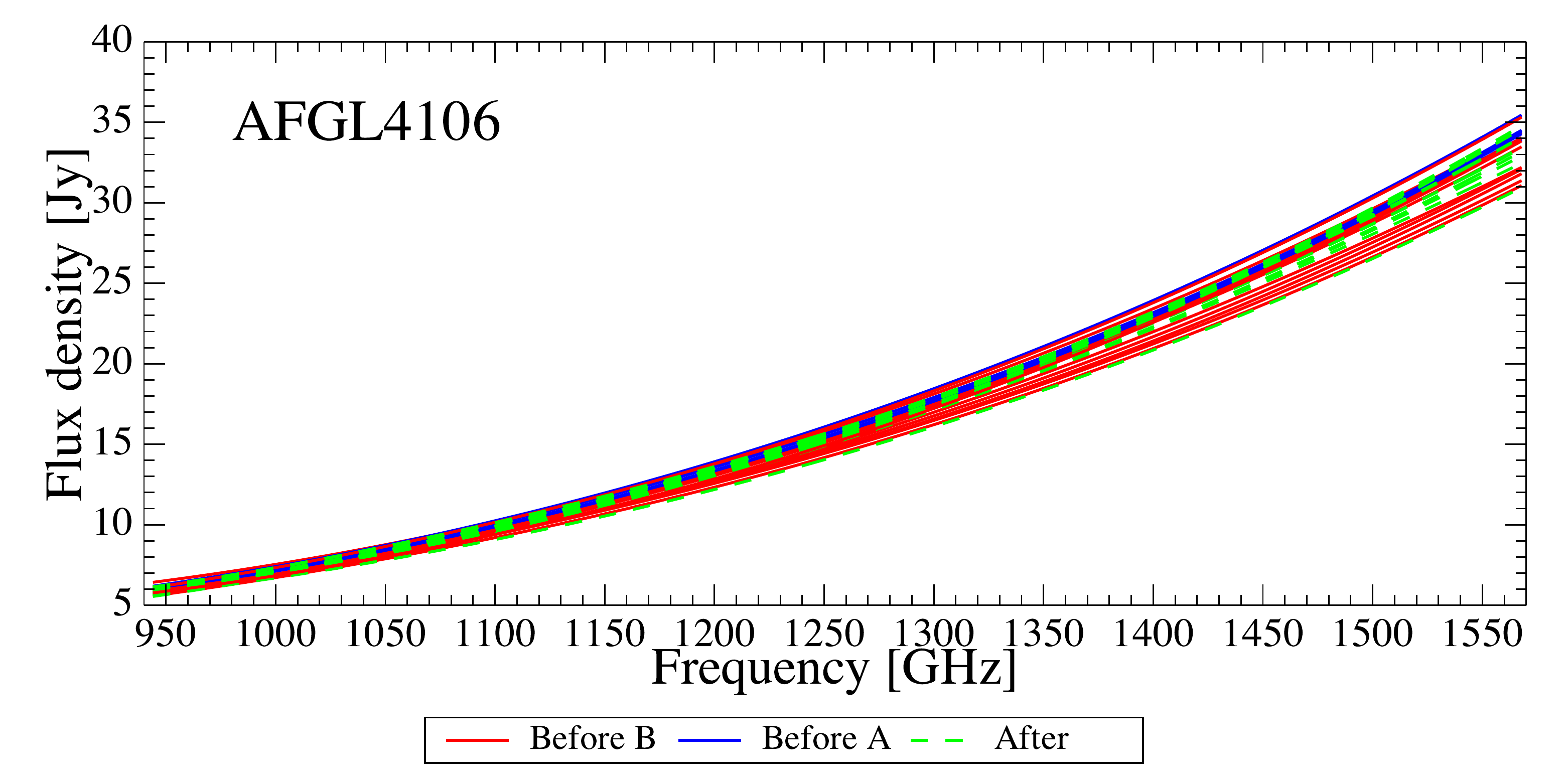}
}
\makebox[\textwidth][l]{
\includegraphics[trim = 8mm 15mm 8mm 5mm, clip,width=0.152\textwidth]{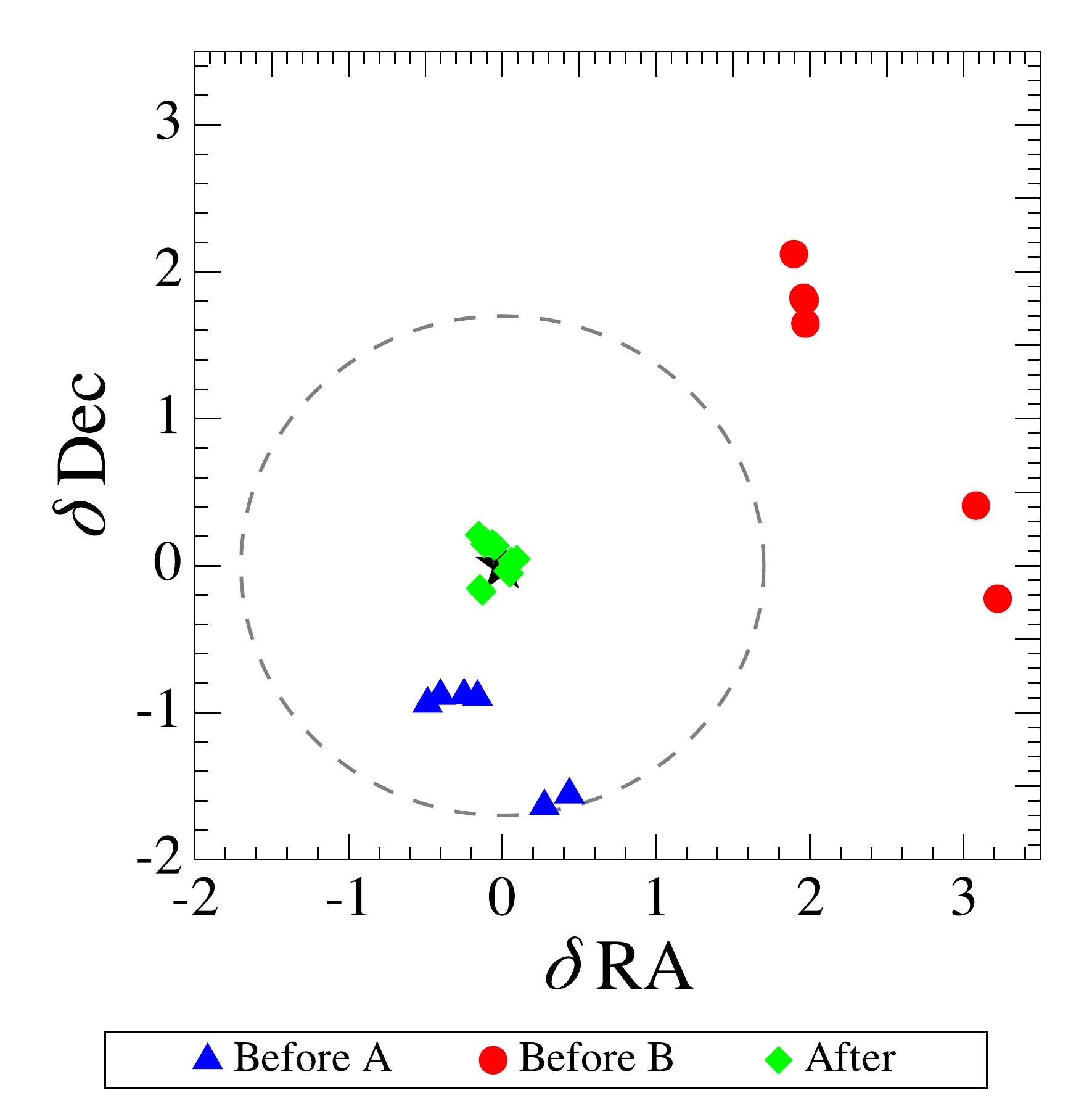}
\includegraphics[trim = 8mm 15mm 8mm 5mm, clip,width=0.333\textwidth]{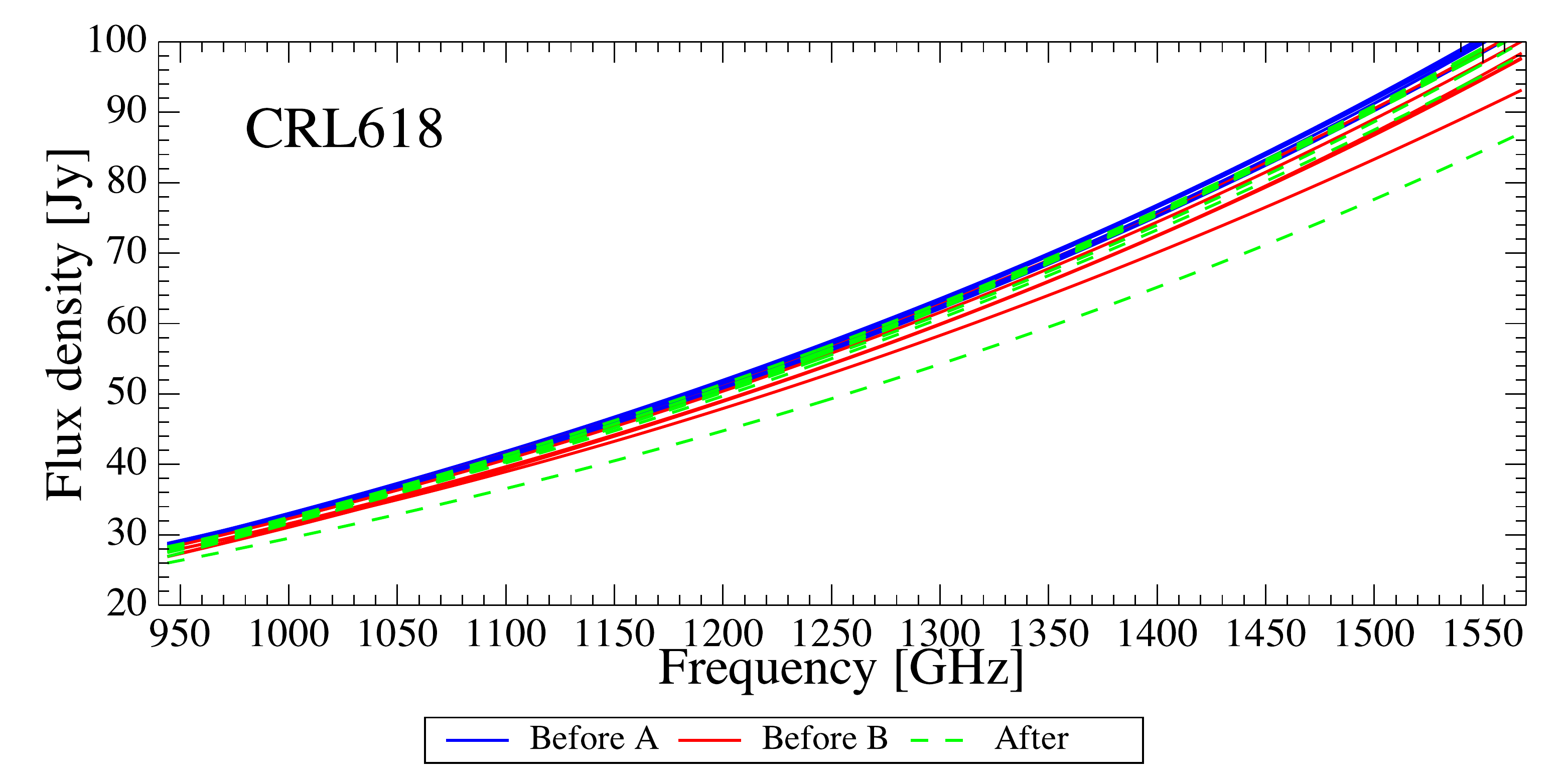}
\includegraphics[trim = 8mm 15mm 8mm 5mm, clip,width=0.152\textwidth]{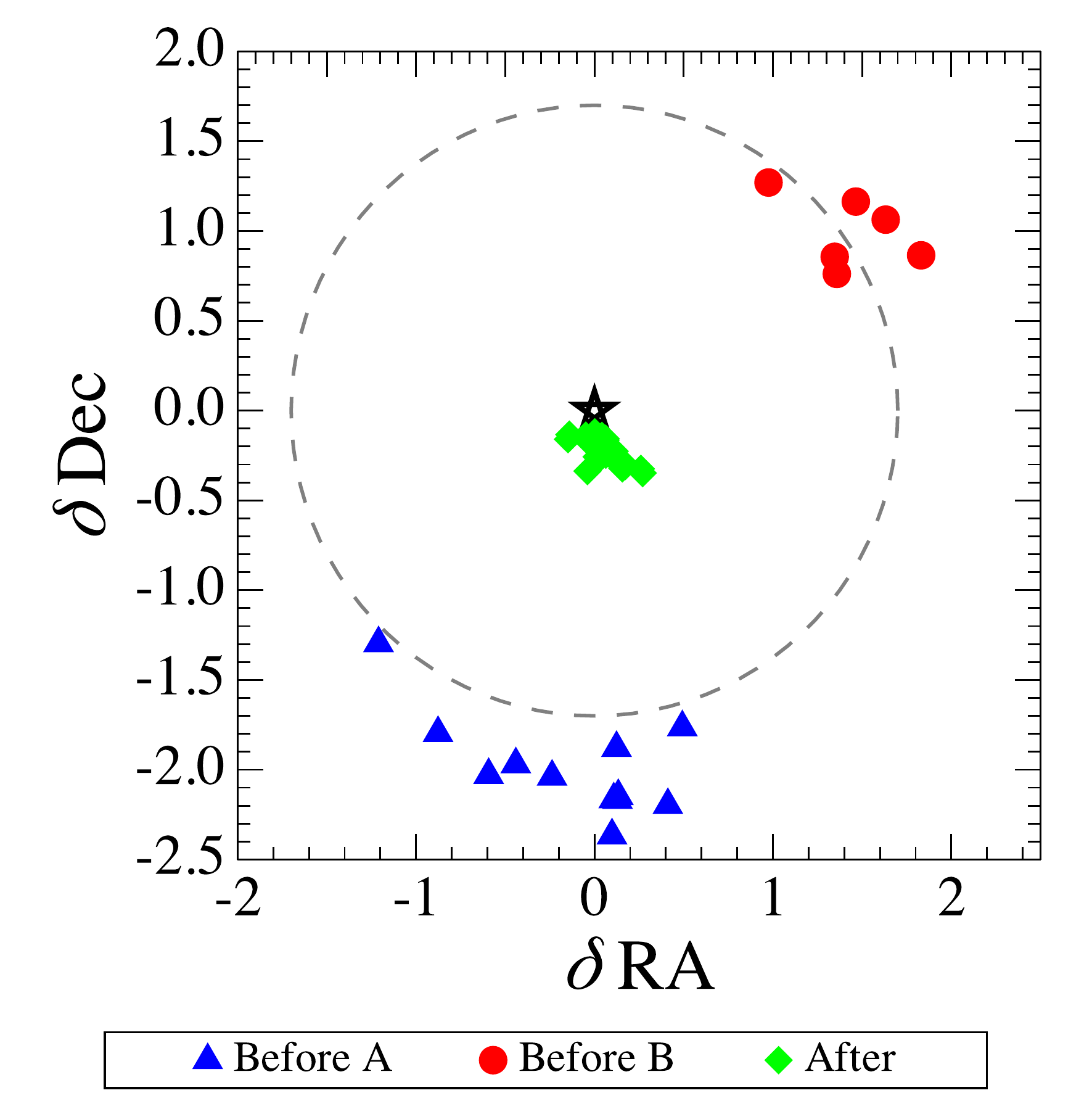}
\includegraphics[trim = 8mm 15mm 8mm 5mm, clip,width=0.333\textwidth]{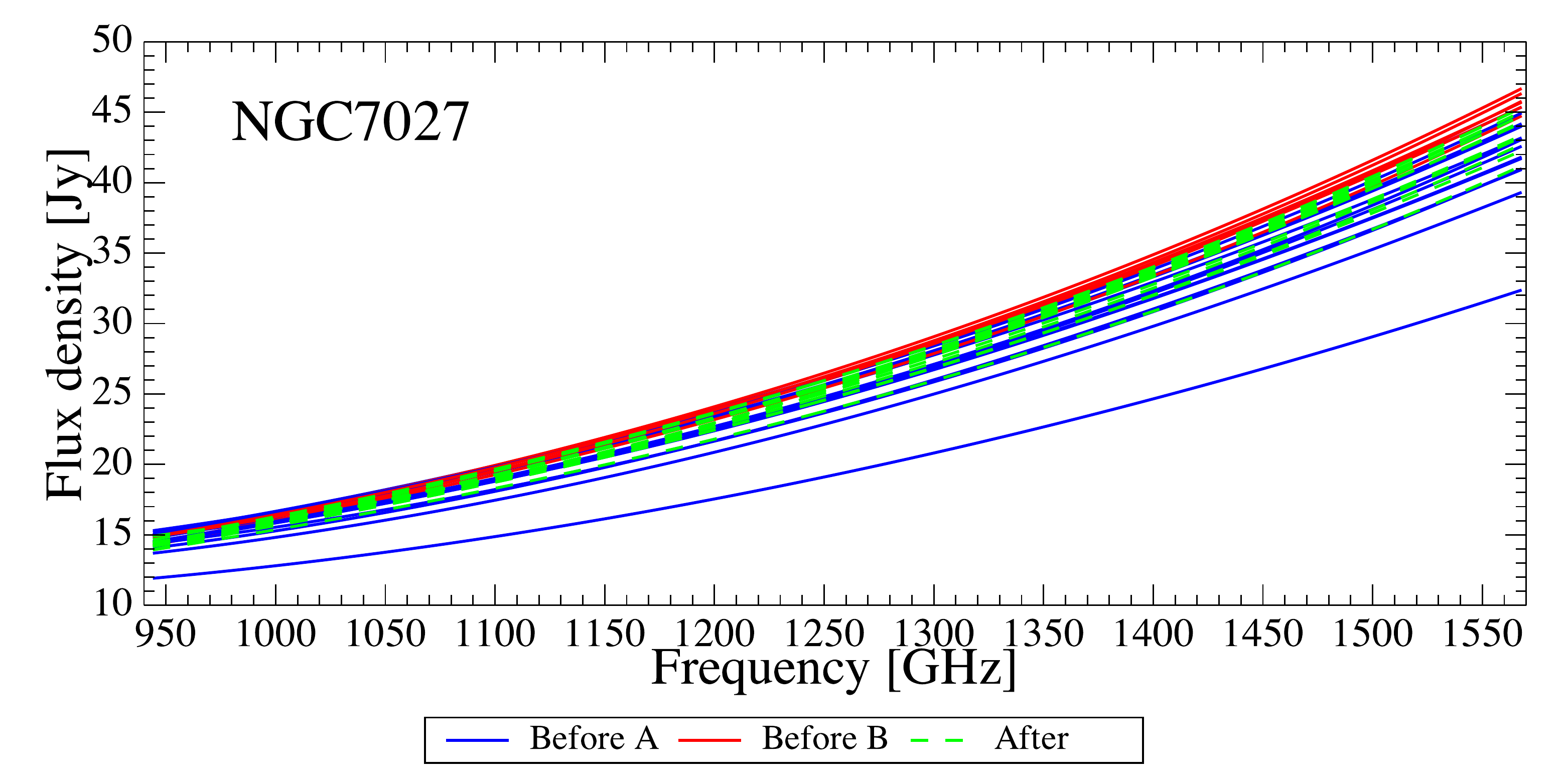}
}
\makebox[\textwidth][l]{
\includegraphics[trim = 8mm 15mm 8mm 5mm, clip,width=0.152\textwidth]{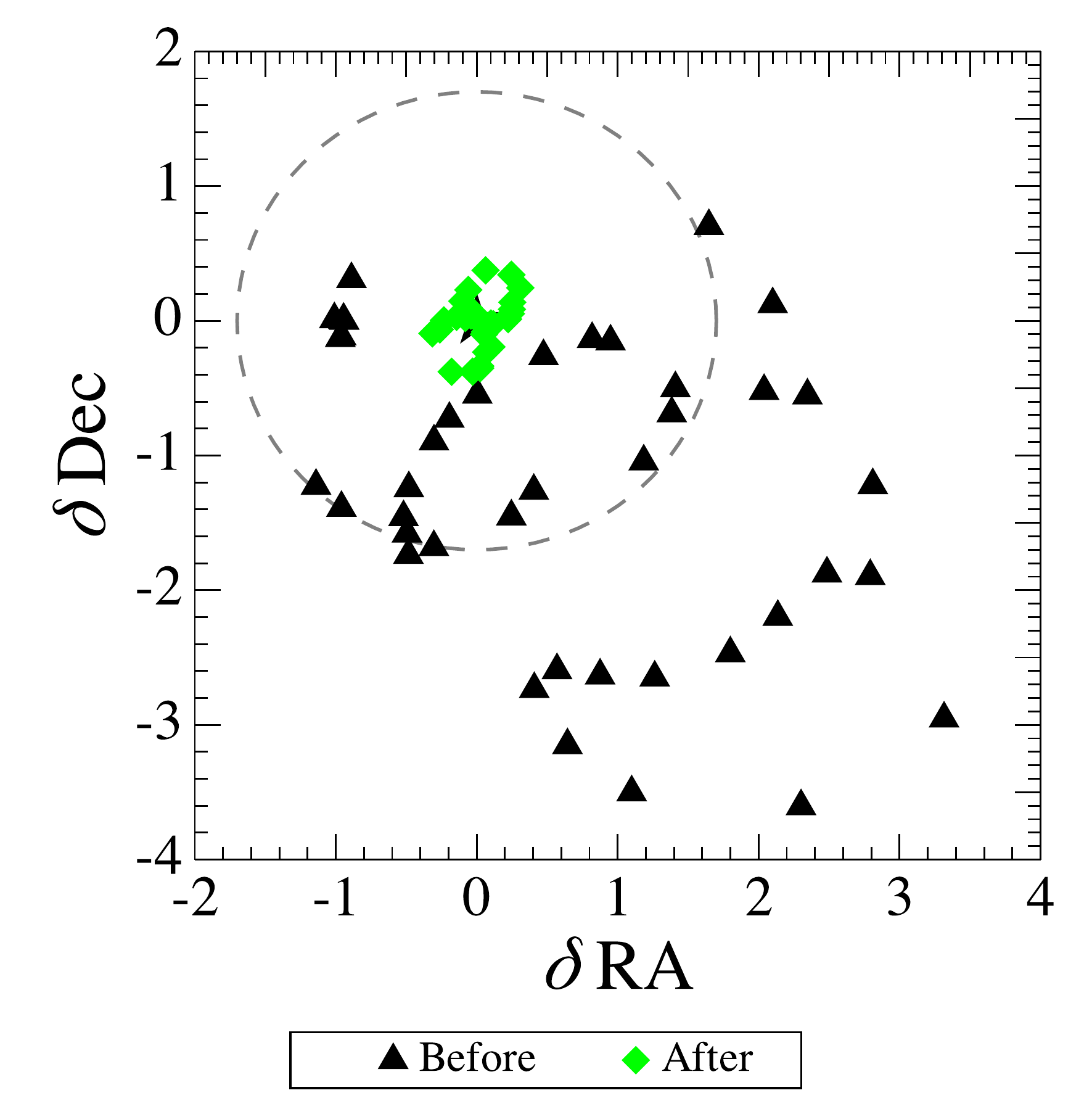}
\includegraphics[trim = 8mm 15mm 8mm 5mm, clip,width=0.333\textwidth]{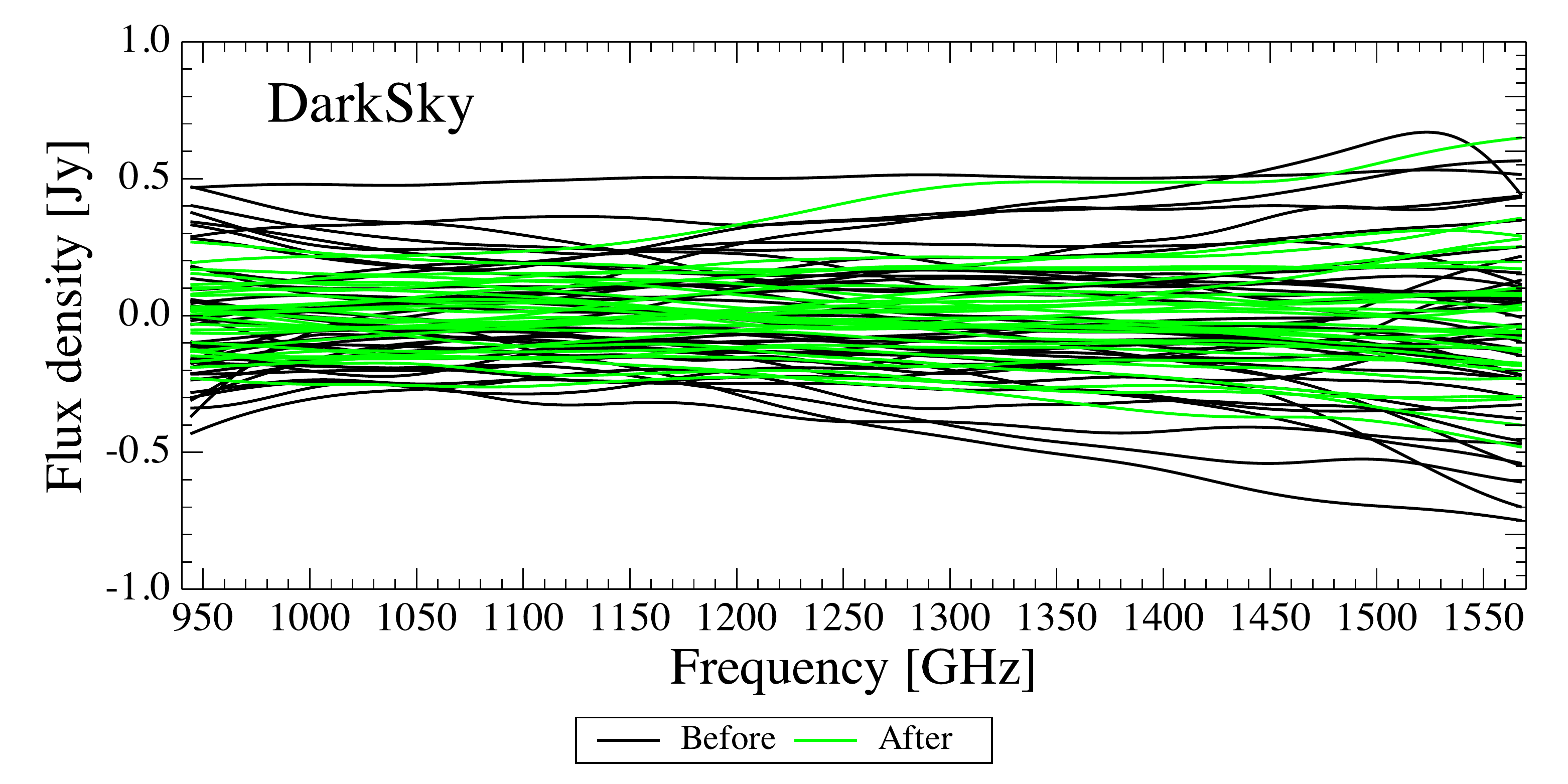}
\includegraphics[trim = 8mm 0mm 8mm 5mm, clip,width=0.5\textwidth]{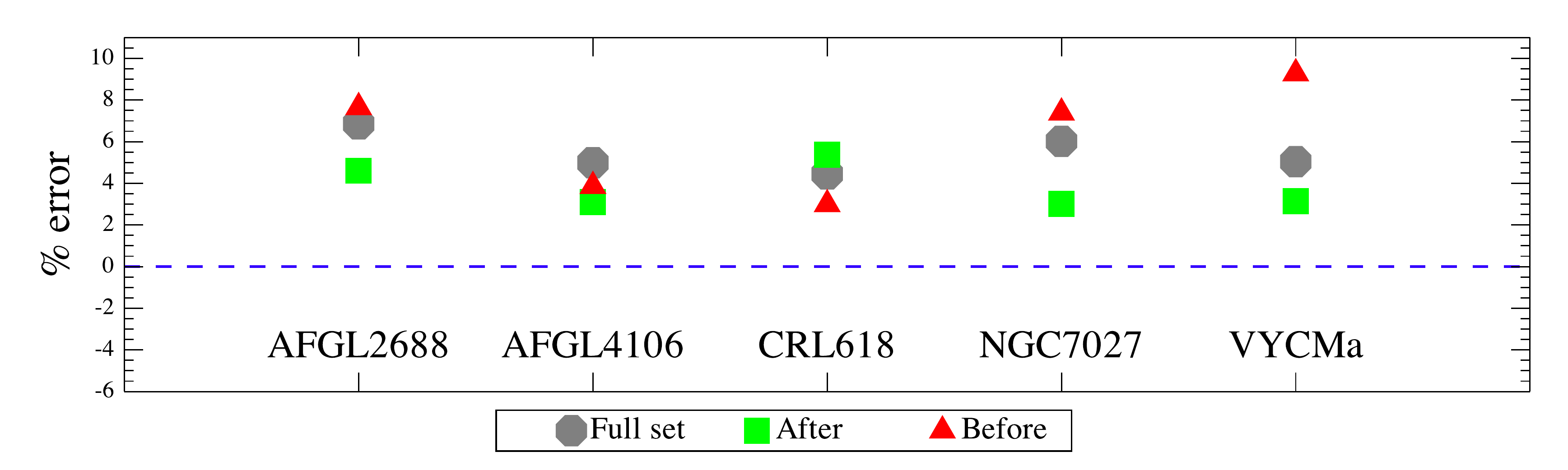}
}
\caption{Difference between the nominated sky position for SSWD4 and the best estimate of actual sky position, along with the respective continuum levels, are plotted for several of the FTS calibration targets. On the left plot of each pair a star marks zero, green diamonds show the $\delta$ RA and Dec for {\it After} observations and red circles and blue triangles represent the {\it Before} observations. The dashed circle has a radius of 1.7\arcsec\,and shows where the {\it Before} coordinates would lie if the pointing had been perfect. The red and blue division for {\it Before} observations is arbitrary to highlight the yearly rotation of the instrument.  The corresponding SSW continuum shown follow the same colour scheme for {\it Before} and {\it After}. Dark sky are shown to the bottom left, but all points are coloured black or grey, due to the more complete rotational coverage. The smoothed spectra are compared for the darks as there is no continuum, with {\it Before} spectra all plotted in black. The bottom right plot shows the spread relative to the mean continuum level at the end of the SSWD4 detector. The scatter of the darks is symmetric around zero flux density, which shows the dark observations are not affected by pointing offset. }
\label{fig:bsm}
\end{figure*}

To examine the impact of having two BSM position epochs for FTS data, the difference between the RA and Dec coordinates of SSWD4 for the actual sky position and the nominal target coordinates were used. The results for several FTS calibration sources, which exhibit no significant intrinsic variability (as discussed in Sections~\ref{sec:continuum} and \ref{sec:lineFlux}), are shown in Fig.~\ref{fig:bsm}, along with the fitted continua (see Section \ref{sec:continuum} for details on how these fits were obtained). In this figure {\it Before} and {\it After} observations are compared. {\it Before} observations also subdivided to illustrate the yearly rotation seen by the instrument, which leads to some grouping of the continua level as the shift is systematic and in a fixed direction in the instrument coordinate system. All {\it After} observations cluster around the nominal target RA and Dec (as expected) and the corresponding continua have a lower spread. The spread relative to the mean continuum level at the high frequency end of the SSW band is shown in the bottom right panel of Fig.~\ref{fig:bsm}. The average spread is 6.2\% for {\it Before} spectra and 3.8\% for {\it After}. Dark sky positions also show a higher scatter of $\delta$\,RA and $\delta$\,Dec for the {\it Before} epoch, although there is no dependency on pointing for the corresponding spectra.

\subsubsection{Pointing offset}\label{sec:pointCorr}

If there are a statistically significant number of observations available of the same source, it is possible to correct the data for pointing offset following the method detailed in \citet[][]{Valtchanov2014}. In this paper we have used a modification of the \citet{Valtchanov2014} method, in order to properly derive the relative pointing offsets for sources that are partially extended within the beam, using the methodology described in \citet{Wu2013}.

Pointing needs to be dealt with differently for observations taken \textit{Before} the BSM change and for those taken \textit{After}. To account for the systematic BSM shift, the {\it Before} observations are corrected back to the 1.7\arcsec\,rest position when applying offsets calculated using the \citet{Valtchanov2014} relative pointing method, rather than the zero-zero position.

In this paper, we obtain offsets for the FTS calibrators sources that were observed a total of eight or more times in high or low resolution modes. The corrections are given in the observation summary tables presented in Appendix~\ref{app:tables}.

\begin{figure}
\centering
\includegraphics[trim = 8mm 0mm 8mm 6mm, clip,width=1.0\hsize]{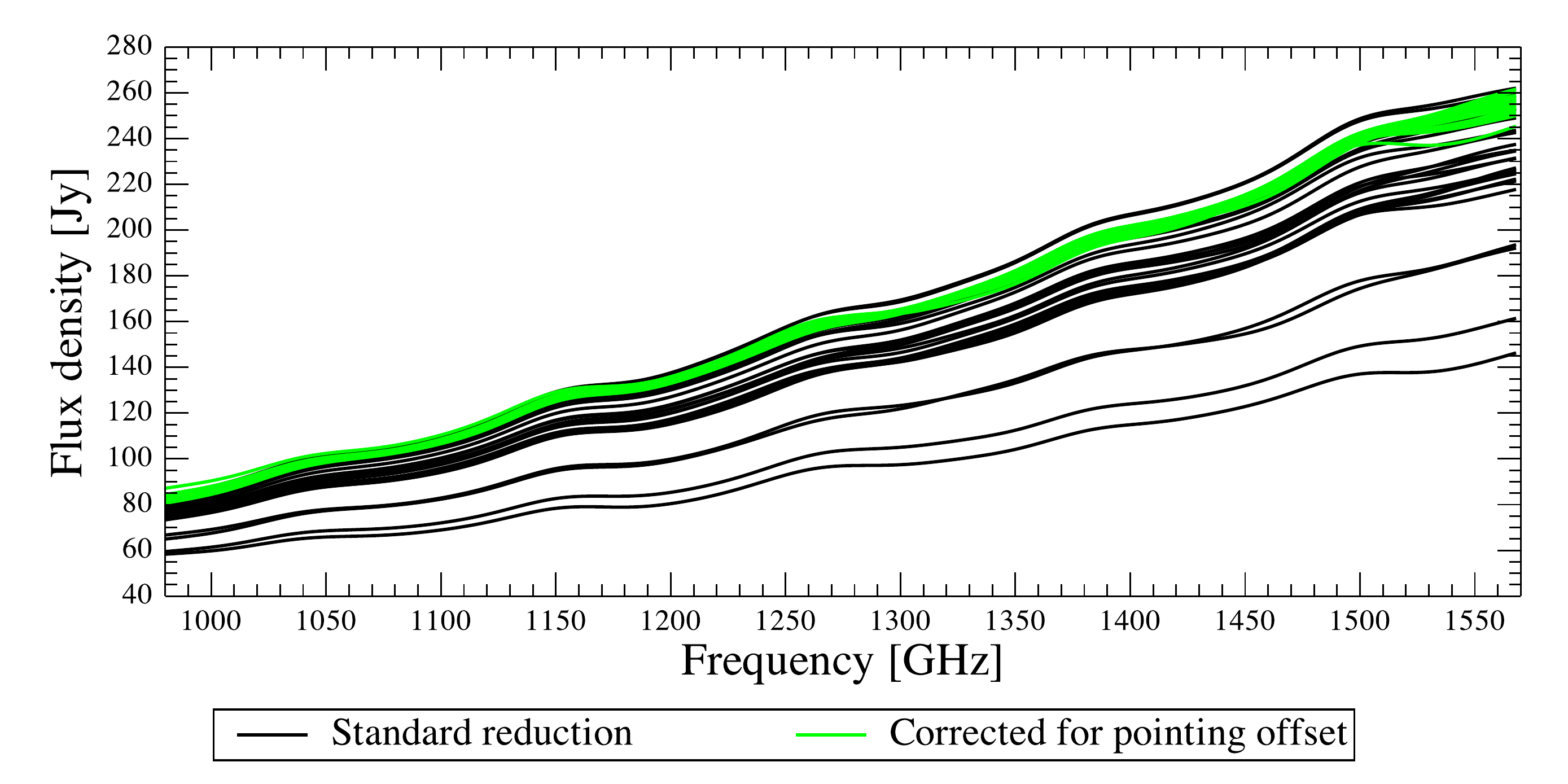}
\caption{Smoothed HR SSWD4 spectra for AFGL2688, before (black) and after (green) pointing correction has been applied. Only SSWD4 is shown as this detector is more sensitive to the effects of pointing offset. After correction for pointing offset, the spread of the spectra is reduced by a factor of 10.}
\label{fig:pointingOffset}
\end{figure}

AFGL2688 is particularly sensitive to pointing offset, as although compact, the core is asymmetric. Most observations of AGL2688 have an estimated offset $>$\,3\arcsec. Three of the AFGL2688 observations were also taken with shifted coordinates to the nominal RA and Dec, leading to large pointing offsets, as high as 7.6\arcsec\,(see Section~\ref{sec:lineSources:AFGL2688}). After correcting the data for pointing offset there is a greatly reduced spread in the spectra, by a factor of 10, which is reflected in the line flux measurements for this source, as discussed in Section~\ref{sec:lineRepeatability}. The full set of smoothed HR observations for this source is shown in Fig.~\ref{fig:pointingOffset}, with and without correcting for pointing offset. The black curves lying above the green in this figure show examples of {\it Before} point-source-calibrated spectra with greater than the true flux density. This is due to a pointing offset back towards the centre of the beam, and therefore the correction lowers the flux density.

\subsection{Fraction of source flux detected by off-axis detectors}\label{sec:outerRings}

Even if observing a compact or point source with the FTS, a small percentage of source flux falls on the detectors lying off the central axis. This percentage depends on the shape of the beam profile for each detector and the extent of the source. Although the point source calibration corrects for this missing fraction, the level of signal in these ``off-axis'' detectors is an important consideration when a background needs to be removed or when assessing whether a source is point-like. The off-axis detectors are arranged in a honeycomb pattern around the centre detectors, with two rings of detectors for SLW and three for SSW (see Fig.~\ref{fig:darkSkyField} and the \citealt{spireHandbook}), and here we determine the fraction of flux in these rings for observations where the source is positioned on the centre detectors. The primary calibrator, Uranus, was not observed on the second ring of SLW or the third ring of SSW, thus these detectors have no reliable point source calibration and are therefore not considered further in this section.

\begin{figure}
\centering
\includegraphics[trim = 8mm 8mm 8mm 0mm, clip,width=0.7\hsize]{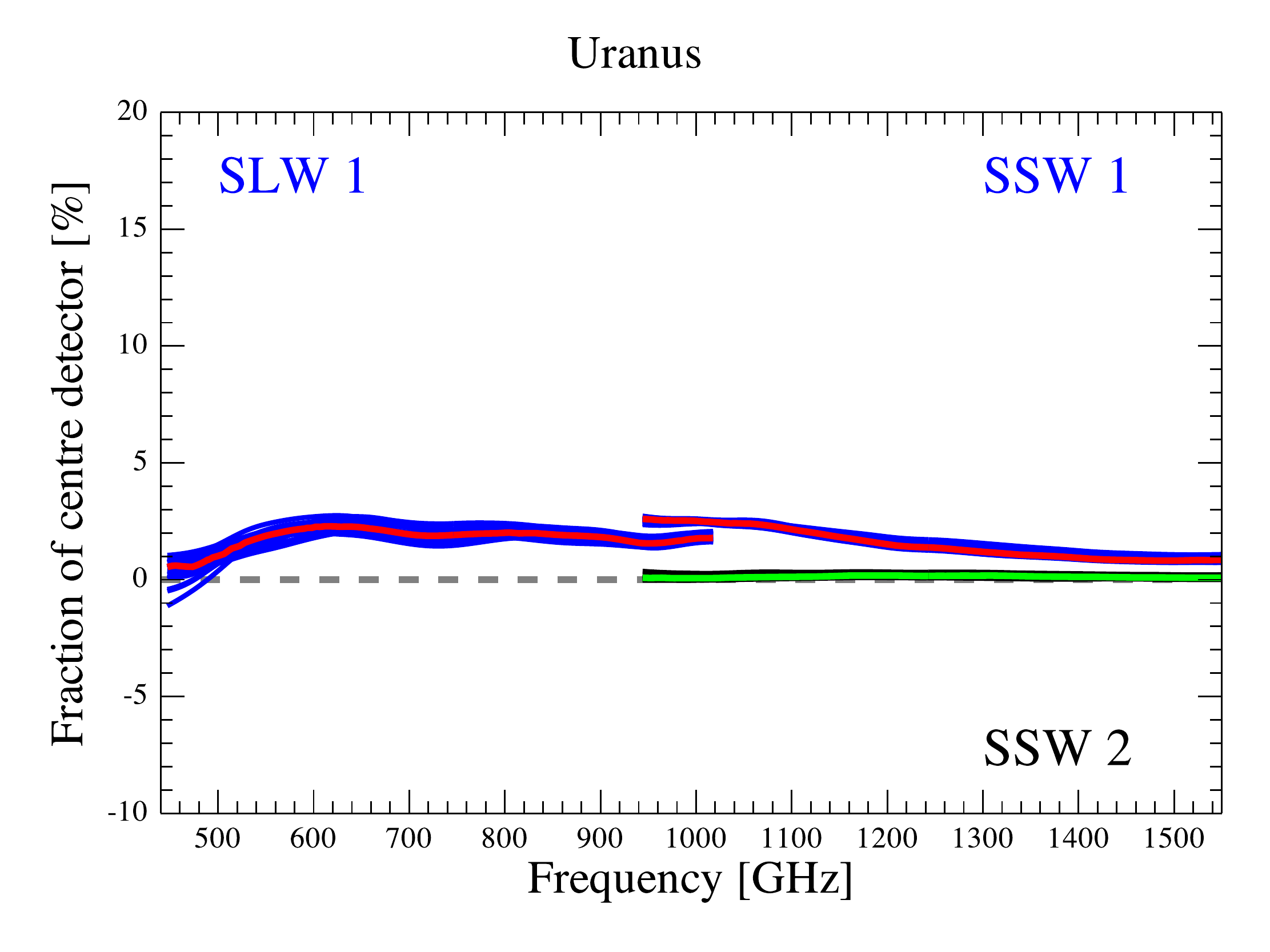}
\includegraphics[trim = 8mm 8mm 8mm 0mm, clip,width=0.7\hsize]{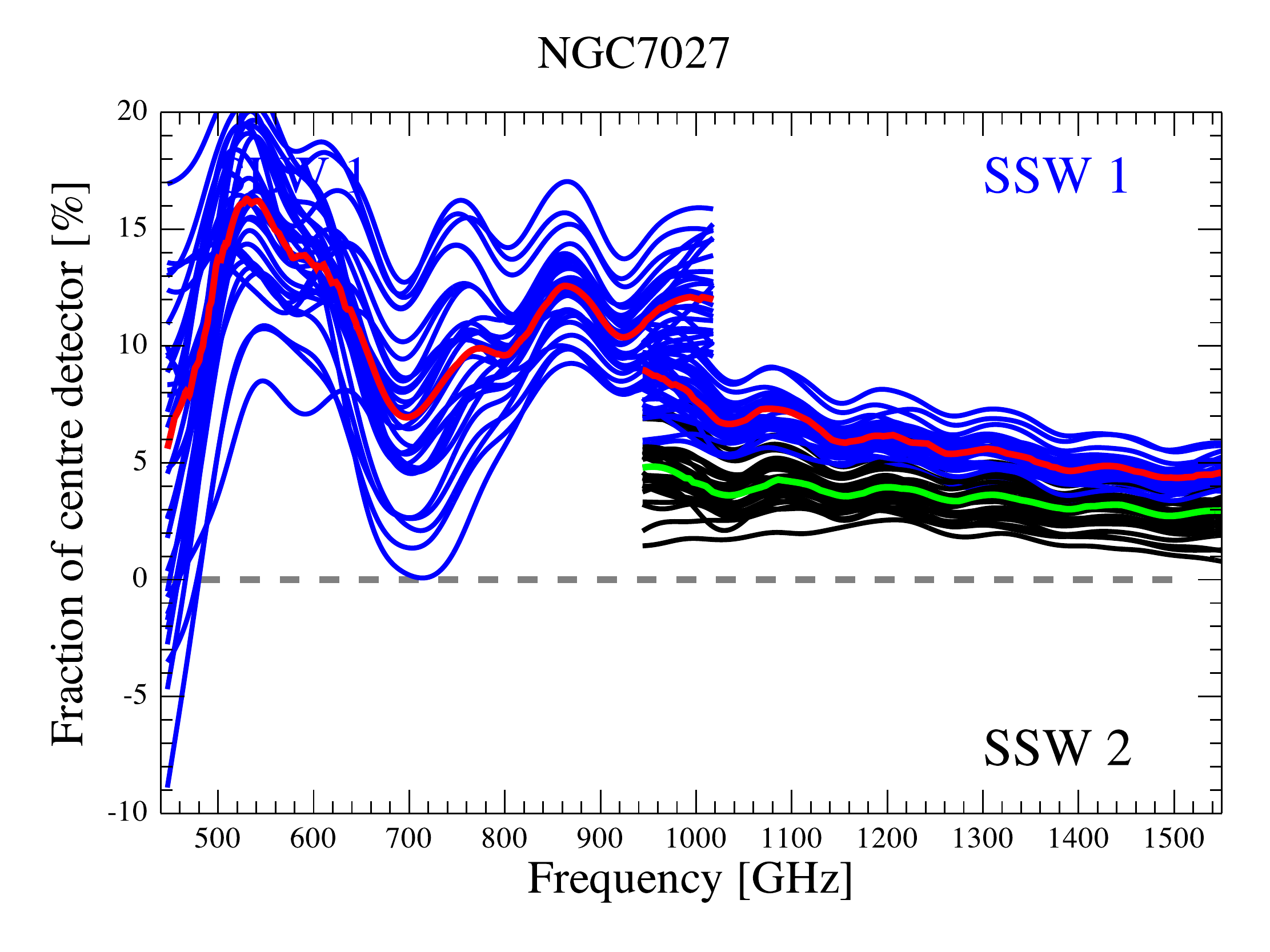}
\caption{Fraction of flux seen by the detectors in the first SLW ring (SLW 1) and those in the first and second SSW detector rings (SSW 1 and SSW 2), with respect to the centre detectors. The averages. are shown by the red and green lines. Results for Uranus (top) illustrate the fraction of flux expected in the first ring of detectors for a point source is roughly 2\% across both frequency bands, whereas there is near to zero flux beyond this. NGC7027 (bottom) is only sightly extended (15\arcsec), but shows a significant excess of flux in the first detector rings, compared to a point-like source.}
\label{fig:fluxInRingsFullFreq}
\end{figure}

\begin{figure}
\centering
\includegraphics[width=0.9\hsize]{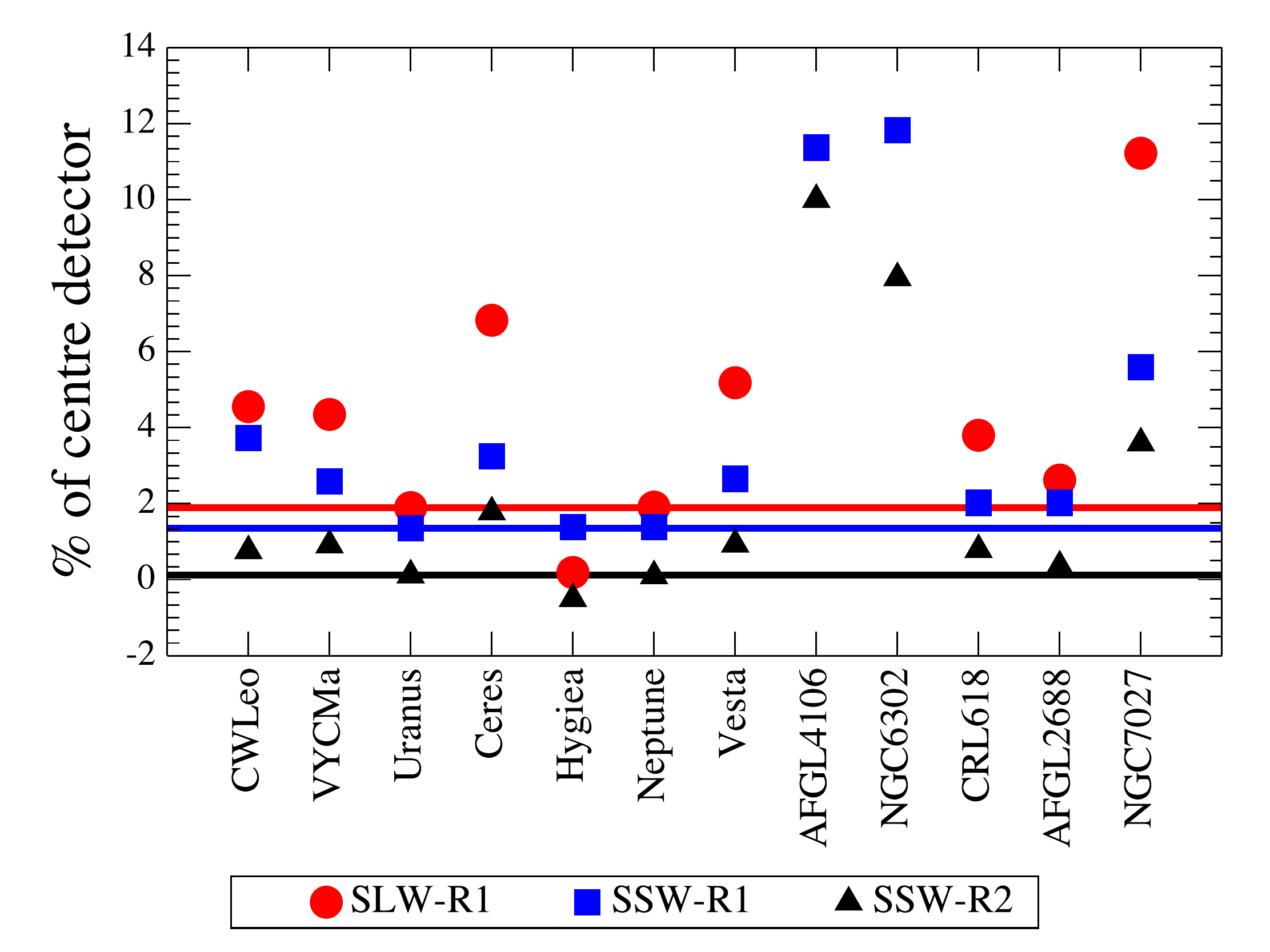}
\caption{Fraction of flux seen by the first SLW (SLW-R1) and the first and second SSW detector rings (SSW-R1 and SSW-R2) with respect to the centre, and averaged in frequency. For easy comparison, the horizontal lines show the results for Uranus and Neptune and correspond in colour to the symbols. AFGL2688, CRL618, CW Leo, VY CMa, and several asteroids can be assessed as near point-like for the FTS. AFGL4106 is considered point-like, but embedded in a significant extended background, and so these fractions are not reliable. NGC6302 and NGC7027 are partially extended, and show significantly greater fractions of flux in the detector rings compared to point-like sources.}
\label{fig:fluxInRingsAve}
\end{figure}

The FTS beam shape was measured using Neptune \citep{Makiwa2013}, but these measurements did not extend sufficiently far into the wings of the beam to accurately predict the signal received by the off-axis detectors. In order to determine the extent of real signal in the off-axis detectors for the repeated calibrators, all HR observations for a given target were smoothed in frequency with a wide Gaussian kernel (21\,GHz). The resulting spectra were averaged per detector ring before dividing by the spectrum from the respective centre detector. Fig.~\ref{fig:fluxInRingsFullFreq} presents the ratios for each detector in the SLW first ring and those in the first and second SSW rings for Uranus and NGC7027, which is a slightly extended source. This figure shows that for sources that are point-like within the FTS beam, e.g. Uranus and Neptune, the expected fraction of total flux is approximately zero in the second SSW ring of detectors (0.1\%), 1.9\% for the first ring of SLW and 1.4\% for the first SSW ring. These ratios were also calculated for the other FTS repeated calibrators, and averaged in frequency (by taking the median), and are shown in Fig.~\ref{fig:fluxInRingsAve}. The ratios for Uranus and Neptune show the expected results for a point-like source, where around 2\% of the source flux lies in the SLW first ring and around 1\% in the SSW first ring, with similar results for AFGL2688 and CRL618, confirming their compact nature at the SPIRE frequencies. Partially extended sources (e.g. NGC6302 and NGC7027) show higher fractions in the detector rings, as do sources embedded in an extended background (e.g. AFLG4106), but in the latter case there is a similar flux fraction found in both SSW rings, and the difference between the values for these two rings is around that expected for a point-like source.

The fraction of source signal in the off-axis detectors is an important consideration if using them to subtract an extended background, because some real signal will be removed (for instance, the fraction of flux given in Fig.~\ref{fig:fluxInRingsAve}). This issue is discussed further in Section~\ref{sec:bgs}.

\subsection{Background subtraction}\label{sec:bgs}

Subtracting a background from the centre detectors of a point source observation may be necessary to correct the continuum shape. The importance of such a subtraction increases for fainter sources, but tends not to have a significant impact for sources with continua of a few Jy or greater. 

\begin{figure}
\centering
\includegraphics[trim = 8mm 0mm 8mm 14mm, clip,width=0.8\hsize]{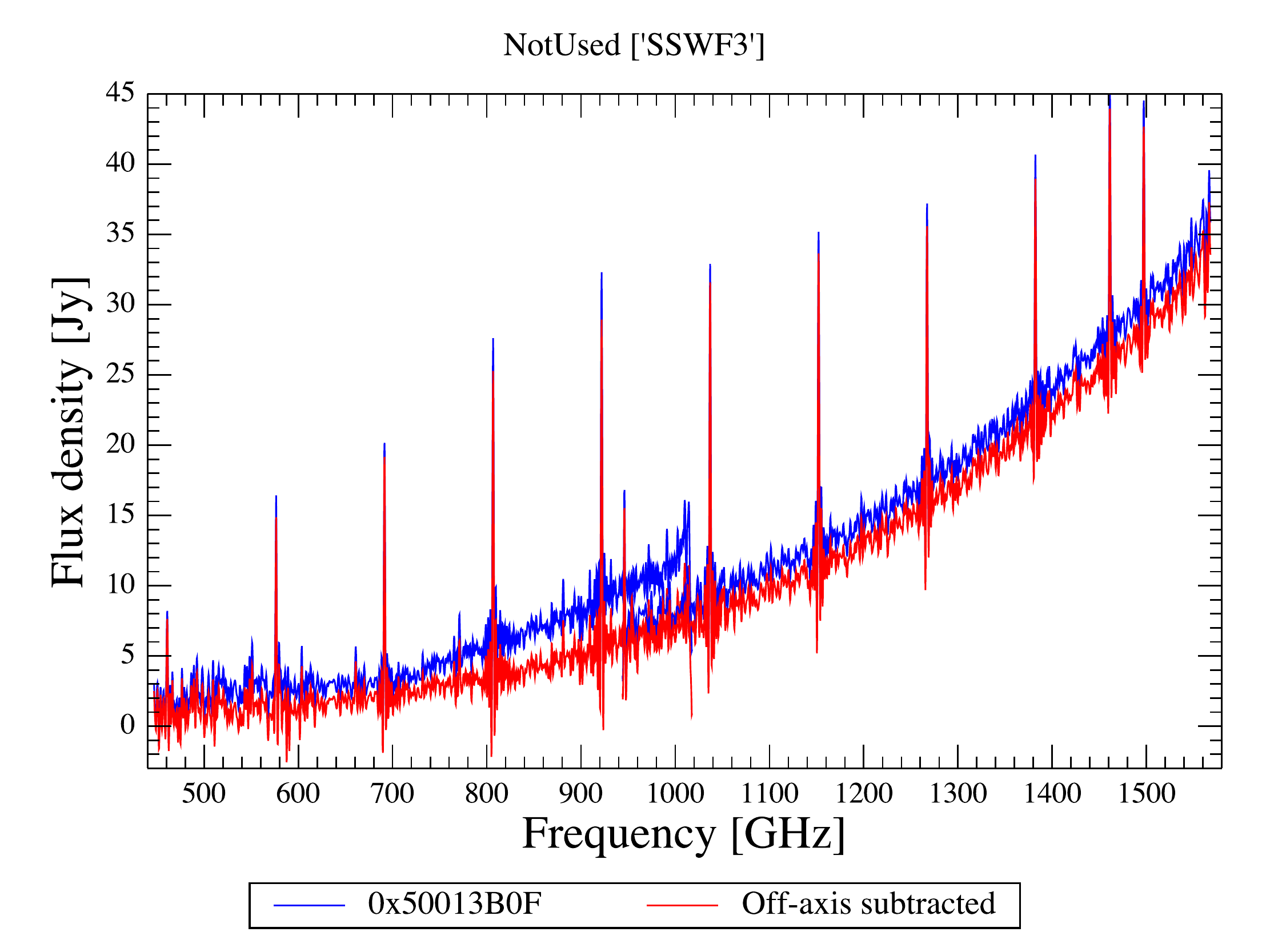}
\caption{An example HR observation of AFGL4106 before and after subtracting the extended background using the average large-scale shape of the off-axis detectors. The corrected spectra (red) show a good join between the bands and improved continuum shape.}
\label{fig:afgl4106bgsEg}
\end{figure}

An extended background can have a pronounced effect on the point-source-calibrated spectra for all strength of sources, as seen for AFGL4106. A background signal tends to be more problematic for the SLW detectors, due to a greater beam size compared to SSW. Figures \ref{fig:footprints} and \ref{fig:lineSourcesEg} include two of the FTS line sources (NGC6302 and AFGL4106) that exhibit a significant step between the SLW and SSW spectra. For NGC6302 this discontinuity is due to a semi-extended morphology, but for AFGL4106, which is compact within the FTS beam, the cause is an extended background. To remove the extended background from the AFGL4106 data, an off-axis subtraction was performed. For each observation, the first ring of SLW and the second rings of SSW detectors were smoothed with a Gaussian kernel, of 21\,GHz width, and examined to check for outliers. Once outliers were discarded, SLW and SSW averages of the remaining smoothed detectors were subtracted from the respective centre detectors, thus removing the large-scale shape due to the background. An example of an AFGL4106 observation before and after background subtraction is shown in Fig.~\ref{fig:afgl4106bgsEg}, where the corrected spectra illustrate the typical improvement seen for all AFGL4106 observations, i.e. an improved spectral shape for SLW and a reduced step between the bands. All asteroid observations were background subtracted using the off-axis detectors, due to high backgrounds. No background subtraction was applied to the other FTS calibrators, to avoid subtracting real signal (as discussed in Section~\ref{sec:outerRings}).

\section{Line source repeatability}\label{sec:repeatabilityLineSources}

This section explores line fitting of FTS data and assesses repeatability using the four main FTS line sources (AFGL2688, AFGL4106, CRL618 and NGC7027). The effects of pointing offset, line asymmetry and the consistency of calibration between the SLW and SSW bands are considered. The strongest lines in the spectra are due to $^{12}$CO rotational transitions from $J$=4--3 to $J$=8--7 for SLW and $J$=9--8 to $J$=13--12 for SSW (see Table~\ref{tab:12co}). These lines were used to assess the line flux and position repeatability throughout the mission.

\subsection{Line fitting}\label{sec:lineFitting}

 The natural instrument line shape is well described by a cardinal sine (ie. a sinc) profile \citep{Spencer10,Naylor10,Naylor2014}, therefore a sinc function, with parameters $p$, is generally used for fitting lines in FTS data and is given by
\begin{eqnarray}
\label{eq:sinc}
f(\nu:p) = p_0 \frac{sin{(u)}}{u},
\end{eqnarray}
where
\begin{eqnarray}
u = \frac{\nu - p_1}{p_2},
\end{eqnarray}
$p_{0}$ is the amplitude in Jy, $p_{1}$ is the line position in GHz, and $p_{2}$ is the sinc width in GHz, which is defined as the distance between the peak and first zero crossing. The uncertainty introduced by assuming the spectral line shape is approximated by a sinc is discussed further in Section~\ref{sec:lineShape}. If lines are partially resolved a more appropriate function is a sinc convolved with a Gaussian
and if standard apodized FTS data are being fitted then a Gaussian profile is a good approximation (see Section~\ref{sec:apodized}).

FTS data with strong spectral features tend to be crowded, and the natural sinc spectral line shape leads to high levels of blending, due to overlapping wings. For this reason, the best approach in fitting an FTS spectrum is to include a sinc function for each line and fit them simultaneously. A simultaneous fit is also advisable for gaining the most accurate match to the continuum, and therefore a polynomial (or modified black body profile) should be included, rather than fitting to a spectrum that has already been baseline subtracted. The fitting method used for all line sources was based on the SPIRE Spectrometer line fitting script available in HIPE \citep{Polehampton14}. The basic approach is to simultaneously fit a low order polynomial (here an order of 3 was used) and a sinc profile for each line. The sinc width is fixed to the width expected for an unresolved line (at best 1.18448\,GHz, but set using the actual resolution of the respective spectrum). 

The fitting algorithm applied uses Levenberg Marquardt minimisation, which is sensitive to the initial parameter guess, and this sensitivity increases with the number of free parameters included. It is therefore essential when fitting a significant number of spectral lines to provide a reasonably close initial guess (particularly for line position), or the fit may converge to an incorrect local minimum. To assist in creating line lists, with optimising the input fitting parameter values in mind, HR observations were co-added for each source, to improve signal-to-noise. The standard pipeline (Fulton et al. in preparation) corrects the frequency scale of FTS data to the LSR frame. The corrected data are interpolated back onto the original frequency grid, which makes co-adding of different observations of the same source trivial. Before co-adding, an initial fit to strong lines and the continuum was performed per observation, and the fitted continuum was then subtracted from each spectrum. The resulting baseline subtracted spectra were then combined by taking a simple mean. The co-added data were used to improve the initial guess positions of faint features during the fitting of individual observations.

\begin{figure}
\centering
\makebox[\textwidth][l]{
\includegraphics[trim = 8mm 3mm 8mm 6mm, clip,width=0.9\hsize]{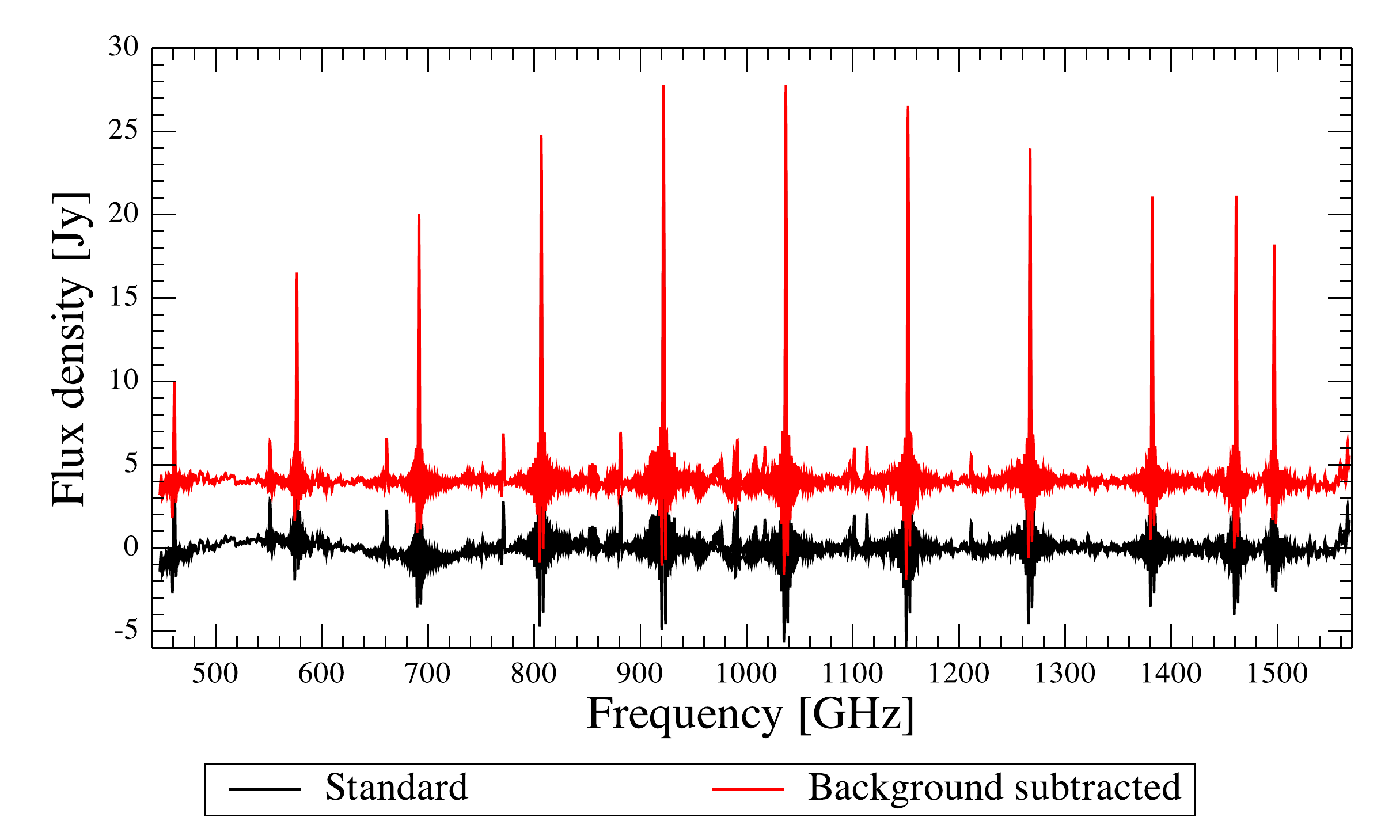}
}
\caption{Co-added AFGL4106 spectra for the centre detectors, with (red) and without (black) correcting the wide-scale spectral shape with a subtraction of the off-axis detectors (see Section~\ref{sec:bgs}). The two results are artificially offset for easy comparison. After removing the extended background, there is an improvement in the polynomial fit to the individual spectra and thus the removal of the continuum, giving a flatter baseline subtracted co-added spectrum.}
\label{fig:afgl4106Coadd}
\end{figure}

Co-adding the observations of AFGL4106 highlights the need for removing the extended background before line fitting. Fig.~\ref{fig:afgl4106Coadd} shows the co-added spectrum for AFGL4106, with and without the background subtracted from the individual spectra, using the off-axis detectors (see Section~\ref{sec:bgs}). The difference between the two results illustrates the worse fit of a polynomial to FTS data when there is high systematic noise present in the continuum. 

\begin{figure}
\centering
\includegraphics[trim = 8mm 2mm 8mm 22mm, clip,width=0.8\hsize]{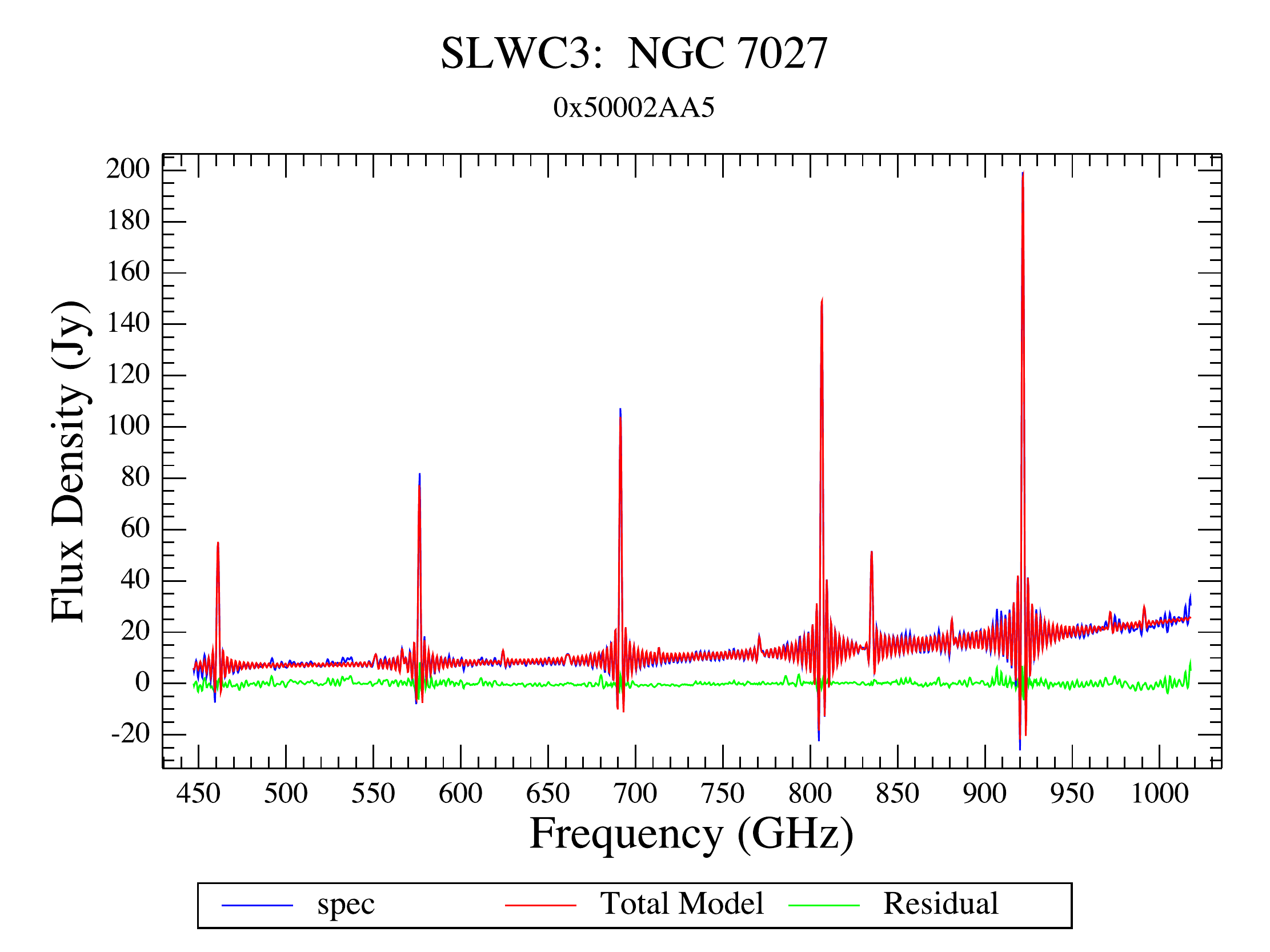}
\caption{Example of the fitted continuum polynomial and line sinc profiles to the SSWC3 spectrum of an observation of NGC7027 during the fitting iteration process detailed in the main text, which is compared to the data and residual on subtraction of the fit.}
\label{fig:egSpecFit}
\end{figure}

To construct the line lists, firstly the $^{12}$CO lines within the FTS frequency bands were included for all sources, then other obvious lines added on a per source basis, e.g., $^{13}$CO and HNC for AFGL2688. The co-added data were then fitted using these lists of high signal-to-noise lines, and the combined fit subtracted. Fainter features visually detectable in the residuals were then added to the lists. While the line profile is well represented by a sinc function the slight asymmetry present (see Section~\ref{sec:lineShape}) does impact the residual after subtracting strong lines, so it is important not to identify this residual as faint lines. Using the co-added data at this stage significantly cuts down on fitting time and fitting failures compared to using individual observations of lower signal-to-noise. Fitting was repeated with the appended line lists, with limits of $\pm$2.0\,GHz applied to prevent large inaccuracies in the fitted line positions. Faint line positions were then adjusted, or unstable lines removed, and the fit repeated until all lines gave a stable fitted position, i.e. close to the input without limits. Finally, the line fitting was run for each observation. Fig.~\ref{fig:egSpecFit} shows an example of the fitting process for NGC7027. For each of the line sources, tables summarising the main species fitted, in addition to the $^{12}$CO lines, are provided in Appendix~\ref{app:lines}. Note that unidentified features are not included in these tables.

\begin{figure}
\centering
\includegraphics[trim = 8mm 2mm 8mm 2mm, clip,width=0.9\hsize]{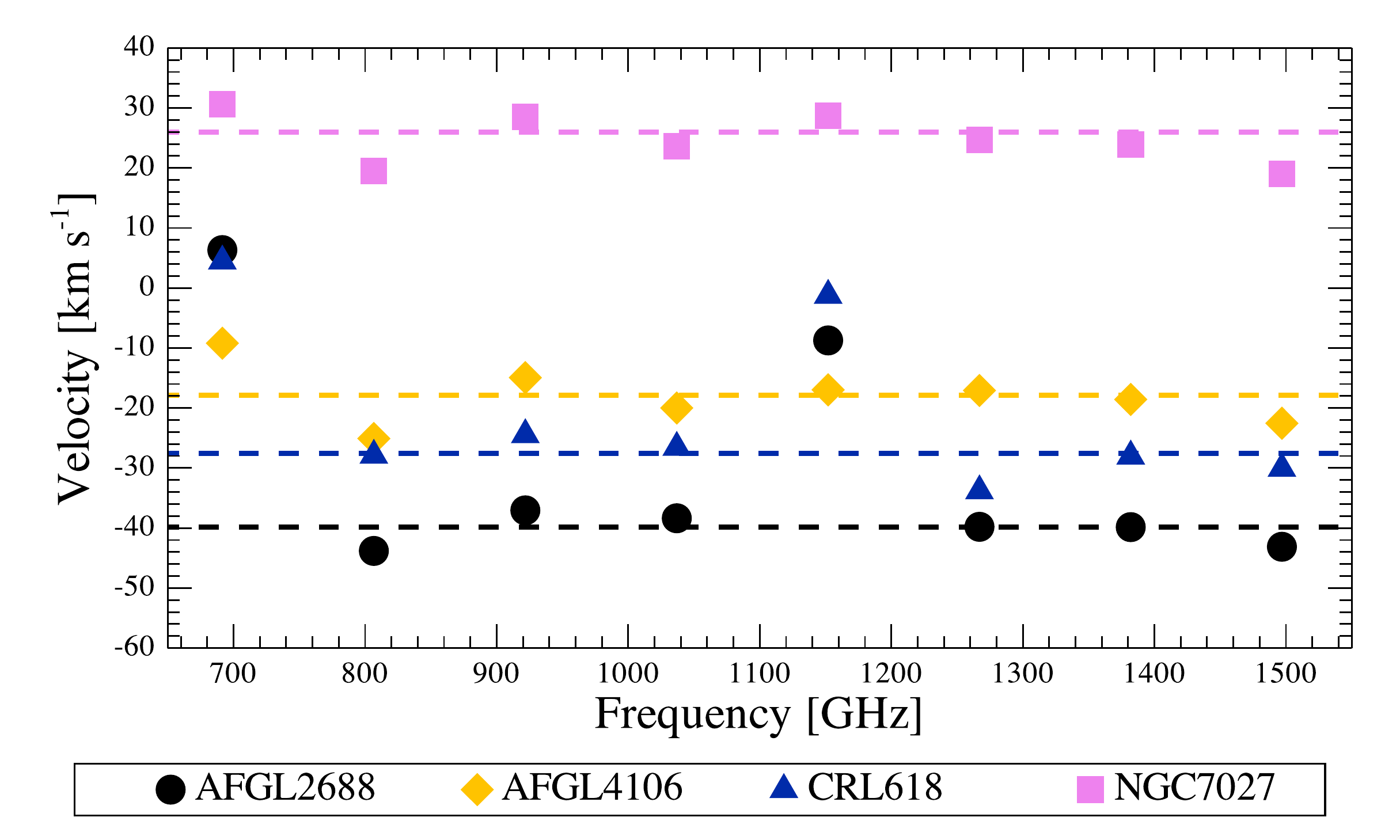}
\caption{Average $^{12}$CO line velocities for the four main FTS line sources for the transitions $J$=6--5 to $J$=13--12. For AFGL2688 and CRL618 there are high levels of blending for the $J$=6--5 and $J$=10--9 lines. Dashed lines show the mean velocity for each source, not including the $^{12}$CO lines $<$\,600\,GHz or those significantly blended.}
\label{fig:tweakingVelocity}
\end{figure}

The intrinsic radial velocity of each source was estimated with respect to the LSR. For the $^{12}$CO lines present above 600\,GHz, the velocity was estimated from the average between the inputted source velocity and that calculated from the fitted results, and adjusted until refitting gave a difference of approximately zero. The average $^{12}$CO velocities are shown in Fig.~\ref{fig:tweakingVelocity} and the final source velocities used are given in Table~\ref{tab:lineVel}, in comparison to previous published values.

\begin{table}
\caption{Intrinsic source velocities with respect to the LSR (v$_{\rm{FTS}}$), estimated from FTS data and used to adjust the initial line positions when fitting the four main FTS line sources. The ``v$_{\rm lit}$'' column gives the published velocity for comparison.}
\medskip
\begin{center}
\begin{tabular}{lccc}
\hline\hline
Name & v$_{\rm{FTS}}$\,[km\,s$^{-1}$] & v$_{\rm lit}$\,[km\,s$^{-1}$] & Ref \\ \hline
AFGL2688 & -39.8$\pm$5.6 & -35.4 & \citet{Herpin02} \\
AFGL4106 & -17.9$\pm$7.4 & -15.8 & \citet{Josselin98} \\
CRL618 & -27.6$\pm$5.8 & -25.0 & \citet{Teyssier06} \\
NGC7027 & +26.0$\pm$5.3 & +25.0 & \citet{Teyssier06} \\
\hline
\end{tabular}
\end{center}
\label{tab:lineVel}
\end{table}

Figures~\ref{fig:linesFittedAFGL2688}--\ref{fig:linesFittedNGC7027}, which can be found in Appendix~\ref{app:lineFittingFigs}, show the lines fitted and example residuals obtained on subtraction of the combined fit for each of the four main line sources. All residuals are over an order of magnitude lower than the data, except towards the high frequency end of SSWD4 for CRL618, which is due to these lines becoming partially resolved (see Section~\ref{sec:lineSources:CRL618}). 
There is a contribution to the residuals for all sources from the asymmetry of the natural line shape, compared to a pure sinc function. This asymmetry is considered in more detail in Section~\ref{sec:lineShape}. One other notable feature is the fringing at the high frequency end of the SLW spectrum for AFGL4106. From HIPE version 12.1 the FTS frequency bands were widened to make all useable data available. However, these additional data tend to be significantly more noisy. AFGL4106 is the faintest of the four sources and although the wide-scale features due to the extended background is corrected for (see Section~\ref{sec:bgs}), the fringing introduced is not. The $^{13}$CO(9--8) line lying in this region requires special attention to improve the consistency of line measurements, which is addressed in Section~\ref{sec:afgl4106NoisyEnds}.

\subsection{Line measurements}\label{sec:lineRepeatability}

The two main quantities that were obtained from fitting spectral lines are the integrated line flux ($I$) and the line centre position. For point-source-calibrated FTS data, $I$ can be calculated in W\,m$^{-2}$ from the fitted sinc function parameters by integrating Equation \ref{eq:sinc} over frequency as
\begin{eqnarray}
I = \displaystyle\int f(\nu:p)\,d\nu = 10^{-26}p_{0}10^9\pi\,p_{2}.
\end{eqnarray}
The signal-to-noise ratio (SNR) was also estimated per line, using the fitted amplitude and the local standard deviation of the residual, obtained after subtracting the combined fit.  

The sinc fit also provides the line centre position, which can be expressed as a radial velocity in the LSR frame, relative to the measured laboratory rest frame line position in km\,s$^{-1}$. Line identifications were checked against \citet{Wesson2010}, with the rest frame line positions taken from the Cologne Database for Molecular Spectroscopy \citep{muller2001}. The main species fitted to each source are given in Appendix~\ref{app:lines}.

The line fitting was carried out for the data before and after correcting for pointing offset, which is described in Section~\ref{sec:pointCorr}.

\subsubsection{Line flux}\label{sec:lineFlux}

\begin{figure}
\centering
\includegraphics[trim = 5mm 0mm 5mm 5mm, clip,width=\hsize]{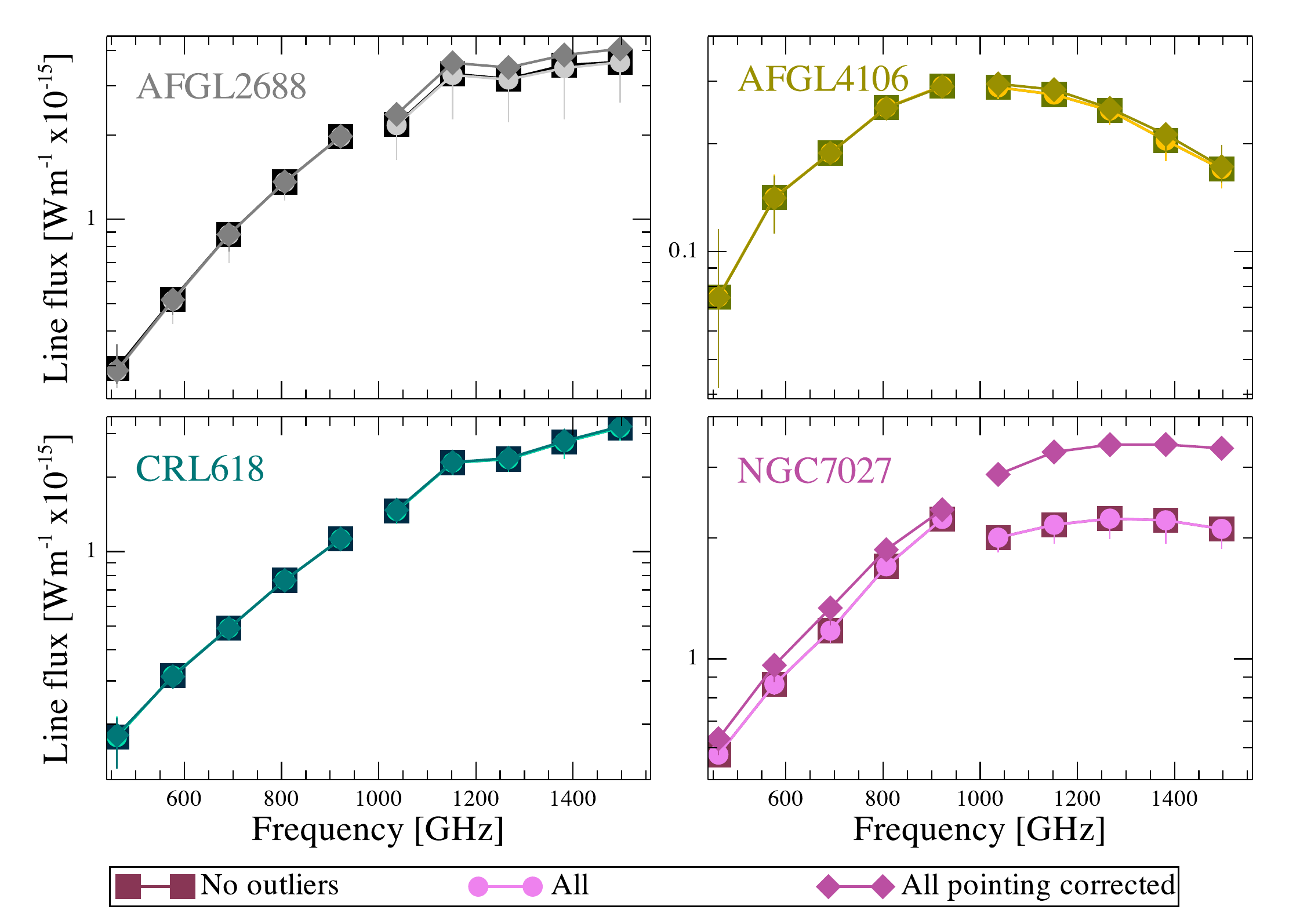}
\caption{Average $^{12}$CO line flux for each of the four main FTS line sources. For AFGL2688, the pointing offset corrected SSW results (diamonds) are more consistent with SLW. The slight angular extent of NGC7027, of 15\arcsec, leads to a more pronounced step between the line flux for SLW and SSW, which sees a marked reduction after correction for pointing offset and the extent. The error bars represent the minimum and maximum fitted line flux. The spread in the measured line flux is presented in Fig.~\ref{fig:lineFluxErr}. The $^{12}$CO transitions included are given in Table~\ref{tab:12co}.}
\label{fig:lineFlux}
\end{figure}

The average integrated $^{12}$CO line flux, taken over all observations per source, are plotted in Fig.~\ref{fig:lineFlux}. No evidence for intrinsic variability of the measured $^{12}$CO flux is seen for any of the sources. NGC7027 is corrected for its slight extent of 15\arcsec\,at the same time as correcting for pointing offset, which leads to a large increase in the the SSW line flux. For the other sources, the pointing corrected data show a slight increase in the average flux for SSW, improving their consistency with the respective SLW values. The increase in SSW line flux is significant for AFGL2688, which exhibits the greatest sensitivity to pointing offsets. The SNR of the fitted $^{12}$CO lines is generally above 100 in SSW and above 50 for SLW ($>$600\,GHz), apart from AFGL4106, which has SNRs $\sim$40 (as shown by Fig.~\ref{fig:snrAve}). The measured spread in line flux over the repeated observations is small, indicated by the error bars in Fig.~\ref{fig:lineFlux} that represent the minimum and maximum values in the distribution. No trend in line flux is seen with OD for any of the lines, showing that the spectral line calibration of the instrument was highly consistent throughout the mission.

\begin{figure}
\centering
\includegraphics[trim = 5mm 0mm 5mm 5mm, clip,width=\hsize]{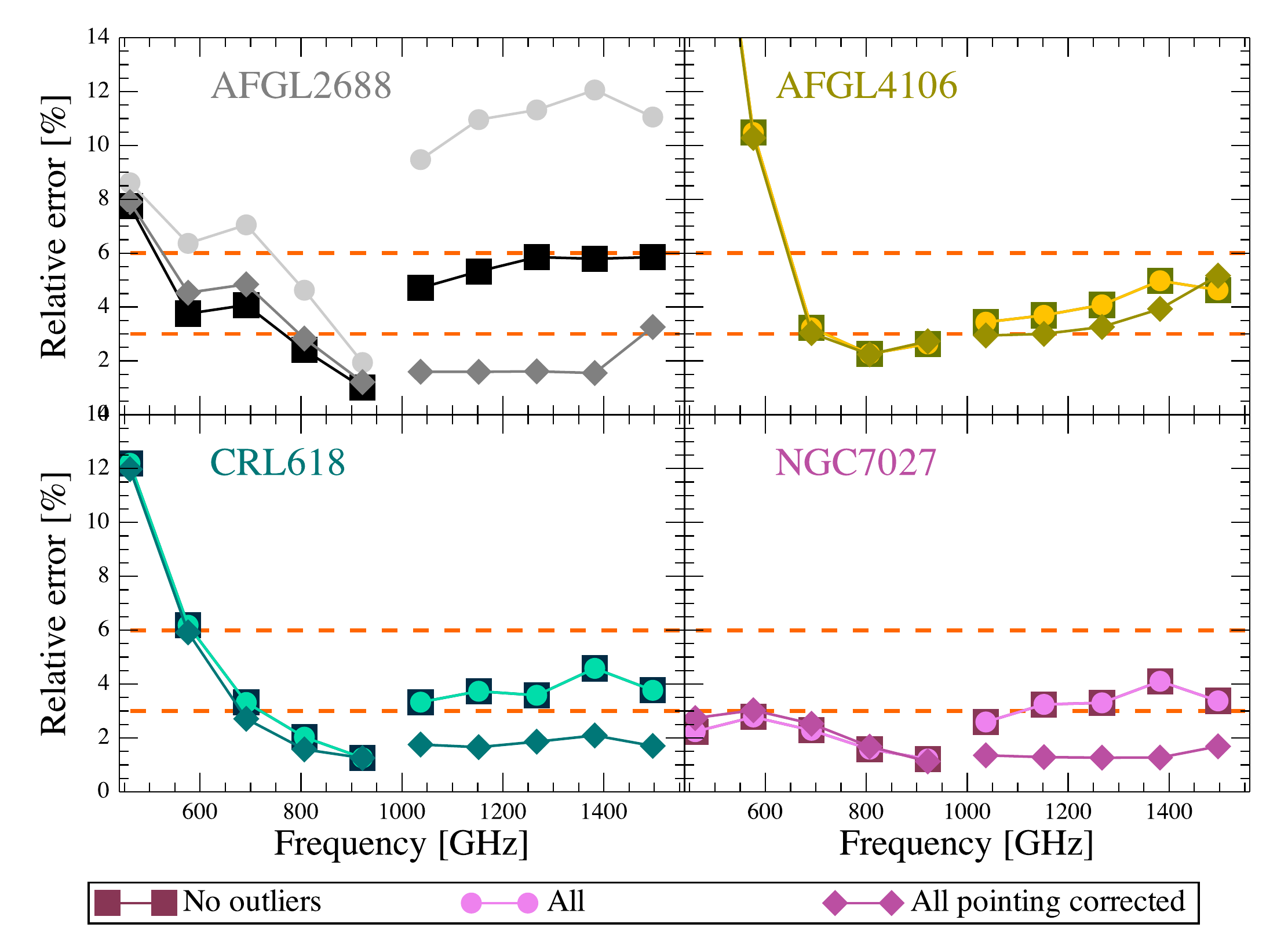}
\caption{The spread in $^{12}$CO line flux for the four main FTS line sources, plotted as a fraction of the average. The ``No outliers'' points are included to compare with the results presented in \citet{Swinyard2014}. Comparing all observations regardless of pointing offset (``All'') with the same observations after pointing correction (``All pointing corrected''), shows a dramatic improvement in the SSW results for AFGL2688 and a good improvement for the other three sources. If the two $^{12}$CO lines below 600\,GHz are discounted, due to significantly lower SNR (as shown in Fig.~\ref{fig:snrAve}), the results for the non-pointing corrected data, with outliers removed is within 6\% and for pointing corrected data, within 3\%. These percentages are indicated by the horizontal dashed lines. The $^{12}$CO transitions included are given in Table~\ref{tab:12co}.}
\label{fig:lineFluxErr}
\end{figure}

The standard deviation in line flux, over the repeated observations, is plotted as a fraction of the average values in Fig.~\ref{fig:lineFluxErr}. This figure shows that for pointing corrected data, the spectral line repeatability was better than 1--2\% for the sources observed with high SNR. For the uncorrected data, the repeatability decreases to between 3--12\% for SSW. As mentioned in Section~\ref{sec:pointCorr}, three of the observations of AFGL2688 were observed with different coordinates. If these known outliers are omitted, the uncorrected repeatability is less than 6\%, which is the case for all four sources. This value is within the expected 1\,$\sigma$ drop in flux (of $\sim$10\% \citep{Swinyard2014}), due to the {\it Herschel} telescope APE.

The source with the highest fitted $^{12}$CO SNRs, and the greatest number of repeated observations, is NGC7027. The spread in line flux in the pointing corrected data are $\sim$1\% for SSW (and 1--2\% for SLW).
However, the SNR of these lines is $\sim$250, indicating that there are systematic effects which limit the repeatability on top of the random spectral noise. These systematic effects are probably related to uncertainties in the determination of the pointing correction rather than due to a limit in the stability of the instrument itself. At the highest frequencies, a small change in pointing can affect the scaling of the spectrum on the level of 1\%.
The spread in line flux significantly increases for the two lowest frequency lines in the SLW band, due to these being the weakest lines, i.e. those with the lowest SNRs, and therefore the repeatability is dominated by random spectral noise and residual from the instrument emission.

In summary, for normal science observations, where it is not possible to know the pointing to a higher accuracy than the {\it Herschel} telescope APE, the repeatability should be taken from the uncorrected data above - i.e. better than 6\%. This value was already presented in \citet{Swinyard2014}, and has not changed from HIPE versions 11 to 13. For observations where pointing offsets are known and corrected, the instrument is capable of much better repeatability, on the order of 1--2\%.

\subsubsection{Line position}

After omitting noisy and blended lines, the distribution of the fitted line centres was reported in \citet{Swinyard2014} as a systematic offset compared to literature values of $<$\,5\,km\,s$^{-1}$, with a spread of $<$\,7\,km\,s$^{-1}$. Here we update those results and present them in more detail.

The mean line centres for each source are determined from $^{12}$CO lines above 600~\,GHz and given in Table~\ref{tab:lineVel}. NGC7027 has the greatest number of observations and highest SNR, and shows a good agreement to within 1\,km\,s$^{-1}$ of the literature value. The other sources show agreement with the literature values of between 2--4\,km\,s$^{-1}$. The spectral resolution of the instrument is 230--800\,km\,s$^{-1}$ and so these values correspond to less than 0.5--2\% of a resolution element.

\begin{figure}
\centering
\includegraphics[trim = 5mm 0mm 5mm 5mm, clip,width=\hsize]{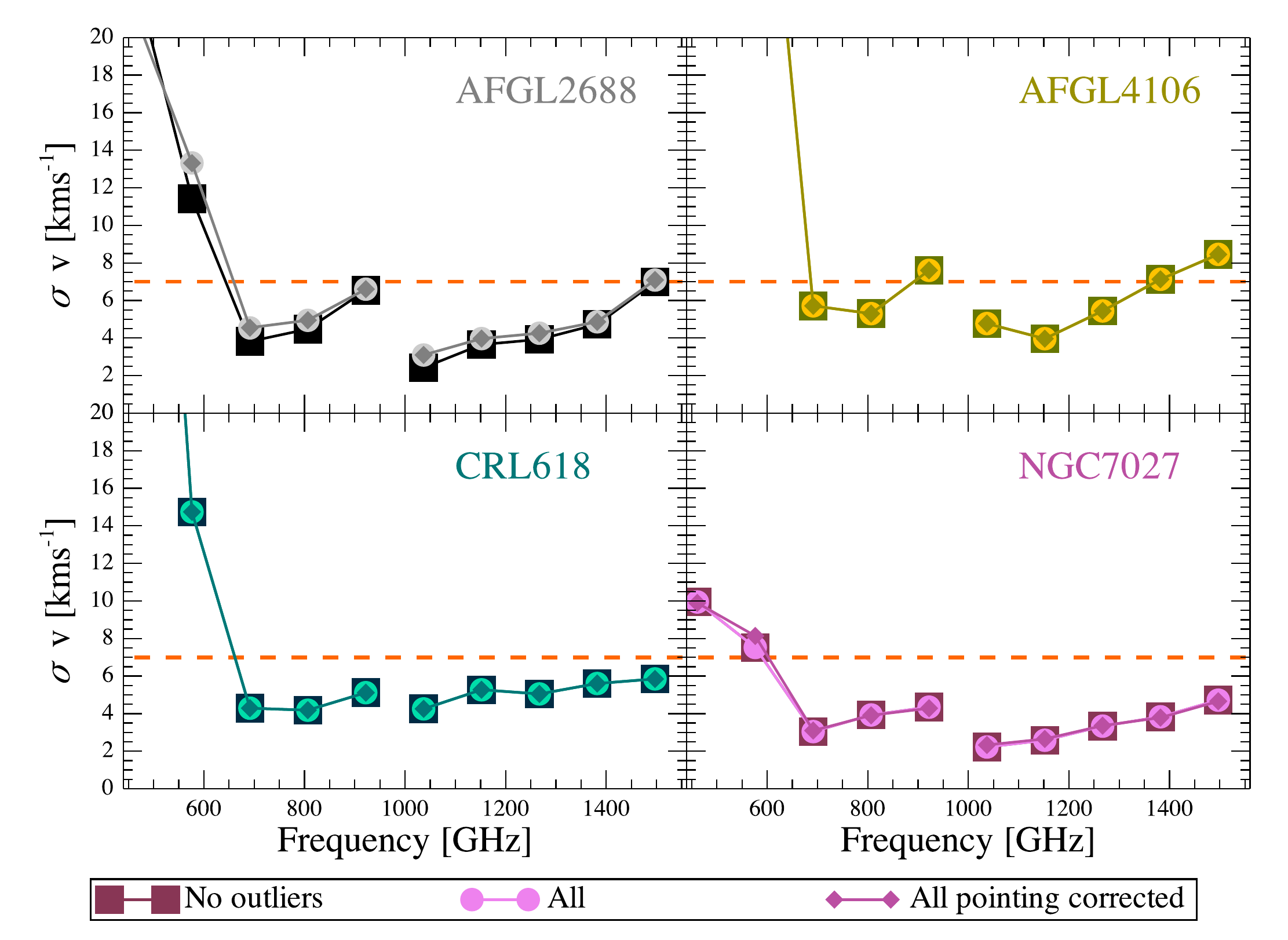}
\caption{The spread in $^{12}$CO line velocity for the four main FTS line sources. If the two $^{12}$CO lines below 600\,GHz are discounted, due low SNR (as shown in Fig.~\ref{fig:snrAve}), the spread in line velocity is $<\,7$km\,s$^{-1}$ regardless of omitting outliers (``No outliers''), using all data (``All''), or using all data after pointing correction (``All pointing corrected''). The $^{12}$CO transitions included are given in Table~\ref{tab:12co}.}
\label{fig:lineVelErr}
\end{figure}

\begin{figure}
\centering
\includegraphics[trim = 5mm 0mm 5mm 5mm, clip,width=\hsize]{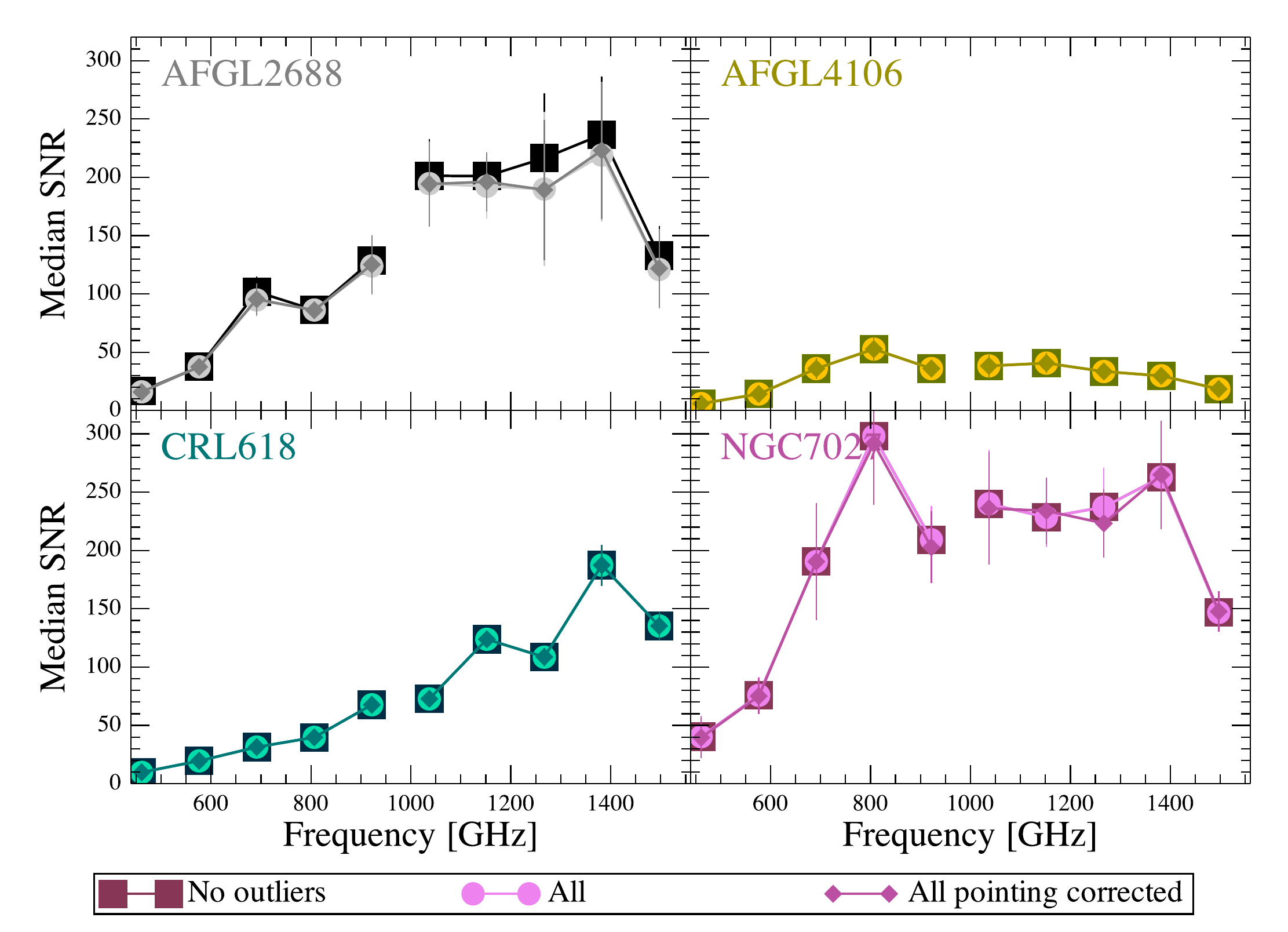}
\caption{The average $^{12}$CO SNR ratio for each of the four main FTS line sources. The SNR ratio roughly drops with decreasing frequency for SLW, and is relatively weak below 600\,GHz. AFGL4106 shows relatively lower SNR across both frequency bands. If known outliers are included (i.e. ``All'' and ``All pointing corrected'') the SNR is lower in SSW for AFGL2688. This is because fringing is increased for observations suffering a significant pointing offset and the correction is applied at the point-source-calibrated stage. As the SSW flux density is scaled up by the correction, so too is the noise. The $^{12}$CO transitions included are given in Table~\ref{tab:12co}.}
\label{fig:snrAve}
\end{figure}

The standard deviation in line velocity measurements over the repeated observations are plotted in Fig.~\ref{fig:lineVelErr} for each source, and show that the repeatability of $^{12}$CO line centres is better than 7\,km\,s$^{-1}$, whether the data are corrected for pointing offset or not. For individual lines, the measured line centre should depend on the SNR of the line being fitted, with the SNR of the measured lines shown in Fig.~\ref{fig:snrAve}. With SNR values of this order, an estimate of the accuracy expected on fitted the line centres can be obtained from the line width divided by the SNR, assuming that the instrumental line shape is well known \citep[e.g.][]{Davis2001}. The measured velocity spread is slightly higher than the expected accuracy by a few km\,s$^{-1}$. This discrepancy may be related to the asymmetry in the line shape (Section~\ref{sec:lineShape}), which increases the uncertainty when the centres are determined using a sinc fit. The most accurate values are for SSW, whereas the velocity spread rapidly increases for SLW below 600\,GHz, where the SNR falls off.

Overall, the results show that the frequency scale is extremely accurate, with systematic difference from literature line centres of $<$5\,km\,s$^{-1}$ (0.5--2\% of a resolution element), and a spread in fitted line centres between observations roughly as expected from the SNR of the measured lines. For lines with a SNR of $\sim$40 and above, a good approximation to quote for the repeatability of line centres is that it is better than 7\,km\,s$^{-1}$ (see table \ref{tab:lineStats}).

The repeatability of measured line centres for other detectors across the array has been investigated by \citet{Benielli2014}, who showed that the spread in measured line centres increases for detectors away from the optical axis, as expected from theory \citep{Davis2001}.

\begin{table}
\caption{Summary of the average line fitting results. All values are the mean of the points presented in Figures~\ref{fig:lineFluxErr}, \ref{fig:lineVelErr} and \ref{fig:snrAve}, but with the first two points of SLW omitted. Results are given for both centre detectors (SLWC3 and SSWD4). The relative error of the line flux ($I_{\rm relErr}$) is given in percent.}
\medskip
\begin{center}
\begin{tabular}{lcccc}
\hline\hline
 & \multicolumn{2}{c}{No outliers} & \multicolumn{2}{c}{All pointing corr.}\\ \hline
AFGL2688 & SLWC3 & SSWD4 & SLWC3 & SSWD4\\ \hline
$I_{\rm relErr}$ & 2.5 & 5.5 & 3.0 & 1.9 \\
$\sigma$\,v [km\,s$^{-1}$] & 4.9 & 4.3 & 5.4 & 4.7
 \\
SNR & 106 & 198 & 102 & 185 \\ \hline\hline
AFGL4106 & SLWC3 & SSWD4 & SLWC3 & SSWD4\\ \hline
$I_{\rm relErr}$  & 2.7 & 4.2 & 2.7 & 3.7 \\
$\sigma$\,v [km\,s$^{-1}$] & 6.2 & 6.0 & 6.2 & 6.0 \\
SNR & 41.5 & 32.1 & 41.5 & 32.1 \\ \hline\hline
CRL618 & SLWC3 & SSWD4 & SLWC3 & SSWD4\\ \hline
$I_{\rm relErr}$ & 2.2 & 3.8 & 1.9 & 1.8 \\
$\sigma$\,v [km\,s$^{-1}$] & 4.5 & 5.2 & 4.5 & 5.2 \\
SNR & 46.3 & 126 & 46.3 & 126 \\ \hline\hline
NG7027 & SLWC3 & SSWD4 & SLWC3 & SSWD4\\ \hline
$I_{\rm relErr}$ & 1.7 & 3.3 & 1.8 & 1.4 \\
$\sigma$\,v [km\,s$^{-1}$] & 3.8 & 3.3 & 3.8 & 3.5 \\
SNR & 233 & 228 & 231 & 221 \\
\hline
\end{tabular}
\end{center}
\label{tab:lineStats}
\end{table}


\subsection{SLW/SSW overlap}\label{sec:overlap}

There is an overlap in frequency between the SLW and SSW detectors, which increased with the introduction of wider bands from HIPE version 12.1, and ranges from 994\,GHz to 1018\,GHz. The fitting results for the four main FTS line sources were used to investigate the consistency of FTS calibration in this overlap region.

\begin{figure}
\centering
\includegraphics[trim = 5mm 24mm 5mm 5mm, clip,width=0.9\hsize]{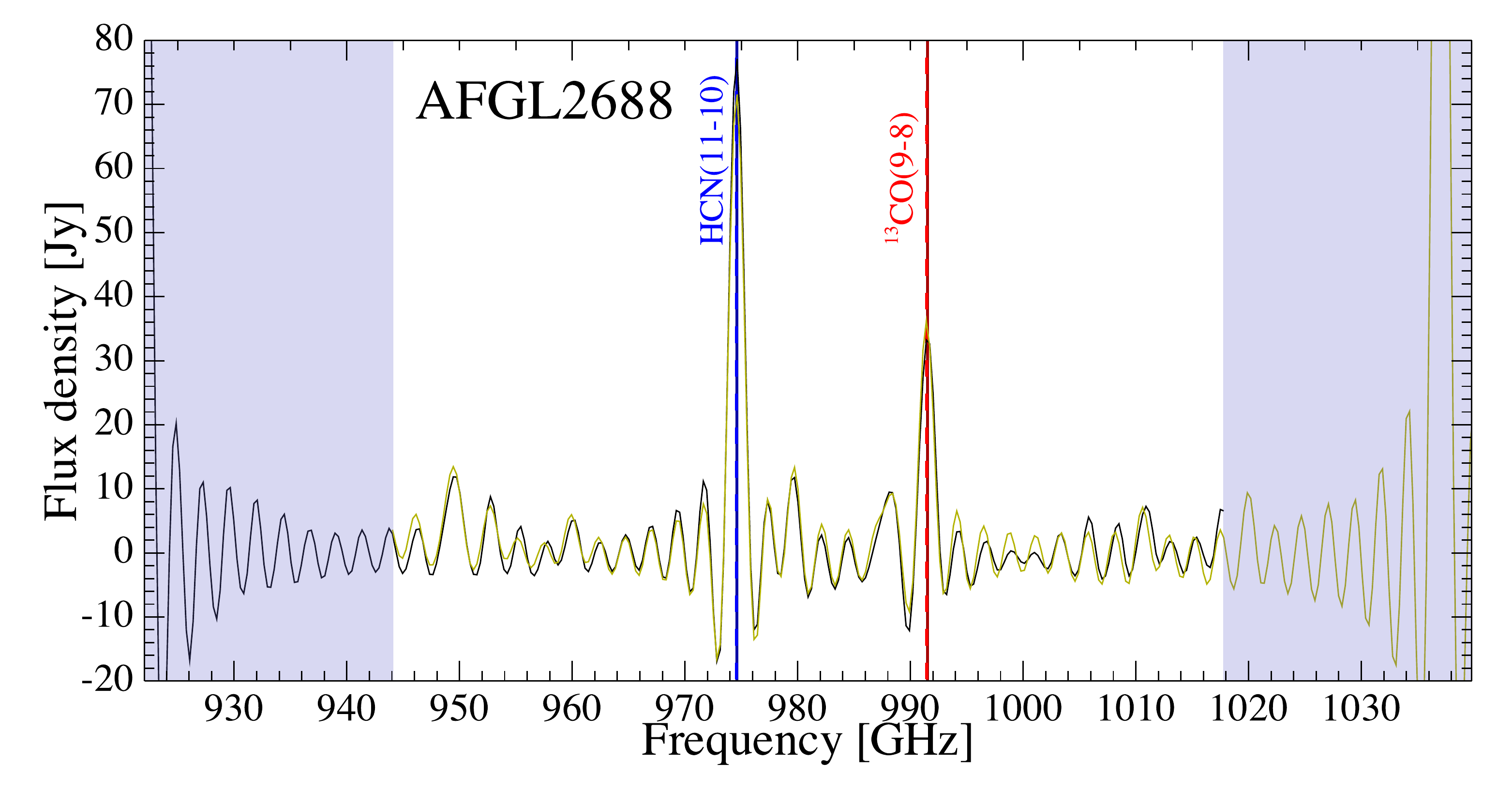}
\includegraphics[trim = 9.5mm 27mm 5mm 5mm, clip,width=0.9\hsize]{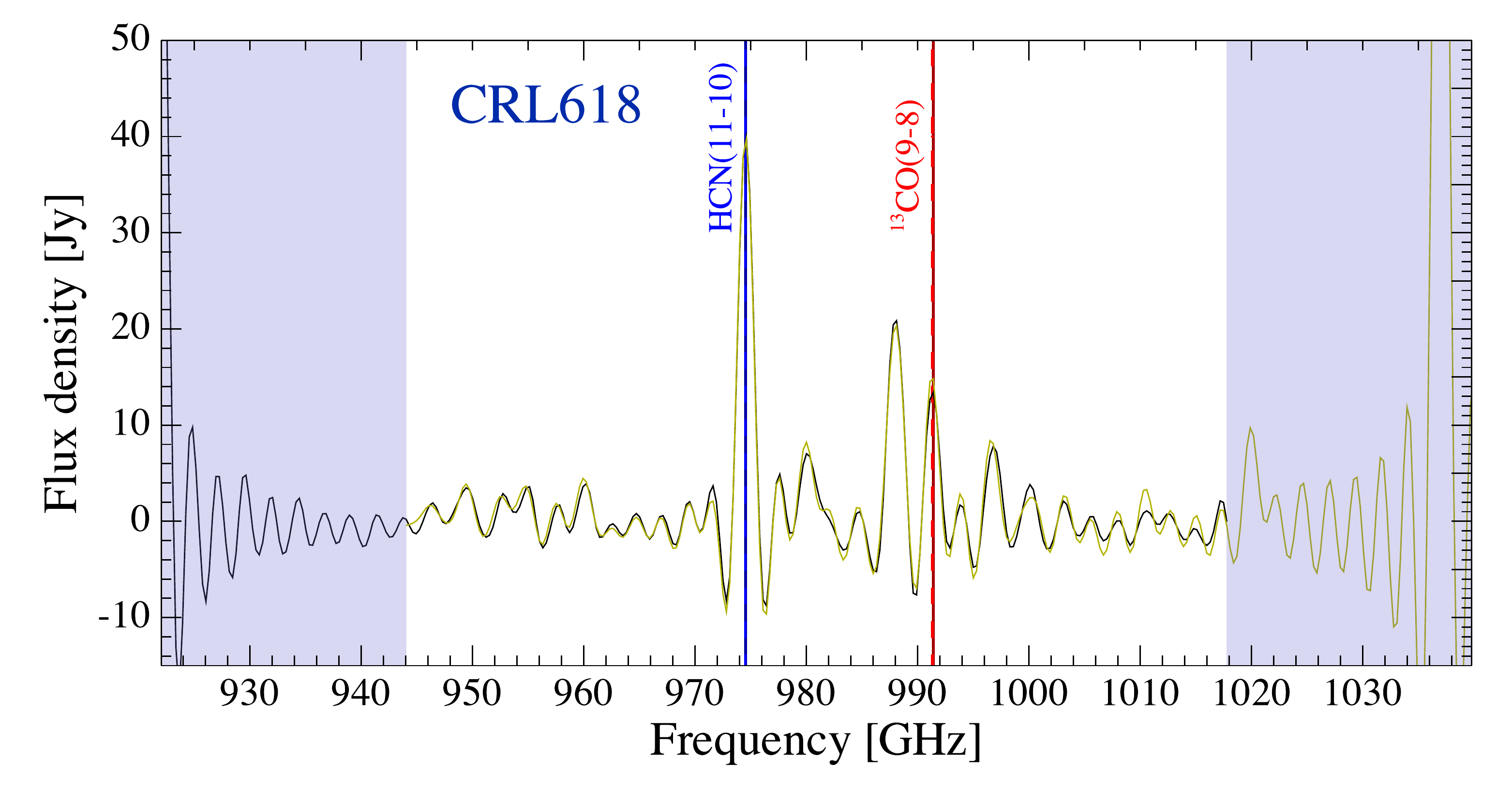}
\includegraphics[trim = 4mm 24mm 5mm 5mm, clip,width=0.9\hsize]{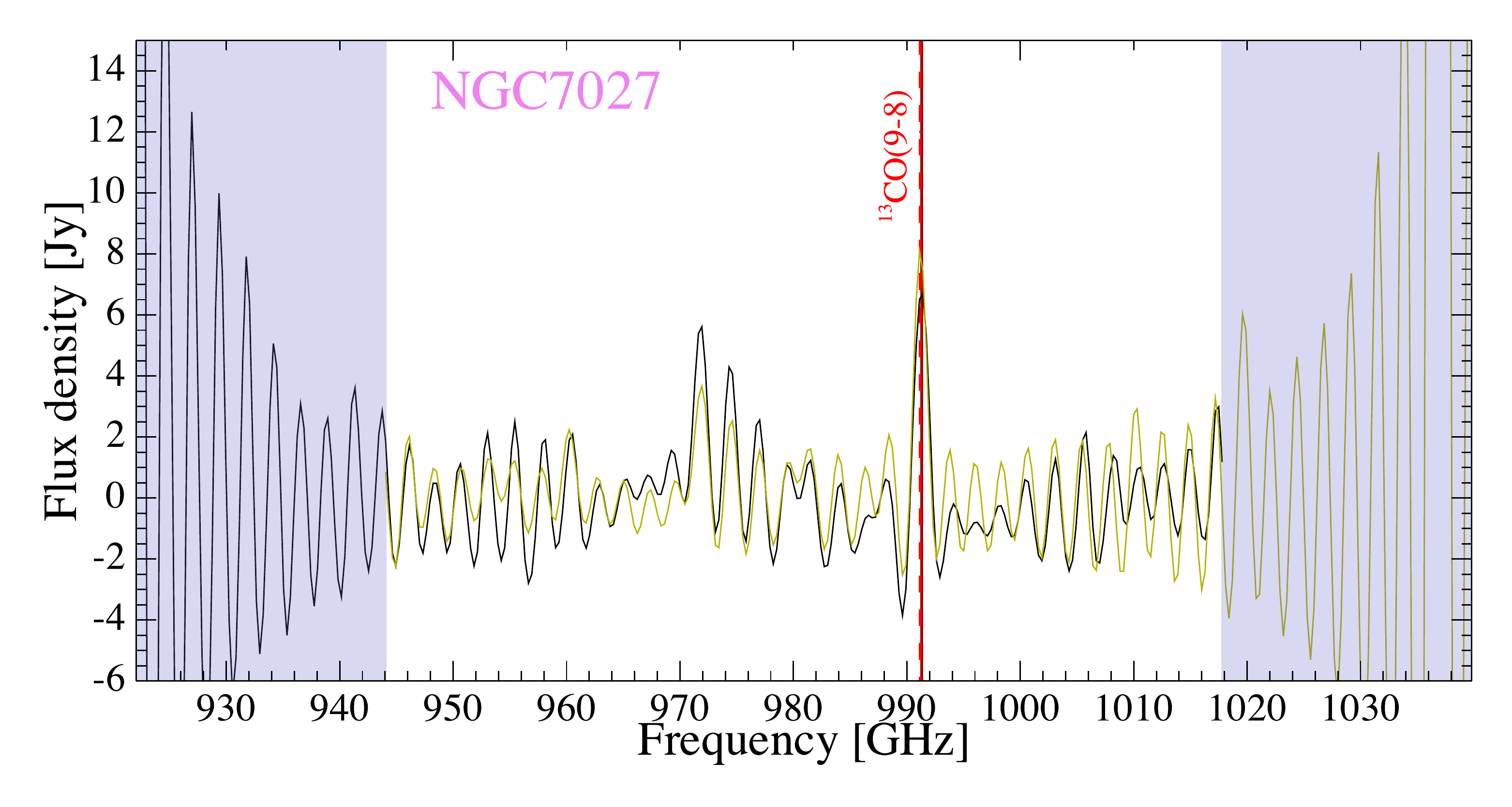}
\includegraphics[trim = 3mm 0mm 5mm 5mm, clip,width=0.9\hsize]{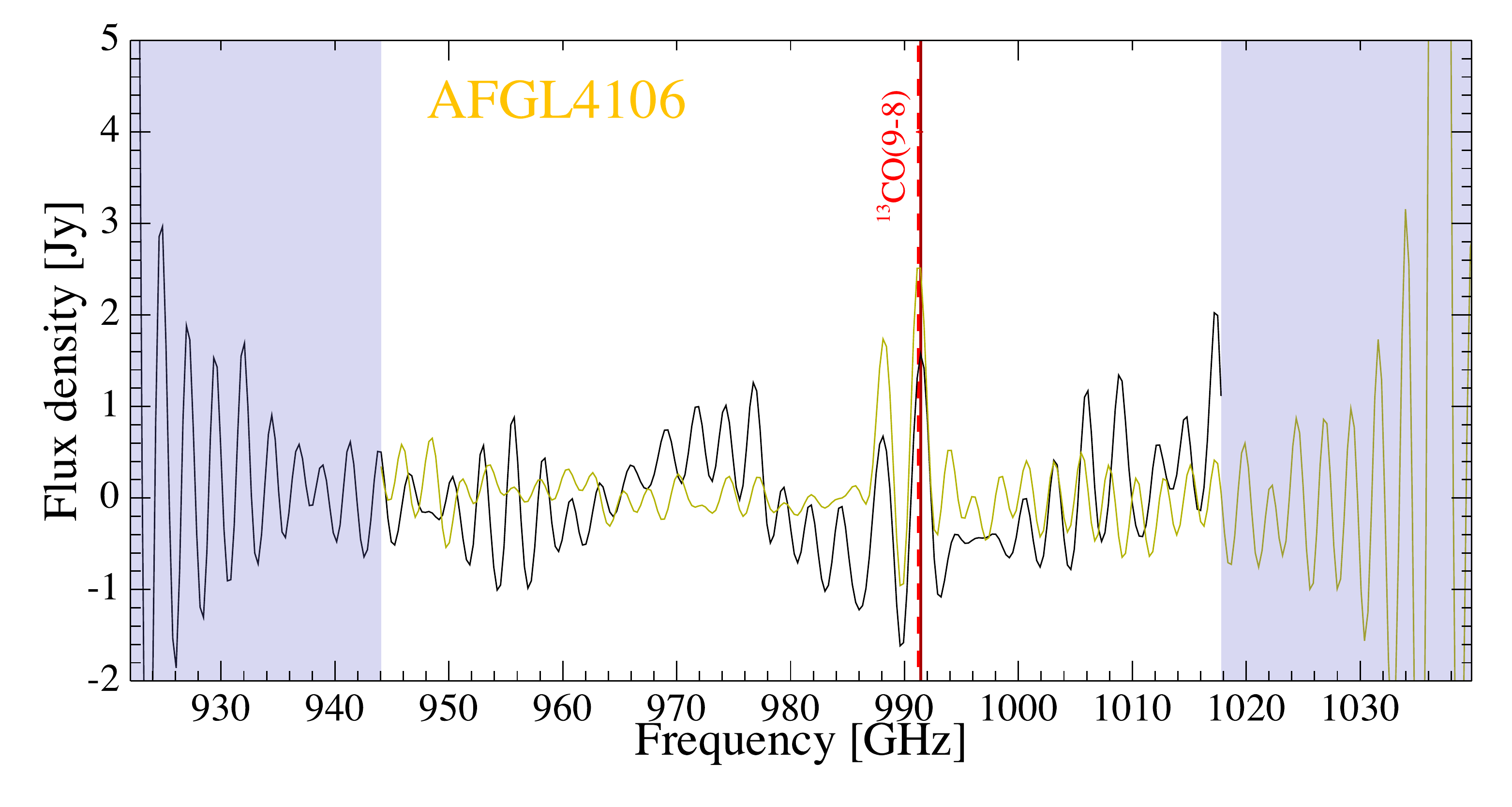}
\caption{Lines fitted in the overlap regions for the four main FTS line sources. The spectra shown are the co-added data, with SLWC3 in black and SSWD4 in brown. The darker solid horizontal lines indicate positions fitted in SLWC3, and the dashed lines the respective SSWD4 fitted positions.}
\label{fig:overlapLines}
\end{figure}

\begin{figure}
\centering
\includegraphics[trim = 2mm 0mm 5mm 5mm, clip,width=1.0\hsize]{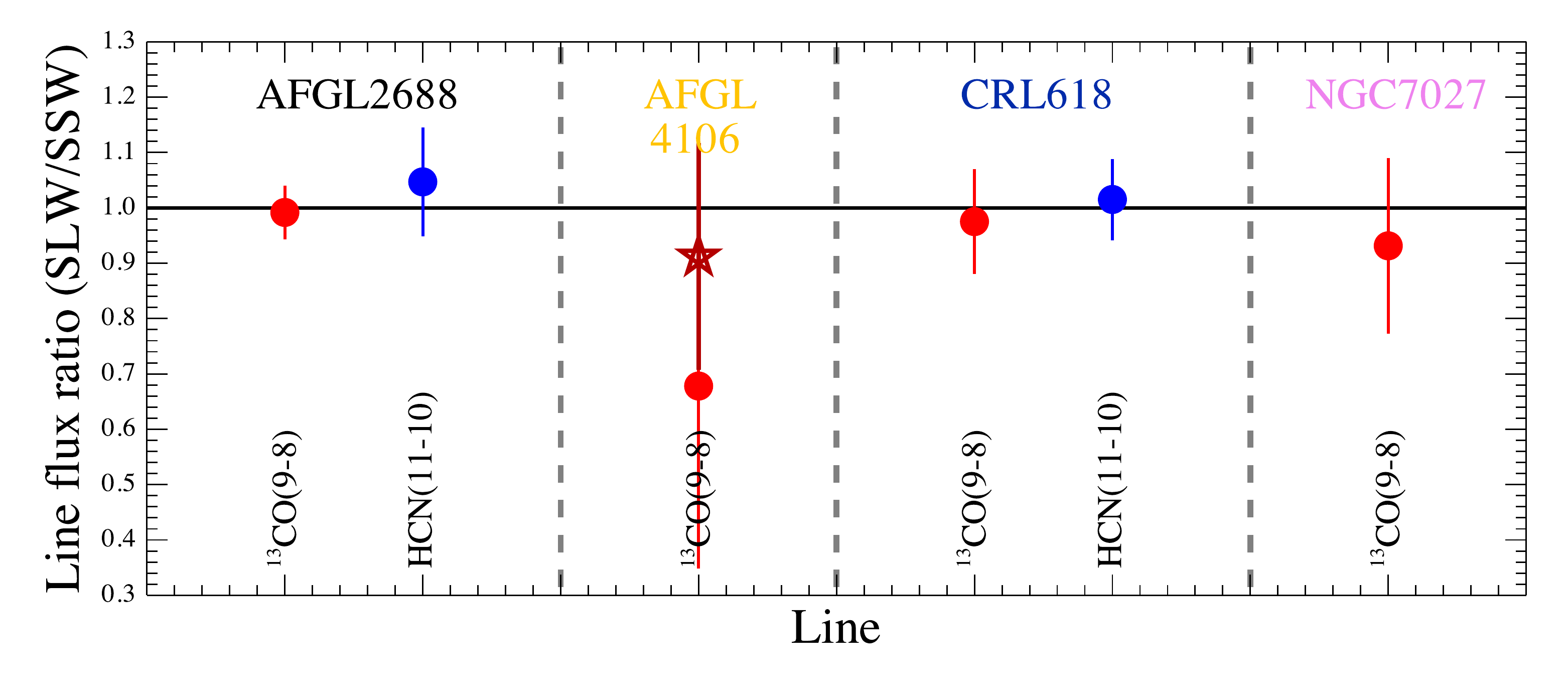}
\caption{SLWC3/SSWD4 line flux ratios for lines fitted in the overlap region. The dark red star represents the revised AFGL4106 ratio after individual refitting of the SLWC3 $^{13}$CO (9-8) line (see Section.~\ref{sec:afgl4106NoisyEnds}). The measured line flux shows good consistency between the bands. }
\label{fig:overlapFluxDiff}
\end{figure}

There are six lines fitted in total within the overlap region, but no $^{12}$CO lines present and only HCN(11-10) in AFGL2688 and CRL618, and $^{13}$CO(9-8) in AFGL2688 could be considered strong, i.e. with a SNR $>$\,10, with the other three having SNR $\sim$\,5. Fig.~\ref{fig:overlapLines} shows the fitted lines that fall within the overlap region, for each of the four sources under consideration. There are only small SLWC3/SSWD4 differences in fitted position ($<<$1\%). Comparing line flux ratios in Fig.~\ref{fig:overlapFluxDiff}, shows the higher signal-to-noise lines are within a couple of percent between the two bands, and all except the $^{13}$CO(9-8) line in AFGL4106 are within $\pm$5\%. Fitting to the line in AFGL4106 is hampered by the high fringing at the end of the SLW band, which gives a poor SLW measurement. Despite the lines available being faint and few, and the relatively higher noise in the overlap region, the results show good consistency between the SLW and SSW calibration.

\subsubsection{Re-fitting AFGL4106}\label{sec:afgl4106NoisyEnds}

The removal of an extended background from the AFGL4106 point-source-calibrated spectra was detailed in Section~\ref{sec:bgs}, but this method of subtraction, using the wide-scale shape of the off-axis detectors, does not address the fringing introduced when the extended background emission is calibrated as a point source. There is pronounced fringing at the high frequency end of SLW, as shown in Section~\ref{sec:overlap}, which leads to a poor fit of the $^{13}$CO (9-8) line located in the overlap region. The line is at 991\,GHz, which lies in the extended frequency band introduced with HIPE 12.1. To see if the line flux measurement could be improved, the line was refitted in the co-added SLWC3 spectrum and resulting line flux compared to the average fit results for SSWD4.

\begin{figure}
\centering
\includegraphics[trim = 0mm 0mm 0mm 0mm, clip, width=1.0\hsize]{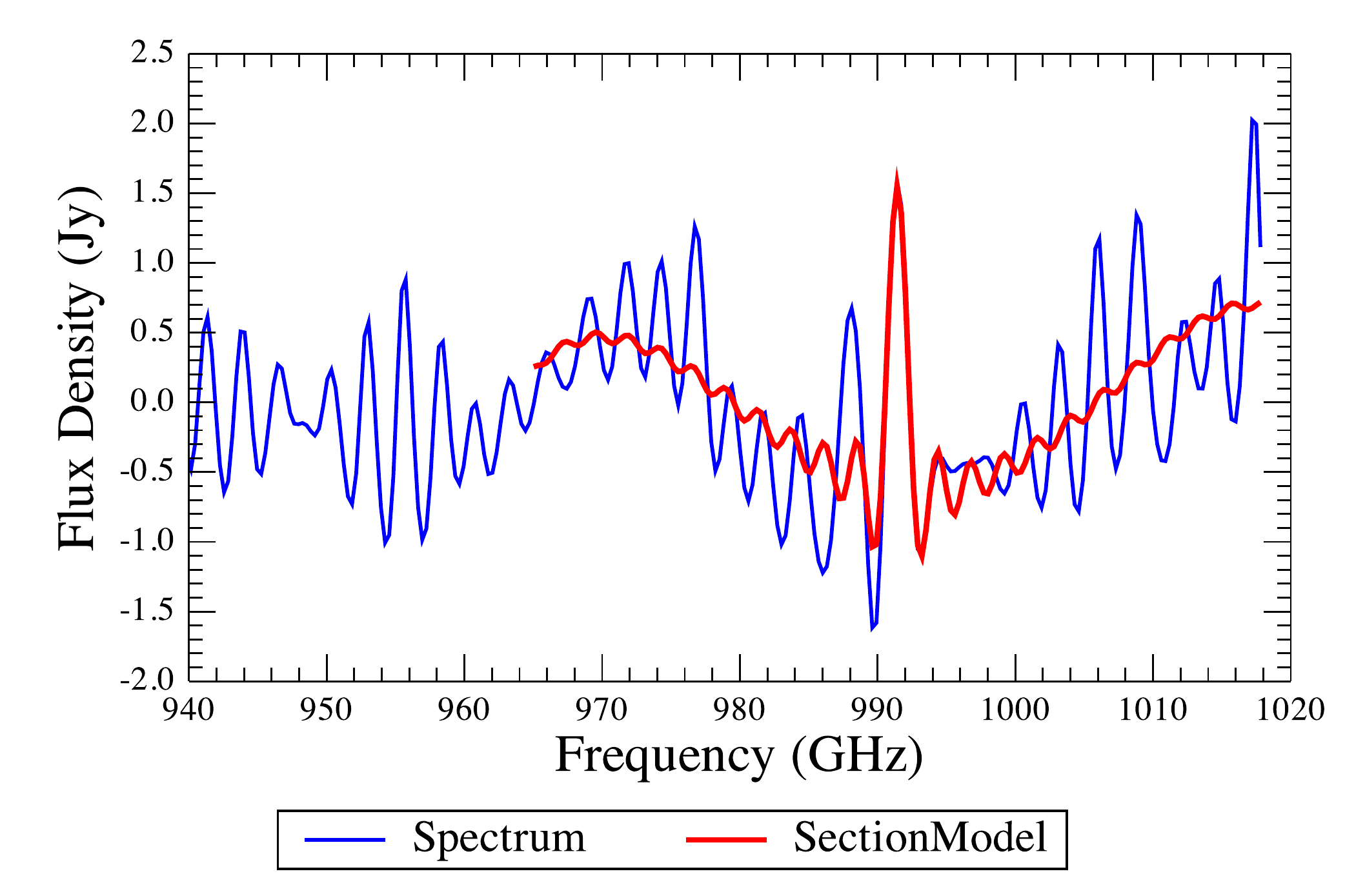}
\caption{Sinc and polynomial fit to a section of the co-added AFGL4106 SLWC3 spectrum, after subtracting all but the $^{13}$CO (9-8) line. The resulting line flux is more consistent with that found for the same line in SSWD4, as shown in Fig.~\ref{fig:overlapFluxDiff}.}
\label{fig:afgl4106refit}
\end{figure}

The co-added data are already baseline subtracted, so the line list for AFGL4106 was fitted without a polynomial. All but the $^{13}$CO (9-8) lines were subtracted using the best fit sinc profiles. A segment of the residual spectrum (965--1025\,GHz) was then fitted with a seventh order polynomial and one sinc profile. The fitted result is shown in Fig.~\ref{fig:afgl4106refit} and the resulting ratio of the line flux for the two bands is plotted as a star in Fig.~\ref{fig:overlapFluxDiff}. The SLWC3 line flux increases from $(1.9\pm0.7)\times10^{-17}$ to $(2.6\pm0.3)\times10^{-17}$ Wm$^{-1}$, which is consistent with the average SSWD4 result of $(2.8\pm0.8)\times10^{-17}$Wm$^{-1}$.

\subsection{Line shape}\label{sec:lineShape}

In order to assess the difference between the measured line shape and a standard sinc function, and the uncertainty introduced when fitting lines, the set of 31 NGC7027 observations was used to build an empirical line shape. The $^{12}$CO lines in the NGC7027 spectra are ideal for this task, due to high SNR and a shape that is less affected by blending compared to the other sources.

\begin{figure}
\centering
\makebox[\hsize][l]{
\includegraphics[trim = 8mm 8mm 8mm 4mm, clip, width=0.516\hsize]{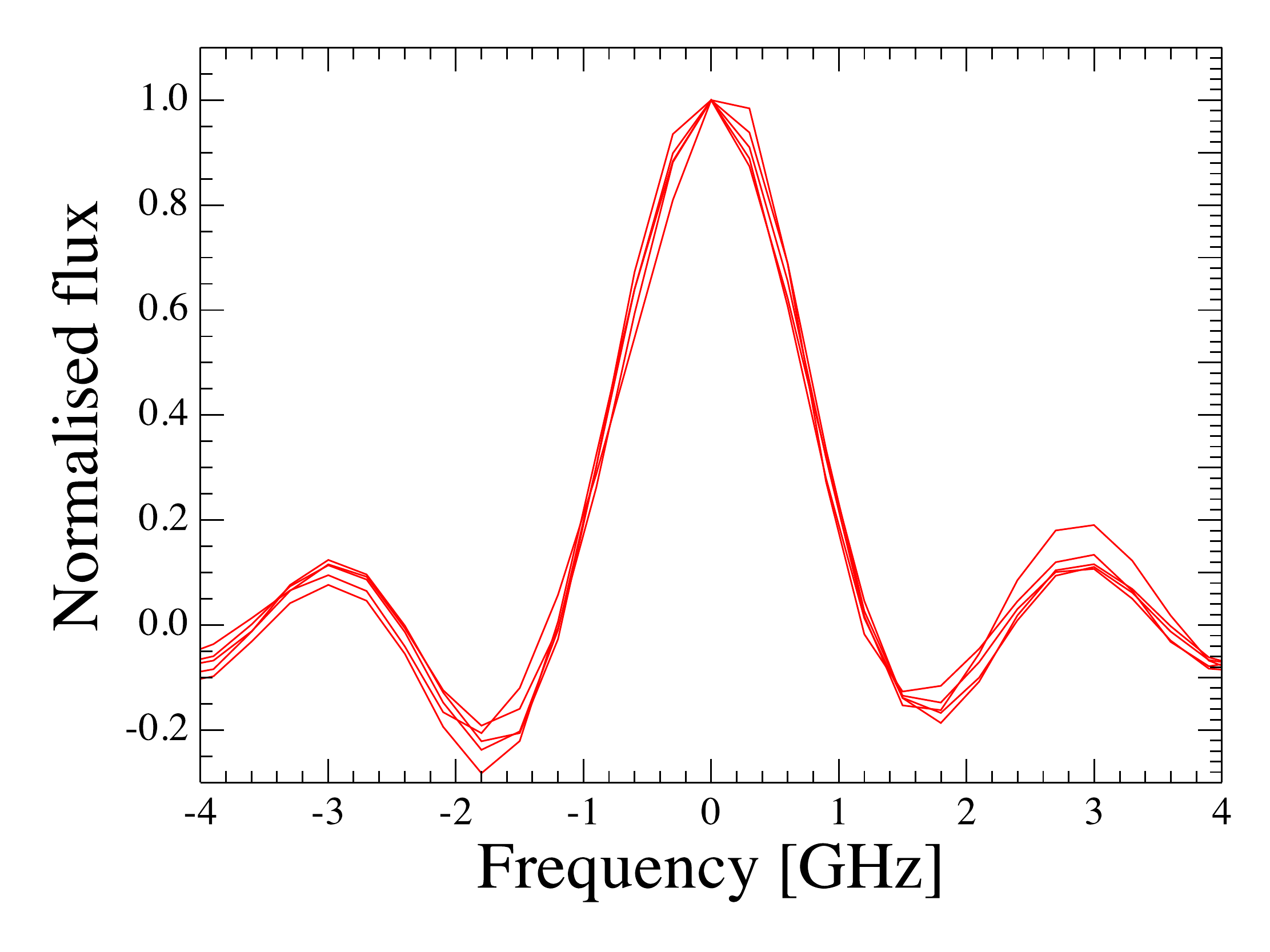}
\includegraphics[trim = 22mm 8mm 7mm 4mm, clip, width=0.484\hsize]{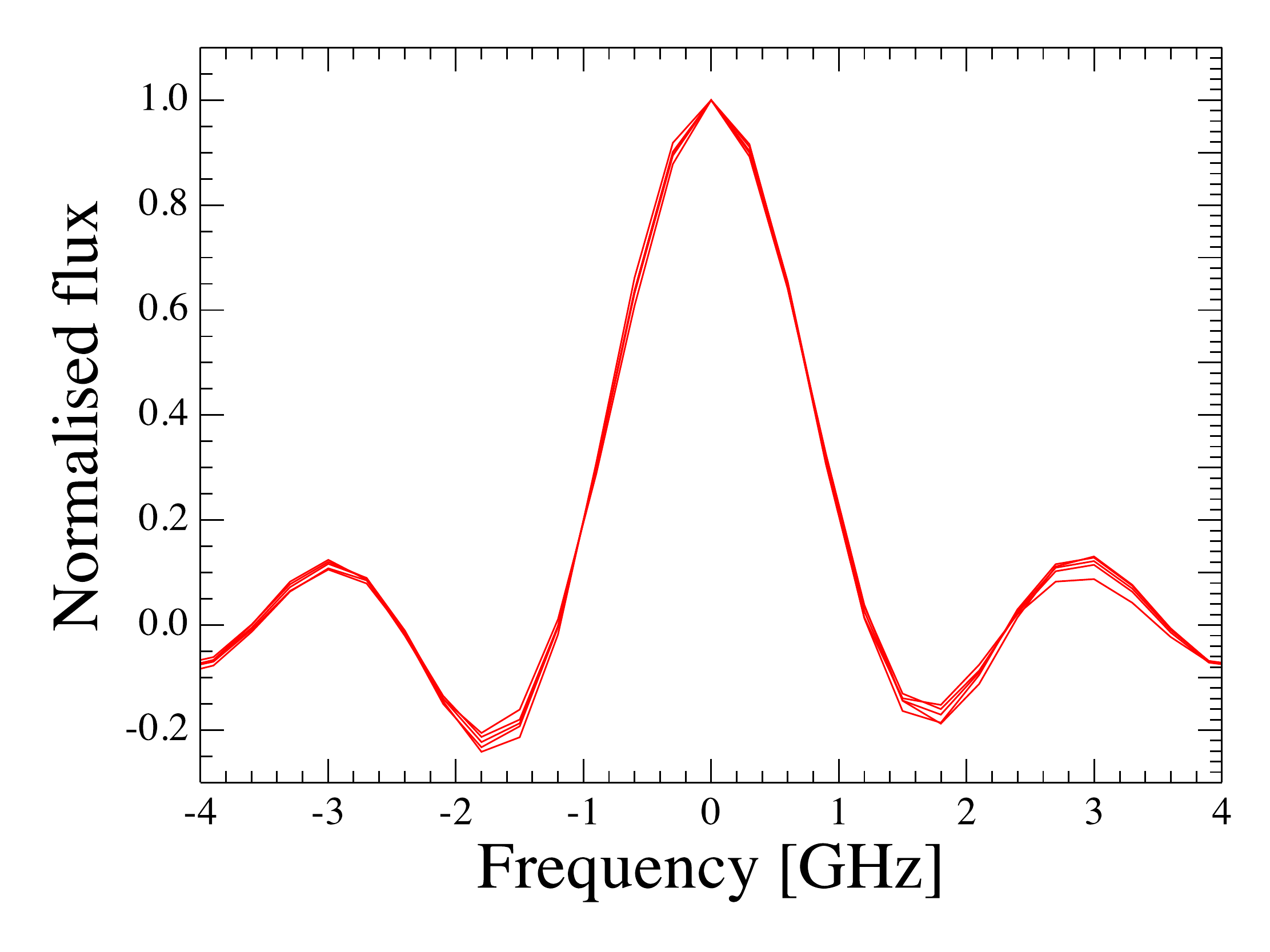}
}
\caption{Example of all the normalised $^{12}$CO lines for one NGC7027 observation, with SLWC3 shown on the left and SSWD4 on the right. Lines from SSW are more consistent.}
\label{fig:empModSpread}
\end{figure}

\begin{figure}
\centering
\includegraphics[width=0.9\hsize]{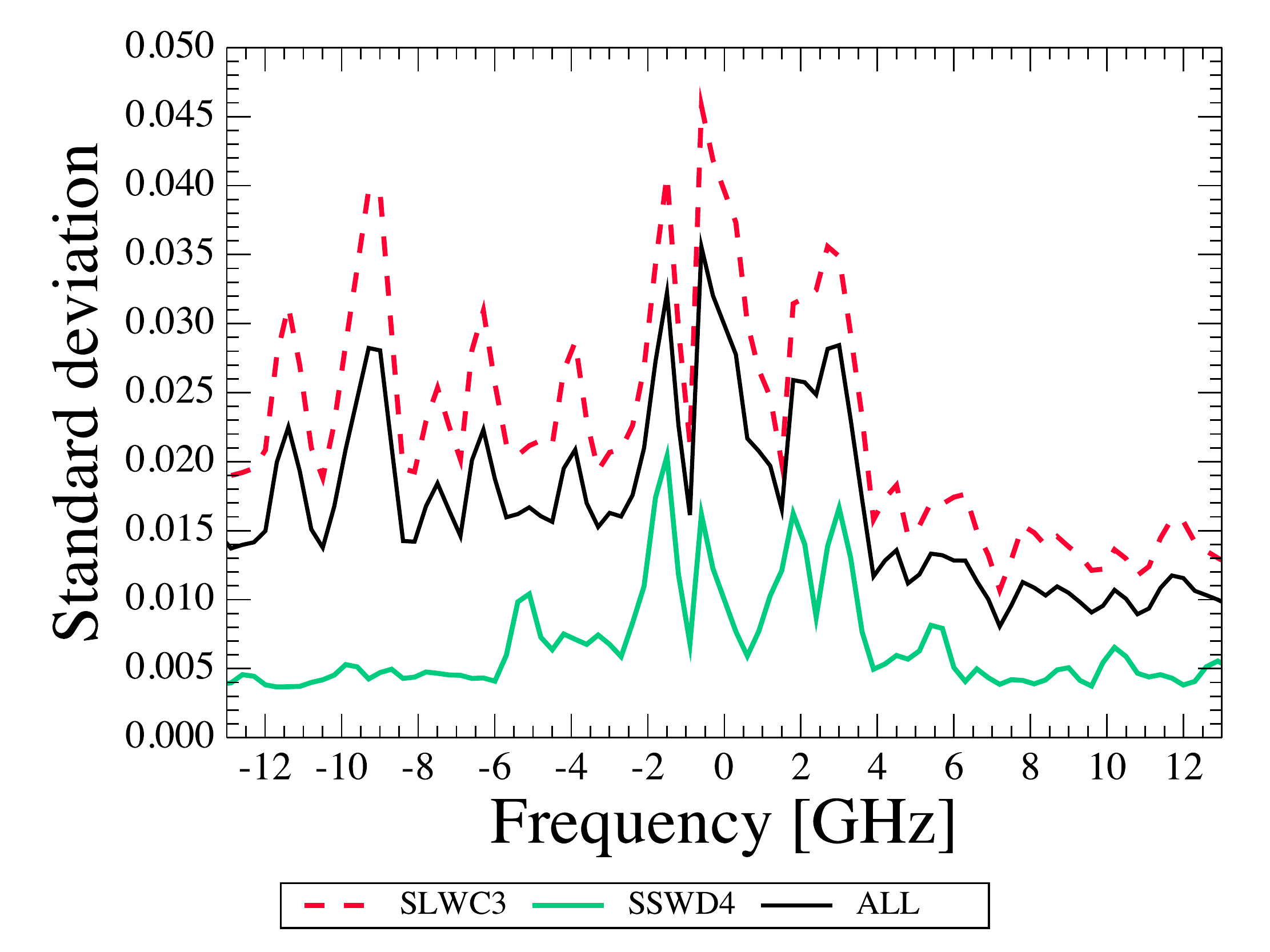}
\caption{Spread in the cropped re-centred and normalised lines from all NGC7027 observations included in the empirical line shape. Using only SSW line optimises the SNR of the resulting line profile.}
\label{fig:empModStddev}
\end{figure}

\begin{figure}
\centering
\includegraphics[trim = 8mm 0mm 8mm 5mm, clip,width=0.8\hsize]{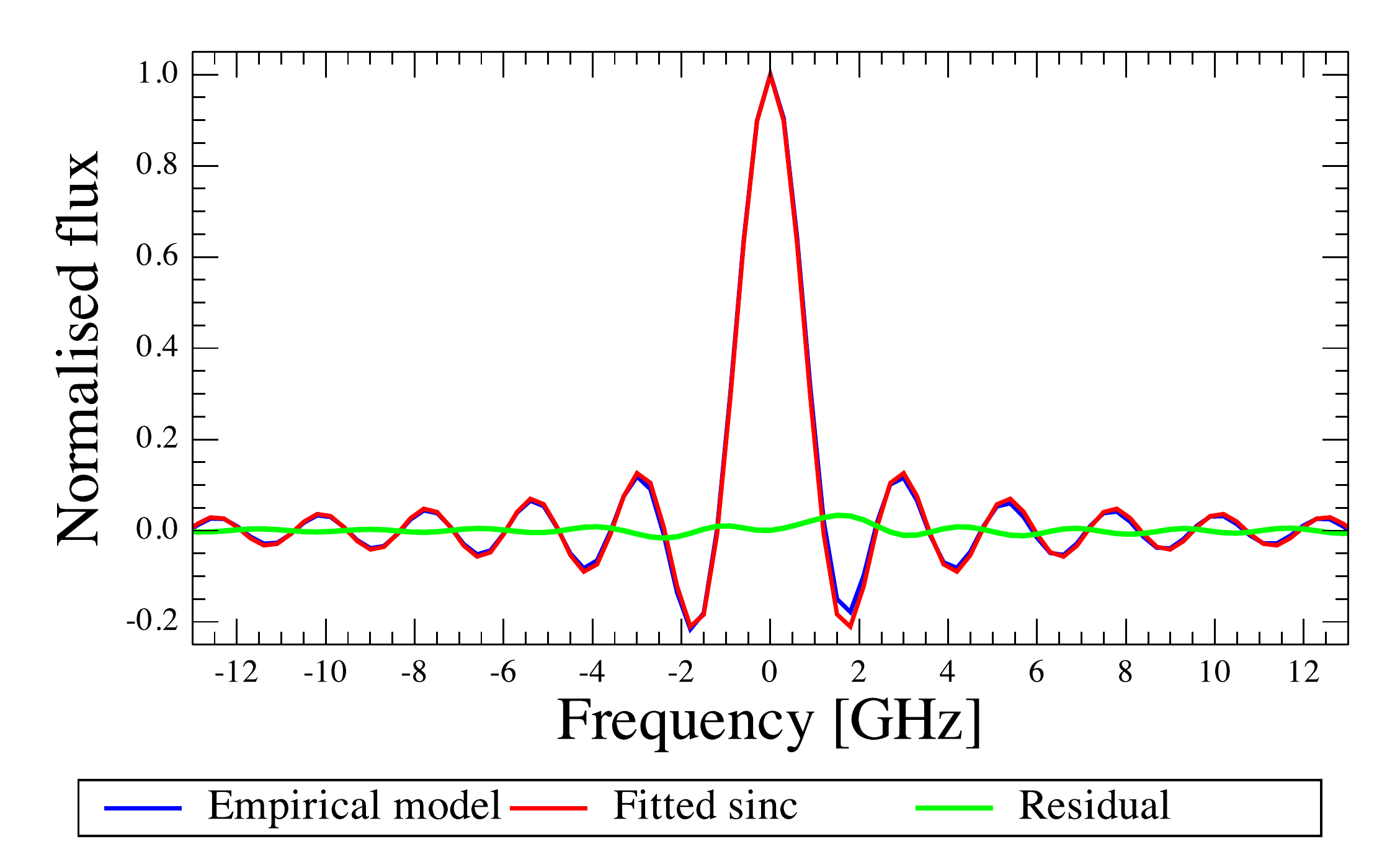}
\includegraphics[trim = 8mm 0mm 8mm 6mm, clip,width=0.8\hsize]{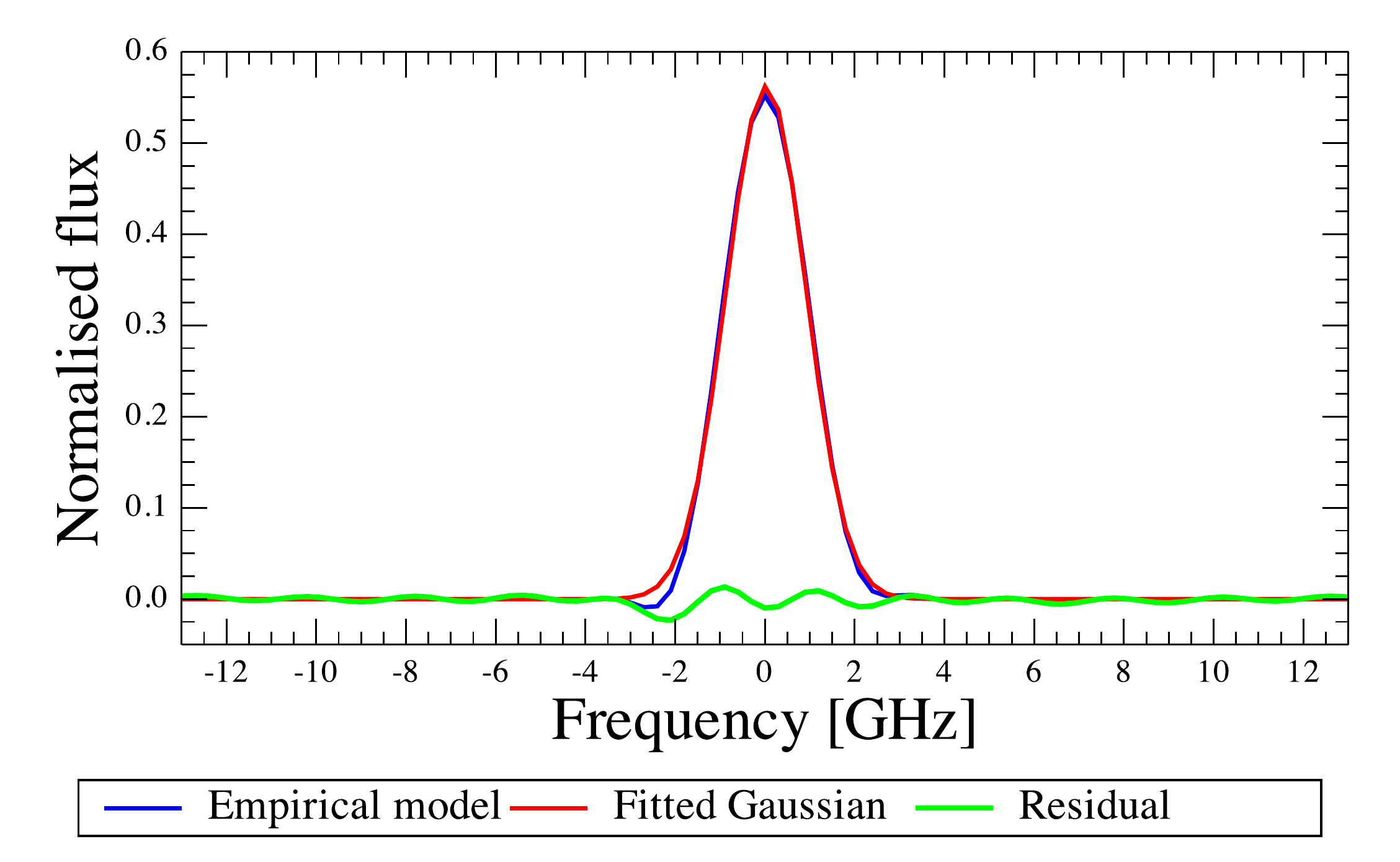}
\caption{Top: empirical line shape compared to a fitted sinc profile and the residual on subtraction of the fit. Bottom: the empirical line shape after apodizing, compared to a fitted Gaussian profile and the residual on subtraction of the fit. Note the apodized line shape was not re-normalised.}
\label{fig:empModFinal}
\end{figure}

Before building the empirical line profile, the NGC7027 data were ``cleaned'' of all but the $^{12}$CO lines. Line fitting was carried out as detailed in Section~\ref{sec:lineFitting}, with a sinc profile fitted to each line and a polynomial for the continuum. For each spectrum the combined fit, not including the $^{12}$CO line profiles, was subtracted to ``clean'' the data of low signal-to-noise lines and the continuum. The ``cleaned'' SSWD4 spectra were cropped around each $^{12}$CO line, and these segments re-centred and normalised to a peak of unity before taking the mean. Only lines observed with SSWD4 are used as they have higher SNR and are more consistent than the lines in SLWC3, as shown by Figures \ref{fig:empModSpread} and \ref{fig:empModStddev}. Fig.~\ref{fig:empModFinal} shows the final empirical line profile fitted with a sinc function to illustrate the FTS line shape asymmetry, which is strongest on the first negative side lobe of the high frequency side. 

When fitting the empirical line profile with a sinc, the width was left as a free parameter to check for any deviation from the expected width. 
All of the NGC7027 observations used for the line profile were made using the highest resolution (1.18448\,GHz), i.e. with a consistent resolution across the set. The intrinsic width of the lines is on the order of 30\,km\,s$^{-1}$ \citep{Herpin02} and so is insignificant compared to the spectral resolution of the instrument. The result of fitting a sinc function to the empirical line shape shows less than 1\% difference at the peak, and less than 0.5\% difference in the width from that expected for the instrument resolution. The curves plotted in Fig.~\ref{fig:empModFinal} were integrated and compared to the line flux from the fitted sinc function.
The ratio of the sinc area to the empirical line shape area shows a 2.6\% shortfall.

\begin{figure}
\centering
\includegraphics[trim = 0mm 5mm 0mm 0mm, clip, width=0.85\hsize]{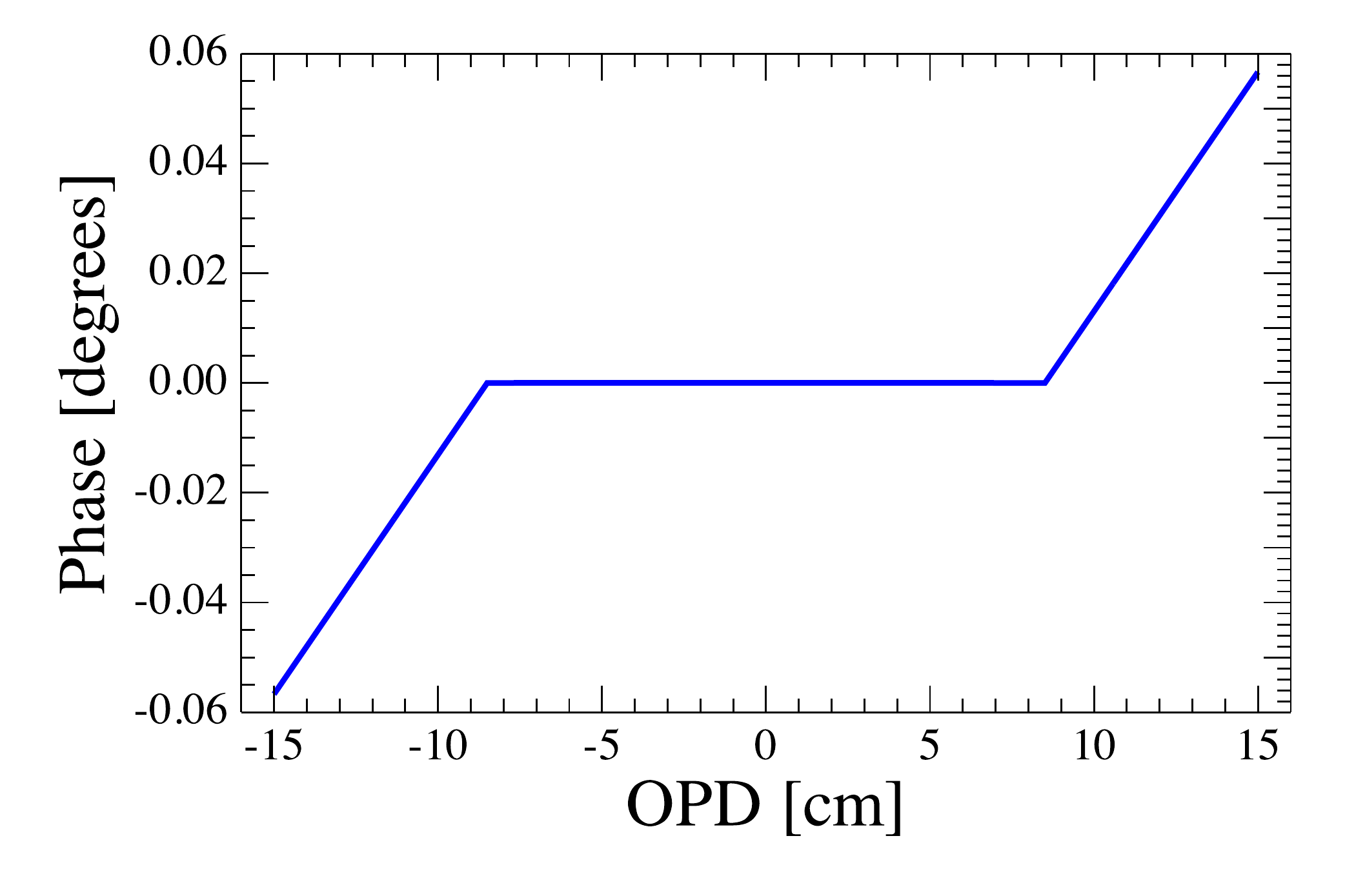}
\includegraphics[trim = 0mm 0mm 0mm 7mm, clip, width=0.85\hsize]{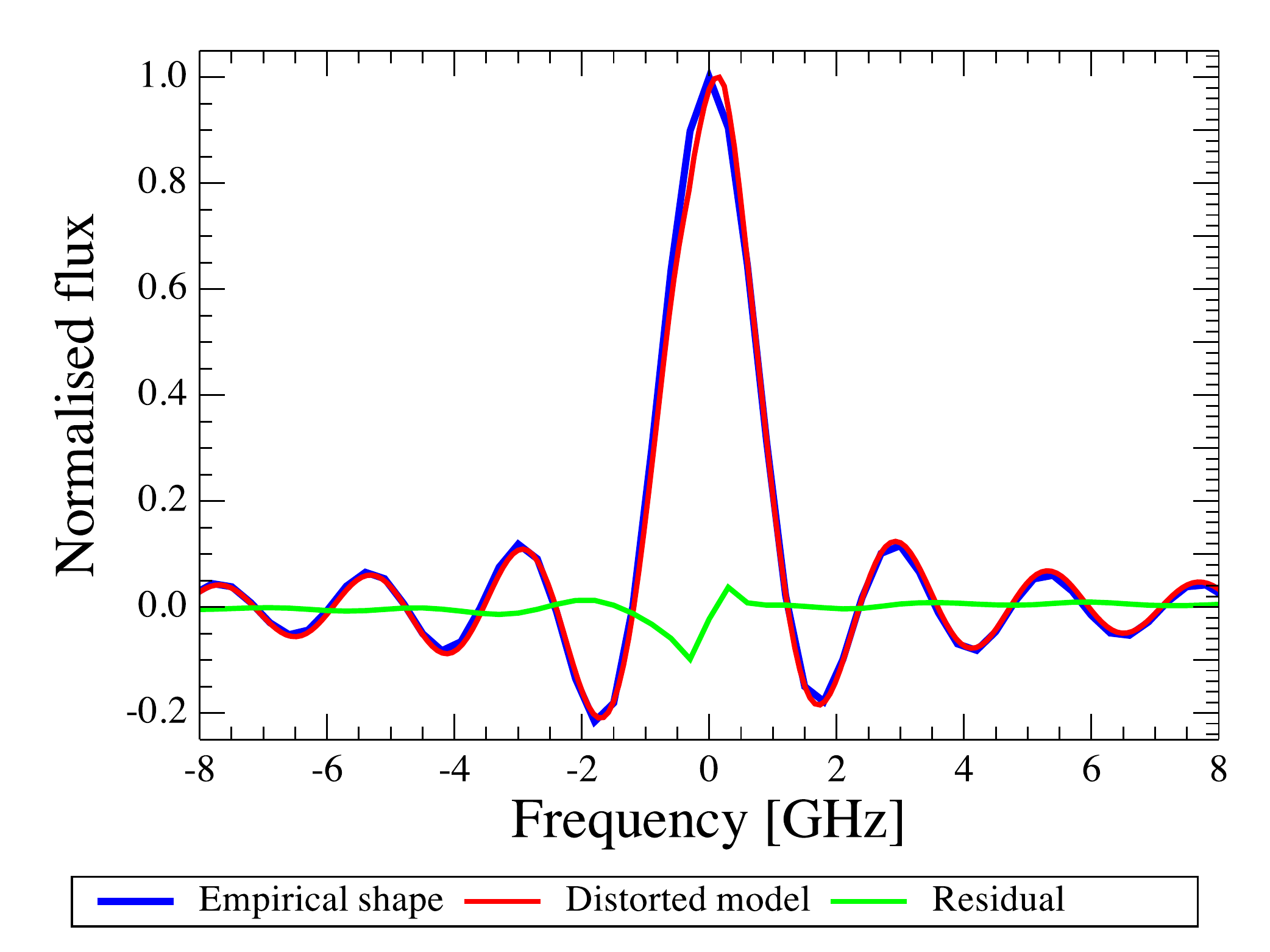}
\caption{A sinc profile with an imposed phase error. The top panel shows the phase is flat over the ``corrected'' section, at the centre of the OPD, and deviates at the longest part. The bottom panel shows the empirical line shape (blue) overlaid with the resulting distorted model (red) and compared to the residual of the two (green).}
\label{fig:phase}
\end{figure}

The asymmetry seen in the FTS line shape is most probably caused by a systematic residual phase shift in the interferogram.  Information at the highest spectral resolution, such as line features, is encapsulated in the interferogram at high OPDs.  However, the phase correction performed by the pipeline is determined only by examining the symmetrical part of the interferogram around zero path difference (Fulton et al. in preparation).  Any residual phase at high path difference, such as that due to small misalignments with respect to the optical axis inside the instrument, could cause systematic phase asymmetry.  
To test this theory we have generated a sinc profile with an imposed phase error versus OPD, as shown in the top panel of Fig.~\ref{fig:phase}.  Here the phase is flat over the ``corrected'' section at the centre of the OPD and deviates at the longest part with an arbitrary slope and magnitude.  The resulting distorted model sinc profile is shown compared to the averaged line profile in the bottom panel of Fig.~\ref{fig:phase}. The slope and magnitude of the residual phase error have been adjusted here to achieve the best fit to the data and we can see that the first order explanation is correct, giving a reasonable fit to the asymmetric part of the profile. Comparing Fig.~\ref{fig:phase} and Fig.~\ref{fig:empModFinal} indicates that although this adjustment improves the fit to the asymmetry, there is some additional discrepancy for the fit to the peak. In future versions of the pipeline we plan to build tools to allow detailed examination of the line profile using this method.  Meanwhile, these simulations show that there is no loss of flux due to the phase shift, but rather a redistribution of flux in the asymmetric line.
Therefore, the 2.6\% difference calculated above by fitting a pure sinc function to the empirical line shape can be taken as an estimate of the systematic errors when measuring line flux with sinc fitting.

\subsection{Fitting apodized data}\label{sec:apodized}

\begin{figure}
\centering
\makebox[\hsize][l]{
\includegraphics[trim = 5mm 5mm 8mm 5mm, clip,width=0.515\hsize]{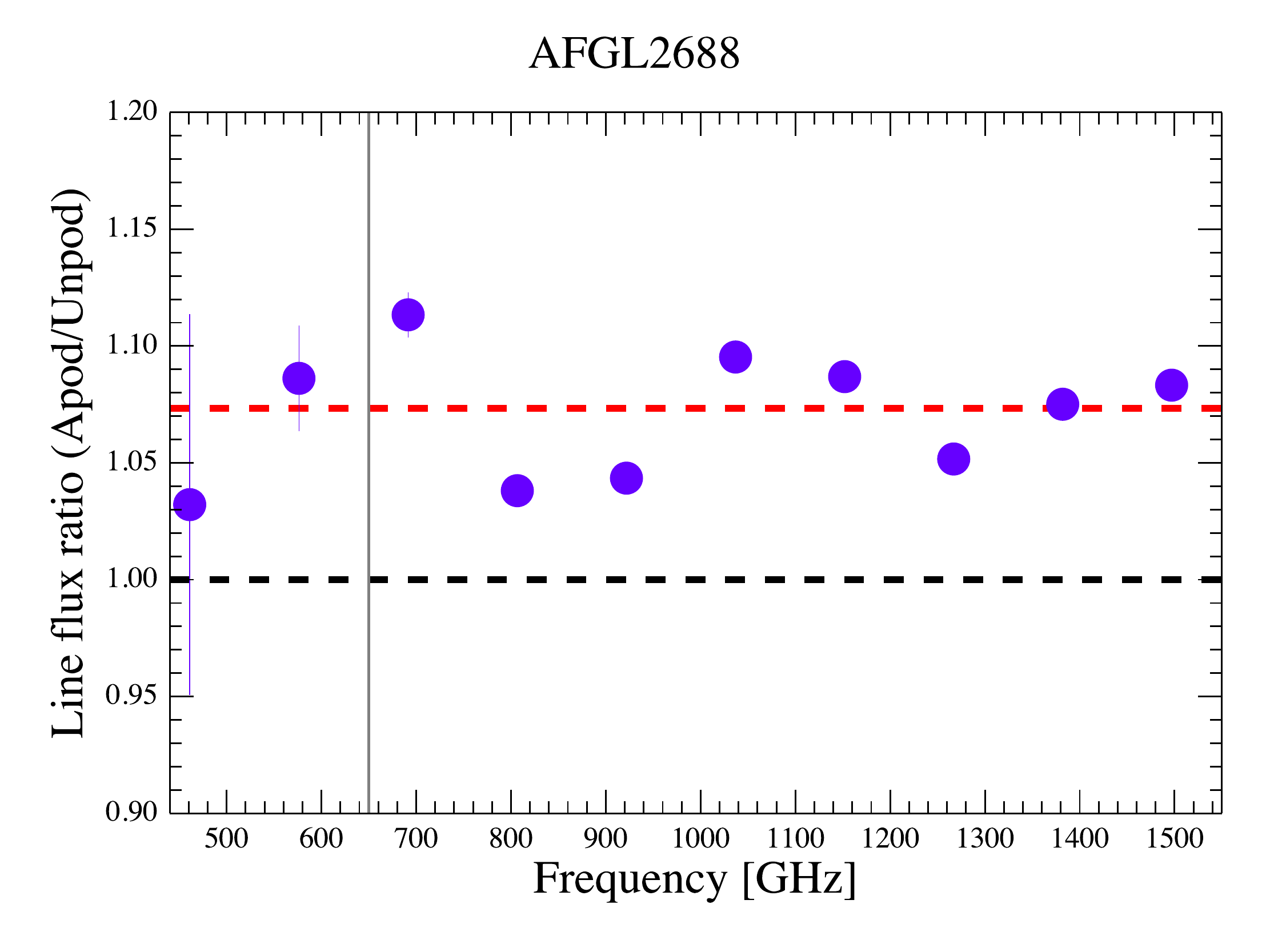}
\includegraphics[trim = 17mm 5mm 8mm 5mm, clip,width=0.485\hsize]{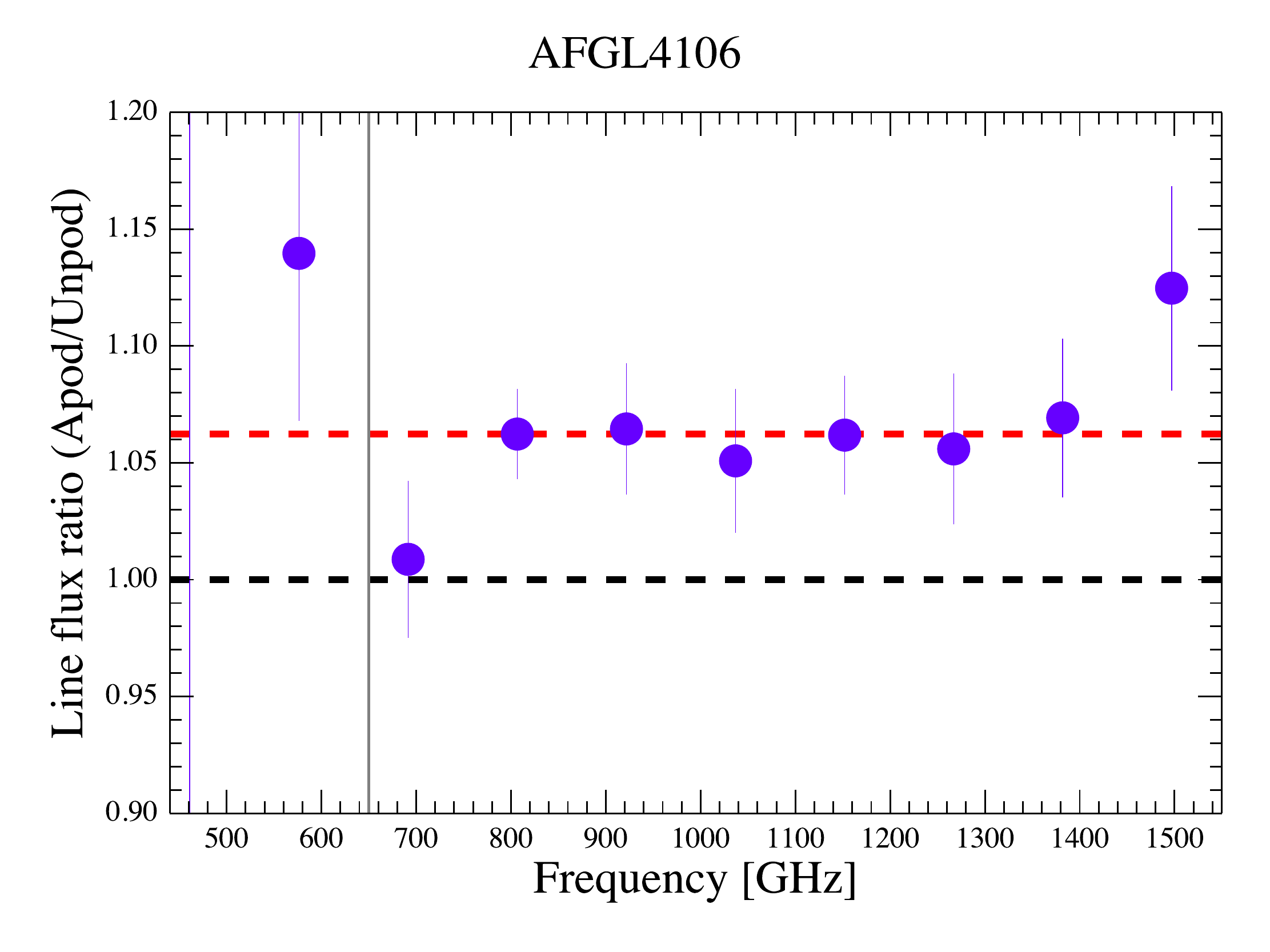}
}
\makebox[\hsize][l]{
\includegraphics[trim = 5mm 5mm 8mm 5mm, clip,width=0.515\hsize]{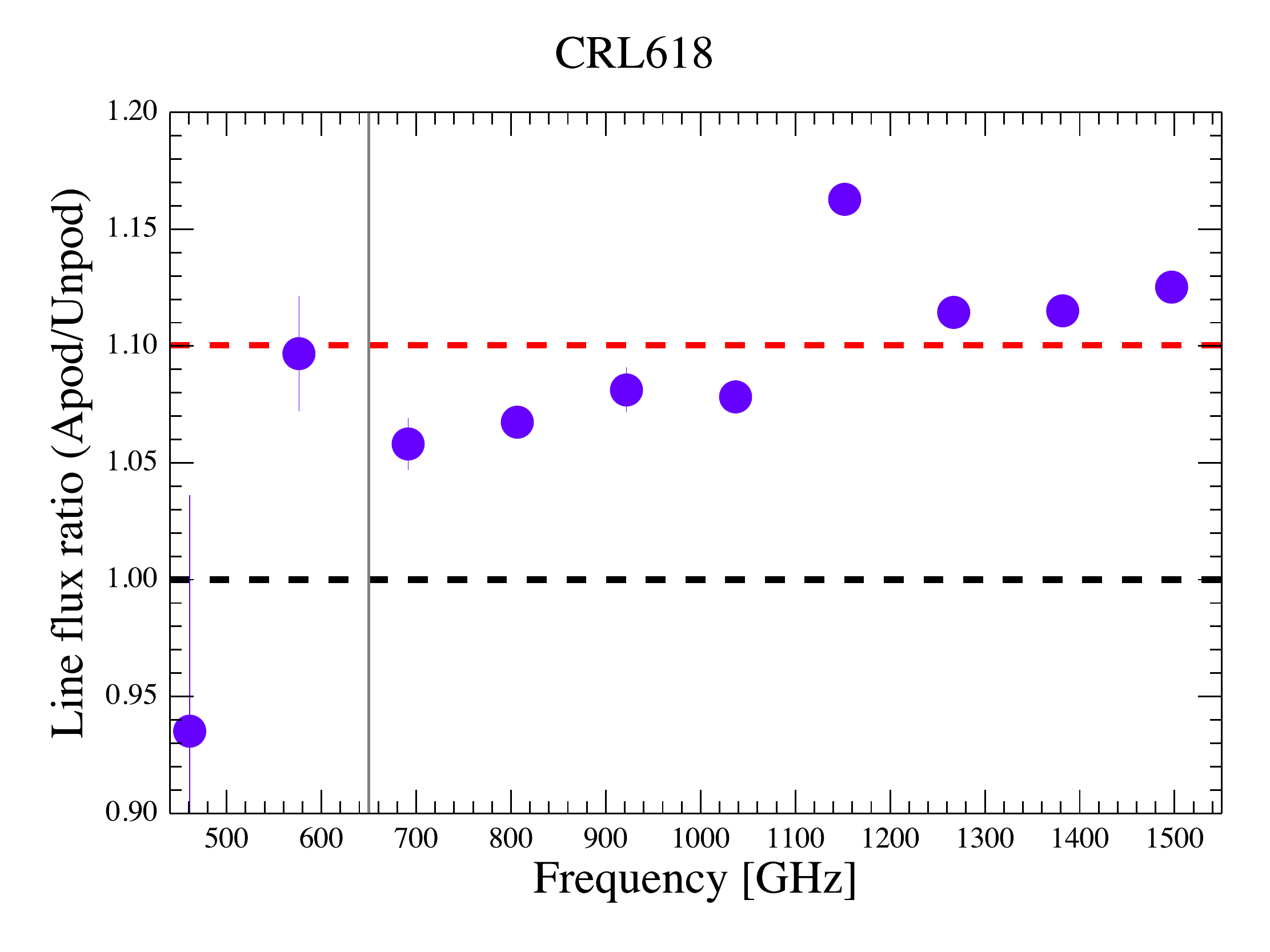}
\includegraphics[trim = 17mm 5mm 8mm 5mm, clip,width=0.485\hsize]{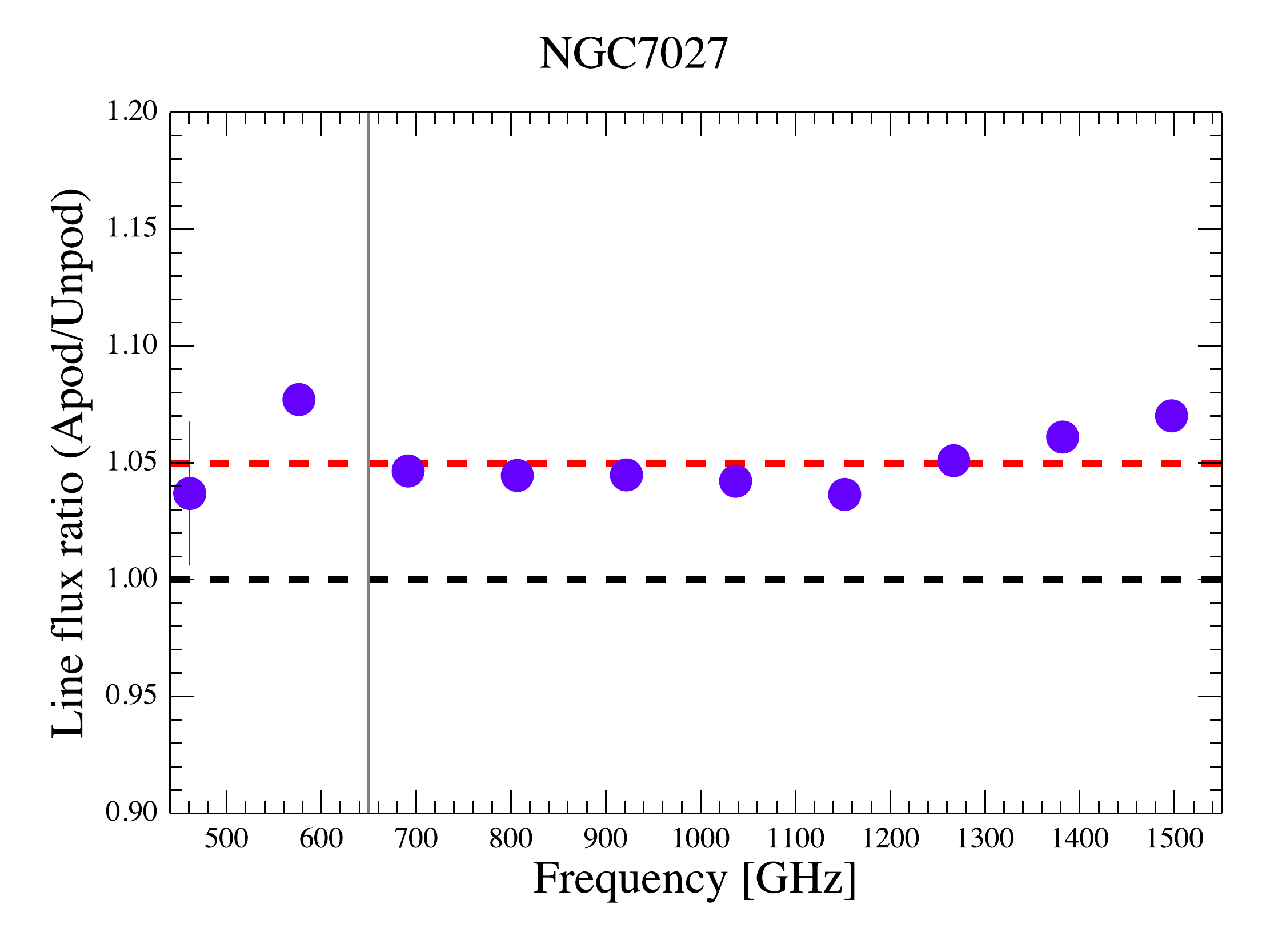}
}
\caption{Fitted $^{12}$CO line flux ratios for apodized to standard data. The horizontal dashed red lines indicate the mean ratio for each source. The points to the left of the vertical black lines are not included in the mean ratio. NGC7027 shows the 5\% excess for Gaussian fitted apodized data, compared to sinc fitted standard data. AFGL4106 and AFGL2688 are consistent with this result. CRL618 shows an increase in the difference, however this also corresponds with the $^{12}$CO lines becoming increasingly more resolved with frequency. The $^{12}$CO transitions included are given in Table~\ref{tab:12co}.}
\label{fig:apodizeLineFlux}
\end{figure}

For data that have been apodized by the standard apodization function \citep[the adjusted Norton-Beer 1.5][]{NaylorTahic07}, the line shape is well represented by a Gaussian profile. Such apodized data have effectively been ``smoothed'', which reduces the sinc wings, but degrades the spectral resolution by a factor of 1.5. The asymmetry seen for standard data, when compared to a sinc function, is also present in apodized data, but is stronger in the lower frequency Gaussian wing. Two different methods were used to assess the difference between line flux obtained from the standard data fitted with a sinc function, compared to that found when fitting a Gaussian profile to apodized data. NGC7027 observations were used for this comparison, after point source calibration. 
The observations of NGC7027 were fitted as previously detailed, i.e., for the standard data, all major lines and the continuum were fitted simultaneously with sinc functions and a polynomial. 
The fitting process was then repeated for the apodized data, but using a Gaussian function and
only fitting to the $^{12}$CO lines. 
Taking the ratio of line flux obtained from the fitted sinc parameters to the equivalent line flux calculated from the Gaussian parameters shows an over estimate of 5\% for the apodized data. 
The empirical line profile was also used to assess this difference. The frequency range of the empirical line shape was extended using a fitted sinc profile, assuming any difference between the pure sinc shape and the true line shape is much less than the uncertainty on the data beyond the first few side lobes. The extended line profile was then apodized, following the steps in the standard pipeline: apply a Fourier transform; apply the standard adjusted Norton-Beer apodization function to the resulting interferogram; apply the reverse Fourier transform; and trim to the standard frequency band edges. The resulting apodized empirical line profile was fitted with a Gaussian profile, and the fit, empirical line profile and residual are plotted in Fig.~\ref{fig:empModFinal}. When comparing line flux obtained from the fitted Gaussian to that with the sinc profile fitted to the line shape before apodizing, and accounting for the estimated shortfall of 2.6\%, the line flux is found to be 4.9\% too great, which agrees with the results from looking at the individual observations of NGC7027. The apodized comparison to standard processing was extended to the other main line sources (AFGL2688, AFGL4106 and CRL618), but the level of blending in these sources leads to a higher scatter in the results, due to the spreading of line flux by apodization. Fig.~\ref{fig:apodizeLineFlux} presents the resulting ratios for all four sources, which gives 6\%, 10\%, 7\% and 5\% for AFGL4106, CRL618, AFGL2688 and NGC7027, respectively.

\section{Continuum repeatability}\label{sec:continuum}

Section~\ref{sec:repeatabilityLineSources} details the repeatability of line features in SPIRE FTS spectra. However, the continuum level can also be investigated to estimate the overall photometric stability across the band. In this section, the continuum level from the four main repeated calibrators is examined. It is important to check for intrinsic source variability, as although there was none found for the $^{12}$CO line measurements in Section~\ref{sec:lineFitting}, this may not hold true for continuum emission.

\begin{figure}
\centering
\includegraphics[trim = 8mm 32mm 8mm 6mm, clip,width=1.00\hsize]{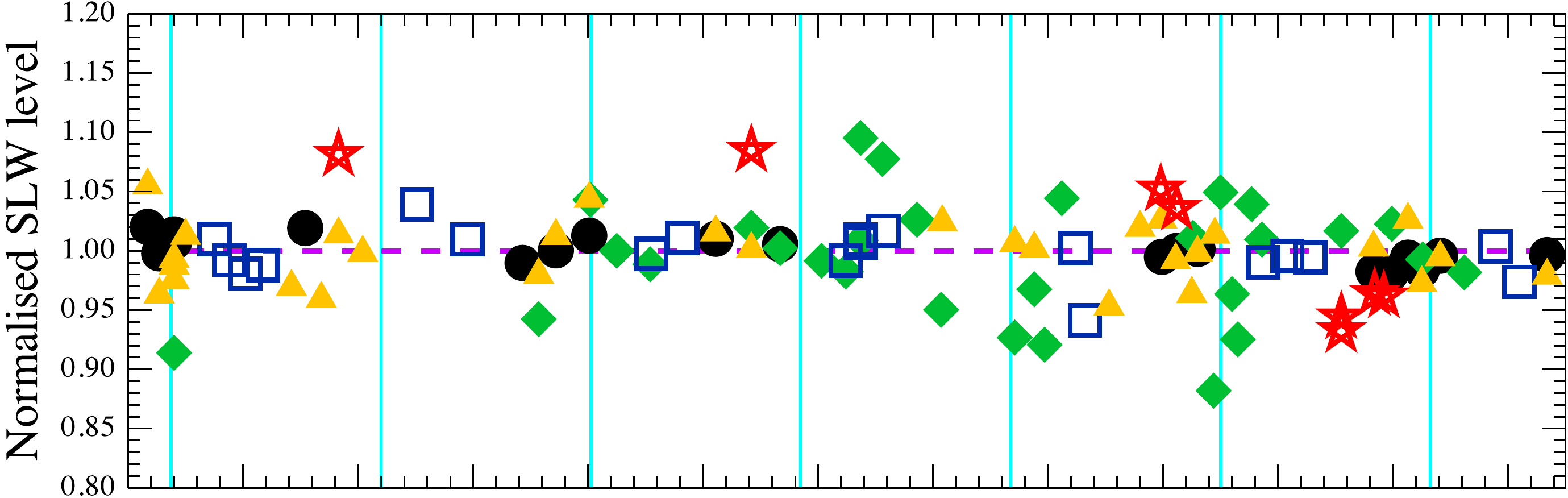}
\includegraphics[trim = 8mm 0mm 8mm 6mm, clip,width=1.0\hsize]{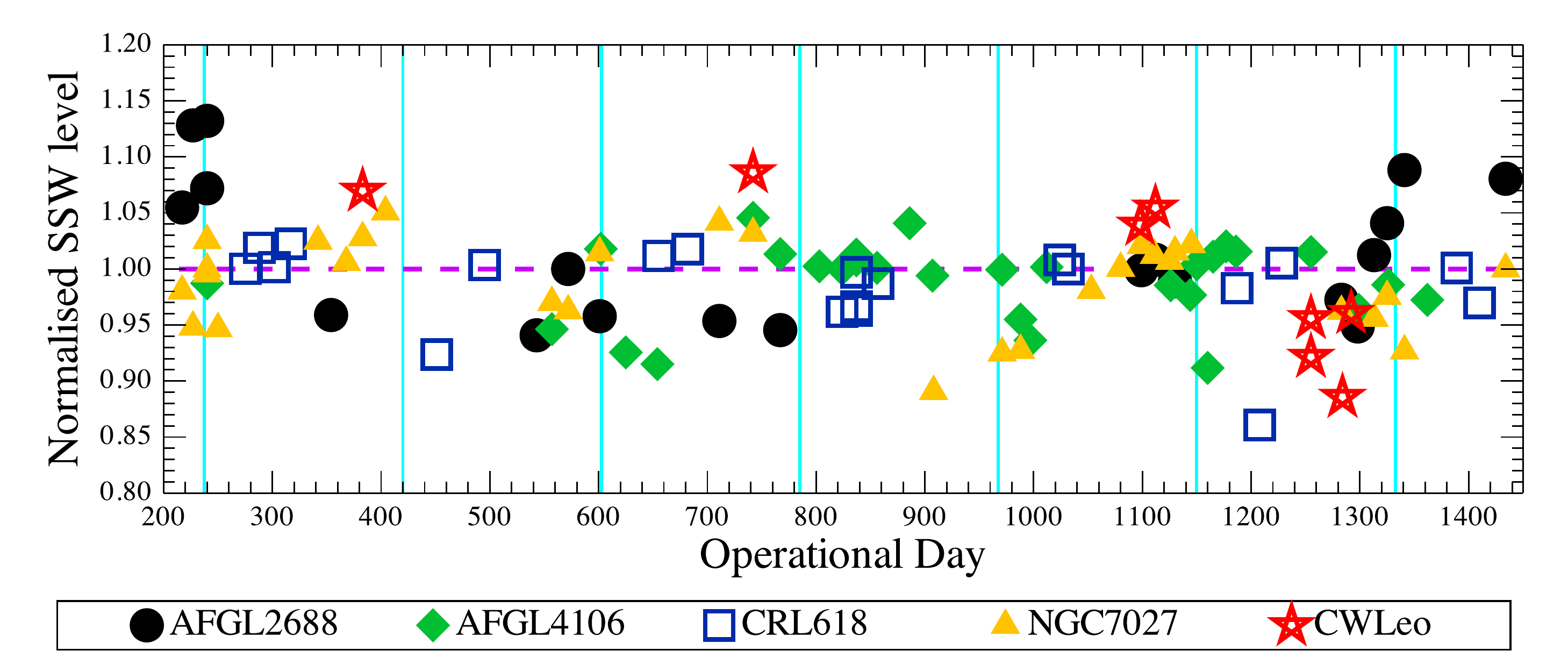}
\caption{Average continuum level as a function of OD for the four main FTS line sources and CW Leo. The values are normalised per source for easier comparison. The level is taken from a polynomial fitted to the continuum, as the mean over 900 to 1000\,GHz for SLW (top) and 1450 to 1550\,GHz for SSW (bottom). Vertical cyan lines indicate six monthly intervals, which can lead to grouping for observations taken before OD\,1011 (see Section~\ref{sec:pointCorr}). AFGL4106 is affected by a background and therefore the SLW points have a relatively high scatter, but the points for the other sources are roughly flat, except for CW Leo. A similar trend is seen for the CW Leo continuum level for SSW and SLW, this source is therefore the only one presented that exhibits significantly intrinsic variable within the FTS uncertainty.}
\label{fig:meanContinua}
\end{figure}

The repeatability of continuum measurements was checked for the four main line sources in comparison to CW Leo, which is a source with significant intrinsic variability \citep[e.g.][]{Becklin1969,Groenewegen2012,Cernicharo2014}. For each HR observation the line fitting process, detailed in Section~\ref{sec:lineFitting}, was used to simultaneously fit sinc functions to the lines and a polynomial to the continuum. Examples of the resulting continua fits can be seen in Fig.~\ref{fig:bsm}, in Section~\ref{sec:pointCorr}. Fig.~\ref{fig:meanContinua} compares the continuum level, calculated as the mean over 900--1000\,GHz for SLW and 1450--1550\,GHz for SSW.

As discussed in Section~\ref{sec:pointCorr}, the effect of pointing offsets manifests more strongly in the SSW spectra, due to the relatively smaller beam size compared to SLW, and an extended background affects SLW more highly for the same reason (e.g. as for AFGL4106). On the other hand, a source with significant intrinsic variability should show a similar time trend in continuum level for both detector arrays, which can be seen for CW Leo in Fig.~\ref{fig:meanContinua}. This Figure shows that there is no significant trend with time for any of the four main line sources, but that there is a larger scatter for SSW. This difference indicates that pointing stability is a more significant source of uncertainty on the continuum than intrinsic variability for these sources, which appears negligible. There is a high scatter for AFGL4106 in SLW, due to the extended background. 

The spread of continua, as a percentage of the mean level is shown to the bottom right of Fig.~\ref{fig:pointingOffset}, and ranges from 5\% to 15\% for data not corrected for pointing offset, with mean values for the four main line sources of 4.4\% for SLW and 13.6\% for SSW. After correcting for pointing offset, these values fall to less than 2\%, however the relative method used to correct the data does inherently correct for intrinsic variability, by assuming the single reference observation for a particular source should match all other observations for that source, regardless of any brightness variation. The spread in continuum levels found are consistent with those for the associated line measurements, which are discussed in Section~\ref{sec:repeatabilityLineSources}.

\section{Planet and asteroid model comparison}\label{sec:ratios}

\subsection{Planets}\label{sec:ratiosPlanets}

Both Uranus and Neptune were regularly observed with the FTS over the whole \textit{Herschel} mission, and both can be considered point sources within the FTS beam \citep{Swinyard2014}.
As mentioned in Section~\ref{sec:calProg} and detailed in \citet{Swinyard2014}, the primary FTS point-source flux calibrator is Uranus. The ESA-4 model of Uranus\footnote{\label{foot:model} The ESA-4 models for Uranus and Neptune are available at\\ ftp://ftp.sciops.esa.int/pub/hsc-calibration/PlanetaryModels/ESA4/.} \citep{Orton14} is used to derive the point-source conversion factor as the ratio of model to observation, after the data have been corrected for pointing offset. The ESA-4 model for Neptune\textsuperscript{\ref{foot:model}} \citep{Moreno98} is also available and is derived independently from the model for Uranus, so an assessment of the repeatability and accuracy of the point-source calibration can be made by comparing data with model predictions for these two planets.

\begin{figure}
\centering
\includegraphics[trim = 0mm 35mm 0mm 0mm, clip,width=0.5\textwidth]{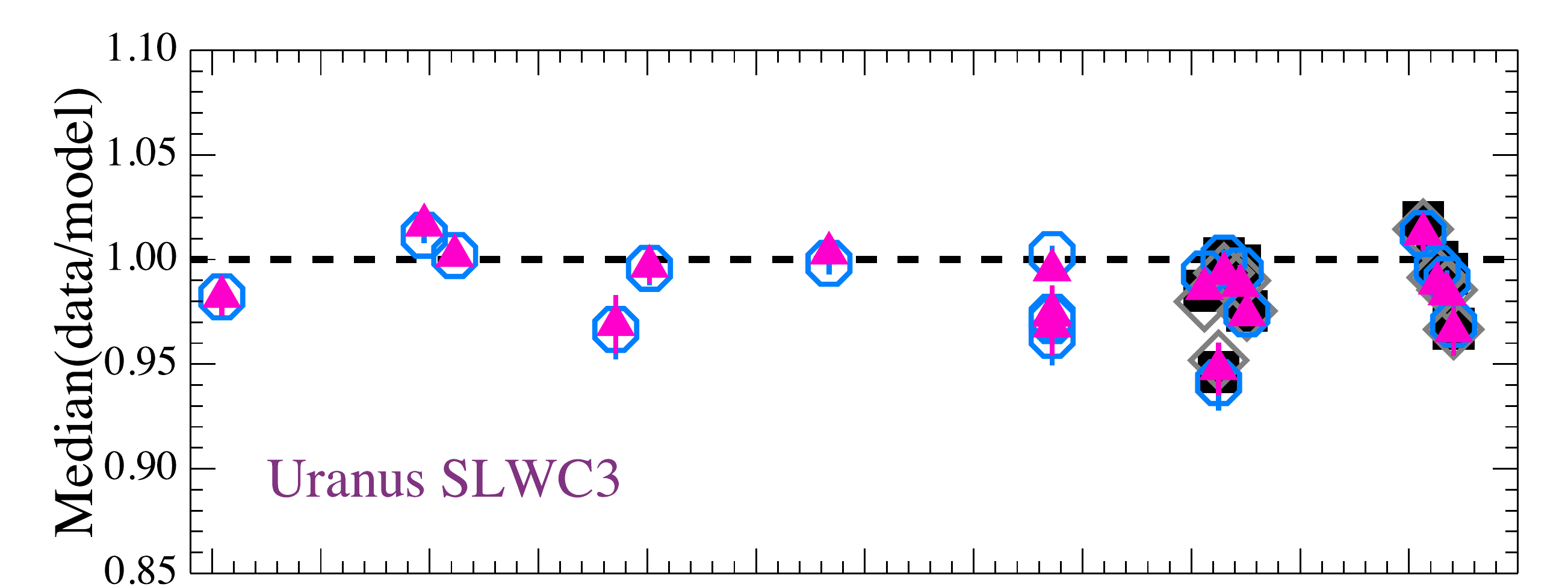}
\includegraphics[trim = 0mm 0mm 0mm 5.5mm, clip,width=0.5\textwidth]{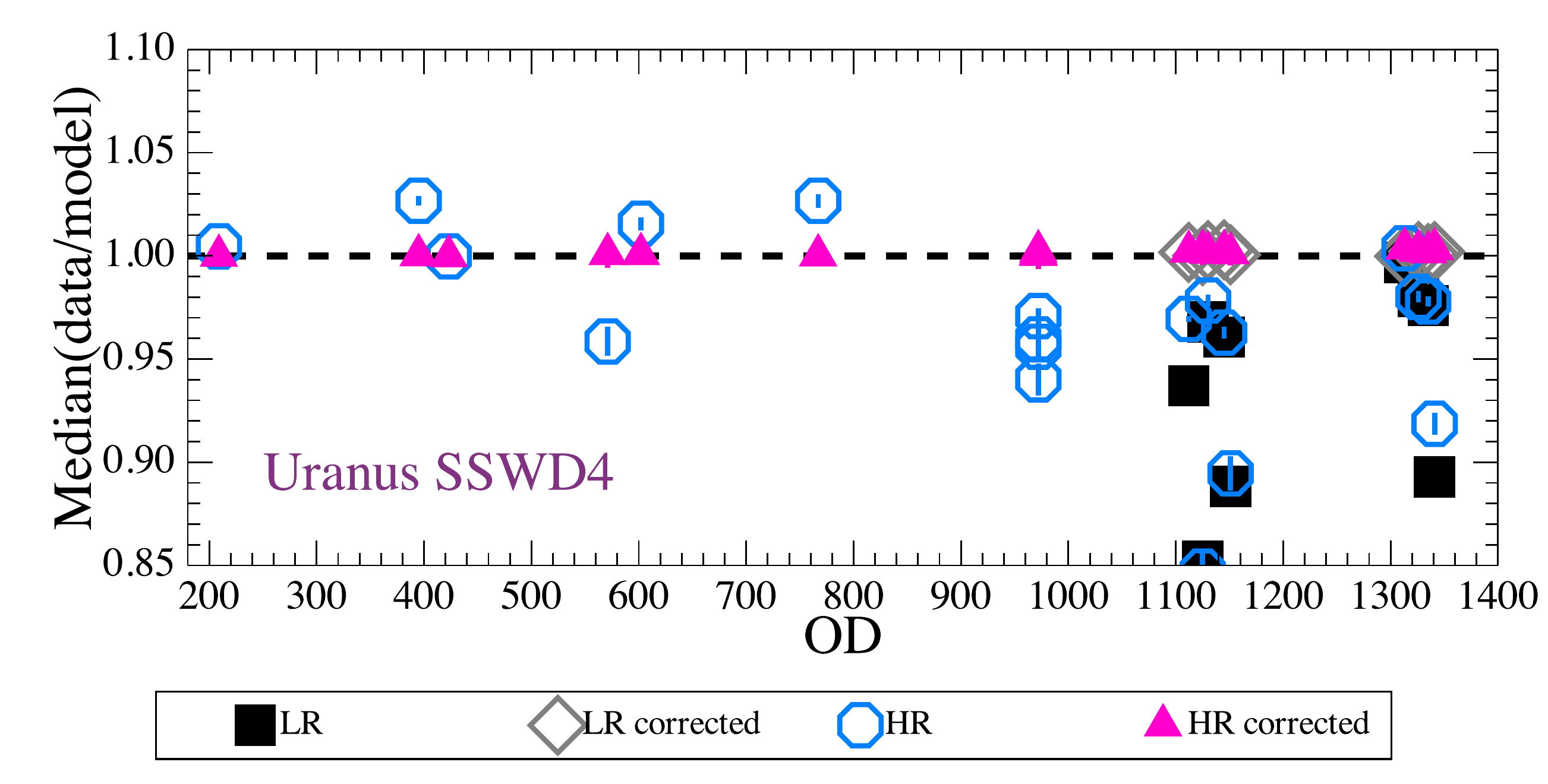}
\includegraphics[trim = 0mm 35mm 0mm 0mm, clip,width=0.5\textwidth]{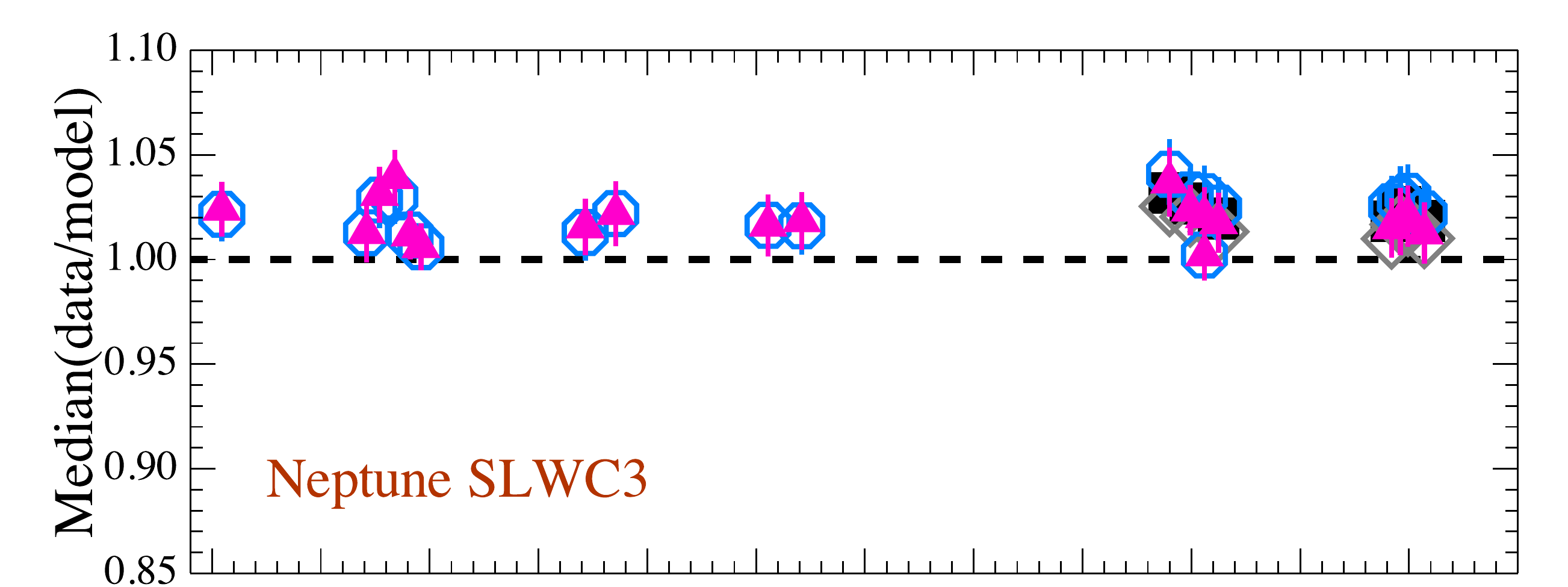}
\includegraphics[trim = 0mm 0mm 0mm 5.5mm, clip,width=0.5\textwidth]{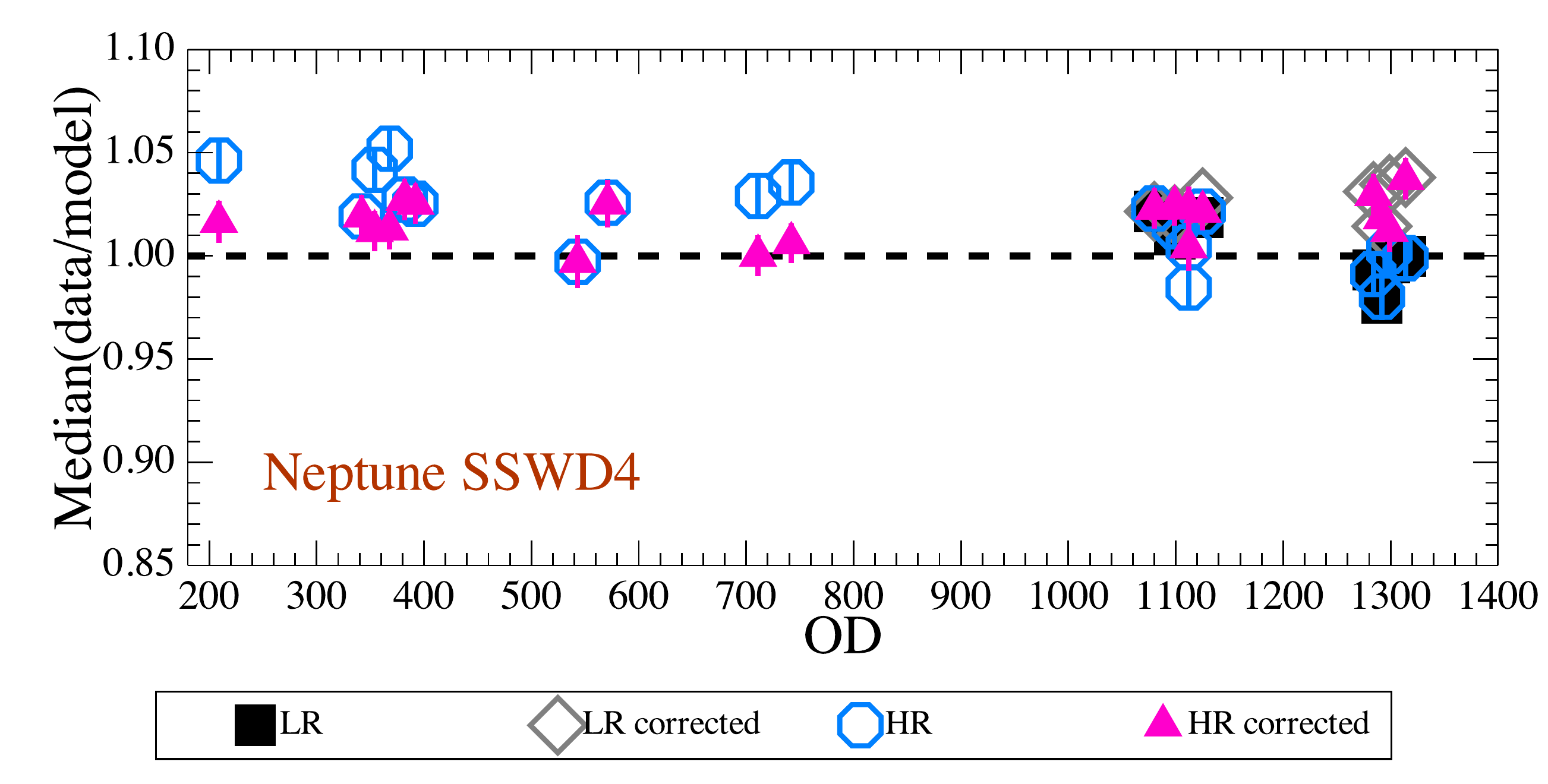}
\caption{Model to data ratios for Uranus and Neptune. There is very low scatter after correcting for pointing offset (``corrected''), although the ratios for Neptune show a systematic offset from 1.0 of 2\%.}
\label{fig:ratiosNeptuneUranus}
\end{figure}

Ratios of data to the respective model were taken for all HR and LR Uranus and Neptune observations. For each ratio a restricted frequency range was used, due to higher noise at the band edges, particularly below 600\,GHz in SLW. The ranges used were 600--900\,GHz for SLW and 1100--1500\,GHz for SSW. To provide a single value per observation, the median ratio was  taken across the frequency range, and the standard error on the mean taken as a 1\,$\sigma$ uncertainty. Fig.~\ref{fig:ratiosNeptuneUranus} shows the resulting ratios with and without correcting the data for pointing offset.

Before correcting for pointing offset, the average HR ratio for Uranus shows an offset from 1.0 of 1\% and 3\% for SLW and SSW, with uncertainties of 2\% and 5\%, respectively (see Table~\ref{tab:ratios}). After correcting for pointing offset, the SLW ratios are approximately unchanged, as expected, whereas HR SSW ratios improve to a mean ratio of 1.0, with $<$1\% spread. These results are consistent with the spectral line flux repeatability presented in Section~\ref{sec:lineRepeatability}. For Neptune the offset from a ratio of 1.0 is 2\% regardless of correcting the data for pointing offset, however the associated spread for SSW improves  from 2\% to 1\% after correction. Before pointing offset is corrected, the average LR Uranus ratio is 0.99$\pm$0.02 for SLW and 0.96$\pm$0.05 for SSW. After correction, the uncertainty of these ratios improve to 1\% for SSW, with a mean ratio of 1.0. For Neptune the average LR ratios are 1.02$\pm$0.01 for SLW and 1.00$\pm$0.02 for SSW. After correcting the LR data for pointing offset, the ratios show a more consistent 2\% shift from 1.0, with the associated SSW scatter reduced to 0.01. The systematic shift in the ratios of 2\%, when comparing to the Neptune model, is consistent with the findings of \citet{Swinyard2014}. There is more consistency between the ``before'' and ``after'' observations and between the two bands for both sets of ratios, compared to those presented in \citet{Swinyard2014}, which is due to improvements in the pointing offset correction folded into the point-source calibration.

\subsection{Asteroids}\label{sec:ratiosAsteroids}

\begin{figure}
\centering
\includegraphics[trim = 5mm 35mm 0mm 0mm, clip, width=0.5\textwidth]{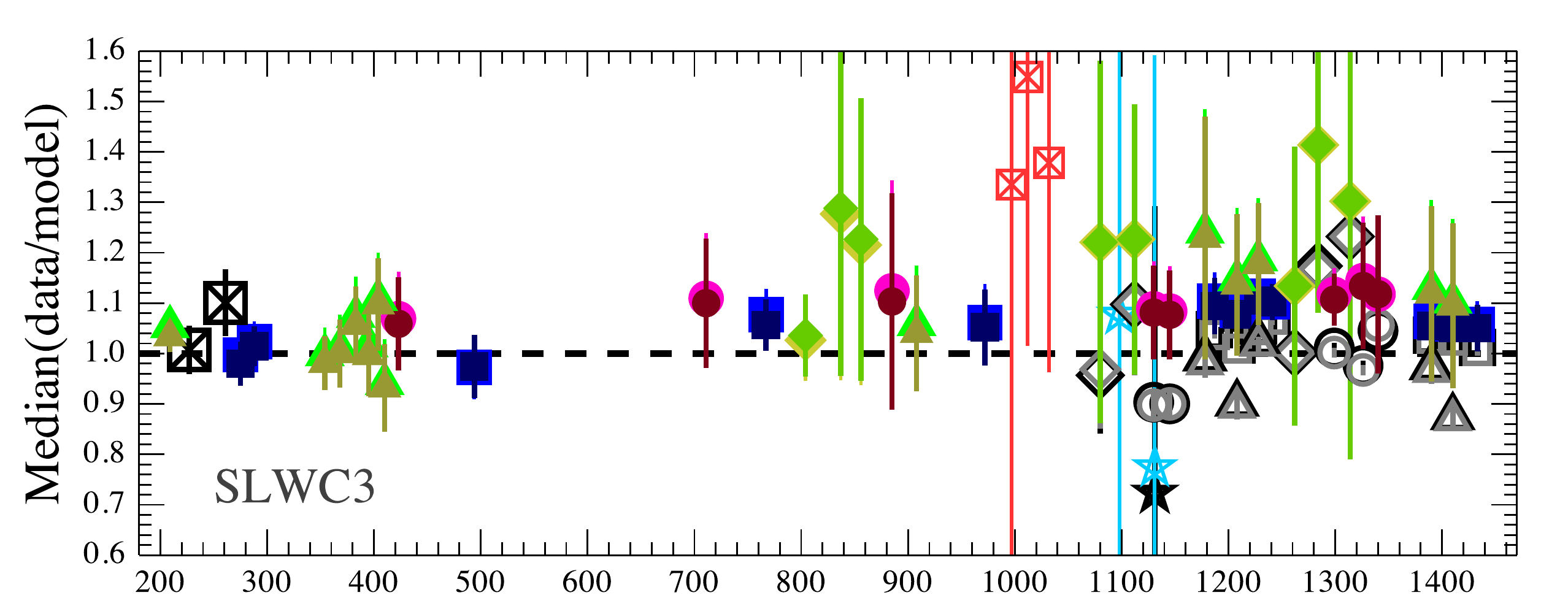}
\includegraphics[trim = 5mm 0mm 0mm 5mm, clip, width=0.5\textwidth]{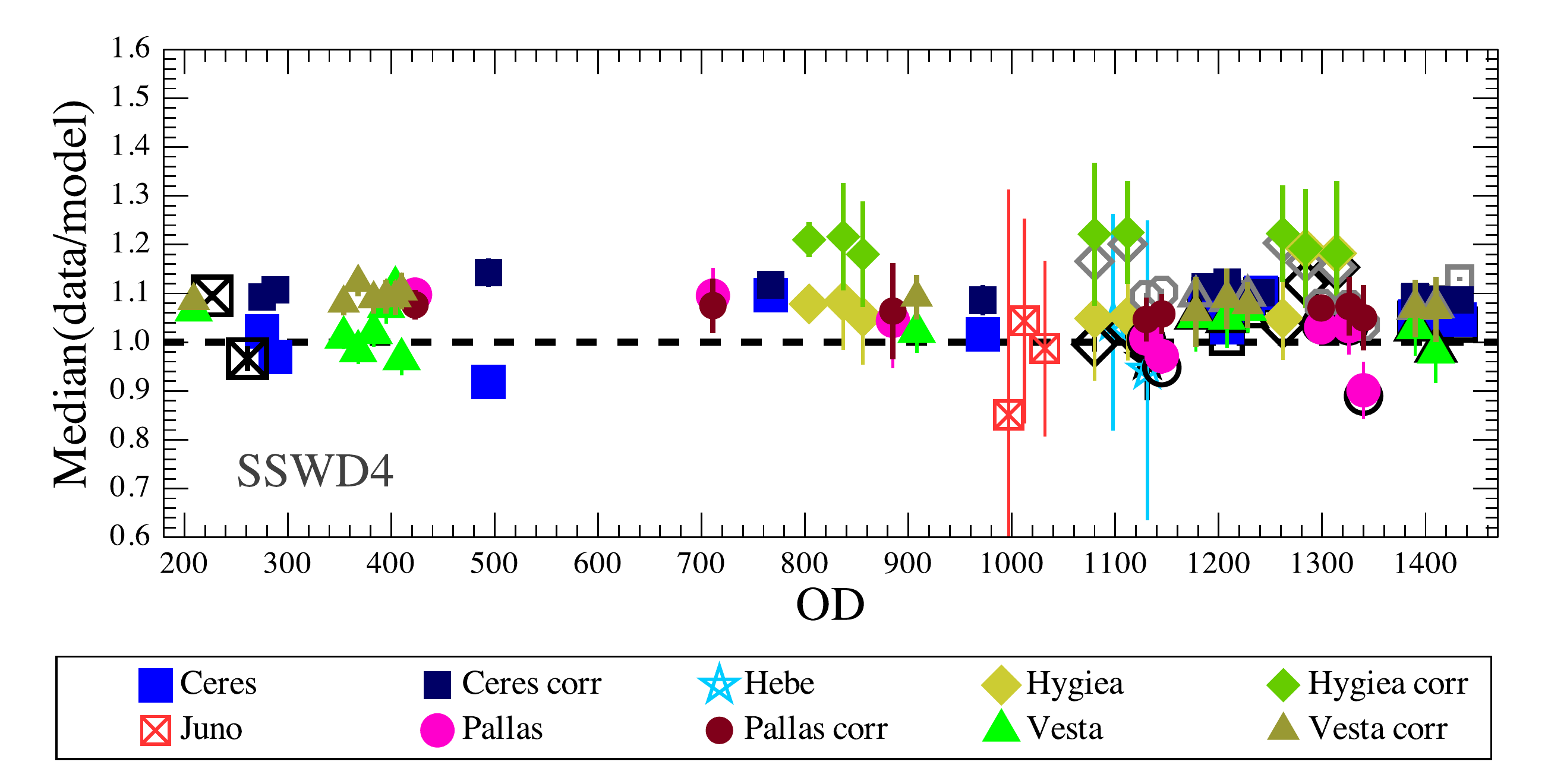}
\caption{Data to model ratios for several asteroids. HR ratios are shown with filled symbols and LR ratios with the corresponding open symbols. The LR symbol for Juno is the same cross filled square, but larger in size (and at an earlier OD to the HR observations). `corr' indicates the ratios for data corrected for pointing offset. Ceres, Pallas and Vesta have ratios consistently nearest to 1.0 due to their more recent models of \citet{Muller13}. The average ratios for these three asteroids after correcting for pointing offset are 1.06$\pm$0.07 for SLW and 1.09$\pm$0.02 for SSW. Note that due to a higher scatter in the ratio values, the y-scaling is wider for these plots than the planet equivalent ratio plots in Fig.~\ref{fig:ratiosNeptuneUranus}.}
\label{fig:ratiosAsteroids}
\end{figure}

Repeated FTS observations of a number of asteroids were compared to the models of \citet{Muller02} and \citet{Muller13}. All observations were background subtracted using the off-axis detectors, as detailed in section~\ref{sec:bgs}. The ratio of data to model was taken following the same method described in Section~\ref{sec:ratiosPlanets}, and the results are presented in Fig.~\ref{fig:ratiosAsteroids} and Table~\ref{tab:ratios}.
The average ratio for each asteroid tends to sit above 1.0, by up to 22\% (i.e. the model falls short of the data), except for Hebe and Juno, which show worse results, but are both faint and have the fewest number of observations available for HR and LR. For the HR ratios, those sources with more recently updated models \citep[Ceres, Pallas and Vesta;][]{Muller13} give an average ratio of 1.06$\pm$0.07 for SLW and 1.04$\pm$0.05 for SSW, for data not corrected for pointing offset. The average ratio for SLW is unchanged after correcting for pointing offset, but for SSW, although the spread in ratios decreases to 2\%, the average ratio increases to 1.09. The pointing corrected result shows there is a consistent offset between the data and the models of $\sim$6--9\%, which is higher than the associated model uncertainties of 5\% quoted by \citet{Muller13}, who explain that such a discrepancy may be due to {\it Herschel} visibility constraints limiting the phase angles tested, or the accuracy with which the asteroid's shapes could be characterised.

\begin{table}
\caption{Model to data ratios for Uranus, Neptune, and the asteroids. Each ratio is the median for all HR or LR observations and presented as ratio$\pm$1\,$\sigma$. The number of observations included in the average ratio is given in brackets in the ``Name(\#)'' column.}

\medskip
\begin{center}
\begin{tabular}{lcccc}
\hline\hline
\multicolumn{5}{c}{\textbf{HR}}\\ \hline\hline
 & \multicolumn{2}{c}{Without pointing corr.} & \multicolumn{2}{c}{With pointing corr.}\\ \hline
Name(\#) & SLWC3 & SSWD4 & SLWC3 & SSWD4 \\ \hline
Uranus(21) & 0.99$\pm$0.02 & 0.97$\pm$0.05 & 0.99$\pm$0.02 & 1.00$\pm$0.00 \\ 
Neptune(21) & 1.02$\pm$0.01 & 1.02$\pm$0.02 & 1.02$\pm$0.01 & 1.02$\pm$0.01 \\ \hline
Ceres(13) & 1.06$\pm$0.04 & 1.05$\pm$0.05 & 1.05$\pm$0.04 & 1.10$\pm$0.02 \\ 
Hebe(3) & 0.92$\pm$0.21 & 0.99$\pm$0.07 & --- & --- \\ 
Hygiea(8) & 1.22$\pm$0.11 & 1.07$\pm$0.06 & 1.23$\pm$0.11 & 1.21$\pm$0.02 \\ 
Juno(3) & 1.38$\pm$0.11 & 0.99$\pm$0.10 & --- & --- \\ 
Pallas(8) & 1.11$\pm$0.02 & 1.03$\pm$0.06 & 1.10$\pm$0.02 & 1.07$\pm$0.01 \\ 
Vesta(13) & 1.07$\pm$0.08 & 1.03$\pm$0.04 & 1.06$\pm$0.08 & 1.09$\pm$0.02 \\ \hline\hline
\multicolumn{5}{c}{\textbf{LR}}\\ \hline\hline
 & \multicolumn{2}{c}{Without pointing corr.} & \multicolumn{2}{c}{With pointing corr.}\\ \hline
Name(\#) & SLWC3 & SSWD4 & SLWC3 & SSWD4 \\ \hline
Uranus(9) & 0.99$\pm$0.02 & 0.96$\pm$0.05 & 0.99$\pm$0.02 & 1.00$\pm$0.01 \\ 
Neptune(7) & 1.02$\pm$0.01 & 1.00$\pm$0.02 & 1.02$\pm$0.01 & 1.03$\pm$0.01 \\ \hline
Ceres(7) & 1.04$\pm$0.02 & 1.05$\pm$0.03 & 1.03$\pm$0.02 & 1.10$\pm$0.01 \\ 
Hebe(2) & 0.72$\pm$0.00 & 0.96$\pm$0.00 & --- & --- \\ 
Hygiea(5) & 1.10$\pm$0.11 & 1.03$\pm$0.07 & 1.10$\pm$0.11 & 1.17$\pm$0.02 \\
Juno(2) & 1.05$\pm$0.07 & 1.03$\pm$0.09 & --- & --- \\ 
Pallas(5) & 0.98$\pm$0.06 & 1.00$\pm$0.06 & 0.97$\pm$0.07 & 1.08$\pm$0.03 \\
Vesta(5) & 0.97$\pm$0.06 & 1.04$\pm$0.03 & 0.96$\pm$0.06 & 1.08$\pm$0.01 \\ 
\hline
\end{tabular}
\end{center}
\label{tab:ratios}
\end{table}

\section{Comparison with the SPIRE photometer measurements}\label{sec:specPhot}

The primary photometer calibrator is Neptune \citep[see][]{Bendo13}, therefore an independent check of FTS calibration can be made with the SPIRE photometer. Two comparisons are made in this section. Firstly, for several of the FTS line sources, synthetic photometry, carried out by integrating the FTS spectra over the SPIRE photometer wavebands, is compared to the average photometry taken from corresponding photometer maps. Secondly, the asteroid ratios discussed in Section~\ref{sec:ratiosAsteroids} are compared to the equivalent ratios taken for the SPIRE photometer, and presented in Lim et al. (in preparation).

For the most accurate photometry of point-like sources with flux density $>$\,20\,mJy, \citet{Pearson2014} recommends the Timeline Fitter \citep{Bendo13}, which is available as a task in HIPE. The task fits a 2D Gaussian to all photometer bolometer timeline readouts, which are near to the source. For each photometer observation included in the comparison, the data were downloaded from the {\it Herschel} Science Archive. All data were reduced using version 11 of the SPIRE pipeline, using the \texttt{spire\_cal\_11\_0} calibration. The timeline fitter task (\textsc{sourceExtractorTimeline}) was run for each of the SPIRE photometer short, medium and long wavebands (250\,$\mu$m PSW, 350\,$\mu$m PMW and 500\,$\mu$m PLW), using the default settings, except for \textsc{useBackInFit} and \textsc{allowVaryBackground}, which were set to \textsc{True}, with an \textsc{rBack} of 300 and 301, for the inner and outer background radius. The Timeline Fitter does not perform a source extraction, so the nominal Spectrometer RA and Dec were used as the coordinate input. The average photometry was taken for each source, and for each band, and the associated uncertainties added in quadrature. Photometer observations of line sources and stars that are used in this paper are summarised in Table~\ref{tab:specMatchPhot}.

\begin{figure}
\centering
\includegraphics[width=\hsize]{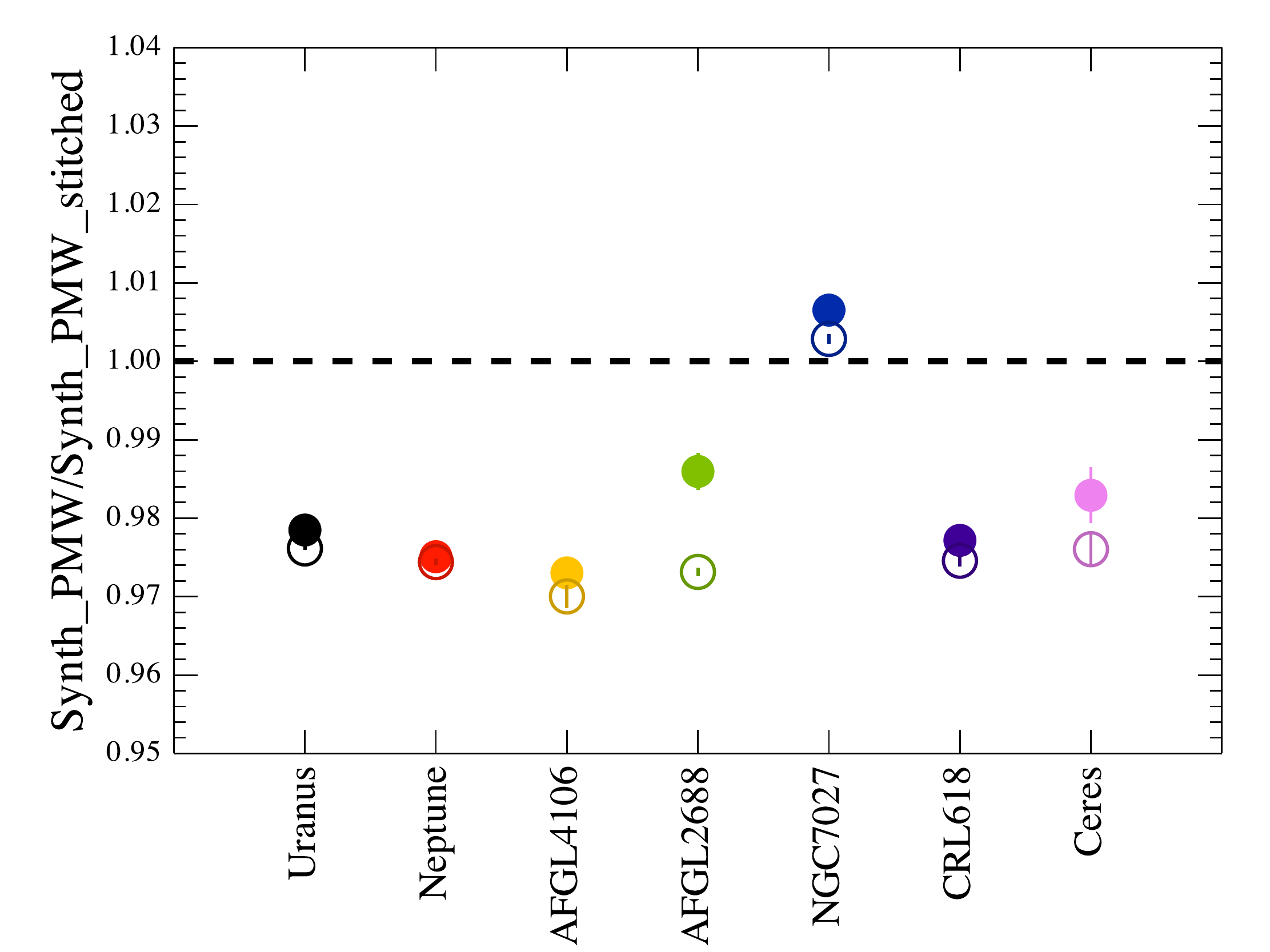}
\caption{Synthetic FTS photometry ratios for the photometer PMW band, comparing results using SLW only with those using SLW and SSW joined together using the mean of the two spectra to replace the overlap region (Synth\_PMW\_stitched). Filled circles show the ratios for data corrected for pointing offset and open circles for uncorrected ratios. NGC7027 is partially extended and AFGL4106 is embedded in a background, so excluding these, it can be seen that approximately 2\% of the synthetic PMW photometry is generated by the region lying within SSW.}
\label{fig:synthPhotCheck}
\end{figure}

To generate synthetic FTS photometry over a SPIRE photometer waveband, a point-source-calibrated spectrum was weighted by the relevant photometer RSRF, including the aperture efficiency, and the result integrated. The method followed is detailed in the \citet{spireHandbook}. 
Taking synthetic photometry for PMW is complicated by a small fraction of the PMW RSRF extending into the SSW frequency range. An SLW spectrum and an SSW spectrum can be joined to ensure the use of the full RSRF range, which increases the synthetic PMW photometry by approximately 2\% compared to only using the SLW spectrum.
This fraction of the photometry from the SSW band for PMW was estimated using all the synthetic photometry for point-like sources and comparing with and without SSWD4 included, see Fig.~\ref{fig:synthPhotCheck}. As already noted, FTS data can often exhibit an offset between the signals in the overlap region of the SLW and SSW bands. This gap can arise due to an imperfect telescope subtraction, a significant background or a semi-extended source. In addition, there is also an increase in noise at the edge of the bands, so the small percentage of the synthetic photometry obtained from SSWD4 is not generally significant when compared to the uncertainty associated with this fraction. The data corrected for pointing offset was used for the comparison with the photometer.

\begin{figure}
\centering
\includegraphics[trim = 8mm 0mm 8mm 5mm, clip,width=1.0\hsize]{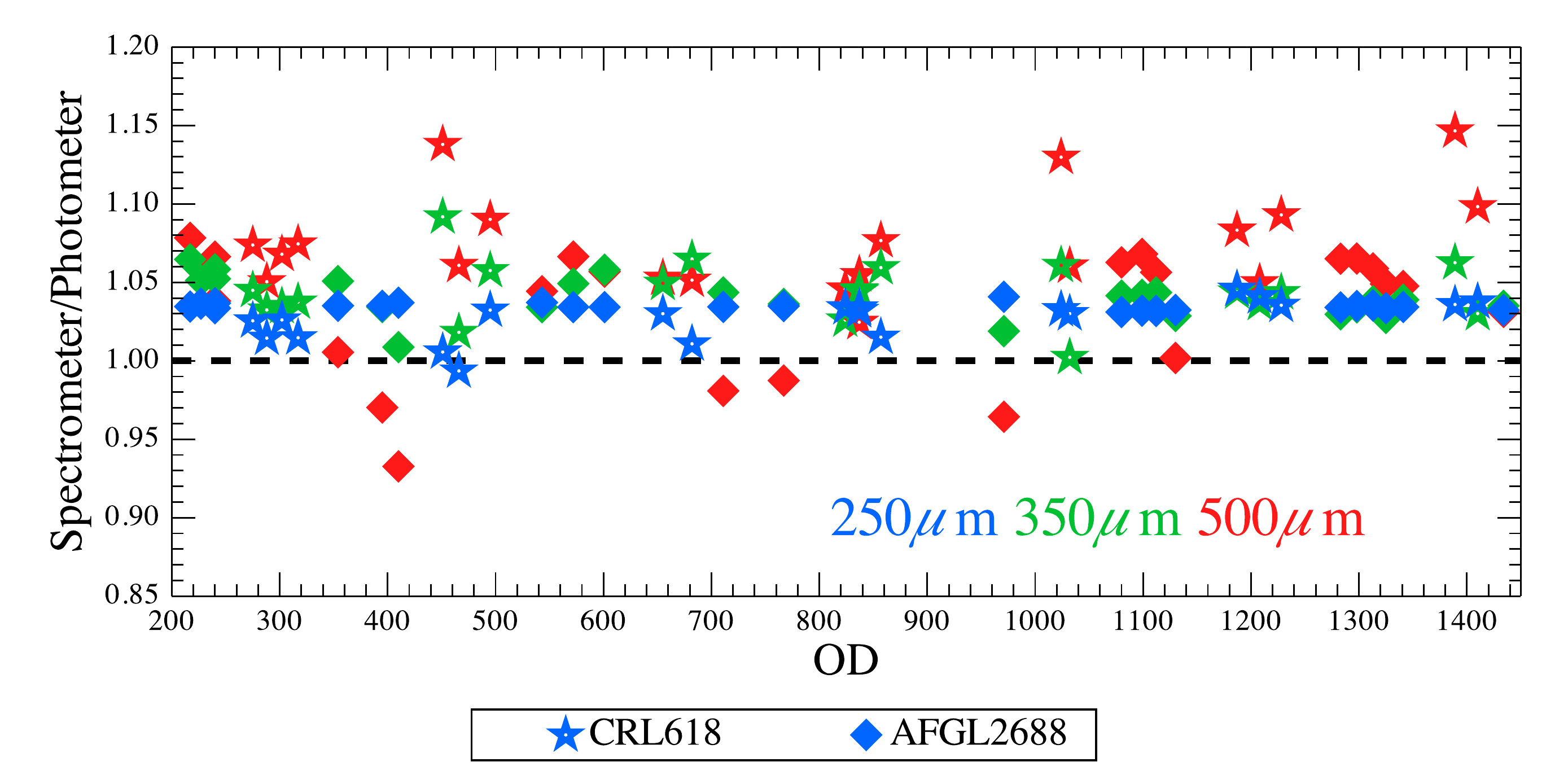}
\caption{Synthetic FTS photometry compared to photometer photometry. The calibration sources included were chosen with reference to Fig.~\ref{fig:fluxInRingsFullFreq} (to identify the most point-like), and restricted to those with available photometer maps and good FTS OD coverage. The PSW points (250\,$\mu$m blue) show the least scatter of 1\% and have an average ratio of 1.03. The average ratio for PMW is 1.04$\pm$0.02. PLW (500\,$\mu$m red) has the highest offset of 5\% and an associated scatter of 4\%.}
\label{fig:synthPhotRatios}
\end{figure}

Using Fig.~\ref{fig:fluxInRingsAve}, CRL618 and AFGL2688 were chosen as the most point-like sources that also have reasonable FTS OD coverage. Both FTS bands were used to derive the synthetic PMW photometry for AFGL2688, as there is no significant step between the bands. The CRL618 PMW synthetic photometry was increased by 2\% to correct for the missing fraction due to the overlap of photometer RSRF into the spectrometer SSW band. The synthetic photometry was converted to the photometer pipeline convention of a monochromatic flux density for a source with a spectrum $S_\nu \sim \nu^{-1}$. The necessary conversion factors (K$_{\rm 4P}$) of 1.0102 (PSW), 1.0095 (PMW) and 1.0056 (PLW) were taken from the \citet{spireHandbook}.
The ratios of FTS synthetic photometry (per observation) to average photometer photometry are shown in Fig.~\ref{fig:synthPhotRatios}.  Overall the PSW ratios are closest to 1, with an increasing systematic offset seen with increasing wavelength (decreasing frequency). 
The mean ratios are 1.03$\pm$0.01 for PSW, 1.04$\pm$0.02 for PMW and 1.05$\pm$0.04 for PLW. Fig.~\ref{fig:ratiosNeptuneUranus} indicates a systematic offset for FTS data when compared to the model of Neptune of 2\%, which when coupled with Fig.~\ref{fig:fluxInRingsAve} (indicating a slight extent of CRL618 and AFGL2688), explains the difference between the photometer and spectrometer photometry, as even a small departure from point-like can lead to flux being missed by the the Timeline fitter.

\begin{figure*}
\centering
\includegraphics[trim = 9mm 29mm 5mm 5mm, clip,width=1.0\hsize]{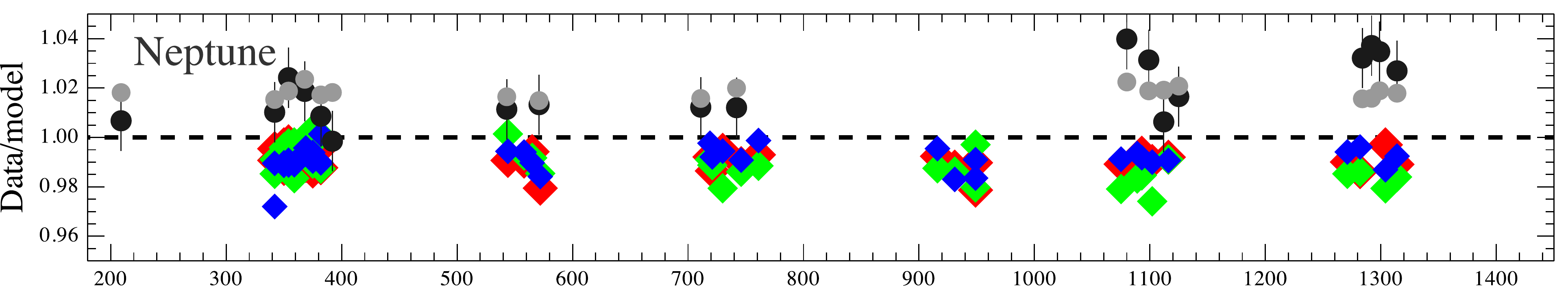}
\includegraphics[trim = 7mm 29mm 5mm 5mm, clip, width=1.0\hsize]{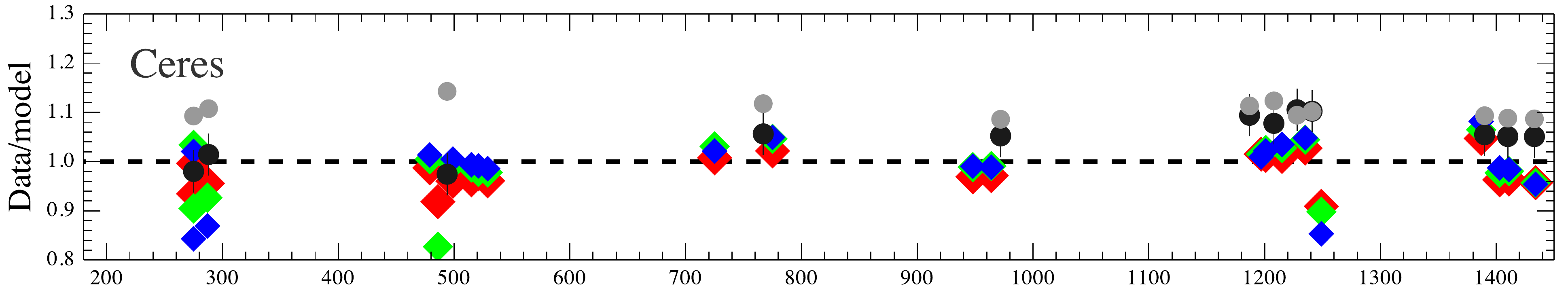}
\includegraphics[trim = 7mm 29mm 5mm 5mm, clip, width=1.0\hsize]{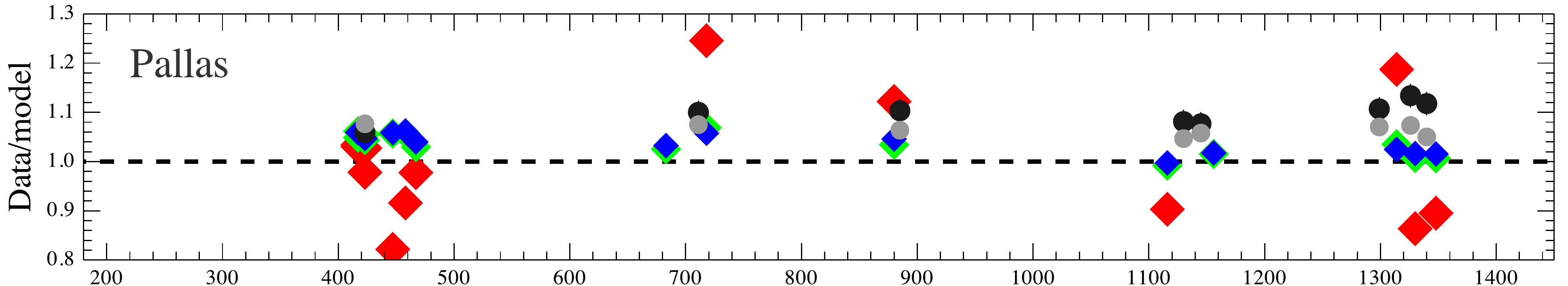}
\includegraphics[trim = 7mm 0mm 5mm 5mm, clip, width=1.0\hsize]{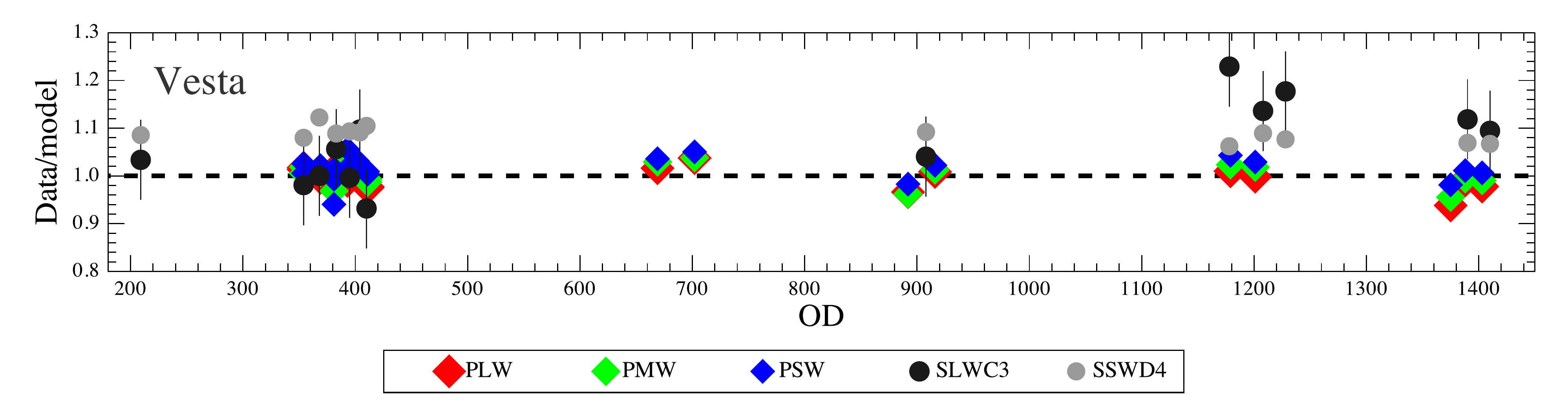}
\caption{Data to model ratios for Neptune, Ceres, Pallas and Vesta. Photometer ratios for PLW, PMW and PSW are represented by diamonds and the spectrometer SLW and SSW ratios by  circles. The FTS data are corrected for pointing offset. Taking into consideration the 2\% difference already found for FTS data, compared to the Neptune model, there is good consistency between the instruments over the whole OD range. Note the y-scale for Neptune is not the same as for the asteroids.}
\label{fig:specPhotAst}
\end{figure*}

The second comparison with the photometer involves the asteroid and Neptune ratios from Lim et al. (in preparation), which are used in Section~\ref{sec:ratios}. Although there is a systematic uncertainty of 6--9\% between the asteroid models and FTS data, a comparison with the photometer is still an interesting exercise to look for consistency over a wide OD range. As already discussed, Neptune is the primary photometer calibrator, and therefore should provide consistent data to model ratios. Fig.~\ref{fig:specPhotAst} shows the ratio comparison for Neptune, Ceres, Pallas and Vesta. All sets of ratios are consistent between the photometer and spectrometer, but do show the 2\% systematic shift that was found in Section~\ref{sec:ratiosPlanets} and attributed to the use of different primary calibrators -- Uranus for the FTS and Neptune for the photometer.

\section{Dark sky observations}\label{sec:darkSky}

The \textit{Herschel} telescope operated at a temperature of 87--90\,K, so its emission dominates nearly all nominal mode FTS observations and requires precise removal during data processing. Observations of dark sky provide a measure of the telescope emission, and are therefore crucial for monitoring FTS calibration accuracy\footnote{A list of all sparse FTS dark sky observations can be found at\\ 
\url{http://herschel.esac.esa.int/twiki/bin/view/Public/SpireDailyDarkObservations}\,.}.
Two significant changes were made in the approach and scheduling of repeated dark sky observations during the mission. From OD\,1079 (27th April 2012) onwards, when the switch away from CR mode occurred, separate HR and LR darks were regularly taken. From OD\,466 (23rd August 2010), a long dark sky observation, at least as long as the longest science observation that day, was taken on each pair of FTS ODs. The primary reason for dedicating a relatively large amount of observing time to dark sky was for the subtraction from science observations, taken the same day (or under similar observing conditions). However, improvements in calibration and understanding of the telescope model \citep{Hopwood2014} mean that a daily dark subtraction is no longer necessary for the majority of observations. The substantial set of dark sky observations available allow an in-depth assessment of overall FTS performance and is used for several key purposes -- for deriving the instrument and telescope RSRFs \citep[see][]{Fulton14a}; for deriving a correction to the telescope model \citep[see][]{Hopwood2014}; to assess FTS sensitivity (see Section~\ref{sec:sensitivity}); to estimate the error on the continuum offset (see Section~\ref{sec:continuumOffset}); and for several other one-off or repeated diagnostic tests, some of which are discussed in this paper.

\subsection{How dark is the dark sky?}\label{sec:stacking}

With an extensive set of dark sky observations available, the ``darkness'' of the nominated SPIRE dark field can be assessed via stacking. All HR dark sky (point-source-calibrated) spectra, with more than five repetitions, were stacked to form spectral cubes. The resulting cubes were then checked for detections.

Although all centred on the same sky coordinates, the set of dark sky observations used were taken at different times over the course of several years, and so their FTS footprints see different rotations (see Fig.~\ref{fig:bsm}, bottom left). Therefore, in order to stack these observations, they were treated as independent measurements at different sky positions, and averaged onto a regular grid using the Na\"{i}ve projection algorithm provided in HIPE. Both an SLW and an SSW spectral cube were generated, equivalent to those obtained from SPIRE FTS mapping observations (Fulton et al. in preparation).

Two versions of cubes were made, one where the large scale shape was removed (i.e. background subtracted) and a second set without this subtraction. The former was checked for spectral line detections at the positions of peaks seen in a stacked PSW map of photometer dark sky, and the latter checked for any clear continua. No significant lines or continua were found. The stacking of the dark sky will be presented in more detail in a separate publication.

\subsection{Uncertainty on the continuum}\label{sec:continuumOffset}

Dark sky observations provide a means to estimate the uncertainty expected on continuum measurements (the continuum offset), which arises from imperfect subtraction of the telescope contribution and, for the lower frequency end of SLW, the instrument contribution. These contributions are fully extended in the FTS beam, and therefore any residual leads to large-scale systematic noise in the continuum of point-source-calibrated spectra.

To assess the uncertainty associated with this residual for both extended and point-source-calibrated spectra, un-averaged HR dark sky observations, with more than 20 repetitions, were used. After excluding extreme outliers, all scans (9724) were smoothed with a Gaussian kernel of FWHM of 21\,GHz, which removes small-scale noise to provide the wide-scale shape. For each frequency bin, the standard deviation was taken across all smoothed scans, to give the 1\,$\sigma$ continuum offset, which is an additive uncertainty.

\begin{figure}
\centering
\includegraphics[trim = 8mm 18mm 8mm 5mm, clip,width=0.9\hsize]{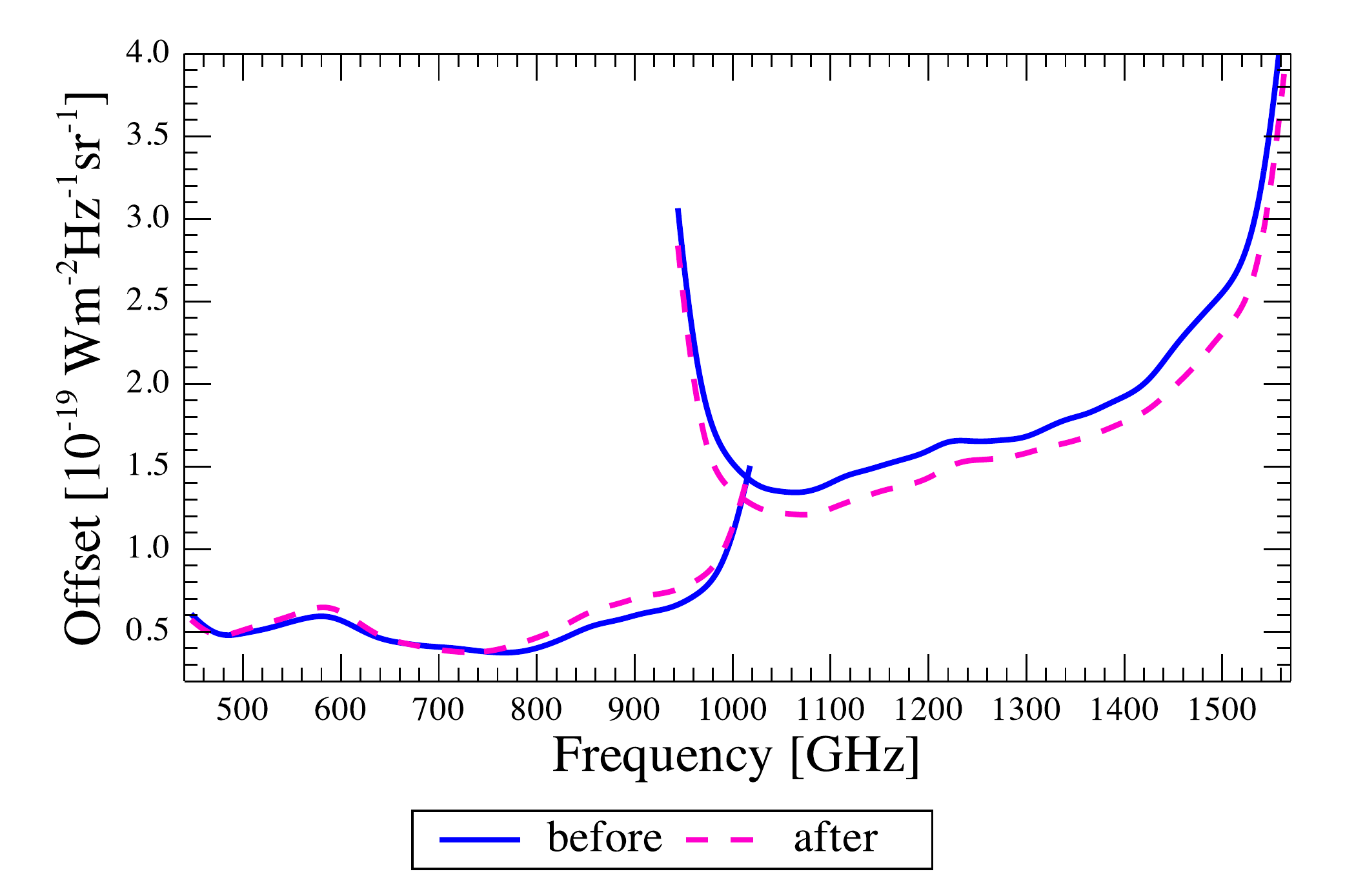}
\includegraphics[trim = 8mm 0mm 8mm 5mm, clip,width=0.9\hsize]{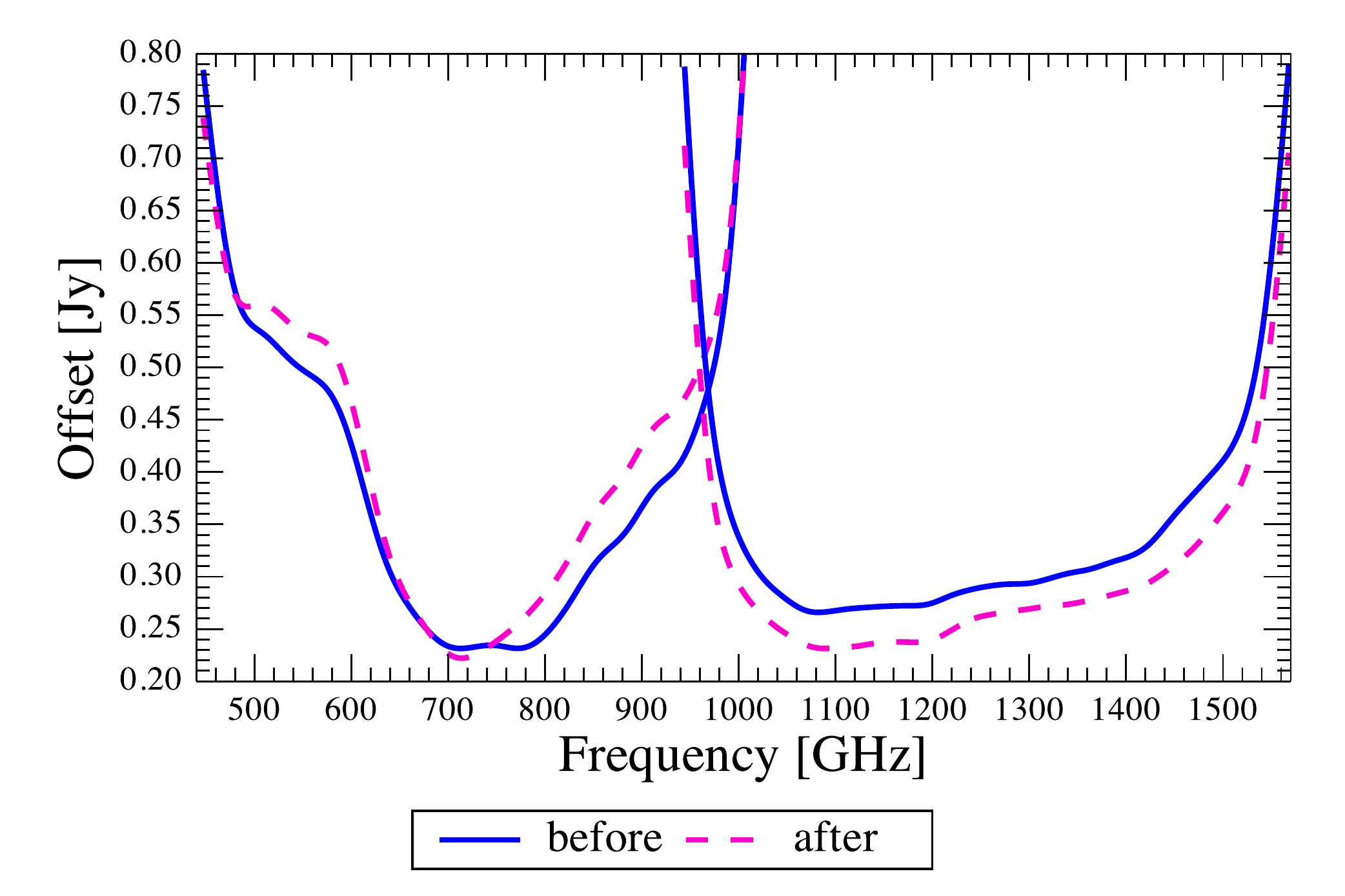}
\caption{Continuum offset for extended-calibrated data (top) and point-source-calibrated data (bottom). The offset is given separately for the two BSM epochs (`before' OD 1011 and on or `after' OD 1011). The median values for the point-source-calibrated data are 0.40\,Jy for SLWC3 and 0.28\,Jy for SSWD4.}
\label{fig:contOffCent}
\end{figure}

\citet{Swinyard2014} present the continuum offset for the centre detectors, using data reduced with HIPE 11. Here we update that result for the wider FTS bands (released with HIPE version 12.1) and improved non-linearity correction introduced in HIPE 13, and present median values for all detectors. Fig.~\ref{fig:contOffCent} shows the point source and extended-calibrated results for the centre detectors, for {\it Before} and {\it After} the BSM correction on OD\,1011. For SSWD4 there is at most a 0.05\,Jy higher uncertainty for {\it Before} data, which is partially due to fewer dark sky observations with $>$\,50 scans available for the estimate, with all HR observations suffering a higher scatter up to the first 20 scans, and due to an improved instrument stability after $\sim$OD\,500. The SLWC3 offset for {\it Before} data is reduced with HIPE 13, due to an improved non-linearity correction, which better calibrates observations taken at the beginning of each pair of FTS ODs. The non-linearity correction for SSWD4 is not significantly changed. The offset is higher at the ends of each frequency band, and the strong influence of the instrument residual can be seen in the lower half of SLWC3 for both calibrations. For point-source-calibrated data, the average uncertainty on the continuum is 0.40\,Jy for SLWC3 and 0.28\,Jy for SSWD4. 

Considering the continuum offset reduction for the centre detectors, since HIPE version 7, there was an average 45\% reduction seen for HIPE 9, due to the introduction of a telescope model correction \citep[see][for details on the derivation of this correction]{Hopwood2014}. Updates to this correction, along with improvements to the RSRFs and point source calibration, have seen this reduction improve to $\sim$60\% for SLW and $\sim$50\% for SSW, for HIPE 10, with a further reduction of up to 72\% and 62\% for SLW and SSW in HIPE 13. This comparison was assessed using a subset of {\it Before} dark sky, so the same set of observations could be used for each version considered. 

The average HIPE 13 continuum offset for all detectors, calculated using the full set of dark sky, are presented in Table~\ref{tab:offsetSigma13}.
The median offsets for all detectors show a similar reduction over evolving versions of HIPE. Except for SSWE2, there is good consistency for offset levels across all detectors, with some scatter for the vignetted SLW detectors. SSWE2 is sensitive to clipping (Fulton et al. in preparation), which causes the higher continuum offset for this detector. 

\subsection{Sensitivity}\label{sec:sensitivity}

\begin{figure}
\begin{center}
\includegraphics[trim = 8mm 16mm 8mm 3mm, clip,width=0.9\hsize]{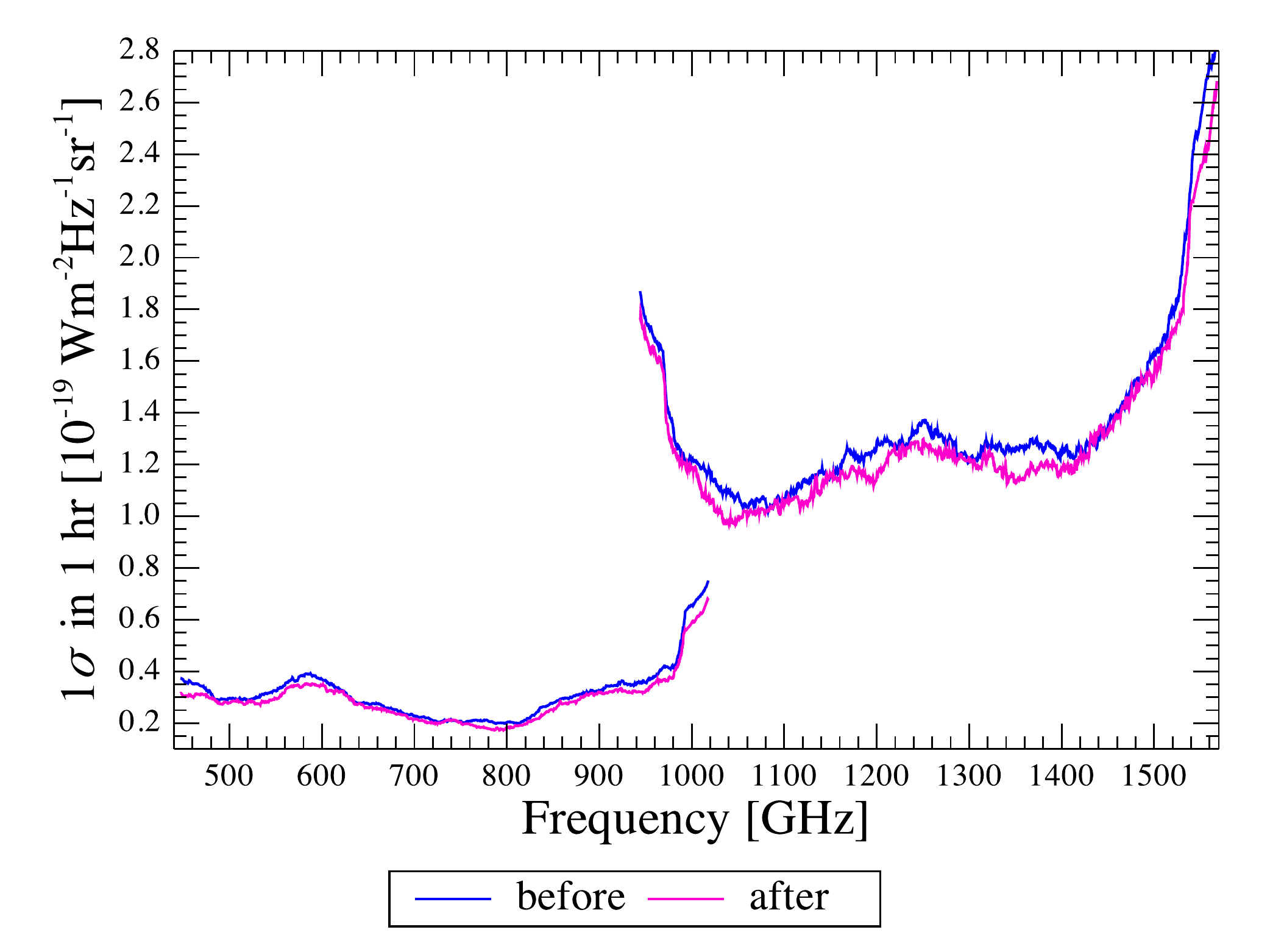}
\includegraphics[trim = 8mm 0mm 8mm 6mm, clip,width=0.9\hsize]{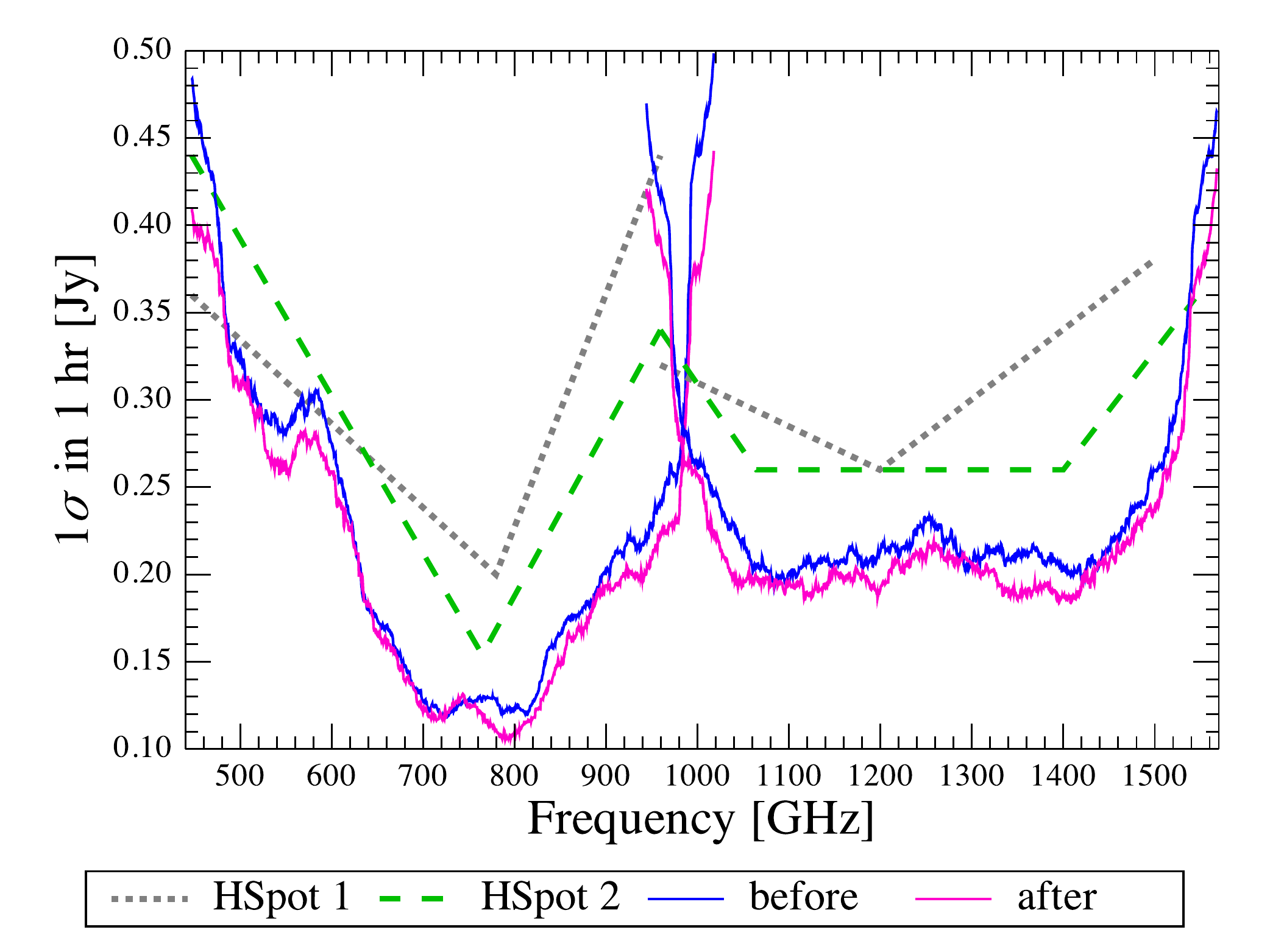}
\end{center}
\caption{FTS sensitivity for extended-calibrated data (top) and point-source-calibrated data (bottom). Curves are shown for both BSM epochs (`before' OD\,1011 and on or `after' OD\, 1011). The point-source-calibrated sensitivity is compared to the initial and revised HSpot values, which give the predicted sensitivity. The average sensitivities for point-source-calibrated data are 0.20\,Jy [1\,$\sigma$; 1 hour] for SSWC3 and 0.21\,Jy [1\,$\sigma$; 1 hour] for SSWD4.}
\label{fig:noiseCentre}
\end{figure}

Spectra of dark sky are also used to assess FTS sensitivity for extended and point-source-calibrated data, where the sensitivity is defined as the expected 1\,$\sigma$ noise in a 1\,hour observation. The ``error'' column provided in averaged FTS data products is the standard error on the mean of the unaveraged scans and is, therefore, assessing the random noise contribution. Except for some very faint sources, the ``error'' does not generally provide a realistic estimate of the total spectral noise. Therefore, to provide a more representative sensitivity, with respect to science observations, HR spectra of Uranus and Ceres were included with the dark sky observations used. For each observation the 1\,$\sigma$ noise is measured directly from the spectrum, using a sliding frequency bin of 50\,GHz. For each frequency sample, a polynomial is fitted over the bin width and subtracted before the standard deviation is taken of the residual within the bin. The bin width is tapered towards the end of the bands. The sensitivity is the median noise, per frequency sample, for all 95 observations included. Fig.~\ref{fig:noiseCentre} shows the results for HIPE 13 point-source-calibrated and extended-calibrated data, for the centre detectors. There is good consistency between the {\it Before} and {\it After} epochs. The average point-source-calibrated sensitivity is 0.20\,Jy [1\,$\sigma$; 1 hour] for SLWC3 and 0.21\,Jy [1\,$\sigma$; 1 hour] for SSWD4.

The improvement in calibration since HIPE version 7 can be expressed as a percentage improvement in sensitivity. There are two significant improvements that can be noted. Firstly, for HIPE 8, a mean telescope RSRF was introduced, constructed from multiple observations rather than a single dark sky, and the sensitivity improved for all frequencies by $\sim$15\%, with respect to HIPE 7. Secondly, another improvement was made, across both bands, when the method to derive the RSRFs was revised \citep[see][for more details]{Fulton14a}, and translates to an enhancement of nearly 40\% from HIPE version 11, compared to 7. Although improvements in calibration were seen for all observations, the greatest impact was on those observations that experienced the highest systematic noise.
This comparison was assessed using a subset of {\it Before} dark sky, so the same set of observations could be used for each version considered. 

The average HIPE 13 sensitivities for all detectors, calculated using the full set of dark sky observations, are presented in Table~\ref{tab:offsetSigma13}. These values shows a good consistency across all detectors, except for SSWB4, which is a significant outlier for both calibrations, suggesting an issue with the detector itself. A closer look shows a significant worsening of the noise for this detector after OD 710, indicative of a sudden event degrading the performance of the detector on this date.

\begin{table*}
\caption{Continuum offset (Offset) and 1\,$\sigma$ in 1 hour sensitivity ($\sigma$). A subscript of PS indicates results for point-source-calibrated data in units of Jy. A subscript of EXT indicated results for extended-calibrated data in units of 10$^{-19}$ Wm$^{-2}$Hz$^{-1}$sr$^{-1}$ for SLW and 10$^{-18}$ Wm$^{-2}$Hz$^{-1}$sr$^{-1}$ for SSW. The values for the centre detectors are shown in bold.}
\medskip
\begin{center}
\begin{tabular}{cccccccccc}
\hline\hline
SLW & Offset$_{\rm EXT}$ & Offset$_{\rm PS}$ & $\sigma_{\rm EXT}$ & $\sigma_{\rm PS}$ & SSW & Offset$_{\rm EXT}$ & Offset$_{\rm PS}$ & $\sigma_{\rm EXT}$ & $\sigma_{\rm PS}$ \\ \hline
SLWA1 & 1.1801 & --- & 0.4791 & --- & SSWA1 & 0.1743 & --- & 0.1225 & --- \\
SLWA2 & 0.6218 & --- & 0.2872 & --- & SSWA2 & 0.1636 & --- & 0.1089 & --- \\
SLWA3 & 0.8581 & --- & 0.3852 & --- & SSWA3 & 0.1783 & --- & 0.1334 & --- \\
SLWB1 & 1.0629 & --- & 0.3331 & --- & SSWA4 & 0.2020 & --- & 0.1151 & --- \\
SLWB2 & 0.5755 & 0.4414 & 0.3050 & 0.2213 & SSWB1 & 0.1603 & --- & 0.1129 & --- \\
SLWB3 & 0.5010 & 0.4018 & 0.2964 & 0.2274 & SSWB2 & 0.1574 & 0.2803 & 0.1121 & 0.1931 \\
SLWB4 & 0.5721 & --- & 0.3496 & --- & SSWB3 & 0.1737 & 0.2975 & 0.1287 & 0.2219 \\
SLWC1 & 1.2752 & --- & 0.6169 & --- & SSWB4 & 0.2378 & 0.4018 & 0.2141 & 0.3557 \\
SLWC2 & 0.5794 & 0.4377 & 0.3155 & 0.2177 & SSWB5 & 0.1704 & --- & 0.1098 & --- \\
{\bf SLWC3} & {\bf 0.5381} & {\bf 0.4038} & {\bf 0.3082} & {\bf 0.2173} & SSWC1 & 0.1726 & --- & 0.1259 & --- \\
SLWC4 & 0.5323 & 0.4127 & 0.2755 & 0.2164 & SSWC2 & 0.1571 & 0.2723 & 0.1146 & 0.1952 \\
SLWC5 & 0.8394 & --- & 0.4280 & --- & SSWC3 & 0.1665 & 0.2873 & 0.1219 & 0.2136 \\
SLWD1 & 0.6354 & --- & 0.3521 & --- & SSWC4 & 0.1864 & 0.3216 & 0.1206 & 0.2054 \\
SLWD2 & 0.5437 & 0.3963 & 0.3164 & 0.2091 & SSWC5 & 0.1731 & 0.2868 & 0.1182 & 0.1955 \\
SLWD3 & 0.5694 & 0.4022 & 0.3060 & 0.2127 & SSWC6 & 0.1763 & --- & 0.1108 & --- \\
SLWD4 & 0.7844 & --- & 0.3113 & --- & SSWD1 & 0.1809 & --- & 0.1256 & --- \\
SLWE1 & 0.6731 & --- & 0.4320 & --- & SSWD2 & 0.1708 & 0.2911 & 0.1247 & 0.2112 \\
SLWE2 & 0.5321 & --- & 0.3221 & --- & SSWD3 & 0.1541 & 0.2667 & 0.1140 & 0.1987 \\
SLWE3 & 0.6006 & --- & 0.4231 & --- & {\bf SSWD4} & {\bf 0.1613} & {\bf 0.2846} & {\bf 0.1262} & {\bf 0.2101} \\
--- & --- & --- & --- & --- & SSWD6 & 0.1672 & 0.2907 & 0.1097 & 0.1848 \\
--- & --- & --- & --- & --- & SSWD7 & 0.1939 & --- & 0.1210 & --- \\
--- & --- & --- & --- & --- & SSWE1 & 0.1681 & --- & 0.1168 & --- \\
--- & --- & --- & --- & --- & SSWE2 & 0.4081 & 0.6818 & 0.1103 & 0.1903 \\
--- & --- & --- & --- & --- & SSWE3 & 0.1550 & 0.2672 & 0.1138 & 0.1929 \\
--- & --- & --- & --- & --- & SSWE4 & 0.1565 & 0.2774 & 0.1217 & 0.2051 \\
--- & --- & --- & --- & --- & SSWE5 & 0.1916 & 0.3407 & 0.1121 & 0.1869 \\
--- & --- & --- & --- & --- & SSWE6 & 0.1724 & --- & 0.1200 & --- \\
--- & --- & --- & --- & --- & SSWF1 & 0.1621 & --- & 0.1161 & --- \\
--- & --- & --- & --- & --- & SSWF2 & 0.2128 & 0.3724 & 0.1051 & 0.1776 \\
--- & --- & --- & --- & --- & SSWF3 & 0.1985 & 0.3557 & 0.1070 & 0.1814 \\
--- & --- & --- & --- & --- & SSWF5 & 0.1635 & --- & 0.1162 & --- \\
--- & --- & --- & --- & --- & SSWG1 & 0.1755 & --- & 0.1414 & --- \\
--- & --- & --- & --- & --- & SSWG2 & 0.1613 & --- & 0.1168 & --- \\
--- & --- & --- & --- & --- & SSWG3 & 0.1885 & --- & 0.1494 & --- \\
--- & --- & --- & --- & --- & SSWG4 & 0.2041 & --- & 0.1606 & --- \\ \hline
\hline
\end{tabular}
\end{center}
\label{tab:offsetSigma13}
\end{table*}

\section{Summary}\label{sec:summary}

An extensive analysis of the FTS systematic programme of calibration observations has been used to assess the performance of the instrument over the entire {\it Herschel} mission. The main results are summarised as follows.
\begin{itemize}
\item The impact of the BSM offset for the first half of the mission is only on sparse observations. An increase in the continuum spread for the repeated calibrators of 2.4\% was found in comparison to data observed after the BSM was set back to the rest position. 
\item The fraction of flux detected by off-axis detectors, for a point source, is 1.9\% for the first ring of SLW, 1.4\% for the first SSW ring and 0.1\% for the second SSW ring. Therefore, for a point source embedded in an extended background the benefits of subtracting this signal should be considered in context of the overall uncertainty.
\item $^{12}$CO line measurements for the four main FTS line sources show a line flux repeatability of $<$\,2\% for well pointed data and $<$\,6\% otherwise. The spread in line velocity was not found to be significantly affected by pointing offset and is $<$\,7\,km\,s$^{-1}$.
\item The SLW and SSW calibration is consistent over the overlap region, with uncertainties of $<$\,5\%.
\item Despite relatively high noise and fringing at the end of the extended band regions, with individual attention, lines can be reliably extracted from this region.
\item The instrument line shape has been measured using unresolved lines, and an empirical profile constructed. The line shape shows a slight asymmetry with respect to a theoretical sinc function.
\item Fitting a sinc to the empirical line profile gives less than a 1\% difference between the peak values and less than 0.5\% difference in the fitted width to that expected for the optimal resolution. There is a 2.6\% systematic shortfall in the fitted integrated line flux, due to the asymmetry. 
\item Fitting the apodized line profile with a Gaussian gives a 5\% systematic excess compared to the equivalent sinc fit to the line profile before apodisation. 
\item For the sources considered when assessing continuum repeatability, there is no significant sign of intrinsic variability, except for CWLeo. The repeatability on the continuum is 4.4\% for SLW and 13.6\% for SSW, although this falls to less than 2\% for data corrected for pointing offset. The continuum repeatability results are inline with those found for spectral lines.
\item The FTS observations of Uranus and Neptune were compared to their models. For Uranus an average ratio of 0.99 was found for SLW, with a spread of 2\%, regardless of correction for pointing offset. For SSW the average ratio improves from 0.97$\pm$0.05 to 1.00$\pm$0.00 after pointing offset is corrected for. The average ratios with the Neptune model show a consistent 2\% systematic offset from a ratio of 1.0 for both SLW and SSW, before and after correction for pointing offset, with a spread of 1\%. This discrepancy is due to the use of different primary calibrators for the photometer and spectrometer.
\item After correcting for pointing offset, the comparison of Ceres, Pallas and Vesta with the respective models shows an average systematic offset and scatter of 1.04$\pm$0.05 for SLW and 1.09$\pm$0.02 for SSW.
\item A comparison of FTS synthetic photometry to SPIRE photometer photometry gives average ratios of 1.03$\pm$0.01 for PSW, 1.04$\pm$0.02 for PMW and 1.05$\pm$0.04 for PLW. The discrepancy between the two sets of photometry is due to the use of a different primary calibrator (Uranus for the FTS and Neptune for the photometer) and a systematic difference between the respective models, and the slight extent of the two sources used for the comparison.
\item FTS and photometer model ratios for Neptune, Ceres, Vesta and Pallas show consistency over the mission, but with the expected 2\% systematic offset introduced by the use of different primary calibrators.
\item Stacking all FTS HR observations of the SPIRE dark field shows no significant continuum or line detection, i.e. the SPIRE dark field is dark.
\item Dark sky data were used to assess the uncertainty on the continuum and the sensitivity. For point-source-calibrated data, the average uncertainty on the continuum is 0.40\,Jy in SLWC3 and 0.28\,Jy in SSWD4. The average point-source-calibrated sensitivity is 0.20\,Jy [1\,$\sigma$; 1 hour] for SLWC3 and 0.21\,Jy [1\,$\sigma$; 1 hour] for SLWD4. The continuum offset and sensitivity are both consistent across the detector arrays.
\item There is no significant difference in sensitivity for the {\it Before} and {\it After} epochs. Compared to {\it After}, the {\it Before} continuum offset is slightly higher for SLWC3 and slightly lower for SSWD4.
\end{itemize}

In conclusion, the existing set of SPIRE FTS calibration observations are sufficient to show the instrument performed with excellent stability throughout the {\it Herschel} mission. The results presented are consistent across both FTS detector arrays, with sensitivity levels that out perform early predictions, as provided by HSpot. There is also good consistency for observations taken `Before' and `After' the correction to the BSM position for sparse observations, and in the overlap region between the long and short wavelength detector bands. One future update that may improve the accuracy of the FTS further is to understand and correct the slight asymmetry in the instrument line shape, which is work in progress for the final pipeline release.

\section*{Acknowledgments}
We thank the referee, J. P. Maillard, for his constructive comments, which helped improve the paper.
SPIRE has been developed by a consortium of institutes led by Cardiff Univ. (UK) and including: Univ. Lethbridge (Canada); NAOC (China); CEA, LAM (France); IFSI, Univ. Padua (Italy); IAC (Spain); Stockholm Observatory (Sweden); Imperial College London, RAL, UCL-MSSL, UKATC, Univ. Sussex (UK); and Caltech, JPL, NHSC, Univ. Colorado (USA). This development has been supported by national funding agencies: CSA (Canada); NAOC (China); CEA, CNES, CNRS (France); ASI (Italy); MCINN (Spain); SNSB (Sweden); STFC, UKSA (UK); and NASA (USA). This research is supported in part by the Canadian Space Agency (CSA) and the Natural Sciences and Engineering Research Council of Canada (NSERC). This research has made use of the NASA/IPAC Infrared Science Archive, which is operated by the Jet Propulsion Laboratory, California Institute of Technology, under contract with the National Aeronautics and Space Administration.

\bibliographystyle{aa}
\bibliography{bib1}

\begin{thebibliography}{55}
\expandafter\ifx\csname natexlab\endcsname\relax\def\natexlab#1{#1}\fi

\bibitem[{{Becklin} {et~al.}(1969){Becklin}, {Frogel}, {Hyland}, {Kristian}, \&
  {Neugebauer}}]{Becklin1969}
{Becklin}, E.~E., {Frogel}, J.~A., {Hyland}, A.~R., {Kristian}, J., \&
  {Neugebauer}, G. 1969, \apjl, 158, L133

\bibitem[{{Bendo} {et~al.}(2013){Bendo}, {Griffin}, {Bock}, {Conversi},
  {Dowell}, {Lim}, {Lu}, {North}, {Papageorgiou}, {Pearson}, {Pohlen},
  {Polehampton}, {Schulz}, {Shupe}, {Sibthorpe}, {Spencer}, {Swinyard},
  {Valtchanov}, \& {Xu}}]{Bendo13}
{Bendo}, G.~J., {Griffin}, M.~J., {Bock}, J.~J., {et~al.} 2013, MNRAS, 433,
  3062

\bibitem[{{Benielli} {et~al.}(2014){Benielli}, {Polehampton}, {Hopwood},
  {Gri{\~n}{\'o}n Mar{\'{\i}}n}, {Fulton}, {Imhof}, {Lim}, {Lu}, {Makiwa},
  {Marchili}, {Naylor}, {Spencer}, {Swinyard}, {Valtchanov}, \& {van der
  Wiel}}]{Benielli2014}
{Benielli}, D., {Polehampton}, E., {Hopwood}, R., {et~al.} 2014, Experimental
  Astronomy, 37, 357

\bibitem[{{Cernicharo} {et~al.}(2014){Cernicharo}, {Teyssier},
  {Quintana-Lacaci}, {Daniel}, {Ag{\'u}ndez}, {Velilla-Prieto}, {Decin},
  {Gu{\'e}lin}, {Encrenaz}, {Garc{\'{\i}}a-Lario}, {de Beck}, {Barlow},
  {Groenewegen}, {Neufeld}, \& {Pearson}}]{Cernicharo2014}
{Cernicharo}, J., {Teyssier}, D., {Quintana-Lacaci}, G., {et~al.} 2014, \apjl,
  796, L21

\bibitem[{Davis {et~al.}(2001)Davis, Abrams, \& J.W.}]{Davis2001}
Davis, S., Abrams, M., \& J.W., B. 2001, Fourier Transform Spectrometry (San
  Diego: Academic Press)

\bibitem[{{Deguchi} {et~al.}(1992){Deguchi}, {Izumiura}, {Nguyen-Q-Rieu},
  {Shibata}, {Ukita}, \& {Yamamura}}]{Deguchi1992}
{Deguchi}, S., {Izumiura}, H., {Nguyen-Q-Rieu}, {et~al.} 1992, \apj, 392, 597

\bibitem[{{Dohlen} {et~al.}(2000){Dohlen}, {Origne}, {Pouliquen}, \&
  {Swinyard}}]{Dohlen2000}
{Dohlen}, K., {Origne}, A., {Pouliquen}, D., \& {Swinyard}, B.~M. 2000, in
  Society of Photo-Optical Instrumentation Engineers (SPIE) Conference Series,
  Vol. 4013, Society of Photo-Optical Instrumentation Engineers (SPIE)
  Conference Series, ed. J.~B. {Breckinridge} \& P.~{Jakobsen}, 119--128

\bibitem[{{Fong} {et~al.}(2006){Fong}, {Meixner}, {Sutton}, {Zalucha}, \&
  {Welch}}]{Fong2006}
{Fong}, D., {Meixner}, M., {Sutton}, E.~C., {Zalucha}, A., \& {Welch}, W.~J.
  2006, \apj, 652, 1626

\bibitem[{{Fulton} {et~al.}(2014){Fulton}, {Hopwood}, {Baluteau}, {Benielli},
  {Imhof}, {Lim}, {Lu}, {Marchili}, {Naylor}, {Polehampton}, {Swinyard}, \&
  {Valtchanov}}]{Fulton14a}
{Fulton}, T., {Hopwood}, R., {Baluteau}, J.-P., {et~al.} 2014, Experimental
  Astronomy, 37, 381

\bibitem[{{Griffin} {et~al.}(2010){Griffin}, {Abergel}, {Abreu}, \& {et
  al.}}]{Griffin10}
{Griffin}, M.~J., {Abergel}, A., {Abreu}, A., \& {et al.} 2010, \aap, 518, L3

\bibitem[{{Groenewegen} {et~al.}(2012){Groenewegen}, {Barlow}, {Blommaert},
  {Cernicharo}, {Decin}, {Gomez}, {Hargrave}, {Kerschbaum}, {Ladjal}, {Lim},
  {Matsuura}, {Olofsson}, {Sibthorpe}, {Swinyard}, {Ueta}, \&
  {Yates}}]{Groenewegen2012}
{Groenewegen}, M.~A.~T., {Barlow}, M.~J., {Blommaert}, J.~A.~D.~L., {et~al.}
  2012, \aap, 543, L8

\bibitem[{{Herpin} {et~al.}(2002){Herpin}, {Goicoechea}, {Pardo}, \&
  {Cernicharo}}]{Herpin02}
{Herpin}, F., {Goicoechea}, J.~R., {Pardo}, J.~R., \& {Cernicharo}, J. 2002,
  ApJ, 577, 961

\bibitem[{{Hopwood} {et~al.}(2014){Hopwood}, {Fulton}, {Polehampton},
  {Valtchanov}, {Benielli}, {Imhof}, {Lim}, {Lu}, {Marchili}, {Pearson}, \&
  {Swinyard}}]{Hopwood2014}
{Hopwood}, R., {Fulton}, T., {Polehampton}, E.~T., {et~al.} 2014, Experimental
  Astronomy, 37, 195

\bibitem[{{Huang} {et~al.}(2010){Huang}, {Hasegawa}, {Dinh-V-Trung}, {Kwok},
  {Muller}, {Hirano}, {Lim}, {Muthu Mariappan}, \& {Lyo}}]{Huang2010}
{Huang}, Z.-Y., {Hasegawa}, T.~I., {Dinh-V-Trung}, {et~al.} 2010, \apj, 722,
  273

\bibitem[{{Jaminet} {et~al.}(1991){Jaminet}, {Danchi}, {Sutton}, {Bieging},
  {Wilner}, {Russell}, \& {Sandell}}]{Jaminet1991}
{Jaminet}, P.~A., {Danchi}, W.~C., {Sutton}, E.~C., {et~al.} 1991, \apj, 380,
  461

\bibitem[{{Jenness} {et~al.}(2002){Jenness}, {Stevens}, {Archibald},
  {Economou}, {Jessop}, \& {Robson}}]{Jenness02}
{Jenness}, T., {Stevens}, J.~A., {Archibald}, E.~N., {et~al.} 2002, \mnras,
  336, 14

\bibitem[{{Josselin} {et~al.}(1998){Josselin}, {Loup}, {Omont}, {Barnbaum},
  {Nyman}, \& {Sevre}}]{Josselin98}
{Josselin}, E., {Loup}, C., {Omont}, A., {et~al.} 1998, A\&AS, 129, 45

\bibitem[{{Kawada} {et~al.}(2007){Kawada}, {Baba}, {Barthel}, {Clements},
  {Cohen}, {Doi}, {Figueredo}, {Fujiwara}, {Goto}, {Hasegawa}, {Hibi}, {Hirao},
  {Hiromoto}, {Jeong}, {Kaneda}, {Kawai}, {Kawamura}, {Kester}, {Kii},
  {Kobayashi}, {Kwon}, {Lee}, {Makiuti}, {Matsuo}, {Matsuura}, {M{\"u}ller},
  {Murakami}, {Nagata}, {Nakagawa}, {Narita}, {Noda}, {Oh}, {Okada}, {Okuda},
  {Oliver}, {Ootsubo}, {Pak}, {Park}, {Pearson}, {Rowan-Robinson}, {Saito},
  {Salama}, {Sato}, {Savage}, {Serjeant}, {Shibai}, {Shirahata}, {Sohn},
  {Suzuki}, {Takagi}, {Takahashi}, {Thomson}, {Usui}, {Verdugo}, {Watabe},
  {White}, {Wang}, {Yamamura}, {Yamauchi}, \& {Yasuda}}]{Kawada2007}
{Kawada}, M., {Baba}, H., {Barthel}, P.~D., {et~al.} 2007, \pasj, 59, 389

\bibitem[{{Kwok} \& {Bignell}(1984)}]{KwokBignell1984}
{Kwok}, S. \& {Bignell}, R.~C. 1984, \apj, 276, 544

\bibitem[{{Lagadec} {et~al.}(2011){Lagadec}, {Verhoelst}, {M{\'e}karnia},
  {Su{\'a}eez}, {Zijlstra}, {Bendjoya}, {Szczerba}, {Chesneau}, {van Winckel},
  {Barlow}, {Matsuura}, {Bowey}, {Lorenz-Martins}, \& {Gledhill}}]{Lagadec2011}
{Lagadec}, E., {Verhoelst}, T., {M{\'e}karnia}, D., {et~al.} 2011, \mnras, 417,
  32

\bibitem[{{Lu} {et~al.}(2014){Lu}, {Polehampton}, {Swinyard}, {Benielli},
  {Fulton}, {Hopwood}, {Imhof}, {Lim}, {Marchili}, {Naylor}, {Schulz},
  {Sidher}, \& {Valtchanov}}]{Lu14}
{Lu}, N., {Polehampton}, E.~T., {Swinyard}, B.~M., {et~al.} 2014, Experimental
  Astronomy, 37, 239

\bibitem[{{Makiwa} {et~al.}(2013){Makiwa}, {Naylor}, {Ferlet}, {Salji},
  {Swinyard}, {Polehampton}, \& {van der Wiel}}]{Makiwa2013}
{Makiwa}, G., {Naylor}, D.~A., {Ferlet}, M., {et~al.} 2013, \ao, 52, 3864

\bibitem[{{Miville-Desch{\^e}nes} \& {Lagache}(2005)}]{MivilleLagache2005}
{Miville-Desch{\^e}nes}, M.-A. \& {Lagache}, G. 2005, \apjs, 157, 302

\bibitem[{{Molster} {et~al.}(1999){Molster}, {Waters}, {Trams}, {Van Winckel},
  {Decin}, {van Loon}, {J{\"a}ger}, {Henning}, {K{\"a}ufl}, {de Koter}, \&
  {Bouwman}}]{Molster1999}
{Molster}, F.~J., {Waters}, L.~B.~F.~M., {Trams}, N.~R., {et~al.} 1999, \aap,
  350, 163

\bibitem[{Moreno(1998)}]{Moreno98}
Moreno, R. 1998, PhD thesis, Universit\'{e} de Paris

\bibitem[{{M{\"u}ller} {et~al.}(2001){M{\"u}ller}, {Thorwirth}, {Roth}, \&
  {Winnewisser}}]{muller2001}
{M{\"u}ller}, H.~S.~P., {Thorwirth}, S., {Roth}, D.~A., \& {Winnewisser}, G.
  2001, \aap, 370, L49

\bibitem[{{M{\"u}ller} {et~al.}(2013){M{\"u}ller}, {Balog}, {Nielbock}, {Lim},
  {Teyssier}, {Olberg}, {Klaas}, {Linz}, {Altieri}, {Pearson}, {Bendo}, \&
  {Vilenius}}]{Muller13}
{M{\"u}ller}, T., {Balog}, Z., {Nielbock}, M., {et~al.} 2013, Experimental
  Astronomy

\bibitem[{{M{\"u}ller} \& {Lagerros}(2002)}]{Muller02}
{M{\"u}ller}, T.~G. \& {Lagerros}, J.~S.~V. 2002, \aap, 381, 324

\bibitem[{{Naylor} {et~al.}(2010){Naylor}, {Baluteau}, {Barlow}, {Benielli},
  {Ferlet}, {Fulton}, {Griffin}, {Grundy}, {Imhof}, {Jones}, {King}, {Leeks},
  {Lim}, {Lu}, {Makiwa}, {Polehampton}, {Savini}, {Sidher}, {Spencer},
  {Surace}, {Swinyard}, \& {Wesson}}]{Naylor10}
{Naylor}, D.~A., {Baluteau}, J.-P., {Barlow}, M.~J., {et~al.} 2010, in Society
  of Photo-Optical Instrumentation Engineers (SPIE) Conference Series, Vol.
  7731, Society of Photo-Optical Instrumentation Engineers (SPIE) Conference
  Series

\bibitem[{{Naylor} {et~al.}(2014){Naylor}, {Baluteau}, {Bendo}, {Benielli},
  {Fulton}, {Gom}, {Griffin}, {Hopwood}, {Imhof}, {Lim}, {Lu}, {Makiwa},
  {Marchili}, {Orton}, {Papageorgiou}, {Pearson}, {Polehampton}, {Schulz},
  {Spencer}, {Swinyard}, {Valtchanov}, {van der Wiel}, {Veenendaal}, \&
  {Wu}}]{Naylor2014}
{Naylor}, D.~A., {Baluteau}, J.-P., {Bendo}, G.~J., {et~al.} 2014, in Society
  of Photo-Optical Instrumentation Engineers (SPIE) Conference Series, Vol.
  9143, Society of Photo-Optical Instrumentation Engineers (SPIE) Conference
  Series, 2

\bibitem[{Naylor \& Tahic(2007)}]{NaylorTahic07}
Naylor, D.~A. \& Tahic, M.~K. 2007, J. Opt. Soc. Am. A., 24, 3644

\bibitem[{{Neugebauer} {et~al.}(1984){Neugebauer}, {Habing}, {van Duinen},
  {Aumann}, {Baud}, {Beichman}, {Beintema}, {Boggess}, {Clegg}, {de Jong},
  {Emerson}, {Gautier}, {Gillett}, {Harris}, {Hauser}, {Houck}, {Jennings},
  {Low}, {Marsden}, {Miley}, {Olnon}, {Pottasch}, {Raimond}, {Rowan-Robinson},
  {Soifer}, {Walker}, {Wesselius}, \& {Young}}]{Neugebauer1984}
{Neugebauer}, G., {Habing}, H.~J., {van Duinen}, R., {et~al.} 1984, \apjl, 278,
  L1

\bibitem[{Orton {et~al.}(2014)Orton, Fletcher, Moses, Mainzer, Hines, Hammel,
  Martin-Torres, Burgdorf, Merlet, \& Line}]{Orton14}
Orton, G.~S., Fletcher, L.~N., Moses, J.~I., {et~al.} 2014, Icarus, in review

\bibitem[{{Ott}(2010)}]{Ott10}
{Ott}, S. 2010, in Astronomical Society of the Pacific Conference Series, Vol.
  434, Astronomical Data Analysis Software and Systems XIX, ed. Y.~{Mizumoto},
  K.-I. {Morita}, \& M.~{Ohishi}, 139

\bibitem[{{Pascale} {et~al.}(2008){Pascale}, {Ade}, {Bock}, {Chapin}, {Chung},
  {Devlin}, {Dicker}, {Griffin}, {Gundersen}, {Halpern}, {Hargrave}, {Hughes},
  {Klein}, {MacTavish}, {Marsden}, {Martin}, {Martin}, {Mauskopf},
  {Netterfield}, {Olmi}, {Patanchon}, {Rex}, {Scott}, {Semisch}, {Thomas},
  {Truch}, {Tucker}, {Tucker}, {Viero}, \& {Wiebe}}]{Pascale2008}
{Pascale}, E., {Ade}, P.~A.~R., {Bock}, J.~J., {et~al.} 2008, \apj, 681, 400

\bibitem[{{Pearson} {et~al.}(2014){Pearson}, {Lim}, {North}, {Bendo},
  {Conversi}, {Dowell}, {Griffin}, {Jin}, {Laporte}, {Papageorgiou}, {Schulz},
  {Shupe}, {Smith}, \& {Xu}}]{Pearson2014}
{Pearson}, C., {Lim}, T., {North}, C., {et~al.} 2014, Experimental Astronomy,
  37, 175

\bibitem[{{Phillips} {et~al.}(1991){Phillips}, {Mampaso}, {Williams}, \&
  {Ukita}}]{Phillips1991}
{Phillips}, J.~P., {Mampaso}, A., {Williams}, P.~G., \& {Ukita}, N. 1991, \aap,
  247, 148

\bibitem[{{Pilbratt} {et~al.}(2010){Pilbratt}, {Riedinger}, {Passvogel}, \& {et
  al}}]{pilbratt10}
{Pilbratt}, G.~L., {Riedinger}, J.~R., {Passvogel}, T., \& {et al}. 2010, \aap,
  518, L1

\bibitem[{{Polehampton}(2014)}]{Polehampton14}
{Polehampton}, E. 2014, in Astronomical Society of the Pacific Conference
  Series, Vol.~0, ADASS XXIV, in press

\bibitem[{{S{\'a}nchez Contreras} {et~al.}(2004){S{\'a}nchez Contreras},
  {Bujarrabal}, {Castro-Carrizo}, {Alcolea}, \&
  {Sargent}}]{SanchezContreras2004}
{S{\'a}nchez Contreras}, C., {Bujarrabal}, V., {Castro-Carrizo}, A., {Alcolea},
  J., \& {Sargent}, A. 2004, \apj, 617, 1142

\bibitem[{{S{\'a}nchez-Portal} {et~al.}(2014){S{\'a}nchez-Portal}, {Marston},
  {Altieri}, {Aussel}, {Feuchtgruber}, {Klaas}, {Linz}, {Lutz}, {Mer{\'{\i}}n},
  {M{\"u}ller}, {Nielbock}, {Oort}, {Pilbratt}, {Schmidt}, {Stephenson}, \&
  {Tuttlebee}}]{SanchezPortal2014}
{S{\'a}nchez-Portal}, M., {Marston}, A., {Altieri}, B., {et~al.} 2014,
  Experimental Astronomy, 37, 453

\bibitem[{{Schulz} {et~al.}(2002){Schulz}, {Huth}, {Laureijs}, {Acosta-Pulido},
  {Braun}, {Casta{\~n}eda}, {Cohen}, {Cornwall}, {Gabriel}, {Hammersley},
  {Heinrichsen}, {Klaas}, {Lemke}, {M{\"u}ller}, {Osip},
  {Rom{\'a}n-Fern{\'a}ndez}, \& {Telesco}}]{Schulz02}
{Schulz}, B., {Huth}, S., {Laureijs}, R.~J., {et~al.} 2002, \aap, 381, 1110

\bibitem[{{Soria-Ruiz} {et~al.}(2013){Soria-Ruiz}, {Bujarrabal}, \&
  {Alcolea}}]{SoriaRuiz2013}
{Soria-Ruiz}, R., {Bujarrabal}, V., \& {Alcolea}, J. 2013, \aap, 559, A45

\bibitem[{Spencer {et~al.}(2010)Spencer, Naylor, \& Swinyard}]{Spencer10}
Spencer, L.~D., Naylor, D.~A., \& Swinyard, B.~M. 2010, Meas. Sci. Technol.,
  21, 065601

\bibitem[{{SPIRE Handbook}(2014)}]{spireHandbook}
{SPIRE Handbook}. 2014, {HERSCHEL-HSC-DOC-0798}, \\accessed from
  http://herschel.esac.esa.int/Documentation.shtml

\bibitem[{{Stansberry} {et~al.}(2007){Stansberry}, {Gordon}, {Bhattacharya},
  {Engelbracht}, {Rieke}, {Marleau}, {Fadda}, {Frayer}, {Noriega-Crespo},
  {Wachter}, {Young}, {M{\"u}ller}, {Kelly}, {Blaylock}, {Henderson},
  {Neugebauer}, {Beeman}, \& {Haller}}]{Stansberry2007}
{Stansberry}, J.~A., {Gordon}, K.~D., {Bhattacharya}, B., {et~al.} 2007, \pasp,
  119, 1038

\bibitem[{{Swinyard} {et~al.}(2014){Swinyard}, {Polehampton}, {Hopwood},
  {Valtchanov}, {Lu}, {Fulton}, {Benielli}, {Imhof}, {Marchili}, {Baluteau},
  {Bendo}, {Ferlet}, {Griffin}, {Lim}, {Makiwa}, {Naylor}, {Orton},
  {Papageorgiou}, {Pearson}, {Schulz}, {Sidher}, {Spencer}, {van der Wiel}, \&
  {Wu}}]{Swinyard2014}
{Swinyard}, B.~M., {Polehampton}, E.~T., {Hopwood}, R., {et~al.} 2014, ArXiv
  e-prints

\bibitem[{{Teyssier} {et~al.}(2006){Teyssier}, {Hernandez}, {Bujarrabal},
  {Yoshida}, \& {Phillips}}]{Teyssier06}
{Teyssier}, D., {Hernandez}, R., {Bujarrabal}, V., {Yoshida}, H., \&
  {Phillips}, T.~G. 2006, A\&A, 450, 167

\bibitem[{{Turner} {et~al.}(2001){Turner}, {Bock}, {Beeman}, {Glenn},
  {Hargrave}, {Hristov}, {Nguyen}, {Rahman}, {Sethuraman}, \&
  {Woodcraft}}]{Turner2001}
{Turner}, A.~D., {Bock}, J.~J., {Beeman}, J.~W., {et~al.} 2001, Appl. Opt., 40,
  4921

\bibitem[{{Valtchanov} {et~al.}(2014){Valtchanov}, {Hopwood}, {Polehampton},
  {Benielli}, {Fulton}, {Imhof}, {Konopczy{\'n}ski}, {Lim}, {Lu}, {Marchili},
  {Naylor}, \& {Swinyard}}]{Valtchanov2014}
{Valtchanov}, I., {Hopwood}, R., {Polehampton}, E., {et~al.} 2014, Experimental
  Astronomy, 37, 207

\bibitem[{{Van Loon} {et~al.}(1999){Van Loon}, {Molster}, {Van Winckel}, \&
  {Waters}}]{vanLoon99}
{Van Loon}, J.~T., {Molster}, F.~J., {Van Winckel}, H., \& {Waters},
  L.~B.~F.~M. 1999, \aap, 350, 120

\bibitem[{{Volk} \& {Kwok}(1997)}]{volkKwok97}
{Volk}, K. \& {Kwok}, S. 1997, \apj, 477, 722

\bibitem[{{Wesson} {et~al.}(2010){Wesson}, {Cernicharo}, {Barlow}, {Matsuura},
  {Decin}, {Groenewegen}, {Polehampton}, {Agundez}, {Cohen}, {Daniel}, {Exter},
  {Gear}, {Gomez}, {Hargrave}, {Imhof}, {Ivison}, {Leeks}, {Lim}, {Olofsson},
  {Savini}, {Sibthorpe}, {Swinyard}, {Ueta}, {Witherick}, \&
  {Yates}}]{Wesson2010}
{Wesson}, R., {Cernicharo}, J., {Barlow}, M.~J., {et~al.} 2010, \aap, 518, L144

\bibitem[{{Wesson} {et~al.}(2011){Wesson}, {Cernicharo}, {Barlow}, {Matsuura},
  {Decin}, {Groenewegen}, {Polehampton}, {Agundez}, {Cohen}, {Daniel}, {Exter},
  {Gear}, {Gomez}, {Hargrave}, {Imhof}, {Ivison}, {Leeks}, {Lim}, {Olofsson},
  {Savini}, {Sibthorpe}, {Swinyard}, {Ueta}, {Witherick}, \&
  {Yates}}]{Wesson2011}
{Wesson}, R., {Cernicharo}, J., {Barlow}, M.~J., {et~al.} 2011, in Astronomical
  Society of the Pacific Conference Series, Vol. 445, Why Galaxies Care about
  AGB Stars II: Shining Examples and Common Inhabitants, ed. F.~{Kerschbaum},
  T.~{Lebzelter}, \& R.~F. {Wing}, 607

\bibitem[{{Wu} {et~al.}(2013){Wu}, {Polehampton}, {Etxaluze}, {Makiwa},
  {Naylor}, {Salji}, {Swinyard}, {Ferlet}, {van der Wiel}, {Smith}, {Fulton},
  {Griffin}, {Baluteau}, {Benielli}, {Glenn}, {Hopwood}, {Imhof}, {Lim}, {Lu},
  {Panuzzo}, {Pearson}, {Sidher}, \& {Valtchanov}}]{Wu2013}
{Wu}, R., {Polehampton}, E.~T., {Etxaluze}, M., {et~al.} 2013, A\&A, 556, 116

\end{thebibliography}

\clearpage

\appendix
\label{app:lineFittingFigs}

\section{Line fitting plots}

\begin{figure*}
\centering
\includegraphics[trim = 0mm 8mm 0mm 8mm, clip,width=\textwidth]{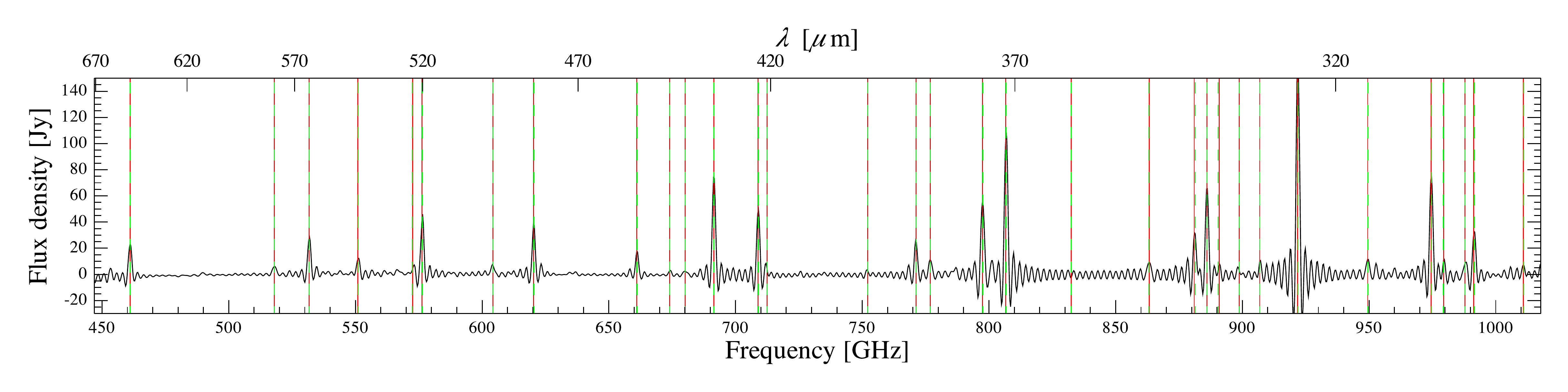}
\includegraphics[trim = 0mm 8mm 0mm 8mm, clip,width=\textwidth]{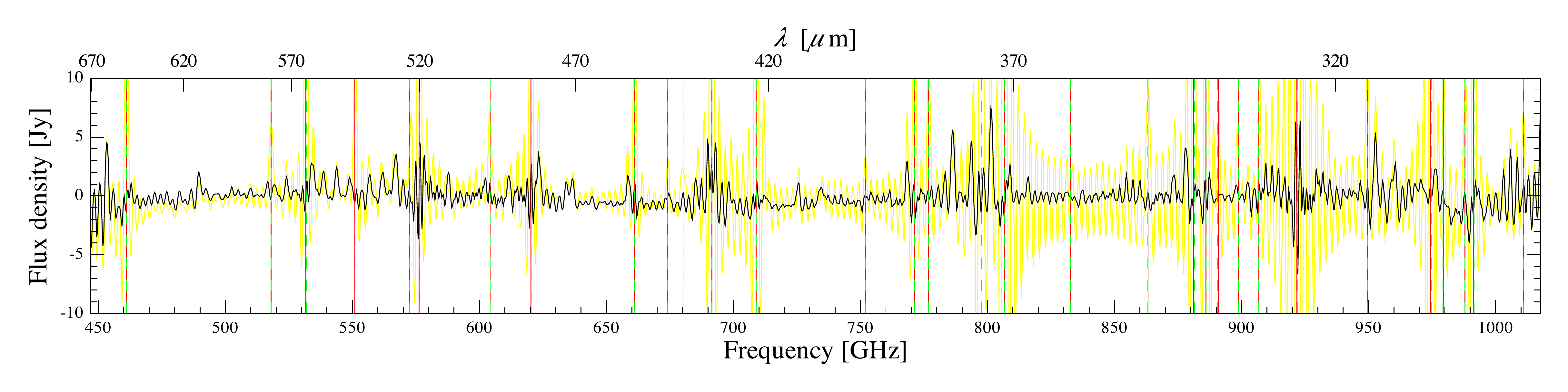}
\includegraphics[trim = 0mm 8mm 0mm 8mm, clip,width=\textwidth]{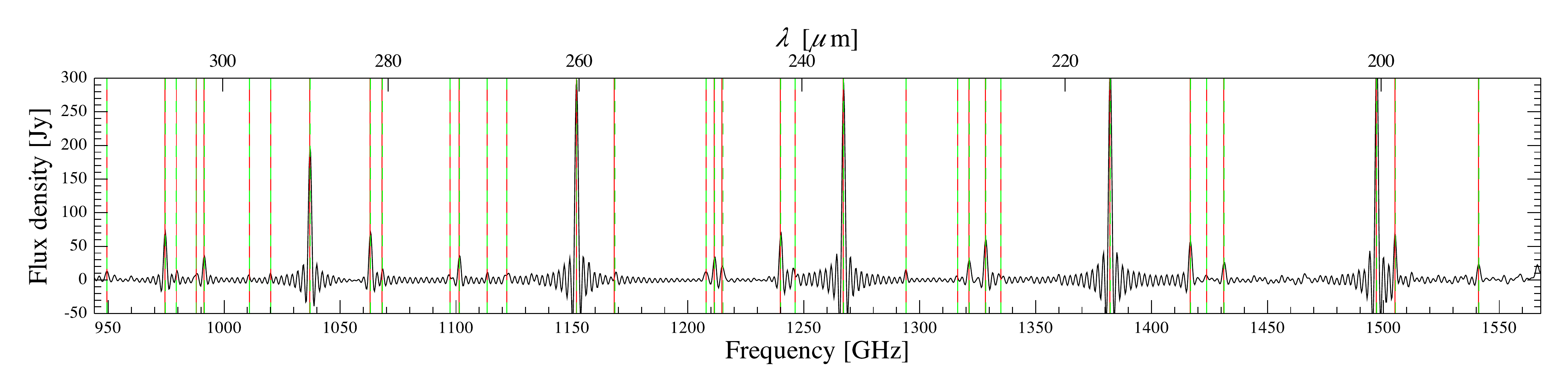}
\includegraphics[trim = 0mm 8mm 0mm 8mm, clip,width=\textwidth]{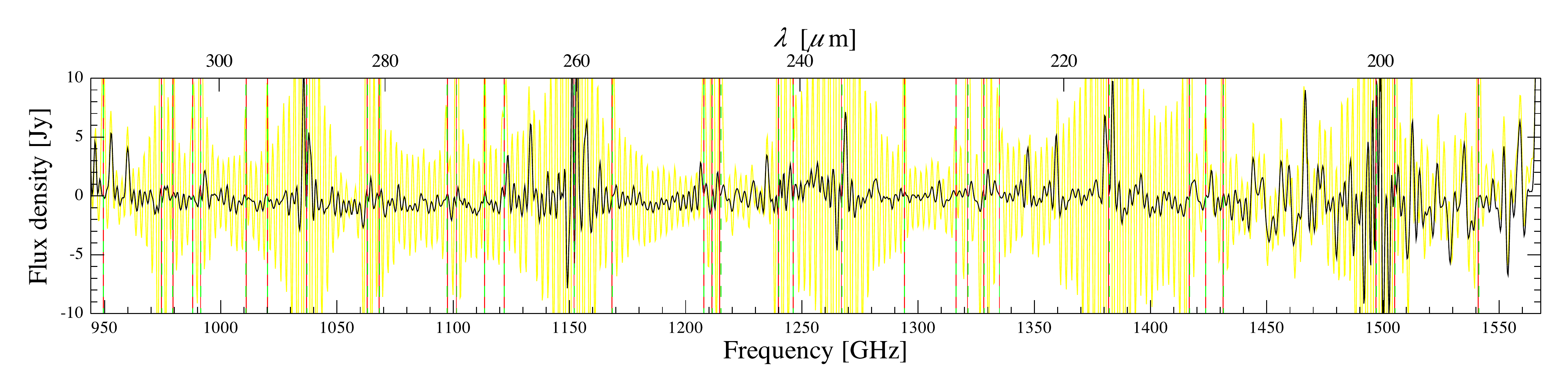}
\caption{Lines fitted to AFGL2688 shown with respect to the co-added spectra (first and third panels, from the top) and the residual on subtraction of the combined fit with the co-added data show in yellow (second and fourth panels, from the top). The top two panels show the results for SLWC3 and the bottom two show SSWD4. Red lines indicate the input line positions and the green dashed lines show the mean fitted positions for all observations.}
\label{fig:linesFittedAFGL2688}
\end{figure*}

\begin{figure*}
\centering
\includegraphics[trim = 0mm 8mm 0mm 8mm, clip,width=\textwidth]{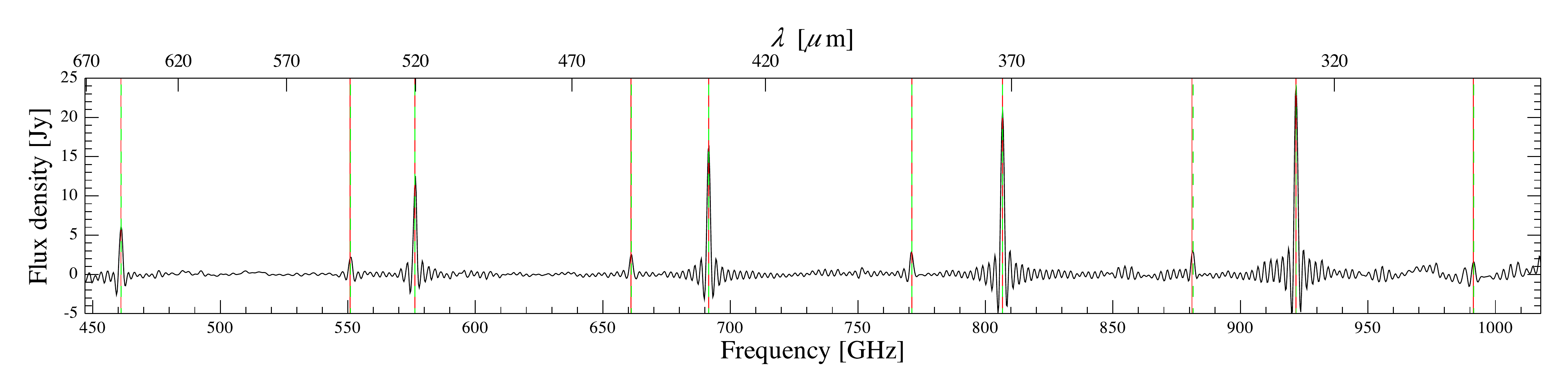}
\includegraphics[trim = 0mm 8mm 0mm 8mm, clip,width=\textwidth]{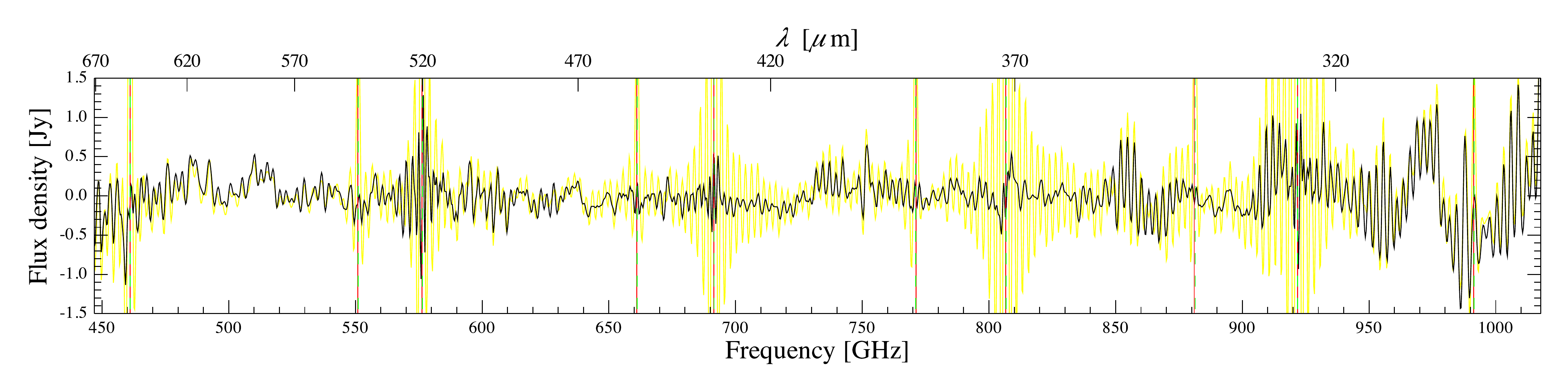}

\includegraphics[trim = 0mm 8mm 0mm 8mm, clip,width=\textwidth]{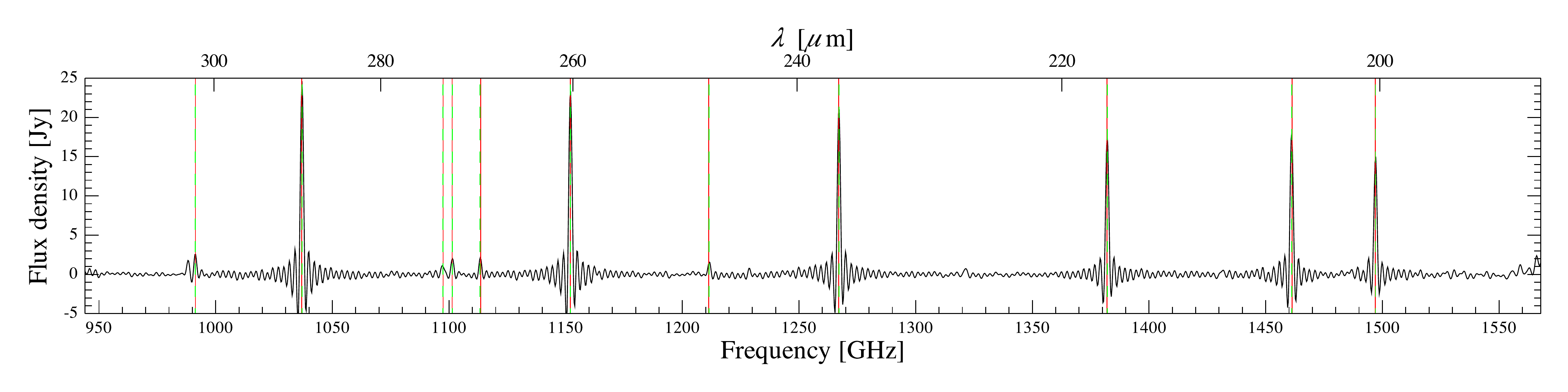}
\includegraphics[trim = 0mm 8mm 0mm 8mm, clip,width=\textwidth]{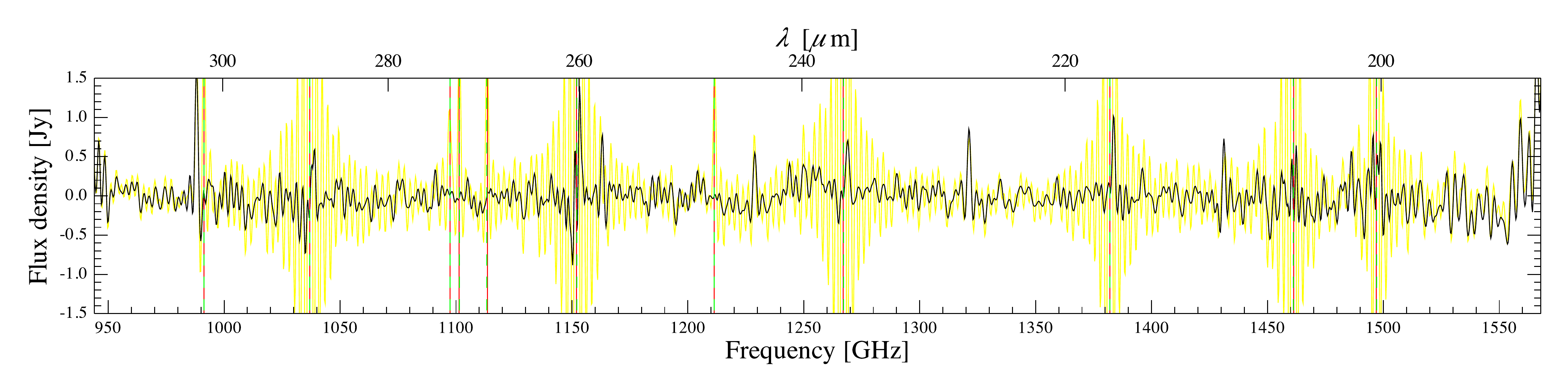}

\caption{Lines fitted to AFGL4106 shown with respect to the co-added spectra (first and third panels, from the top) and the residual on subtraction of the combined fit with the co-added data show in yellow (second and fourth panels, from the top). The top two panels show the results for SLWC3 and the bottom two show SSWD4. Red lines indicate the input line positions and the green dashed lines show the mean fitted positions for all observations. }
\label{fig:linesFittedAFGL4106}
\end{figure*}

\begin{figure*}
\centering
\includegraphics[trim = 0mm 8mm 0mm 8mm, clip,width=\textwidth]{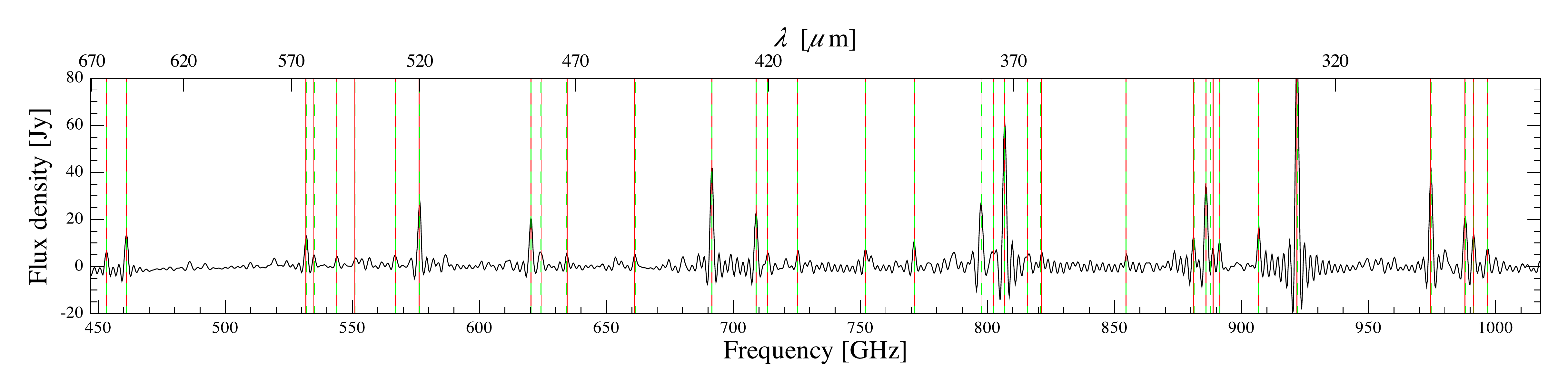}
\includegraphics[trim = 0mm 8mm 0mm 8mm, clip,width=\textwidth]{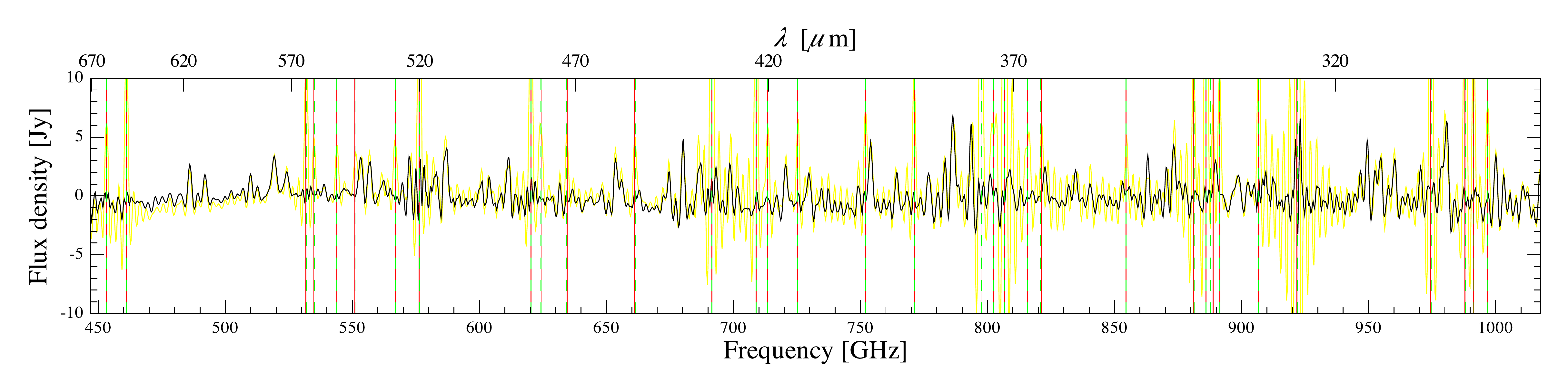}
\includegraphics[trim = 0mm 8mm 0mm 8mm, clip,width=\textwidth]{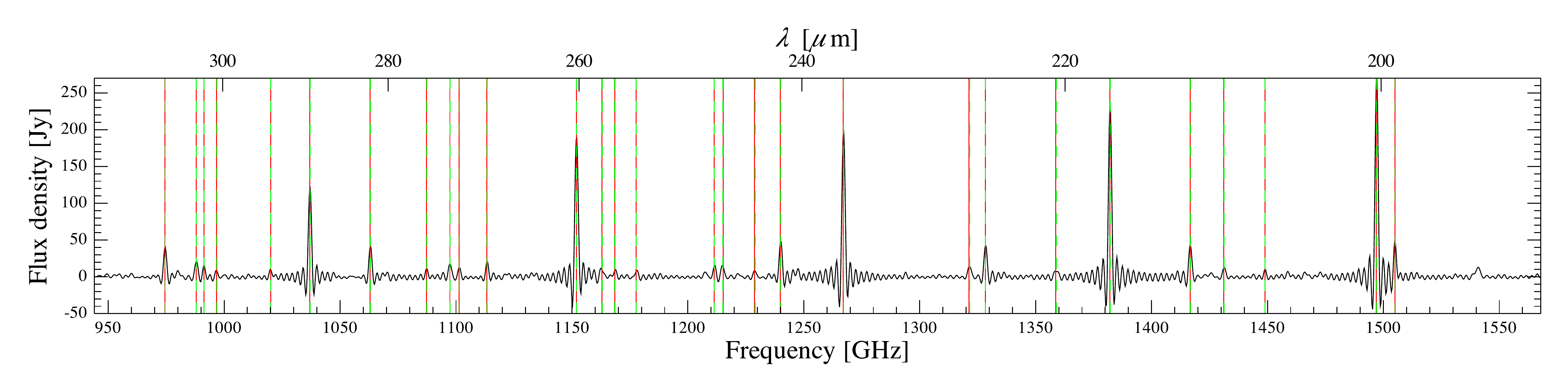}
\includegraphics[trim = 0mm 8mm 0mm 8mm, clip,width=\textwidth]{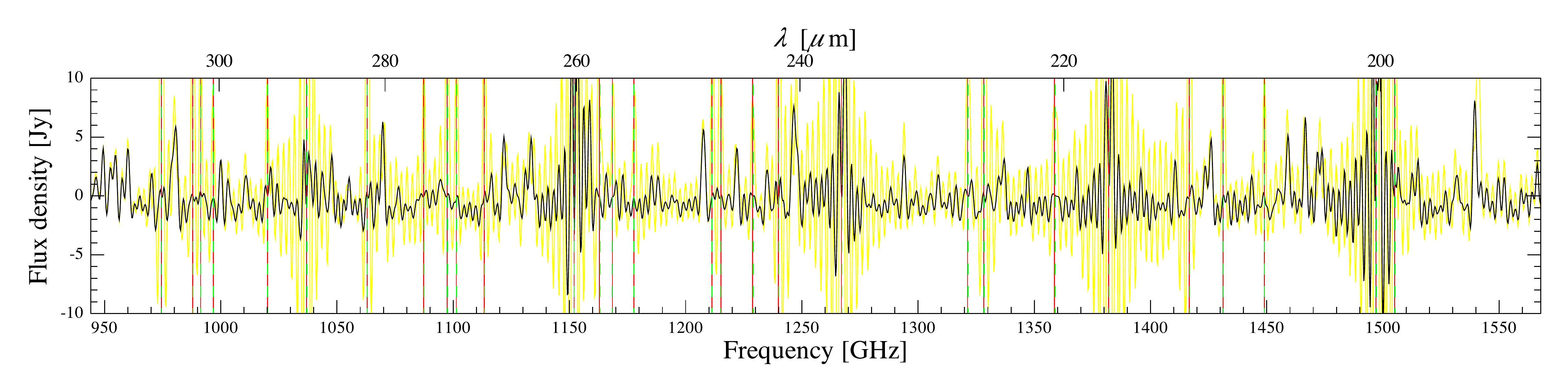}
\caption{Lines fitted to CRL618 shown with respect to the co-added spectra (first and third panels, from the top) and the residual on subtraction of the combined fit with the co-added data show in yellow (second and fourth panels, from the top). The top two panels show the results for SLWC3 and the bottom two show SSWD4. Red lines indicate the input line positions and the green dashed lines show the mean fitted positions for all observations.}
\label{fig:linesFittedCRL618}
\end{figure*}

\begin{figure*}
\centering
\includegraphics[trim = 0mm 8mm 0mm 8mm, clip,width=\textwidth]{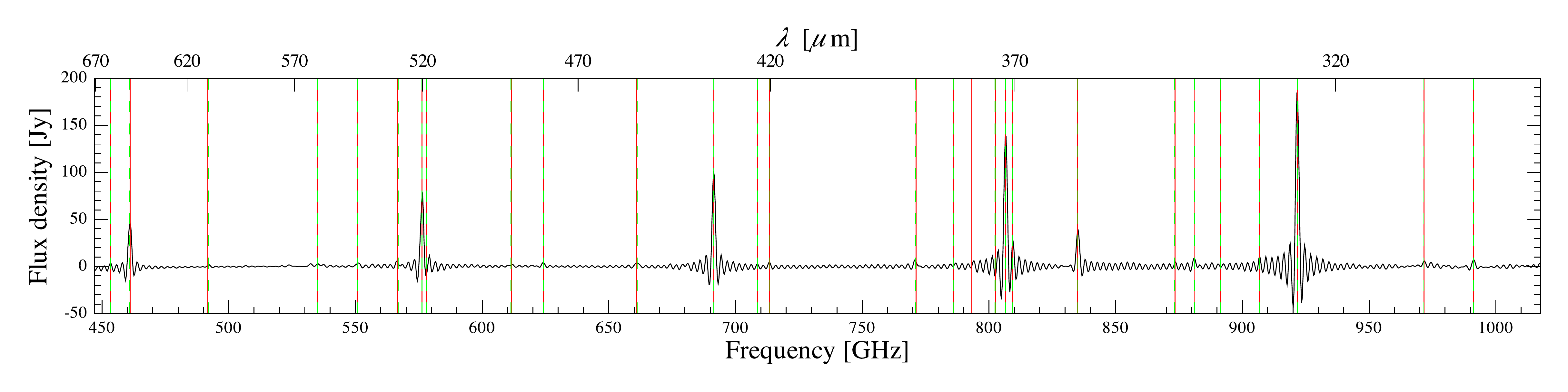}
\includegraphics[trim = 0mm 8mm 0mm 8mm, clip,width=\textwidth]{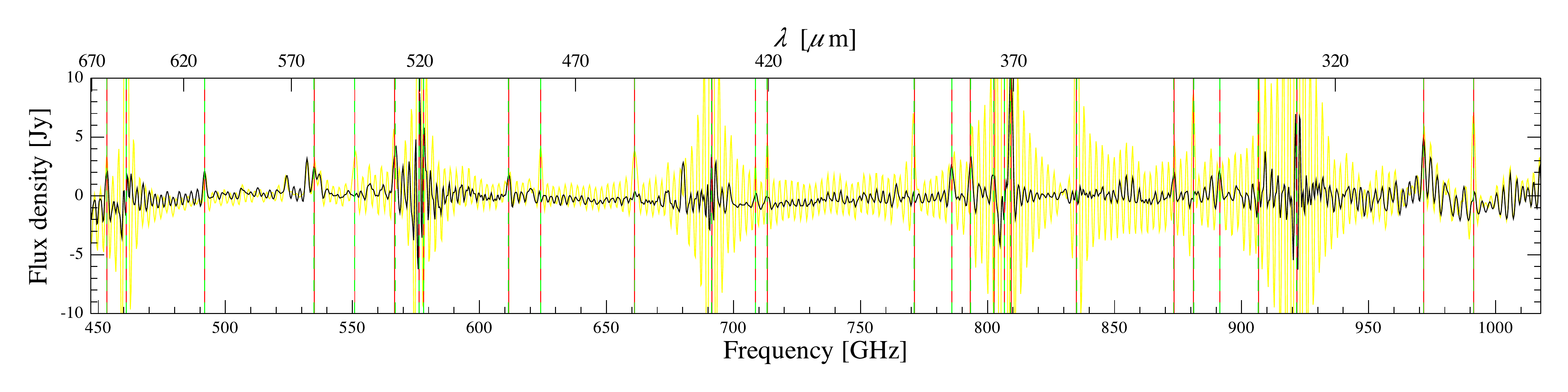}
\includegraphics[trim = 0mm 8mm 0mm 8mm, clip,width=\textwidth]{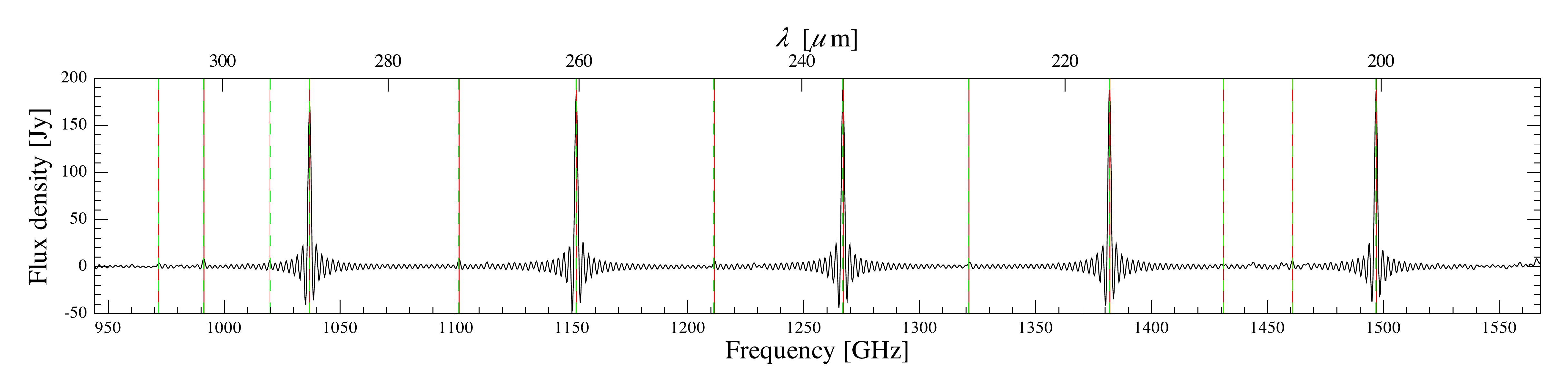}
\includegraphics[trim = 0mm 8mm 0mm 8mm, clip,width=\textwidth]{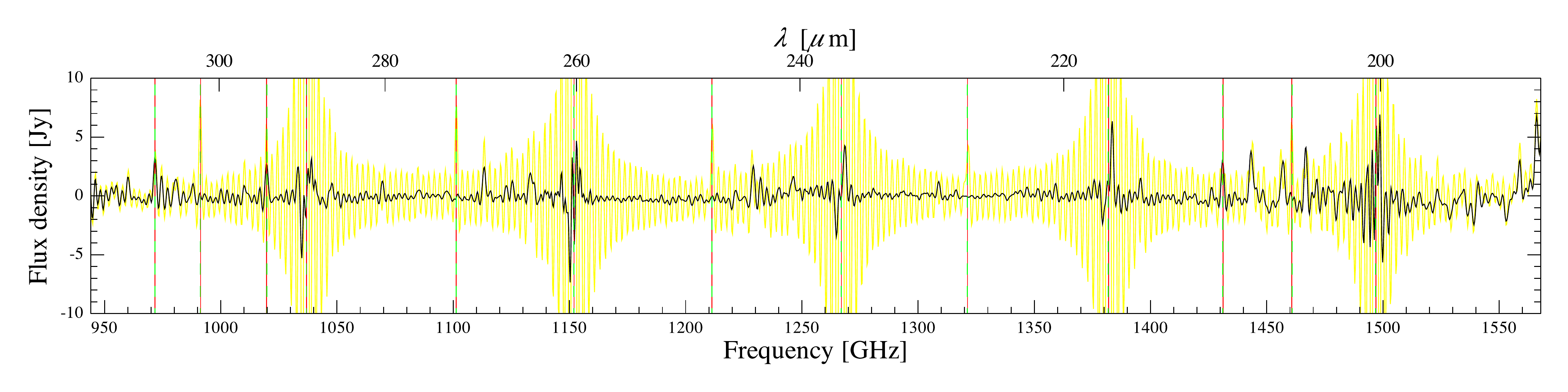}
\caption{Lines fitted to NGC7027 shown with respect to the co-added spectra (first and third panels, from the top) and the residual on subtraction of the combined fit with the co-added data show in yellow (second and fourth panels, from the top). The top two panels show the results for SLWC3 and the bottom two show SSWD4. Red lines indicate the input line positions and the green dashed lines show the mean fitted positions for all observations. }
\label{fig:linesFittedNGC7027}
\end{figure*}

Figures to illustrate the spectral lines fitted to the four main line sources AFGL2688, AFGL4106, CRL618 and NGC7027. For each of Figures~\ref{fig:linesFittedAFGL2688}--\ref{fig:linesFittedNGC7027}, the first and third panels from the top, show the co-added spectra and the second and fourth panels from the top show the residual on subtraction of the combined fit, where the un-subtracted data is shown in yellow. The top two panels show the results for SLWC3 and the bottom two show those for SSWD4. Red lines indicate the input line positions and the green dashed lines show the mean fitted positions for all observations.

\clearpage

\section{Details of FTS repeated calibration observations}
\label{app:tables}

Tables summarising observations comprising the FTS calibration monitoring programme. There is one table of observations per source, unless otherwise stated below. Observations made prior to OD209 are not included.
AFGL2688, AFGL4106, CRL618 and NGC7027 observations are given in Tables~\ref{tab:AFGL2688Obs}--\ref{tab:ngc7027Obs}. Tables~\ref{tab:NGC6302Obs}--\ref{tab:vycmaObs} present the remaining line sources, apart from those with few observations, which are grouped together in Table~\ref{tab:othersObs}. The observations of Uranus are in Table~\ref{tab:UranusObs} and those of Neptune are in Table~\ref{tab:NeptuneObs}. Mars and Saturn are presented in Tables~\ref{tab:MarsObs} and \ref{tab:SaturnObs}. Observations of the asteroids are in Tables~\ref{tab:CeresObs}--\ref{tab:mrObs}.

\begin{table}
\caption{AFGL2688 observations taken after OD\,189. Bias mode and spatial sampling are indicated above each table section. Number of repetitions are provided in the ``Reps'' column and the commanded resolution is given in the ``Res'' column. Pointing offset is given in the final column (P$_{\rm off}$).}
\medskip
\begin{center}
\begin{tabular}{lcclcc}
\hline\hline
\multicolumn{6}{c}{\textbf{AFGL2688}}\\ \hline
OD & dd-mm-yy & Reps & Obsid & Res & P$_{\rm off}$[\arcsec] \\ \hline
\multicolumn{6}{c}{\textbf{HR/CR nominal sparse}}\\ \hline
217 & 17-12-09 & 5 & 1342188198 & CR & 2.6$\pm$0.3 \\ 
227 & 27-12-09 & 5 & 1342188669$^{\rm P_0}$ & CR & 0.0$\pm$0.3 \\ 
240 & 09-01-10 & 5 & 1342189122 & CR & 2.4$\pm$0.3 \\ 
240 & 09-01-10 & 17 & 1342189123$^{\rm *}$ & HR & 0.0$\pm$0.4 \\ 
354 & 03-05-10 & 5 & 1342195770 & CR & 4.5$\pm$0.3 \\ 
395 & 13-06-10 & 5 & 1342198270$^{\rm \dagger}$ & CR & 6.2$\pm$0.3 \\ 
410 & 27-06-10 & 5 & 1342199251$^{\rm \dagger}$ & CR & 7.6$\pm$0.3 \\ 
543 & 08-11-10 & 4 & 1342208387 & CR & 4.5$\pm$0.3  \\ 
572 & 06-12-10 & 4 & 1342210856 & CR & 3.6$\pm$0.3 \\ 
601 & 05-01-11 & 4 & 1342212325 & CR & 4.3$\pm$0.3 \\ 
711 & 25-04-11 & 4 & 1342219570 & CR & 4.6$\pm$0.3 \\ 
767 & 20-06-11 & 4 & 1342222872 & CR & 4.6$\pm$0.3 \\ 
971 & 10-01-12 & 4 & 1342237008$^{\rm \dagger}$ & CR & 8.2$\pm$0.3 \\ 
1080 & 27-04-12 & 4 & 1342245076$^{\rm \dagger}$ & HR & 6.0$\pm$0.3 \\ 
1099 & 16-05-12 & 4 & 1342245863 & HR & 3.3$\pm$0.3 \\ 
1112 & 30-05-12 & 4 & 1342246287 & HR & 3.3$\pm$0.3 \\ 
1130 & 17-06-12 & 4 & 1342247105 & HR & 3.6$\pm$0.3 \\ 
1283 & 17-11-12 & 4 & 1342255262 & HR & 3.7$\pm$0.3 \\ 
1298 & 01-12-12 & 4 & 1342256356 & HR & 4.1$\pm$0.3 \\ 
1313 & 17-12-12 & 4 & 1342257342 & HR & 3.1$\pm$0.3 \\ 
1325 & 29-12-12 & 4 & 1342257916 & HR & 2.5$\pm$0.3 \\ 
1341 & 14-01-13 & 4 & 1342259590 & HR & 1.2$\pm$0.3 \\ 
1434 & 16-04-13 & 4 & 1342270191 & HR & 1.6$\pm$0.3 \\ \hline
\multicolumn{6}{c}{\textbf{LR nominal sparse}}\\ \hline
1080 & 27-04-12 & 4 & 1342245077 & LR & 6.1$\pm$0.3 \\
1099 & 16-05-12 & 4 & 1342245862 & LR & 3.1$\pm$0.3 \\
1112 & 30-05-12 & 4 & 1342246286 & LR & 3.1$\pm$0.3 \\
1130 & 17-06-12 & 4 & 1342247104 & LR & 3.6$\pm$0.3 \\
1283 & 17-11-12 & 4 & 1342255261 & LR & 3.7$\pm$0.3 \\
1298 & 01-12-12 & 4 & 1342256355 & LR & 3.8$\pm$0.3 \\
1313 & 17-12-12 & 4 & 1342257341 & LR & 2.8$\pm$0.3 \\
1325 & 29-12-12 & 4 & 1342257917 & LR & 2.4$\pm$0.3 \\
1341 & 14-01-13 & 4 & 1342259589 & LR & 0.3$\pm$0.4 \\
1434 & 16-04-13 & 4 & 1342270192 & LR & 1.3$\pm$0.3 \\ \hline
\hline
\end{tabular}
\end{center}
\begin{tablenotes}[normal,flushleft]
\item $^{\rm *}${Science observation}
\item $^{\rm \dagger}${Known outlier}
\item $^{\rm P_0}${Pointing offset reference observation}
\end{tablenotes}
\label{tab:AFGL2688Obs}
\end{table}

\begin{table}
\caption{AFGL4106 observations taken after OD\,189. Bias mode and spatial sampling are indicated above each table section. Number of repetitions are provided in the ``Reps'' column and the commanded resolution is given in the ``Res'' column. Pointing offset is given in the final column (P$_{\rm off}$). Two pointing offsets are given for the special calibration observation, which is comprised of two pointings -- one with the BSM set at zero-zero and one with it is set to the {\it Before} offset position.}
\medskip
\begin{center}
\begin{tabular}{lcclcc}
\hline\hline
\multicolumn{6}{c}{\textbf{AFGL4106}}\\ \hline
OD & dd-mm-yy & Reps & Obsid & Res & P$_{\rm off}$[\arcsec] \\ \hline
\multicolumn{6}{c}{\textbf{HR/CR nominal sparse}}\\ \hline
240 & 09-01-10 & 17 & 1342189115$^{\rm *}$ & HR & 2.6$\pm$0.4 \\ 
543 & 07-11-10 & 4 & 1342208380$^{\rm \dagger}$ & CR & 15.8$\pm$0.4 \\ 
557 & 22-11-10 & 4 & 1342209854 & CR & 3.4$\pm$0.3 \\ 
602 & 05-01-11 & 4 & 1342212341 & CR & 0.3$\pm$0.5 \\ 
625 & 29-01-11 & 4 & 1342213382 & CR & 3.1$\pm$0.4 \\ 
654 & 26-02-11 & 4 & 1342214814 & CR & 3.8$\pm$0.3 \\ 
742 & 26-05-11 & 4 & 1342221710 & CR & 0.5$\pm$0.8 \\ 
767 & 19-06-11 & 4 & 1342222861 & CR & 1.9$\pm$0.3 \\ 
803 & 25-07-11 & 4 & 1342224747 & CR & 2.2$\pm$0.3 \\ 
824 & 16-08-11 & 4 & 1342227457 & CR & 2.4$\pm$0.4 \\ 
837 & 28-08-11 & 4 & 1342227777 & CR & 2.1$\pm$0.4 \\ 
837 & 28-08-11 & 4 & 1342227778$^{\rm \dagger\dagger}$ & CR & 2.1$\pm$0.3 \\  
 &  & & & & 0.7$\pm$0.5 \\ 
856 & 16-09-11 & 4 & 1342228698 & CR & 2.2$\pm$0.4 \\ 
886 & 16-10-11 & 4 & 1342231058 & CR & 1.4$\pm$0.5 \\ 
907 & 06-11-11 & 4 & 1342231972 & CR & 2.9$\pm$0.4 \\ 
971 & 09-01-12 & 4 & 1342236994 & CR & 2.4$\pm$0.4 \\ 
988 & 26-01-12 & 4 & 1342238240 & CR & 3.0$\pm$0.3 \\ 
997 & 04-02-12 & 4 & 1342238696 & CR & 3.4$\pm$0.3 \\ 
1012 & 20-02-12 & 4 & 1342239362 & CR & 1.2$\pm$0.3\\ 
1126 & 12-06-12 & 4 & 1342246991 & HR & 1.9$\pm$0.4 \\ 
1144 & 30-06-12 & 4 & 1342247560 & HR & 2.4$\pm$0.4 \\ 
1150 & 07-07-12 & 4 & 1342247748 & HR & 1.2$\pm$0.3 \\ 
1160 & 16-07-12 & 4 & 1342248228 & HR & 3.3$\pm$0.3 \\ 
1165 & 22-07-12 & 4 & 1342248419 & HR & 1.7$\pm$0.4 \\ 
1177 & 03-08-12 & 4 & 1342249072$^{\rm P_0}$ & HR & 0.0$\pm$0.3 \\ 
1186 & 12-08-12 & 4 & 1342249459 & HR & 1.0$\pm$0.6 \\
1255 & 20-10-12 & 4 & 1342253667 & HR & 1.2$\pm$0.3 \\
1299 & 03-12-12 & 4 & 1342256379 & HR & 2.3$\pm$0.3 \\
1326 & 30-12-12 & 4 & 1342257935 & HR & 2.0$\pm$0.3 \\
1362 & 04-02-13 & 4 & 1342262911 & HR & 2.0$\pm$0.3 \\ \hline
\multicolumn{6}{c}{\textbf{LR nominal sparse}}\\ \hline
1126 & 12-06-12 & 4 & 1342246990 & LR & 1.7$\pm$0.4 \\
1144 & 30-06-12 & 4 & 1342247561 & LR & 2.2$\pm$0.4 \\
1150 & 07-07-12 & 4 & 1342247749 & LR & 1.5$\pm$0.4 \\
1160 & 16-07-12 & 4 & 1342248229 & LR & 3.4$\pm$0.3 \\
1165 & 22-07-12 & 4 & 1342248420 & LR & 1.2$\pm$0.4 \\
1177 & 03-08-12 & 4 & 1342249071 & LR & 0.0$\pm$0.3 \\
1186 & 12-08-12 & 4 & 1342249458 & LR & 0.0$\pm$0.4 \\
1255 & 20-10-12 & 4 & 1342253666 & LR & 0.3$\pm$0.5 \\
1299 & 03-12-12 & 4 & 1342256378 & LR & 2.4$\pm$0.3 \\
1326 & 30-12-12 & 4 & 1342257934 & LR & 1.9$\pm$0.3 \\
1362 & 04-02-13 & 4 & 1342262912 & LR & 2.3$\pm$0.3 \\ \hline
\hline
\end{tabular}
\end{center}
\begin{tablenotes}[normal,flushleft]
\item $^{\rm *}${Science observation}
\item $^{\rm \dagger}${Known outlier}
\item $^{\rm \dagger\dagger}${Special calibration}
\item $^{\rm P_0}${Pointing offset reference observation}
\end{tablenotes}
\label{tab:AFGL4106Obs}
\end{table}

\begin{table}
\caption{CRL618 observations taken after OD\,189. Bias mode and spatial sampling are indicated above each table section. Number of repetitions are provided in the ``Reps'' column and the commanded resolution is given in the ``Res'' column. Pointing offset is given in the final column (P$_{\rm off}$). Two pointing offsets are given for the special calibration observation, which is comprised of two pointings -- one with the BSM set at zero-zero and one with it is set to the {\it Before} offset position.}
\medskip
\begin{center}
\begin{tabular}{lcclcc}
\hline\hline
\multicolumn{6}{c}{\textbf{CRL618}}\\ \hline
OD & dd-mm-yy & Reps & Obsid & Res & P$_{\rm off}$[\arcsec] \\ \hline
\multicolumn{6}{c}{\textbf{HR/CR nominal sparse}}\\ \hline
275 & 13-02-10 & 5 & 1342190681 & CR & 1.9$\pm$0.3 \\ 
288 & 26-02-10 & 5 & 1342191215 & CR & 1.3$\pm$0.3 \\ 
302 & 12-03-10 & 5 & 1342192178 & CR & 1.8$\pm$0.4 \\ 
317 & 27-03-10 & 5 & 1342192837 & CR & 1.3$\pm$0.4 \\ 
415 & 08-08-10 & 4 & 1342202259 & CR & 3.6$\pm$0.3 \\ 
466 & 22-08-10 & 4 & 1342204023$^{\rm \dagger}$ & CR & 4.4$\pm$0.3 \\ 
495 & 21-09-10 & 4 & 1342204927 & CR & 1.8$\pm$0.3 \\ 
655 & 28-02-11 & 4 & 1342214858 & CR & 2.0$\pm$0.3 \\ 
682 & 27-03-11 & 4 & 1342216888 & CR & 1.5$\pm$0.4 \\ 
824 & 15-08-11 & 4 & 1342227452 & CR & 2.8$\pm$0.3 \\ 
837 & 29-08-11 & 4 & 1342227784 & CR & 2.8$\pm$0.3 \\ 
837 & 29-08-11 & 4 & 1342227785$^{\rm \dagger\dagger}$ & CR & 1.2$\pm$0.3 \\ 
 & &  &  & & 2.8$\pm$0.3 \\ 
857 & 18-09-11 & 4 & 1342228743 & CR & 2.2$\pm$0.3 \\ 
1024 & 02-03-12 & 4 & 1342240019$^{\rm P_0}$ & CR & 0.0$\pm$0.3 \\ 
1032 & 10-03-12 & 4 & 1342242592 & CR & 1.1$\pm$0.3 \\ 
1187 & 13-08-12 & 4 & 1342249478 & HR & 1.8$\pm$0.3 \\ 
1208 & 03-09-12 & 4 & 1342250533 & HR & 3.9$\pm$0.3 \\ 
1228 & 23-09-12 & 4 & 1342251299 & HR & 1.0$\pm$0.4 \\ 
1389 & 03-03-13 & 4 & 1342265828 & HR & 0.9$\pm$0.3 \\ 
1410 & 24-03-13 & 4 & 1342268302 & HR & 2.0$\pm$0.3 \\ \hline
\multicolumn{6}{c}{\textbf{LR nominal sparse}}\\ \hline
1187 & 13-08-12 & 4 & 1342249479 & LR & 1.9$\pm$0.3 \\
1208 & 03-09-12 & 4 & 1342250531 & LR & 4.0$\pm$0.3 \\
1228 & 23-09-12 & 4 & 1342251298 & LR & 0.0$\pm$0.6 \\
1389 & 03-03-13 & 4 & 1342265827 & LR & 0.1$\pm$0.4 \\
1410 & 24-03-13 & 4 & 1342268301 & LR & 1.6$\pm$0.4 \\ \hline
\multicolumn{6}{c}{\textbf{H+LR nominal sparse}}\\ \hline
1208 & 03-09-12 & 8 & 1342250532 & LR & --- \\ \hline
\multicolumn{6}{c}{\textbf{LR nominal full}}\\ \hline
302 & 12-03-10 & 4 & 1342192179 & LR & --- \\
\hline
\end{tabular}
\end{center}
\begin{tablenotes}[normal,flushleft]
\item $^{\rm \dagger}${Known outlier}
\item $^{\rm \dagger\dagger}${Special calibration}
\item $^{\rm P_0}${Pointing offset reference observation}
\end{tablenotes}
\label{tab:crl618Obs}
\end{table}

\begin{table}
\caption{NGC7027 observations taken after OD\,189. Bias mode and spatial sampling are indicated above each table section. Number of repetitions are provided in the ``Reps'' column and the commanded resolution is given in the ``Res'' column. Pointing offset is given in the final column (P$_{\rm off}$).}
\medskip
\begin{center}
\begin{tabular}{lcclcc}
\hline\hline
\multicolumn{6}{c}{\textbf{NGC7027}}\\ \hline
OD & dd-mm-yy & Reps & Obsid & Res & P$_{\rm off}$[\arcsec] \\ \hline
\multicolumn{6}{c}{\textbf{HR/CR nominal sparse}}\\ \hline
217 & 17-12-09 & 10 & 1342188197 & CR & 2.6$\pm$0.4 \\ 
227 & 27-12-09 & 4 & 1342188670 & CR & 3.5$\pm$0.4 \\ 
240 & 09-01-10 & 10 & 1342189121 & CR & 1.4$\pm$0.4 \\
240 & 09-01-10 & 17 & 1342189124$^{*}$ & CR & 2.4$\pm$0.3 \\
240 & 09-01-10 & 17 & 1342189125$^{*}$ & CR & 2.4$\pm$0.3 \\
250 & 19-01-10 & 10 & 1342189543 & CR & 3.7$\pm$0.3 \\
326 & 05-04-10 & 5 & 1342193812$^{\rm \dagger}$ & CR & 7.3$\pm$0.3 \\
342 & 21-04-10 & 5 & 1342195347 & CR &1.5$\pm$0.3 \\
368 & 17-05-10 & 5 & 1342196614 & CR & 2.5$\pm$0.3 \\
383 & 01-06-10 & 5 & 1342197486 & CR & 2.2$\pm$0.3 \\
409 & 22-06-10 & 5 & 1342198921 & CR & 0.9$\pm$0.3 \\
557 & 22-11-10 & 4 & 1342209856 & CR & 3.3$\pm$0.3 \\
572 & 06-12-10 & 4 & 1342210858 & CR & 3.4$\pm$0.3 \\
601 & 05-01-11 & 4 & 1342212324 & CR & 1.7$\pm$0.4 \\
711 & 25-04-11 & 4 & 1342219571 & CR & 1.3$\pm$0.3 \\
742 & 26-05-11 & 4 & 1342221697 & CR & 1.8$\pm$0.3 \\
908 & 08-11-11 & 4 & 1342231993 & CR  & 4.9$\pm$0.3 \\
971 & 10-01-12 & 4 & 1342237007 & CR & 4.4$\pm$0.3 \\
988 & 27-01-12 & 4 & 1342238245 & CR & 4.3$\pm$0.3 \\
1053 & 01-04-12 & 4 & 1342243593 & HR & 2.3$\pm$0.4 \\
1080 & 27-04-12 & 4 & 1342245075 & HR &1.8$\pm$0.3 \\
1099 & 16-05-12 & 4 & 1342245861 & HR & 0.4$\pm$0.9 \\
1111 & 28-05-12 & 4 & 1342246255 & HR & 1.5$\pm$0.3 \\
1125 & 11-06-12 & 4 & 1342246971 & HR & 1.7$\pm$0.3 \\
1130 & 17-06-12 & 4 & 1342247106 & HR & 1.3$\pm$0.4 \\
1145 & 02-07-12 & 4 & 1342247623$^{\rm P_0}$ & HR & 0.0$\pm$0.3 \\
1283 & 17-11-12 & 4 & 1342255260 & HR & 3.0$\pm$0.3 \\
1313 & 17-12-12 & 4 & 1342257340 & HR & 3.0$\pm$0.3 \\
1325 & 29-12-12 & 4 & 1342257918 & HR & 2.6$\pm$0.3 \\
1341 & 14-01-13 & 4 & 1342259592 & HR & 3.7$\pm$0.3 \\
1434 & 16-04-13 & 4 & 1342270193 & HR & 1.9$\pm$0.3 \\ \hline
\multicolumn{6}{c}{\textbf{LR nominal sparse}}\\ \hline
1080 & 27-04-12 & 4 & 1342245074 & LR & 2.1$\pm$0.4 \\
1099 & 16-05-12 & 4 & 1342245860  & LR & 0.6$\pm$1.0 \\
1111 & 28-05-12 & 4 & 1342246254  & LR & 1.9$\pm$0.4 \\
1125 & 11-06-12 & 4 & 1342246972 & LR & 2.6$\pm$0.3 \\
1130 & 17-06-12 & 4 & 1342247107 & LR & 1.3$\pm$0.5 \\
1145 & 02-07-12 & 4 & 1342247624 & LR & 0.8$\pm$0.5 \\
1283 & 17-11-12 & 4 & 1342255259 & LR & 3.7$\pm$0.4 \\
1313 & 17-12-12 & 4 & 1342257341 & LR & 3.7$\pm$0.4 \\
1325 & 29-12-12 & 4 & 1342257919 & LR & 2.9$\pm$0.4 \\
1341 & 14-01-13 & 4 & 1342259591 & LR & 4.1$\pm$0.4 \\
1434 & 16-04-13 & 4 & 1342270194 & LR & 2.1$\pm$0.4 \\ \hline
\multicolumn{6}{c}{\textbf{HR/CR nominal full}}\\ \hline
742 & 26-05-11 & 4 & 1342221698$^{*}$ & HR & --- \\
\hline
\end{tabular}
\end{center}
\begin{tablenotes}[normal,flushleft]
\item $^{\rm *}${Science observation}
\item $^{\rm \dagger}${Known outlier}
\item $^{\rm P_0}${Pointing offset reference observation}
\end{tablenotes}
\label{tab:ngc7027Obs}
\end{table}

\begin{table}
\caption{NGC6302 observations taken after OD\,189. Bias mode and spatial sampling are indicated above each table section. Number of repetitions are provided in the ``Reps'' column and the commanded resolution is given in the ``Res'' column. Pointing offset is given in the final column (P$_{\rm off}$).}
\medskip
\begin{center}
\begin{tabular}{lcclcc}
\hline\hline
\multicolumn{6}{c}{\textbf{NGC6302}}\\ \hline
OD & dd-mm-yy & Reps & Obsid & Res & P$_{\rm off}$[\arcsec] \\ \hline
\multicolumn{6}{c}{\textbf{HR/CR nominal sparse}}\\ \hline
288 & 26-02-10 & 17 & 1342191224$^{\rm *}$ & HR & 4.0$\pm$0.4 \\ 
1012 & 20-02-12 & 4 & 1342239361 & CR & 6.6$\pm$0.4 \\ 
1024 & 02-03-12 & 4 & 1342240014 & CR & 5.5$\pm$0.4 \\ 
1208 & 03-09-12 & 4 & 1342250537 & HR & 5.7$\pm$0.4 \\ 
1229 & 24-09-12 & 4 & 1342251322 & HR & 5.0$\pm$0.3 \\ 
1242 & 06-10-12 & 4 & 1342252895$^{**}$ & HR & 0.0$\pm$0.3 \\ 
1242 & 06-10-12 & 4 & 1342252896 & HR & 4.5$\pm$0.3 \\ 
1389 & 02-03-13 & 4 & 1342265809 & HR & 5.1$\pm$0.4 \\ 
1410 & 23-03-13 & 4 & 1342268288 & HR & 5.9$\pm$0.4 \\ \hline
\multicolumn{6}{c}{\textbf{LR nominal sparse}}\\ \hline
1208 & 03-09-12 & 4 & 1342250536 & LR & 6.1$\pm$0.4 \\
1229 & 24-09-12 & 4 & 1342251323 & LR & 5.0$\pm$0.4 \\
1242 & 06-10-12 & 4 & 1342252894 & LR & 4.4$\pm$0.4 \\
1389 & 02-03-13 & 4 & 1342265808 & LR & 5.2$\pm$0.4 \\
1410 & 23-03-13 & 4 & 1342268289 & LR & 5.7$\pm$0.4 \\ \hline
\hline
\end{tabular}
\end{center}
\begin{tablenotes}[normal,flushleft]
\item $^{\rm *}${Science observation}
\item $^{**}${Continuum peak and pointing offset reference observation}
\item $^{\rm \dagger}${Known outlier}
\end{tablenotes}
\label{tab:NGC6302Obs}
\end{table}

\begin{table}
\caption{R Dor observations taken after OD\,189. Bias mode and spatial sampling are indicated above each table section. Number of repetitions are provided in the ``Reps'' column and the commanded resolution is given in the ``Res'' column. Pointing offset is given in the final column (P$_{\rm off}$).}
\medskip
\begin{center}
\begin{tabular}{lcclcc}
\hline\hline
\multicolumn{6}{c}{\textbf{R DOR}}\\ \hline
OD & dd-mm-yy & Reps & Obsid & Res & P$_{\rm off}$[\arcsec] \\ \hline
\multicolumn{6}{c}{\textbf{HR/CR nominal sparse}}\\ \hline
971 & 09-01-12 & 16 & 1342236993 & CR & 3.2$\pm$0.3 \\ 
1079 & 27-04-12 & 17 & 1342245114$^{\rm *}$ & HR & 0.0$\pm$0.3 \\ 
1228 & 23-08-12 & 4 & 1342251293 & HR & 1.6$\pm$0.4 \\ 
1255 & 20-10-12 & 4 & 1342253670 & HR & 1.8$\pm$0.4 \\ \hline
\multicolumn{6}{c}{\textbf{LR nominal sparse}}\\ \hline
1228 & 23-08-12 & 4 & 1342251292 & LR & 1.1$\pm$0.4 \\
1255 & 20-10-12 & 4 & 1342253669 & LR & 2.2$\pm$0.3 \\ \hline
\hline
\end{tabular}
\end{center}
\begin{tablenotes}[normal,flushleft]
\item $^{\rm *}${Science observation and pointing offset reference observation}
\item $^{\rm \dagger}${Known outlier}
\end{tablenotes}
\label{tab:rdorObs}
\end{table}

\begin{table}
\caption{CW Leo observations taken after OD\,189. Bias mode and spatial sampling are indicated above each table section. Number of repetitions are provided in the ``Reps'' column and the commanded resolution is given in the ``Res'' column. No pointing offset is given, due to CW Leo's intrinsic variability, which precludes the v14 relative method. CW Leo is also known as IRC+10216, with FIR/submm domain variability \citep{Cernicharo2014}.}
\medskip
\begin{center}
\begin{tabular}{lcclcc}
\hline\hline
\multicolumn{5}{c}{\textbf{CW LEO}}\\ \hline
OD & dd-mm-yy & Reps & Obsid & Res \\ \hline
\multicolumn{5}{c}{\textbf{HR/CR nominal sparse}}\\ \hline
383 & 31-05-10 & 17 & 1342197466$^{\rm *}$  & HR \\ 
742 & 26-05-11 & 4 & 1342221713$^{\rm *}$ & HR \\ 
908 & 07-11-11 & 4 & 1342231984 & CR \\ 
1098 & 16-05-12 & 4 & 1342245849$^{\rm *}$ & HR \\ 
1112 & 29-05-12 & 4 & 1342246269 & HR \\ 
1255 & 20-10-12 & 4 & 1342253660 & HR \\ 
1255 & 20-10-12 & 4 & 1342253661$^{\rm *}$ & HR \\ 
1284 & 18-11-12 & 4 & 1342255284 & HR \\ 
1292 & 26-11-12 & 4 & 1342256105$^{\rm *}$ & HR \\ \hline
\multicolumn{5}{c}{\textbf{LR nominal sparse}}\\ \hline
1112 & 29-05-12 & 4 & 1342246270 & LR \\
1255 & 20-10-12 & 4 & 1342253662 & LR \\
1284 & 18-11-12 & 4 & 1342255283 & LR \\
\hline
\end{tabular}
\end{center}
\begin{tablenotes}[normal,flushleft]
\item $^{\rm *}${Science observation}
\end{tablenotes}
\label{tab:cwleoObs}
\end{table}

\begin{table}
\caption{VY CMa observations taken after OD\,189. Bias mode and spatial sampling are indicated above each table section. Number of repetitions are provided in the ``Reps'' column and the commanded resolution is given in the ``Res'' column. Pointing offset is given in the final column (P$_{\rm off}$).}
\medskip
\begin{center}
\begin{tabular}{lcclcc}
\hline\hline
\multicolumn{6}{c}{\textbf{VY CMA}}\\ \hline
OD & dd-mm-yy & Reps & Obsid & Res & P$_{\rm off}$[\arcsec] \\ \hline
\multicolumn{6}{c}{\textbf{HR/CR nominal sparse}}\\ \hline
317 & 27-03-10 & 17 & 1342192834$^{\rm *}$ & HR & 1.9$\pm$0.4 \\ 
907 & 06-11-11 & 4 & 1342231976 & CR & 3.9$\pm$0.3 \\ 
1054 & 02-04-12 & 4 & 1342243639 & CR & 2.9$\pm$0.3 \\ 
1079 & 27-04-12 & 4 & 1342245116 & HR & 1.1$\pm$0.3 \\ 
1098 & 15-05-122 & 4 & 1342245843 & HR & 1.1$\pm$0.3 \\ 
1241 & 06-10-12 & 4 & 1342252292 & HR & 0.2$\pm$0.4 \\ 
1255 & 20-10-12 & 4 & 1342253665 & HR & 0.5$\pm$0.3 \\ 
1433 & 16-04-13 & 4 & 1342270034$^{\rm P_0}$ & HR & 0.0$\pm$0.3 \\ \hline
\multicolumn{6}{c}{\textbf{LR nominal sparse}}\\ \hline
1079 & 27-04-12 & 4 & 1342245115 & LR & 0.8$\pm$0.6 \\
1098 & 15-05-122 & 4 & 1342245842 & LR & 0.6$\pm$0.7 \\
1241 & 06-10-12 & 4 & 1342252291 & LR & 0.0$\pm$1.0 \\
1255 & 20-10-12 & 4 & 1342253664 & LR & 0.5$\pm$0.8 \\
1433 & 16-04-13 & 4 & 1342270033 & LR & 0.0$\pm$0.9 \\ \hline
\hline
\end{tabular}
\end{center}
\begin{tablenotes}[normal,flushleft]
\item $^{\rm *}${Science observation}
\item $^{\rm P_0}${Pointing offset reference observation}
\end{tablenotes}
\label{tab:vycmaObs}
\end{table}

\begin{table}
\caption{A number of calibration observations taken after OD\,189. Bias mode and spatial sampling are indicated above each table section. Number of repetitions are provided in the ``Reps'' column and the commanded resolution is given in the ``Res'' column.}
\medskip
\begin{center}
\begin{tabular}{lcccc}
\hline\hline
\multicolumn{5}{c}{\textbf{OTHER LINE SOURCES}}\\ \hline 
Name & dd-mm-yy & OD & Reps & Obsid \\ \hline
\multicolumn{5}{c}{\textbf{HR nominal sparse}}\\ \hline 
IK Tau & 12-03-10 & 302 & 17 & 1342192176$^{\rm *}$ \\ 
Omi Cet & 19-01-10 & 250 & 17 & 1342189546$^{\rm *}$ \\ 
Omi Cet & 03-02-13 & 1361 & 12 & 1342262863 \\
W Hya & 09-01-10 & 240 & 17 & 1342189116$^{\rm *}$ \\ \hline
\multicolumn{5}{c}{\textbf{LR nominal sparse}}\\ \hline
Omi Cet & 03-02-13 & 1361 & 12 & 1342262862 \\
\hline
\end{tabular}
\end{center}
\begin{tablenotes}[normal,flushleft]
\item $^{\rm *}${Science observation}
\end{tablenotes}
\label{tab:othersObs}
\end{table}

\clearpage
\begin{table}
\caption{Uranus observations taken after OD\,189. Bias mode and spatial sampling are indicated above each table section. Number of repetitions are provided in the ``Reps'' column and the commanded resolution is given in the ``Res'' column. Pointing offset is given in the final column (P$_{\rm off}$), which has an associated 0.3\arcsec\,uncertainty. Two offsets are given for the special calibration observation, which comprises of two pointings. Pointing offset are given for sparse mode, except the the MR observations.}
\medskip
\begin{center}
\begin{tabular}{lcclcc}
\hline\hline
\multicolumn{6}{c}{\textbf{URANUS}}\\ \hline
OD & dd-mm-yy & Reps & Obsid & Res & P$_{\rm off}$[\arcsec] \\ \hline
\multicolumn{6}{c}{\textbf{HR/CR nominal sparse}}\\ \hline
209 & 09-12-09 & 4 & 1342187880 & CR & 1.6$\pm$0.3 \\ 
383 & 01-06-10 & 22 & 1342197472 & CR & 0.5$\pm$0.3 \\ 
383 & 01-06-10 & 6 & 1342197473 & CR & 2.9$\pm$0.5 \\ 
383 & 01-06-10 & 6 & 1342197474 & CR & 3.4$\pm$0.5 \\ 
383 & 01-06-10 & 6 & 1342197475 & CR & 3.3$\pm$0.5 \\ 
383 & 01-06-10 & 6 & 1342197476 & CR & 2.0$\pm$0.7 \\ 
383 & 01-06-10 & 6 & 1342197477 & CR & 3.0$\pm$0.6 \\ 
383 & 01-06-10 & 6 & 1342197478 & CR & 2.3$\pm$0.6 \\ 
395 & 13-06-10 & 4 & 1342198273 & CR & 0.4$\pm$0.3 \\ 
423 & 10-07-10 & 64 & 1342200175$^{\rm *}$ & HR & 1.7$\pm$0.3 \\
571 & 05-12-10 & 4 & 1342210844 & CR & 2.8$\pm$0.3 \\  
602 & 05-01-11 & 4 & 1342212338 & CR & 1.2$\pm$0.3 \\ 
767 & 19-06-11 & 4 & 1342222864 & CR & 0.4$\pm$0.4 \\ 
767 & 19-06-11 & 4 & 1342222865 & CR & 6.0$\pm$0.4 \\ 
767 & 19-06-11 & 4 & 1342222866 & CR & 6.7$\pm$0.4 \\ 
767 & 19-06-11 & 4 & 1342222867 & CR & 5.3$\pm$0.4 \\ 
767 & 19-06-11 & 4 & 1342222868 & CR & 5.2$\pm$0.4 \\ 
767 & 19-06-11 & 4 & 1342222869 & CR & 3.6$\pm$0.5 \\ 
972 & 10-01-12 & 4 & 1342237013 & CR & 2.8$\pm$0.3 \\ 
972 & 10-01-12 & 6 & 1342237014 & HR & 3.1$\pm$0.3 \\ 
972 & 10-01-12 & 6 & 1342237017$^{\rm \dagger\dagger}$ & CR & 1.8$\pm$0.4 \\
 & &  & &  & 2.7$\pm$0.3 \\
972 & 10-01-12 & 4 & 1342237019 & CR & 2.8$\pm$0.6 \\ 
972 & 10-01-12 & 4 & 1342237020 & CR & 2.5$\pm$0.5 \\ 
972 & 10-01-12 & 4 & 1342237021 & CR & 7.1$\pm$0.4 \\ 
972 & 10-01-12 & 4 & 1342237022 & CR & 4.4$\pm$0.5 \\ 
972 & 10-01-12 & 4 & 1342237023 & CR & 6.1$\pm$0.4 \\ 
1112 & 30-05-12 & 22 & 1342246285$^{\rm \dagger\dagger}$ & HR & 1.9$\pm$0.3 \\
 &  &  & & & 3.6$\pm$0.3 \\
1125 & 11-06-12 & 6 & 1342246974 & HR & 4.2$\pm$0.3 \\
1130 & 17-06-12 & 6 & 1342247100 & HR & 1.6$\pm$0.3 \\ 
1145 & 02-07-12 & 6 & 1342247618 & HR & 2.1$\pm$0.3 \\ 
1151 & 08-07-12 & 6 & 1342247767 & HR & 3.5$\pm$0.3 \\ 
1313 & 16-12-12 & 4 & 1342257307$^{\rm P_0}$ & HR & 0.0$\pm$0.3 \\  
1326 & 29-12-12 & 4 & 1342257926 & HR & 1.5$\pm$0.3 \\  
1335 & 07-01-13 & 4 & 1342258697 & HR & 1.6$\pm$0.3 \\  
1341 & 14-01-13 & 4 & 1342259588 & HR & 3.1$\pm$0.3 \\ \hline
\multicolumn{6}{c}{\textbf{LR nominal sparse}}\\ \hline
972 & 10-01-12 & 6 & 1342237016 & LR & 2.8$\pm$0.3 \\
1112 & 30-05-12 & 22 & 1342246283$^{\rm \dagger\dagger}$ & LR & 2.7$\pm$0.3 \\
 & &  & & & 4.2$\pm$0.3 \\
1125 & 11-06-12 & 6 & 1342246975 & LR & 4.1$\pm$0.3 \\
1130 & 17-06-12 & 6 & 1342247101 & LR & 1.9$\pm$0.3 \\ 
1145 & 02-07-12 & 6 & 1342247619 & LR & 2.1$\pm$0.3 \\
1151 & 08-07-12 & 6 & 1342247766 & LR & 3.5$\pm$0.3 \\
1313 & 16-12-12 & 4 & 1342257305 & LR & 0.5$\pm$0.4 \\
1326 & 29-12-12 & 4 & 1342257927 & LR & 1.5$\pm$0.3 \\
1335 & 07-01-13 & 4 & 1342258696 & LR & 1.7$\pm$0.3 \\
1341 & 14-01-13 & 4 & 1342259587 & LR & 3.5$\pm$0.3 \\ \hline
\end{tabular}
\end{center}
\label{tab:UranusObs}
\end{table}

\addtocounter{table}{-1}
\begin{table}
\caption{}
\medskip
\begin{center}
\begin{tabular}{lcclcc}
\hline
\multicolumn{6}{c}{\textbf{URANUS cont.}}\\ \hline
\multicolumn{6}{c}{\textbf{MR nominal sparse}}\\ \hline
972 & 10-01-12 & 6 & 1342237015 & MR & --- \\  \hline
\multicolumn{6}{c}{\textbf{HR/CR bright sparse}}\\ \hline
209 & 09-12-09 & 4 & 1342187879 & CR & --- \\ 
383 & 01-06-10 & 5 & 1342197485 & CR & --- \\
395 & 13-06-10 & 4 & 1342198274 & CR & --- \\
410 & 28-06-10 & 4 & 1342199261 & CR & --- \\
423 & 10-07-10 & 4 & 1342200174 & CR & --- \\
972 & 10-01-12 & 6 & 1342237018 & CR & --- \\ \hline
\multicolumn{6}{c}{\textbf{LR nominal intermediate}}\\ \hline
383 & 01-06-10 & 4 & 1342197481 & LR & --- \\
383 & 01-06-10 & 4 & 1342197482 & LR & --- \\
383 & 01-06-10 & 4 & 1342197483 & LR & --- \\
383 & 01-06-10 & 4 & 1342197484 & LR & --- \\
\multicolumn{6}{c}{\textbf{LR nominal special calibration map}}\\ \hline
1112 & 30-05-12 & 4 & 1342246284$^{\rm \dagger\dagger}$ & LR & --- \\
1313 & 16-12-12 & 4 & 1342257306$^{\rm \dagger\dagger}$ & LR & --- \\
\hline
\end{tabular}
\end{center}
\begin{tablenotes}[normal,flushleft]
\item $^{\rm *}${Science observation}
\item $^{\rm P_0}${Pointing offset reference observation}
\item $^{\rm \dagger\dagger}${Special calibration}
\end{tablenotes}
\label{tab:UranusObs}
\end{table}

\begin{table}
\caption{Neptune observations taken after OD\,189. Bias mode and spatial sampling are indicated above each table section. Number of repetitions are provided in the ``Reps'' column and the commanded resolution is given in the ``Res'' column. Pointing offset is given in the final column (P$_{\rm off}$), which has an associated 0.3\arcsec\,uncertainty. Two pointing offsets are given for the special calibration observation, which is comprised of two pointings -- one with the BSM set at zero-zero and one with it is set to the {\it Before} offset position. Pointing offset are given for sparse mode, except for off-axis detector observations.}
\medskip
\begin{center}
\begin{tabular}{lcclcc}
\hline\hline
\multicolumn{6}{c}{\textbf{NEPTUNE}}\\ \hline
OD & dd-mm-yy & Reps & Obsid & Res & P$_{\rm off}$[\arcsec] \\ \hline
\multicolumn{6}{c}{\textbf{HR/CR nominal sparse}}\\ \hline
209 & 09-12-09 & 4 & 1342187883 & HR & --- \\ 
209 & 09-12-09 & 4 & 1342187884 & HR & --- \\ 
209 & 09-12-09 & 4 & 1342187885 & HR & --- \\ 
209 & 09-12-09 & 4 & 1342187887 & CR & 0.7 \\ 
342 & 21-04-10 & 6 & 1342195348 & CR & 1.8 \\ 
354 & 03-05-10 & 6 & 1342195771 & CR & 1.0 \\ 
368 & 17-05-10 & 6 & 1342196617$^{\rm P_0}$ & CR & 0.0 \\ 
381 & 31-05-10 & 6 & 1342197362 & CR & --- \\ 
381 & 31-05-10 & 6 & 1342197363 & CR & --- \\ 
381 & 31-05-10 & 6 & 1342197364 & CR & --- \\ 
381 & 31-05-10 & 6 & 1342197365 & CR & --- \\ 
381 & 31-05-10 & 6 & 1342197366 & CR & --- \\ 
381 & 31-05-10 & 6 & 1342197367 & CR & --- \\  
381 & 31-05-10 & 22 & 1342197368 & CR & 1.6 \\ 
392 & 09-06-10 & 100 & 1342198429$^{\rm *}$ & HR & 1.6 \\ 
543 & 08-11-10 & 4 & 1342208385 & CR & 2.3 \\ 
557 & 22-11-10 & 4 & 1342209855$^{\rm \dagger}$ & CR & 4.0 \\ 
571 & 05-12-10 & 4 & 1342210841 & CR & 1.5 \\ 
711 & 25-04-11 & 4 & 1342219564 & CR & 1.5 \\ 
742 & 26-05-11 & 4 & 1342221703 & CR & 1.3 \\ 
908 & 08-11-11 & 4 & 1342231992$^{\rm \dagger}$ & CR & 3.2 \\ 
1080 & 27-04-12 & 16 & 1342245081 & HR & 0.3 \\ 
1099 & 16-05-12 & 16 & 1342245866 & HR & 1.0 \\ 
1112 & 30-05-12 & 22 & 1342246280$^{\rm \dagger\dagger}$ & HR & 2.0 \\
 & & & & & 2.2 \\
1125 & 11-06-12 & 6 & 1342246977 & HR & 0.5 \\ 
1284 & 18-11-12 & 4 &1342255278 & HR & 1.8 \\ 
1292 & 25-11-12 & 4 & 1342256096 & HR & 2.1 \\ 
1299 & 02-12-12 & 4 & 1342256369 & HR & 1.4 \\ 
1314 & 17-12-12 & 4 & 1342257352 & HR & 1.6 \\ \hline
\multicolumn{6}{c}{\textbf{LR nominal sparse}}\\ \hline
1080 & 27-04-12 & 16 & 1342245080 & LR & 0.4 \\
1099 & 16-05-12 & 16 & 1342245865 & LR & 1.2 \\
1112 & 30-05-12 & 22 & 1342246278$^{\rm \dagger\dagger}$ & LR & --- \\
1125 & 11-06-12 & 6 & 1342246976 & LR & 0.8 \\
1284 & 18-11-12 & 4 &1342255277 & LR & 1.8 \\
1292 & 25-11-12 & 4 & 1342256095 & LR & 2.2 \\
1299 & 02-12-12 & 4 & 1342256367 & LR & 1.7 \\
1314 & 17-12-12 & 4 & 1342257350 & LR & 1.6 \\ \hline
\multicolumn{6}{c}{\textbf{CR/HR bright sparse}}\\ \hline
382 & 31-05-10 & 2 & 1342197369 & CR & --- \\
1099 & 17-05-12 & 4 & 1342245867 & HR & --- \\ \hline
\multicolumn{6}{c}{\textbf{LR nominal full}}\\ \hline
354 & 03-05-10 & 4 & 1342195774 & LR & --- \\
354 & 03-05-10 & 4 & 1342195775 & LR & --- \\ \hline
\multicolumn{6}{c}{\textbf{LR nominal special calibration map}}\\ \hline
1112 & 30-05-12 & 4 & 1342246279 & LR & --- \\
1299 & 02-12-12 & 4 & 1342256368 & LR & --- \\
1314 & 17-12-12 & 4 & 1342257351 & LR & --- \\
\hline
\end{tabular}
\end{center}
\begin{tablenotes}[normal,flushleft]
\item $^{\rm *}${Science observation}
\item $^{\rm \dagger}${Known outlier}
\item $^{\rm \dagger\dagger}${Special calibration}
\item $^{\rm P_0}${Pointing offset reference observation}
\end{tablenotes}
\label{tab:NeptuneObs}
\end{table}

\begin{table}
\caption{Mars observations taken after OD\,189. All Mars observations were taken in bright mode. Number of repetitions are provided in the ``Reps'' column and the commanded resolution is given in the ``Res'' column.}
\medskip
\begin{center}
\begin{tabular}{lcclc}
\hline\hline
\multicolumn{5}{c}{\textbf{MARS}}\\ \hline
OD & dd-mm-yy & Reps & Obsid & Res \\ \hline
\multicolumn{5}{c}{\textbf{HR/CR bright sparse}}\\ \hline
327 & 06-04-10 & 2 & 1342193677 & HR \\
327 & 06-04-10 & 2 & 1342193676 & HR \\
327 & 06-04-10 & 2 & 1342193675 & HR \\
327 & 06-04-10 & 2 & 1342193674 & HR \\
383 & 31-05-10 & 2 & 1342197462 & HR \\ 
404 & 22-06-10 & 4 & 1342198930 & CR \\ 
886 & 16-10-11 & 4 & 1342231056 & CR \\ 
886 & 16-10-11 & 4 & 1342231059 & CR \\ 
886 & 16-10-11 & 4 & 1342231062 & CR \\ 
886 & 17-10-11 & 4 & 1342231070 & CR \\ 
886 & 17-10-11 & 4 & 1342231079 & CR \\ 
886 & 17-10-11 & 4 & 1342231076 & CR \\ 
886 & 17-10-11 & 4 & 1342231085 & CR \\ 
1098 & 16-05-12 & 4 & 1342245848 & HR \\ 
1144 & 30-06-12 & 4 & 1342247563 $^{\rm \dagger\dagger}$ & HR \\ 
1144 & 30-06-12 & 4 & 1342247571 $^{\rm \dagger\dagger}$ & HR \\ \hline
\hline
\end{tabular}
\end{center}
\begin{tablenotes}[normal,flushleft]
\item $^{\rm \dagger\dagger}${Special calibration}
\end{tablenotes}
\label{tab:MarsObs}
\end{table}

\begin{table}
\caption{Saturn observations taken after OD\,189. All Saturn observations were taken in bright mode. Number of repetitions are provided in the ``Reps'' column and the commanded resolution is given in the ``Res'' column.}
\medskip
\begin{center}
\begin{tabular}{lcclc}
\hline\hline
\multicolumn{5}{c}{\textbf{SATURN}}\\ \hline
OD & dd-mm-yy & Reps & Obsid & Res \\ \hline
\multicolumn{5}{c}{\textbf{HR/CR bright sparse}}\\ \hline
327 & 25-07-11 & 10 & 1342224754$^{\rm *}$ & HR \\
395 & 13-06-10 & 4 & 1342198279 & CR \\
1150 & 07-07-12 & 8 & 1342247750 & HR \\ \hline
\hline
\end{tabular}
\end{center}
\begin{tablenotes}[normal,flushleft]
\item $^{\rm *}${Science observation}
\end{tablenotes}
\label{tab:SaturnObs}
\end{table}

\begin{table}
\caption{Ceres observations taken after OD\,189. Bias mode and spatial sampling are indicated above each table section. Number of repetitions are provided in the ``Reps'' column and the commanded resolution is given in the ``Res'' column. Pointing offset is given in the final column (P$_{\rm off}$).}
\medskip
\begin{center}
\begin{tabular}{lcclcc}
\hline\hline
\multicolumn{6}{c}{\textbf{CERES}}\\ \hline
OD & dd-mm-yy & Reps & Obsid & Res & P$_{\rm off}$[\arcsec] \\ \hline
\multicolumn{6}{c}{\textbf{HR/CR nominal sparse}}\\ \hline
275 & 13-02-10 & 10 & 1342190673 & CR & 3.2$\pm$0.3 \\ 
288 & 26-02-10 & 10 & 1342191220 & CR & 4.0$\pm$0.3 \\ 
327 & 06-04-10 & 10 & 1342193667$^{\rm \dagger}$ & CR & 6.4$\pm$0.3 \\ 
494 & 19-09-10 & 4 & 1342204878 & CR & 4.7$\pm$0.3 \\ 
767 & 19-06-11 & 4 & 1342222862 & CR & 1.8$\pm$0.4 \\ 
972 & 10-01-12 & 4 & 1342237010 & CR & 3.3$\pm$0.4 \\ 
1187 & 12-08-12 & 4 & 1342249467 & HR & 1.6$\pm$0.4 \\ 
1208 & 03-09-12 & 4 & 1342250535 & HR & 2.6$\pm$0.4 \\ 
1228 & 23-09-12 & 4 & 1342251297 & HR & 1.2$\pm$0.4 \\ 
1241 & 06-10-12 & 4 & 1342252294$^{\rm P_0}$ & HR & 0.0$\pm$0.3 \\ 
1390 & 04-03-13 & 4 & 1342265862 & HR & 2.2$\pm$0.3 \\ 
1410 & 04-03-13 & 4 & 1342268300 & HR & 2.3$\pm$0.3 \\ 
1433 & 16-04-13 & 4 & 1342270037 & HR & 2.3$\pm$0.3 \\ \hline
\multicolumn{6}{c}{\textbf{LR nominal sparse}}\\ \hline
1187 & 12-08-12 & 4 & 1342249466 & LR & 1.9$\pm$0.4 \\
1208 & 03-09-12 & 4 & 1342250534 & LR & 3.0$\pm$0.3 \\
1228 & 23-09-12 & 4 & 1342251296 & LR & 1.1$\pm$0.4 \\
1241 & 06-10-12 & 4 & 1342252293 & LR & 0.4$\pm$0.5 \\
1390 & 04-03-13 & 4 & 1342265861 & LR & 2.3$\pm$0.3 \\
1410 & 24-03-13 & 4 & 1342268299 & LR & 2.3$\pm$0.3 \\
1433 & 16-04-13 & 4 & 1342270036 & LR & 2.5$\pm$0.3 \\ \hline
\multicolumn{6}{c}{\textbf{LR nominal special calibration map}}\\ \hline
1390 & 04-03-13 & 4 & 1342265860 & LR & ---\\
1410 & 24-03-13 & 4 & 1342268298 & LR & --- \\
1433 & 16-04-13 & 4 & 1342270038 & LR & --- \\
\hline
\end{tabular}
\end{center}
\begin{tablenotes}[normal,flushleft]
\item $^{\rm \dagger}${Known outlier}
\item $^{\rm P_0}${Pointing offset reference observation}
\end{tablenotes}
\label{tab:CeresObs}
\end{table}

\begin{table}
\caption{Hebe observations taken after OD\,189. Bias mode and spatial sampling are indicated above each table section. Number of repetitions are provided in the ``Reps'' column and the commanded resolution is given in the ``Res'' column.}
\medskip
\begin{center}
\begin{tabular}{lcclc}
\hline\hline
\multicolumn{5}{c}{\textbf{HEBE}}\\ \hline
OD & dd-mm-yy & Reps & Obsid & Res \\ \hline
\multicolumn{5}{c}{\textbf{HR/CR nominal sparse}}\\ \hline
1098 & 16-05-12 & 12 & 1342245850 & HR \\ 
1131 & 17-06-12 & 12 & 1342247115 & HR \\ 
1361 & 03-02-13 & 12 & 1342262857 & HR \\ \hline
\multicolumn{5}{c}{\textbf{LR nominal sparse}}\\ \hline
1131 & 17-06-12 & 12 & 1342247116 & LR \\
1361 & 03-02-13 & 12 & 1342262856 & LR \\ \hline
\hline
\end{tabular}
\end{center}
\label{tab:HebeObs}
\end{table}

\begin{table}
\caption{Hygiea observations taken after OD\,189. Bias mode and spatial sampling are indicated above each table section. Number of repetitions are provided in the ``Reps'' column and the commanded resolution is given in the ``Res'' column. Pointing offset is given in the final column (P$_{\rm off}$).}
\medskip
\begin{center}
\begin{tabular}{lcclcc}
\hline\hline
\multicolumn{6}{c}{\textbf{HYGIEA}}\\ \hline
OD & dd-mm-yy & Reps & Obsid & Res & P$_{\rm off}$[\arcsec] \\ \hline
\multicolumn{6}{c}{\textbf{HR/CR nominal sparse}}\\ \hline
804 & 26-07-11 & 14 & 1342224763 & CR & 3.8$\pm$0.5 \\ 
837 & 29-08-11 & 4 & 1342227794 & CR & 3.7$\pm$0.5 \\ 
856 & 17-09-11 & 4 & 1342228705 & CR & 4.1$\pm$0.5 \\ 
1080 & 28-04-12 & 4 & 1342245095 & HR & 3.8$\pm$0.5 \\ 
1112 & 30-05-12 & 4 & 1342246282 & HR & 3.8$\pm$0.5 \\ 
1262 & 27-10-12 & 4 & 1342253968 & HR & 3.7$\pm$0.5 \\ 
1284 & 18-11-12 & 4 & 1342255276$^{\rm P_0}$ & HR & 0.0$\pm$0.3 \\ 
1314 & 17-12-12 & 4 & 1342257349 & HR & 0.1$\pm$3.2 \\ \hline
\multicolumn{6}{c}{\textbf{LR nominal sparse}}\\ \hline
1080 & 28-04-12 & 4 & 1342245096 & LR & 4.4$\pm$0.5 \\
1112 & 30-05-12 & 4 & 1342246281 & LR & 4.1$\pm$0.5 \\
1262 & 27-10-12 & 4 & 1342253969 & LR & 4.0$\pm$0.5 \\
1284 & 18-11-12 & 4 & 1342255275 & LR & 2.5$\pm$0.6 \\
1314 & 17-12-12 & 4 & 1342257348 & LR & 0.2$\pm$3.7 \\ \hline
\multicolumn{6}{c}{\textbf{MR nominal sparse}}\\ \hline
342 & 21-04-10 & 20 & 1342195342 & MR & --- \\
\hline
\end{tabular}
\end{center}
\begin{tablenotes}[normal,flushleft]
\item $^{\rm P_0}${Pointing offset reference observation}
\end{tablenotes}
\label{tab:HygieaObs}
\end{table}

\begin{table}
\caption{Juno observations taken after OD\,189. Bias mode and spatial sampling are indicated above each table section. Number of repetitions are provided in the ``Reps'' column and the commanded resolution is given in the ``Res'' column.}
\medskip
\begin{center}
\begin{tabular}{lcclc}
\hline\hline
\multicolumn{5}{c}{\textbf{JUNO}}\\ \hline
OD & dd-mm-yy & Reps & Obsid & Res \\ \hline
\multicolumn{5}{c}{\textbf{HR/CR nominal sparse}}\\ \hline
997 & 04-02-12 & 4 & 1342238698 & CR \\ 
1012 & 20-02-12 & 12 & 1342239360 & CR \\ 
1032 & 11-02-12 & 12 & 1342242599 & CR \\ \hline
\multicolumn{5}{c}{\textbf{MR nominal sparse}}\\ \hline
227 & 27-12-09 & 30 & 1342188674 & MR \\ 
261 & 30-01-10 & 30 & 1342189901$^{\rm *}$ & MR \\ 
\hline
\end{tabular}
\end{center}
\begin{tablenotes}[normal,flushleft]
\item $^{\rm *}${Science observation}
\end{tablenotes}
\label{tab:JunoObs}
\end{table}

\begin{table}
\caption{Pallas observations taken after OD\,189. Bias mode and spatial sampling are indicated above each table section. Number of repetitions are provided in the ``Reps'' column and the commanded resolution is given in the ``Res'' column. Pointing offset is given in the final column (P$_{\rm off}$).}
\medskip
\begin{center}
\begin{tabular}{lcclcc}
\hline\hline
\multicolumn{6}{c}{\textbf{PALLAS}}\\ \hline
OD & dd-mm-yy & Reps & Obsid & Res & P$_{\rm off}$[\arcsec] \\ \hline
\multicolumn{6}{c}{\textbf{HR/CR nominal sparse}}\\ \hline
423 & 10-07-10 & 10 & 1342200177 & CR & 0.0$\pm$1.2 \\
711 & 25-04-11 & 14 & 1342219569 & CR & 0.0$\pm$0.3 \\ 
885 & 16-10-11 & 4 & 1342231051$^{\rm P_0}$ & CR & 2.2$\pm$0.8 \\ 
1130 & 17-06-12 & 12 & 1342247103  & HR & 2.4$\pm$0.4 \\ 
1145 & 02-07-11 & 12 & 1342247621 & HR & 3.2$\pm$0.4 \\ 
1299 & 02-12-12 & 12 & 1342256371 & HR & 1.8$\pm$0.5 \\ 
1326 & 29-12-12 & 4 & 1342257929 & HR & 1.7$\pm$0.6 \\ 
1340 & 12-01-13 & 4 & 1342259571 & HR & 4.2$\pm$0.4 \\ \hline
\multicolumn{6}{c}{\textbf{LR nominal sparse}}\\ \hline
1130 & 17-06-12 & 12 & 1342247102 & LR & 2.6$\pm$0.4 \\
1145 & 02-07-11 & 12 & 1342247620 & LR & 3.6$\pm$0.4 \\
1299 & 02-12-12 & 12 & 1342256372 & LR & 1.8$\pm$0.5 \\
1326 & 29-12-12 & 4 & 1342257928 & LR & 1.9$\pm$0.5 \\
1340 & 12-01-13 & 4 & 1342259572 & LR & 4.4$\pm$0.4 \\ \hline
\hline
\end{tabular}
\end{center}
\begin{tablenotes}[normal,flushleft]
\item $^{\rm P_0}${Pointing offset reference observation}
\end{tablenotes}
\label{tab:PallasObs}
\end{table}

\begin{table}
\caption{Vesta observations taken after OD\,189. Bias mode and spatial sampling are indicated above each table section. Number of repetitions are provided in the ``Reps'' column and the commanded resolution is given in the ``Res'' column. Pointing offset is given in the final column (P$_{\rm off}$). No offset is given when the source is positioned in an off-axis detector or for the bright mode observation. }
\medskip
\begin{center}
\begin{tabular}{lcclcc}
\hline\hline
\multicolumn{6}{c}{\textbf{VESTA}}\\ \hline
OD & dd-mm-yy & Reps & Obsid & Res & P$_{\rm off}$[\arcsec] \\ \hline
\multicolumn{6}{c}{\textbf{HR/CR nominal sparse}}\\ \hline
209 & 09-12-09 & 50 & 1342187897 & CR & 2.0$\pm$0.4 \\ 
209 & 09-12-09 & 40 & 1342187898 & CR & --- \\ 
209 & 09-12-09 & 40 & 1342187899 & CR & --- \\ 
209 & 09-12-09 & 40 & 1342187900 & CR & --- \\ 
354 & 03-05-10 & 10 & 1342195769 & CR & 3.2$\pm$0.3 \\ 
368 & 17-05-10 & 10 & 1342196620 & CR & 3.5$\pm$0.3 \\ 
383 & 31-05-10 & 10 & 1342197465 & CR & 3.0$\pm$0.4 \\ 
395 & 13-06-10 & 10 & 1342198265 & CR & 1.9$\pm$0.4 \\ 
404 & 22-06-10 & 10 & 1342198924$^{\rm P_0}$ & CR & 0.0$\pm$0.3 \\ 
410 & 27-06-10 & 10 & 1342199247 & CR & 3.9$\pm$0.4 \\ 
908 & 08-11-11 & 4 & 1342231991 & CR & 3.0$\pm$0.4 \\ 
1178 & 04-08-12 & 4 & 1342249049 & HR & 1.3$\pm$1.3 \\ 
1208 & 02-09-12 & 4 & 1342250525 & HR & 1.9$\pm$0.5 \\ 
1228 & 23-09-12 & 4 & 1342251295 & HR & 0.6$\pm$1.1 \\ 
1390 & 04-03-13 & 4 & 1342265847 & HR & 2.2$\pm$0.4 \\ 
1410 & 24-03-13 & 4 & 1342268297 & HR & 3.2$\pm$0.4 \\ \hline
\multicolumn{6}{c}{\textbf{LR nominal sparse}}\\ \hline
1178 & 04-08-12 & 4 & 1342249048 & LR & 1.5$\pm$0.8 \\
1208 & 02-09-12 & 4 & 1342250524 & LR & 1.9$\pm$0.5 \\
1228 & 23-09-12 & 4 & 1342251294 & LR & 1.5$\pm$0.6 \\
1390 & 04-03-13 & 4 & 1342265846 & LR & 2.2$\pm$0.5 \\
1410 & 24-03-13 & 4 & 1342268296 & LR & 3.2$\pm$0.4 \\ \hline
\multicolumn{6}{c}{\textbf{HR/CR bright sparse}}\\ \hline
209 & 09-12-09 & 10 & 1342187901 & CR & --- \\
\hline
\end{tabular}
\end{center}
\begin{tablenotes}[normal,flushleft]
\item $^{\rm P_0}${Pointing offset reference observation}
\end{tablenotes}
\label{tab:VestaObs}
\end{table}

\begin{table}
\caption{Observations of other asteroids taken after OD\,189. Bias mode and spatial sampling are indicated above each table section. Number of repetitions are provided in the ``Reps'' column and the commanded resolution is given in the ``Res'' column.}
\medskip
\begin{center}
\begin{tabular}{lcccc}
\hline\hline
\multicolumn{5}{c}{\textbf{OTHER ASTEROIDS}}\\ \hline
Name & OD & dd-mm-yy & Reps & Obsid \\ \hline
\multicolumn{5}{c}{\textbf{CR nominal sparse}}\\ \hline
Europa & 1054 & 02-04-12 & 12 & 1342243626 \\ \hline
\multicolumn{5}{c}{\textbf{MR nominal sparse}}\\ \hline
Thisbe & 227 & 27-12-09 & 30 & 1342188676 \\ 
Cybele & 227 & 27-12-09 & 30 & 1342188675 \\
Europa & 302 & 12-03-10 & 30 & 1342192177 \\
Europa & 317 & 27-03-10 & 30 & 1342192836 \\
\hline
\end{tabular}
\end{center}
\label{tab:mrObs}
\end{table}

\clearpage

\section{Lines fitted to the FTS line sources}
\label{app:lines}

Tables of the line catalogues fitted to the main four FTS line sources, AFGL2688, AFGL4106, CRL618 and NGC7027. The $^{12}$CO lines within the FTS frequency bands, which were included in the fit for each source, are given in Table~\ref{tab:12co}. The additional main species included for each source are provided in Tables~\ref{tab:AFGL2688Lines}--\ref{tab:NGC7027Lines}. Unidentified lines are not included in these catalogues.

\begin{table}
\caption{$^{12}$CO lines}
\medskip
\begin{center}
\begin{tabular}{lc}
\hline\hline
\multicolumn{2}{c}{\textbf{SLWC3}}\\ \hline
Transition & Frequency [GHz] \\ \hline
4 - 3  & 461.041 \\
5 - 4  & 576.268 \\
6 - 5  & 691.473 \\
7 - 6  & 806.652 \\
8 - 7  & 921.800 \\ \hline
\multicolumn{2}{c}{\textbf{SSWD4}}\\ \hline
Transition & Frequency [GHz] \\ \hline
9 - 8 & 1036.912 \\
10 - 9 & 1151.985 \\
11 - 10 & 1267.014 \\
12 - 11 & 1381.995 \\
13 - 12 & 1496.923 \\
\hline
\end{tabular}
\end{center}
\label{tab:12co}
\end{table}

\begin{table}
\caption{Lines fitted to AFGL2688 in addition to the $^{12}$CO lines listed in Table~\ref{tab:12co}. Unidentified features are not included.}
\medskip
\begin{center}
\begin{tabular}{lcc}
\hline\hline
\multicolumn{3}{c}{\textbf{SLWC3}}\\ \hline
Species & Transition & Frequency [GHz] \\ \hline
$^{13}$CO & 5 - 4 & 550.926 \\
$^{13}$CO & 6 - 5 & 661.067 \\
$^{13}$CO & 7 - 6 & 771.184 \\
$^{13}$CO & 8 - 7 & 881.000 \\
$^{13}$CO & 9 - 8 & 991.329 \\
HCN & 6 - 5 & 531.710 \\
HCN & 7 - 6 & 620.300 \\
HCN & 8 - 7 & 708.877 \\
HCN & 9 - 8 & 797.450 \\
HCN & 10 - 9 & 885.980 \\
HCN & 11 - 10 & 974.488 \\ \hline
\multicolumn{3}{c}{\textbf{SSWD4}}\\ \hline
Species & Transition & Frequency [GHz] \\ \hline
$^{13}$CO & 9 - 8 & 991.329 \\
$^{13}$CO & 10 - 9 & 1101.348 \\
$^{13}$CO & 11 - 10 & 1211.329 \\
$^{13}$CO & 12 - 11 & 1321.265 \\
$^{13}$CO & 13 - 12 & 1431.154 \\
$^{13}$CO & 14 - 13 & 1540.988 \\
HCN & 11 - 10 & 974.488 \\
HCN & 12 - 11 & 1062.97 \\
HCN & 14 - 13 & 1239.88 \\
HCN & 15 - 14 & 1328.29 \\
HCN & 16 - 15 & 1416.67 \\
HCN & 17 - 16 & 1505.041 \\ \hline
\hline
\end{tabular}
\end{center}
\label{tab:AFGL2688Lines}
\end{table}

\begin{table}
\caption{Lines fitted to AFGL4106 in addition to the $^{12}$CO lines listed in Table~\ref{tab:12co}. Unidentified features are not included.}
\medskip
\begin{center}
\begin{tabular}{lcc}
\hline\hline
\multicolumn{3}{c}{\textbf{SLWC3}}\\ \hline
Species & Transition & Frequency [GHz] \\ \hline
$^{13}$CO & 5 - 4 & 550.926 \\
$^{13}$CO & 6 - 5 & 661.067 \\
$^{13}$CO & 7 - 6 & 771.184 \\
$^{13}$CO & 8 - 7 & 881.000 \\
$^{13}$CO & 9 - 8 & 991.329 \\ \hline
\multicolumn{3}{c}{\textbf{SSWD4}}\\ \hline
Species & Transition & Frequency [GHz] \\ \hline
$^{13}$CO & 9 - 8 & 991.329 \\
$^{13}$CO & 10 - 9 & 1101.348 \\
$^{13}$CO & 11 - 10 & 1211.329 \\
\hline
\end{tabular}
\end{center}
\label{tab:AFGL4106Lines}
\end{table}

\begin{table}
\caption{Lines fitted to CRL618 in addition to the $^{12}$CO lines listed in Table~\ref{tab:12co}. Unidentified features are not included.}
\medskip
\begin{center}
\begin{tabular}{lcc}
\hline\hline
\multicolumn{3}{c}{\textbf{SLWC3}}\\ \hline
Species & Transition & Frequency [GHz] \\ \hline
$^{13}$CO & 5 - 4 & 550.926 \\
$^{13}$CO & 6 - 5 & 661.067 \\
$^{13}$CO & 7 - 6 & 771.184 \\
$^{13}$CO & 8 - 7 & 881.000 \\
$^{13}$CO & 9 - 8 & 991.329 \\
HCN & 6 - 5 & 531.710 \\
HCN & 7 - 6 & 620.300 \\
HCN & 8 - 7 & 708.877 \\
HCN & 9 - 8 & 797.450 \\
HCN & 11 - 10 & 974.488 \\
HNC & 5 - 4 & 453.269 \\
HNC & 6 - 5 & 543.897 \\
HNC & 7 - 6 & 634.51 \\
HNC & 8 - 7 & 725.106 \\
HNC & 9 - 8 & 815.683 \\
HNC & 11 - 10 & 996.77 \\
H$_2$O & 2$_{11}$ - 2$_{02}$ & 752.032 \\
H$_2$O & 2$_{02}$- 1$_{11}$ & 987.918 \\ \hline
\multicolumn{3}{c}{\textbf{SSWD4}}\\ \hline
Species & Transition & Frequency [GHz] \\ \hline
$^{13}$CO & 9 - 8 & 991.329 \\
$^{13}$CO & 10 - 9 & 1101.348 \\
$^{13}$CO & 11 - 10 & 1211.329 \\
$^{13}$CO & 12 - 11 & 1321.265 \\
$^{13}$CO & 13 - 12 & 1431.154 \\
HCN & 11 - 10 & 974.488 \\
HCN & 12 - 11 & 1062.97 \\
HCN & 14 - 13 & 1239.88 \\
HCN & 15 - 14 & 1328.29 \\
HCN & 16 - 15 & 1416.67 \\
HCN & 17 - 16 & 1505.041 \\
HNC & 11 - 10 & 996.77 \\
HNC & 12 - 11 & 1087.275 \\
HNC & 13 - 12 & 1177.751 \\
HNC & 15 - 14 & 1358.607 \\
HNC & 16 - 15 & 1448.982 \\
H$_2$O & 2$_{02}$ - 1$_{11}$ & 987.918 \\
H$_2$O & 3$_{12}$ - 3$_{03}$ & 1097.365 \\
H$_2$O & 1$_{11}$ - 0$_{00}$ & 1113.34 \\
H$_2$O & 3$_{21}$ - 3$_{12}$ & 1162.933 \\
H$_2$O & 2$_{20}$ - 2$_{11}$ & 1228.799 \\
\hline
\end{tabular}
\end{center}
\label{tab:CRL618Lines}
\end{table}

\begin{table}
\caption{Lines fitted to NGC7027 in addition to the $^{12}$CO lines listed in Table~\ref{tab:12co}. Unidentified features are not included.}
\medskip
\begin{center}
\begin{tabular}{lcc}
\hline\hline
\multicolumn{3}{c}{\textbf{SLWC3}}\\ \hline
Species & Transition & Frequency [GHz] \\ \hline
$^{13}$CO & 5 - 4 & 550.926 \\
$^{13}$CO & 6 - 5 & 661.067 \\
$^{13}$CO & 7 - 6 & 771.184 \\
$^{13}$CO & 8 - 7 & 881.000 \\
$^{13}$CO & 9 - 8 & 991.329 \\
HCO$^{+}$ & 6 - 5 & 535.023 \\
HCO$^{+}$ & 7 - 6 & 624.144 \\
HCO$^{+}$ & 8 - 7 & 713.294 \\
HCO$^{+}$ & 10 - 9 & 891.475 \\
$[$CI$]$ & 2 - 1 & 809.280 \\
CH$^{+}$ & 1 - 0 & 834.999 \\
\hline
\multicolumn{3}{c}{\textbf{SSWD4}}\\ \hline
Species & Transition & Frequency [GHz] \\ \hline
$^{13}$CO & 9 - 8 & 991.329 \\
$^{13}$CO & 11 - 10 & 1211.329 \\
$^{13}$CO & 12 - 11 & 1321.265 \\
$^{13}$CO & 13 - 12 & 1431.154 \\
HCO$^{+}$ & 13 - 12 & 1158.800 \\
 \hline
\end{tabular}
\end{center}
\label{tab:NGC7027Lines}
\end{table}

\clearpage
\section{Details of SPIRE photometer observations}
\label{app:photometerTables}

Table~\ref{tab:specMatchPhot} details the SPIRE photometer observations used for the brightness comparison in Table~\ref{tab:sourceCount} (for line sources and stars)
and for the comparison with FTS synthetic photometry for AFGL2688 and CRL618 in Section~\ref{sec:specPhot}.

\begin{table}
\caption{SPIRE photometer observations used for Table~\ref{tab:sourceCount} (for line sources and stars) and for the comparison with FTS synthetic photometry for AFGL2688 and CRL618 in Section~\ref{sec:specPhot}. All photometer observations were taken in \textsc{SpirePhotoLargeScan}, nominal mode.}
\medskip
\begin{center}
\begin{tabular}{llll}
\hline\hline
OD & Obsid & obsMode & Reps \\ \hline
\multicolumn{4}{c}{\textbf{AFGL2688}}\\ \hline
180 & 1342186836  & Large Map & 3 \\
217 & 1342188168  & Large Map & 3 \\ \hline
\multicolumn{4}{c}{\textbf{CRL618}}\\ \hline
134 & 1342184387 & Large Map & 5 \\
275 & 1342190661 & Large Map & 4 \\ \hline
\multicolumn{4}{c}{\textbf{NGC7027}}\\ \hline
217 & 1342188172 & Large Map & 2 \\
226 & 1342188597 & Large Map & 2 \\
232 & 1342188832 & Large Map & 2 \\ \hline
\multicolumn{4}{c}{\textbf{CW Leo}}\\ \hline
164 & 1342186293 & Large Map & 3 \\
181 & 1342186943 & Large Map & 3 \\
529 & 1342207040 & Large Map & 3 \\
746 & 1342221902 & Large Map & 3 \\
893 & 1342231352 & Large Map & 3 \\
1117 & 1342246623 & Large Map & 3 \\
1265 & 1342254051 & Large Map & 3 \\ \hline
\multicolumn{4}{c}{\textbf{NGC6302}}\\ \hline
134 & 1342184379 & Large Map & 4 \\
648 & 1342214573 & Small Map & 4 \\ \hline
\multicolumn{4}{c}{\textbf{R Dor}}\\ \hline
216 & 1342188164 & Large Map & 3 \\ \hline
\multicolumn{4}{c}{\textbf{VY CMa}}\\ \hline
181 & 1342186941 & Large Map & 3 \\
904 & 1342231847 & Small Map & 4 \\ \hline
\multicolumn{4}{c}{\textbf{Omi Cet}}\\ \hline
249 & 1342189423 & Large Map & 3 \\ \hline
\multicolumn{4}{c}{\textbf{IK Tau}}\\ \hline
287 & 1342191180 & Large Map & 3 \\ \hline
\multicolumn{4}{c}{\textbf{W Hya}}\\ \hline
250 & 1342189519 & Large Map & 3 \\ \hline
\end{tabular}
\end{center}
\label{tab:specMatchPhot}
\end{table}

\label{lastpage}
\end{document}